\documentclass[graybox, secnum]{svmult}

\usepackage{mathptmx}       % selects Times Roman as basic font
\usepackage{helvet}         % selects Helvetica as sans-serif font
\usepackage{courier}        % selects Courier as typewriter font
\usepackage{type1cm}        % activate if the above 3 fonts are
\usepackage{makeidx}         % allows index generation
\usepackage{graphicx}        % standard LaTeX graphics tool
\usepackage{multicol}        % used for the two-column index
\usepackage[bottom]{footmisc}% places footnotes at page bottom
\usepackage{soul}            % for high-lighting of text
\newcommand{\HBarXiv}[2]
{#2}%(arXiv version)

\usepackage{amsfonts}
\usepackage{amstext}
\usepackage{amsmath}
\usepackage{amssymb}

\usepackage[dvipdfmx]{hyperref}        % for hyperlinks
\hypersetup{
hidelinks,
    colorlinks=true,
    urlcolor=blue,
}

\newcommand{\red}[1]{\textcolor{black}{#1}}

\newcommand{\rf}[1]{(\ref{#1})}

\newcommand{\e}{{\rm e}}

\renewcommand{\a}{\alpha{\halftinyspace}}

\newcommand{\tr}{{\rm tr}}
\newcommand{\Tr}{{\rm Tr}}

\newcommand{\cH}{{\cal H}}

\newcommand{\cW}{{\cal W}}

\newcommand{\Ham}{{\cal H}}%{{\mathscr H}}

\newcommand{\mbar}[1]{\overline {#1} \hskip 1pt{}}
\newcommand{\Hop}{\Ham}%{\cH}
\newcommand{\Hopstar}{\Hop^\star}

\newcommand{\Hkin}{{\cal H}_{\rm kin}}
\newcommand{\tHkin}{\tilde{\cal H}_{\rm kin}}

\newcommand{\T}{T} %D}

\newcommand{\bW}{\mbar W}

\newcommand{\Hc}{\cH_{\rm c}}	%{\cH}

\newcommand{\Bf}[1]{\mbox{{\boldmath $#1$}}}
\newcommand{\Bfscript}[1]{\mbox{{\scriptsize \boldmath $#1$}}}

\newcommand{\define}{\stackrel{\rm def}{\equiv}}

\font\fourteenmsbm = msbm10 scaled\magstep2
\font\twelvemsbm = msbm10 scaled\magstep1
\font\tenmsbm = msbm10
\font\eightmsbm = msbm8
\font\sixmsbm = msbm6
\newcommand{\DBL}[1]{\leavevmode\raise-.10ex\hbox{\fourteenmsbm #1}}
\newcommand{\Dbl}[1]{\leavevmode\raise-.10ex\hbox{\twelvemsbm #1}}
\newcommand{\dbl}[1]{\leavevmode\raise-.00ex\hbox{\tenmsbm #1}}
\newcommand{\dblsmall}[1]{\leavevmode\raise-.05ex\hbox{\eightmsbm #1}}
\newcommand{\dbltiny}[1]{\leavevmode\raise-.05ex\hbox{\sixmsbm #1}}

\newcommand{\bra}[1]{\langle #1 |}

\newcommand{\ket}[1]{| #1 \rangle}

\newcommand{\bracket}[2]{ \langle #1 | #2 \rangle}

\newcommand{\vac}{\bra{{\rm vac}}}
\newcommand{\cuum}{\ket{{\rm vac}}}
\newcommand{\vacuumNorm}{\bracket{{\rm vac}}{{\rm vac}}}
\newcommand{\expect}[1]{\langle #1 \tinyspace\rangle}

\newcommand{\Norderingbig}[1]{
 {\scriptsize \substack{\circ \\ \circ}}{\dbltinyspace}
  #1
 {\dbltinyspace} {\scriptsize \substack{\circ \\ \circ}}}

\newcommand{\combi}[2]{\left( \!\! \begin{array}{c} 
	\raise0.5ex\hbox{$#1$} \\ \lower0.5ex\hbox{$#2$} \\ 
	\end{array} \!\! \right)}

\newcommand{\commutator}[2]{[\, #1 {\dbltinyspace}, #2 \,]}

\newcommand{\anticommutator}[2]{\{\tinyspace #1 \;\!,\>\! #2 \tinyspace\}}

\newcommand{\Pbracket}[2]{\anticommutator{#1}{#2}}

\newcommand{\dd}{{\rm d}}
\newcommand{\pder}[1]{\frac{\partial}{\partial #1}}
\newcommand{\pdder}[1]{\frac{\partial^2}{\partial #1^2}}

\newcommand{\GG}{G}%{a}

\newcommand{\ope}[1]{#1}%{\hat{#1}}

\newcommand{\halftinyspace}{\hspace{0.0278em}} % = 1/36quad
\newcommand{\tinyspace}{\hspace{0.0556em}}%\hspace{0.8pt}}%\>\!}%
\newcommand{\trehalftinyspace}{\hspace{0.0834em}}%\hspace{1.2pt}}%\>\!}%
\newcommand{\dbltinyspace}{\hspace{0.1112em}}%\hspace{1.6pt}}%\>\!}%
\newcommand{\femhalftinyspace}{\hspace{0.1390em}}%\hspace{2.0pt}}%\>\!}%
\newcommand{\trpltinyspace}{\hspace{0.1668em}}%\hspace{2.4pt}}%\>\!}%
\newcommand{\qdrpltinyspace}{\hspace{0.2224em}}%\hspace{3.2pt}}%\>\!}%
\newcommand{\neghalftinyspace}{\hspace{-0.0278em}}
\newcommand{\negtinyspace}{\hspace{-0.0556em}}%\;\!\!}%\hspace{-0.8pt}}
\newcommand{\negtrehalftinyspace}{\hspace{-0.0834em}}%\hspace{-1.2pt}}%\>\!}%
\newcommand{\negdbltinyspace}{\hspace{-0.1112em}}%\hspace{-1.6pt}}%\>\!}%
\newcommand{\negtrpltinyspace}{\hspace{-0.1668em}}%\hspace{-2.4pt}}%\>\!}%
\newcommand{\negqdrpltinyspace}{\hspace{-0.2224em}}%\hspace{-3.2pt}}%\>\!}%
\newcommand{\ddim}{{d{\trehalftinyspace}}}%{N}	%粒子数Nとかぶることあり

\newcommand{\half}{\frac{1}{2}}
\newcommand{\onethird}{\frac{1}{3}}

\newcommand{\quarter}{\frac{1}{4}}

\newcommand{\eight}{8}%{0}

\newcommand{\cc}{\mu} %\Lambda

\newcommand{\oIndex}{\varnothing}	%0		%\emptyset
\newcommand{\OIndex}{\varnothing}	%[{\tinyspace}0{\tinyspace}]
\newcommand{\pIndex}{+}				%[+]
\newcommand{\mIndex}{-}				%[-]
\newcommand{\pmIndex}{\pm}			%[\pm]
\newcommand{\CoeffSinglet}{\kappa_{{\tinyspace}0}}	%{\bar{\kappa}}		%C

\allowdisplaybreaks[1]	%The value should be 0, 1, 2, 3 or 4.

\HBarXiv{
\newcommand{\Article}{chapter}
\newcommand{\Chapter}{section}
\newcommand{\Section}{subsection}
\newcommand{\Subsection}{subsubsection}
}{
\newcommand{\Article}{article}		%review
\newcommand{\Chapter}{chapter}
\newcommand{\Section}{section}
\newcommand{\Subsection}{subsection}
}

\newcommand{\age}{age}			%age, era, period
\newcommand{\period}{period}	%age, era, period

\newcommand{\TPindent}{\hspace{12pt}}

\begin{document}
%\tableofcontents{}
\title*{%
\HBarXiv{
The causality road from dynamical triangulations 
to quantum gravity%
\\
that describes our Universe%
}{
\centerline{
The causality road from dynamical triangulations 
}
\centerline{
to quantum gravity %${}^{{\raisebox{-1.8ex}{\mbox{}}}}$
that describes our Universe%
${}^{{\raisebox{-1.8ex}{\mbox{$\dag$}}}}$
}
}
}
\titlerunning{%
\HBarXiv{%												%\HBarXiv command
The causality road %from DT 
to quantum gravity 
that describes our Universe%
}{\mbox{}}% 											%\HBarXiv command
}
% Use \titlerunning{Short Title} for an abbreviated version of
% your contribution title if the original one is too long
\author{%
\HBarXiv{%												%\HBarXiv command
Yoshiyuki Watabiki%
}{\centerline{\large\it Yoshiyuki Watabiki}}%			%\HBarXiv command
}% \thanks{corresponding author} and Second Author}
\HBarXiv{}{\authorrunning{\mbox{}}}%					%\HBarXiv command
% Use \authorrunning{Short Title} for an abbreviated version of
% your contribution title if the original one is too long
\institute{%
\HBarXiv{%												%\HBarXiv command
Yoshiyuki Watabiki \at Tokyo Institute of Technology, Tokyo, Japan, \email{watabiki@th.phys.titech.ac.jp}
%\and Second Author \at Institute 2, Address of Institute 2 \email{name2@email.address}
}{% 													%\HBarXiv command
$\dag$\hspace{2pt}% 	 								%\HBarXiv command
This is a contribution to 								%\HBarXiv command
the Handbook of Quantum Gravity 						%\HBarXiv command
which will be published in %the beginning of 			%\HBarXiv command
2023.  It will appear as a chapter in the section of 	%\HBarXiv command
the handbook denoted 									%\HBarXiv command
``Causal Dynamical Triangulations''.					%\HBarXiv command
}%														%\HBarXiv command
}
%
% Use the package "url.sty" to avoid
% problems with special characters
% used in your e-mail or web address
%
\maketitle
\HBarXiv{}{%											%\HBarXiv command
\vspace{-108pt}\mbox{}\\%								%\HBarXiv command
\centerline{Tokyo Institute of Technology,}\\%			%\HBarXiv command
\centerline{Department of Physics,%  					%\HBarXiv command
 High Energy Theory Group,}\\%							%\HBarXiv command
\centerline{Oh-okayama 2-12-1, Meguro-ku,%				%\HBarXiv command
 Tokyo 152-8551, Japan}\\%								%\HBarXiv command
\centerline{{email: watabiki@th.phys.titech.ac.jp}}\\%	%\HBarXiv command
\vspace{36pt}\\%										%\HBarXiv command
}%							 							%\HBarXiv command
\HBarXiv{}{\hspace{147pt}\vspace{4pt}}%   			   	%\HBarXiv command
\abstract{%
\HBarXiv{}{\\{\TPindent}}%							   	%\HBarXiv command
It is shown how one, 
guided by causality, 
starting from so-called dynamical triangulations, 
is led to a candidate of quantum gravity that describes our Universe. 
This theory is based on $W$-- and Jordan algebras. 
It explains 
how our Universe was created, 
how cosmic inflation began and ended, 
how the topology and the geometry of our Universe was formed, 
and 
what was the origin of Big Bang energy. 
The theory also leads to a modified Friedmann equation 
which explains the present accelerating expansion of our Universe 
without appealing to the cosmological constant.}

\HBarXiv{%												%\HBarXiv command
\section*{Keywords} 
Quantum Gravity (QG); 
Dynamical Triangulation (DT); 
Causal Dynamical \\
Triangulation (CDT); 
Non-critical String Field Theory; 
Conformal Field Theory; 
\\
$W$ Algebra; 
Jordan Algebra; 
String Theory;
Cosmic Inflation;
Big Bang Theory
%Please provide keywords required to facilitate search of chapter on web; maximum 10 keywords.
}{}%													%\HBarXiv command

\HBarXiv{}{
%%\renewcommand{\contentsname}{\vspace*{100pt}Contents}
%\tableofcontents
%\addtocontents{toc}{\vskip40pt}
\newpage
}%														%\HBarXiv command

\section{Introduction}
\label{sec:intro}

%\section{Introduction and Motivation of quantum gravity}
%\label{sec:Motivation}

%なぜ量子重力理論を考えるのか？\ まずはその動機について考えてみよう。

\subsection{Determinism in Physics (Laplace's demon)}
\label{sec:Determinism}

The world state
${\cal X}{\negtinyspace}(\Bf{x},t)$
[\,where $\Bf{x}$ \red{denotes} spatial coordinates 
\red{and} $t$ time\,]
is expressed by 
matter fields 
$\psi_\alpha^{(i)}{\negtinyspace}(\Bf{x},t)$ 
[\,$i \!=\! 1$, $2$, \ldots\,], 
gauge fields 
$A_\rho^{(a)}{\negtinyspace}(\Bf{x},t)$ 
[\,$a \!=\! 1$, $2$, \ldots\,], 
and 
the metric %of spacetime 
$g_{\mu\nu}(\Bf{x},t)$. 
%世界の状態%
%${\cal X}{\negtinyspace}(\Bf{x},t)$
%[\,$\Bf{x}$は空間座標，$t$は時刻\,]
%は
%各種物質場$\psi_\alpha^{(i)}(\Bf{x},t)$ [\,$i \!=\! 1$, \ldots\,]と
%各種ゲージ場$A_\rho^{(a)}(\Bf{x},t)$ [\,$a \!=\! 1$, \ldots\,]と
%計量$g_{\mu\nu}(\Bf{x},t)$で表される。
Namely, %In other words, 
${\cal X}{\negtinyspace}(\Bf{x},t) \define
 \big\{
 \{ \psi_\alpha^{(i)}{\negtinyspace}(\Bf{x},t)\},
 \{ A_\rho^{(a)}{\negtinyspace}(\Bf{x},t)\},
 % \psi_\alpha^{(1)}(t),\psi_\alpha^{(2)}(t),\ldots,
 % A_\rho^{(1)}(\Bf{x},t),A_\rho^{(2)}(\Bf{x},t),\ldots,
  g_{\mu\nu}(\Bf{x},t)
 \big\}$%
. 
Understanding the mechanism that determines ${\cal X}{\negtinyspace}(\Bf{x},t)$ is 
one of the most important problems in physics. 
%そして，この
%${\cal X}{\negtinyspace}(\Bf{x},t)$%
%を決定するメカニズムを理解することは物理学の最も重要な課題の一つである。

In the classical theory, 
if one %can perfectly 
knows the world state ${\cal X}{\negtinyspace}(\Bf{x},t_0)$ 
at a certain time $t_0$,%
\footnote{%{\dbltinyspace}%
$t_0$ in this {\Chapter} has nothing to do with 
the present time $t_0$ used in 
{\Chapter}s \ref{sec:BasicTheory} and \ref{sec:MFE}. 
}
one %can know
knows the world state ${\cal X}{\negtinyspace}(\Bf{x},t)$ 
at any time $t$ in the past and future.
%古典論では，
%ある時刻$t_0$の世界の状態${\cal X}{\negtinyspace}(\Bf{x},t_0)$%
%を完璧に知ることができれば，
%過去や未来の任意の時刻$t$の世界の状態${\cal X}{\negtinyspace}(\Bf{x},t)$%
%を知ることができる。
This is because the classical theory is a determination theory 
that uniquely determines the past and future states from the current state.%
%古典論は現在の状態から過去や未来の世界の状態を
%一意的に決める決定論だからである。
\footnote{%{\dbltinyspace}%
The problem of whether it is possible to perfectly know 
the world state ${\cal X}{\negtinyspace}(\Bf{x},t_0)$, 
the so-called observation problem, 
is difficult and very deep but can be separated, 
so we will not treat it in this {\Article}. 
%ある時刻$t_0$の世界の状態${\cal X}{\negtinyspace}(\Bf{x},t_0)$%
%を完璧に知ることが可能かどうかの問題，%議論の余地はあるが，
%いわゆる観測問題は，
%難しく奥が深い問題だが切り離すことが可能なので，
%ここでは問わないことにする。
}
However, the question remains, 
``How was the state 
${\cal X}{\negtinyspace}(\Bf{x},t_0)$ 
chosen from countless possibilities?''%
%しかし，%たとえそれができたとしても
%これでは
%「無数に存在する可能性の中から，
%なぜ状態${\cal X}{\negtinyspace}(\Bf{x},t_0)$が選ばれたのか？」
%という疑問が残る。
\footnote{%{\dbltinyspace}%
There %is a similar question, 
are similar questions, 
``Where do we come from? What are we? Where are we going?''
%「なぜ我々は存在するのか？」という疑問と同じ類の疑問である。
}
This 
%is a problem that cannot be solved in classical theory, 
%but can be `partially' solved in 
\red{question
cannot be answered in the classical theory, 
but a `partial' answer exists in a}
quantum theory where 
the probability that ${\cal X}{\negtinyspace}(\Bf{x},t)$ is chosen 
proportional to 
the square of the wave function 
$\Psi{\negtinyspace}\big({\cal X}{\negtinyspace}(\Bf{x},t);t\big)$, i.e.\  
$\big|\Psi{\negtinyspace}\big({\cal X}{\negtinyspace}(\Bf{x},t);t\big)\big|^2$. 
%これは%古典論最大の問題で，%ある。%この問題は
%古典論では解決できない問題なのだが，
%「時刻$t$の世界の状態が${\cal X}{\negtinyspace}(\Bf{x},t)$となる確率は，
%波動関数$\Psi{\negtinyspace}\big({\cal X}{\negtinyspace}(\Bf{x},t);t\big)$%
%の大きさの2乗$\big|\Psi{\negtinyspace}\big({\cal X}{\negtinyspace}(\Bf{x},t);t\big)\big|^2$に比例する」
%とする量子論によって「部分的に」解決される。
In \red{a} quantum theory, 
the world state ${\cal X}{\negtinyspace}(\Bf{x},t)$ is chosen by probability 
from all possibilities 
using the wave function $\Psi{\negtinyspace}\big({\cal X}{\negtinyspace}(\Bf{x},t);t\big)$. 
%量子論では時刻$t$の世界の状態${\cal X}{\negtinyspace}(\Bf{x},t)$は
%考えられるすべての可能性の中から
%波動関数$\Psi{\negtinyspace}\big({\cal X}{\negtinyspace}(\Bf{x},t);t\big)$%
%によって確率的に選ばれると考えるのである。
\red{As in a classical theory, also in a quantum theory} 
%As in classical theory, in quantum theory, 
the wave function 
$\Psi{\negtinyspace}\big({\cal X}{\negtinyspace}(\Bf{x},t);t\big)$ at any time $t$ 
will be uniquely determined from the wave function 
$\Psi{\negtinyspace}\big({\cal X}{\negtinyspace}(\Bf{x},t_0);t_0\big)$ at a certain time $t_0$. 
%そして，古典論と同様に量子論でも，
%任意の時刻$t$の波動関数$\Psi{\negtinyspace}\big({\cal X}{\negtinyspace}(\Bf{x},t);t\big)$%
%は
%ある時刻$t_0$の波動関数$\Psi{\negtinyspace}\big({\cal X}{\negtinyspace}(\Bf{x},t_0);t_0\big)$%
%から一意的に決定されるとする。

In this way, 
the world state ${\cal X}{\negtinyspace}(\Bf{x},t)$ at any time $t$ 
is uniquely and probabilistically determined from the wave function 
$\Psi{\negtinyspace}\big({\cal X}{\negtinyspace}(\Bf{x},t_0);t_0\big)$, 
but the question still remains, 
``How is the wave function 
$\Psi{\negtinyspace}\big({\cal X}{\negtinyspace}(\Bf{x},t_0);t_0\big)$ chosen?''
%このようにして
%任意の時刻$t$の世界の状態${\cal X}{\negtinyspace}(\Bf{x},t)$は
%$\Psi{\negtinyspace}\big({\cal X}{\negtinyspace}(\Bf{x},t_0);t_0\big)$から一意的に決まるのだが，
%「%
%%無数に存在する可能性の中から${\cal X}{\negtinyspace}(\Bf{x},t_0)$が
%%選ばれる理由はわかったが，
%なぜ波動関数$\Psi{\negtinyspace}\big({\cal X}{\negtinyspace}(\Bf{x},t_0);t_0\big)$が選ばれたのか？」
%という疑問は未だ残る。
This is the meaning of 
\red{the word `partial' used above, 
and a quantum theory cannot answer this question} 
%`partially', 
%and quantum theory cannot solve this problem 
completely. 
%これが先ほど「部分的に」と言った意味で，
%量子論でも完全解決には至らないのである。
However, 
\red{there {\it is} an answer if 
``the universe starts from a point.''}
%this problem can be solved easily in a simple way, 
%``The universe starts from a point.''
%ところが，この問題は
%「宇宙は点から始まる」という簡単な方法であっさりと解決する。
If the space is not a point, 
matter fields, gauge fields, and the metric are distributed 
in various ways in space, 
but 
if the space is a point, 
the distribution is unique, 
namely, 
the world state ${\cal X}{\negtinyspace}(\Bf{x},t_0)$ is unique, 
so 
the wave function $\Psi{\negtinyspace}\big({\cal X}{\negtinyspace}(\Bf{x},t_0);t_0\big)$ 
is also unique. 
%物質やゲージ場，計量は拡がった空間中にいろいろな状態で分布しうるが，
%「点」の空間座標は唯一しかないため，
%このときの状態${\cal X}{\negtinyspace}(\Bf{x},t_0)$は唯一となり，よって，
%$\Psi{\negtinyspace}\big({\cal X}{\negtinyspace}(\Bf{x},t_0);t_0\big)$も唯一となるからである。
%\footnote{%{\dbltinyspace}%
One should also note that 
this state is prohibited in classical theory 
which is a determination theory. 
%これは決定論に従う古典論ではありえない状態である。
%The idea that the universe becomes a point state is completely different 
%between classical theory and quantum theory.
%宇宙が点状態になるという考えは，
%古典論と量子論では全く違うのである。
The quantum theory plays an important role in order to realize this idea. 
The quantum theory can generate a nontrivial state 
${\cal X}{\negtinyspace}(\Bf{x},t)$ 
even if the universe starts from a point. 
%}
However, 
if the space is a point, 
the problem arises that all physical densities are infinite, 
and new questions arise, 
``What was the state of the universe before the universe was a point?'', 
``Was the universe born from nothing?'' 
%しかし，点という状態はあらゆる物理量の密度が無限大になるという問題を孕み，
%さらには，
%「点の状態の前の宇宙はどのような状態だったのか？」
%「宇宙は無から生じたのか？」
%という新たな疑問が生じる。
These questions are basic problems of quantum gravity theory, 
\red{and we will discuss them in detail later.} 
%so let us give a careful discussion later. 
%これらの問題は量子重力理論の基本的な問題になるので，
%後々，丁寧な議論を与えることにしよう。
%%（解決すべき問題なのかも含めて）
Though new questions such as these arise, %but
quantum theory thus provides an answer to %the 
\red{some of the conceptual} %decisive 
problems in determinism 
that classical theory had. 
%このように量子論は古典論が持っていた決定論の問題に一つの解答を与える。

\red{The quantum theory has succeeded in 
describing the strong, weak and electromagnetic} 
%Moreover, 
%the quantum theory is applied to electromagnetics, 
%and has succeeded in describing strong and weak} 
interactions. 
%そして，量子論は電磁気学に適用され，
%さらには強い力や弱い力を記述する量子論が完成する。
This is 
${\rm SU}(3) {\negtrpltinyspace}\times{\negtrpltinyspace}
 {\rm SU}(2) {\negtrpltinyspace}\times{\negtrpltinyspace}
 {\rm U}(1)$
Yang-Mills gauge theory, 
\red{the so-called Standard Model. 
It is a renormalizable quantum field theory.} 
%so-called the standard model. 
%いわゆる
%${\rm SU}(3) {\negtrpltinyspace}\times{\negtrpltinyspace}
% {\rm SU}(2) {\negtrpltinyspace}\times{\negtrpltinyspace}
% {\rm U}(1)$%
%の対称性を持つYang-Millsゲージ理論，標準理論である。
On the other hand, 
the quantization of gravity 
\red{has been  difficult 
because it is not renormalizable quantum field theory. 
String theory went beyond quantum field theory 
and succeeded in curing the UV problems 
related to quantum gravity. 
It became a candidate for a quantum gravity theory.} 
%was extremely difficult 
%because of renormalization, 
%but string theory succeeded in this and 
%became a candidate for quantum gravity theory.
%The quantization of the Einstein gravity theory 
%is extremely difficult. 
%ところが，重力を記述するEinstein理論の量子化については困難を極める。
However, 
string theory has not yet 
addressed the mechanism of the birth of the universe. 
%clarified the mechanism involved in the birth of the universe. 
The reason for this seems to be related 
not only to the problem of quantization 
but also to the existence of the wave function 
$\Psi{\negtinyspace}\big({\cal X}{\negtinyspace}(\Bf{x},t);t\big)$. 
%However, 
%the reason why string theory cannot yet address 
%the mechanism related to the birth of the universe 
%seems to be related to problems 
%%peculiar to gravity other than renormalization. 
%of the existence of the world's wave function in addition to renormalization.
%Renormalization is one of the reasons 
%why quantization of gravity is difficult, 
%but that is not the only problem.
%The quantization of gravity has a conceptual aspect 
%that is related to the significance of the existence of 
%the wave function 
%$\Psi{\negtinyspace}\big({\cal X}{\negtinyspace}(\Bf{x},t);t\big)$ of the world.
%重力の量子化が困難な理由には，
%繰り込みという技術的な側面はあるが，問題はそれだけではなく，
%世界の波動関数%
%$\Psi{\negtinyspace}\big({\cal X}{\negtinyspace}(\Bf{x},t);t\big)$%
%の存在意義にもかかわる
%概念的な側面もあり，視点や考え方を大きく変える必要があると思われる。
%This may be the reason why string theory has failed.  
%despite being said to be a candidate for quantum gravity theory. 
%弦理論が量子重力理論の候補と言われながらも頓挫している
%理由がここにあるのではないだろうか。
Before considering the time evolution of this wave function, 
we will give a \red{sketchy} %rough 
definition of the quantum gravity theory 
and consider the problems of this theory in the next {\Section}.
%そこで，次の節では，量子重力理論の大雑把な定義を与え，
%この理論が持つ問題について考えてみよう。

\subsection{Overview of Quantum Gravity %Theory 
(QG)}
\label{sec:QuantumGravityOverview}

Let us give a \red{short} %rough 
definition of the quantum gravity theory (QG) 
and explain some problems it poses.
%%%量子重力理論の詳しい議論をする前に，
%量子重力理論の大雑把な定義を与え，
%この理論が持つ問題について少し説明しよう。
The string theory is discussed at the end of this {\Section}.
%ただし，弦理論についてはこの節の最後に議論する。
%subsubsection\ref{sec:PositionOfStringTheory}で行う。

\subsubsection{\red{Formal definition} %Rough definition 
of QG}
\label{sec:QuantumGravityRoughDef}

\red{Quantization is described by summing over 
all possible field configurations with 
the weight of each configuration 
being the exponential of the (classical) action.} 
%Quantization is described by summing up all possible configurations. 
This is the so-called path integral. 
%量子論は，全ての可能な状態を足し上げる操作，
%いわゆる，経路積分によって記述される。
If we apply \red{the prescription} %this 
to QG, 
the path integral becomes an operation 
that sums over all possible configurations of spacetime, 
and in terms of the partition function, 
it %roughly 
has the following form. 
%これを量子重力理論に適用するなら，経路積分は
%全ての可能な時空を足し上げる操作となり，
%分配関数で言えば，%$\ddim$次元の量子重力理論は
%大雑把に次の形になる。%
%%\footnote{%{\dbltinyspace}%
%%In order to perform the integration
%%$\int{\negdbltinyspace}
%% {\cal D}g_{\mu\nu} {\cal D}A_\rho {\cal D}\psi_\alpha$, 
%%delicate regularization is inevitable 
%%%The expression \rf{PartitionFunQGfixedTopology} is ambiguous 
%%if the configulation 
%%$[{\dbltinyspace}
%% g_{\mu\nu}(x), A_\rho(x),\psi_\alpha(x)
%%{\dbltinyspace}]$ 
%%is not discrete. 
%%}
\begin{eqnarray}
&&
Z \,=\,
\sum_{{\cal T}\rule{0pt}{5.5pt}} 
  C_{{\negtinyspace}{\cal T}} {\tinyspace} Z_{{\cal T}}
\,,
\label{PartitionFunQG}
\\
&&
Z_{{\cal T}} \,=\,
\int\limits_{\hbox{\scriptsize{\dbltinyspace}topology ${\cal T}$ is fixed}}%
\hspace{-25pt}{\cal D}g_{\mu\nu} {\cal D}A_\rho 
{\cal D}\psi_\alpha {\tinyspace}
\exp{\negdbltinyspace}\bigg\{
  i \!\int\! \dd^\ddim{\negdbltinyspace}x {\trpltinyspace}
  {\cal L}{\tinyspace}[{\dbltinyspace}g_{\mu\nu}(x),
                       A_\rho(x),\psi_\alpha(x){\dbltinyspace}]
\bigg\}
\,,
\qquad
\label{PartitionFunQGfixedTopology}
\end{eqnarray}
where 
$\ddim$ is the dimension of spacetime, 
$g_{\mu\nu}$ is the metric of spacetime, 
$A_\rho$ is gauge fields, 
$\psi_\alpha$ is matter fields, 
${\cal L}$ is the Lagrangian density.%
\footnote{%{\dbltinyspace}%
The supersymmetric fields should be added 
if the supersymmetry (SUSY) exists. 
%$g_{mn}$ includes graviton and gravitino.
%$A_\rho$ includes gauge and gaugino fields.
%$\psi_\alpha$ includes scalar fields fermion fields and gauge fields.
}
In eq.\ \rf{PartitionFunQG} 
all spacetime with different topologies ${\cal T}$ are summed over 
with weights $C_{{\negtinyspace}{\cal T}}$. %of spacetime with a topology ${\cal T}$. 
In eq.\ \rf{PartitionFunQGfixedTopology} 
the metric of spacetime $g_{\mu\nu}$ with a fixed topology ${\cal T}$, 
gauge fields $A_\rho$, matter fields $\psi_\alpha$ 
are path-integrated. 
%Note that 
%in \rf{PartitionFunQG} and \rf{PartitionFunQGfixedTopology} 
%all possible configurations are summed over.
%\footnote{%{\dbltinyspace}%
%For example, %in the covariant expression %of spacetime 
%the spaces with Euclidean metric will appear anytime and anywhere 
%if the metric at the beginning of the universe is Euclidean. 
%%So, this viewpoint restricts the theory of QG. 
%%もし宇宙誕生時の計量がユークリッド計量になるのなら，
%%ユークリッド計量の空間いつでもどこでも出現することになる。
%}

The topology ${\cal T}$, %of $Z_{{\cal T}}$ 
which contributes most to the sum on the rhs of \rf{PartitionFunQG}, 
is the topology of spacetime in the classical theory. 
The metric $g_{\mu\nu}$, gauge fields $A_\rho$ 
and matter fields $\psi_\alpha$, 
which contribute most to the path integral 
on the rhs of \rf{PartitionFunQGfixedTopology}, 
determine the shape of the spacetime, 
the state of gauge fields and matter fields, 
respectively, in the classical theory. 
%\rf{PartitionFunQG}の右辺の和の中で最も大きな寄与をする$Z_{{\cal T}}$の
%${\cal T}$が現実世界の時空のトポロジーとなり，
%\rf{PartitionFunQGfixedTopology}の右辺の経路積分の中で最も大きな寄与をする
%計量$g_{\mu\nu}$とゲージ場$A_\rho$と物質場$\psi_\alpha$が
%現実世界の時空の形とゲージ場・物質場の状態を決める。
The states that contribute most to the path integral 
are the states that satisfy the equation of motion (on-shell 
\red{states). 
The other states that do not satisfy 
the equation of motion are denoted off-shell states.} 
%states), 
%and 
%other states are the states that do not satisfy 
%the equation of motion (off-shell states). 
%経路積分に最も大きな寄与をする状態が
%運動方程式を満たす状態(on-shell状態)となり，
%それ以外の状態が
%運動方程式を満たさない状態(off-shell状態)となるのである。
\red{The quantum averages, 
which include off-shell states, 
will in general differ from the classical solutions, 
and will in this way reveal  quantum effects. 
However, in the case of QG there is a new aspect: 
universes may split and merge,}%
%The difference from the solution obtained from 
%the equation of motion in classical theory is the quantum effect. 
%The difference with the equation of motion in classical theory 
%appears as quantum effects. 
%そして，古典論の運動方程式からのズレが量子効果として現れる。
%This is the relationship between the classical theory 
%and the quantum theory, 
%but in the case of QG, the universes split and merge,%
%これが古典論と量子論の関係なのだが，
%QGの場合は宇宙の分離や融合が起きるため，%
\footnote{%{\dbltinyspace}%
Fractal structure of space, wormholes, and baby universes 
appear by this process. 
%時空にフラクタル構造が現れたり，
%wormhole や baby universe が現れたりする。
}
\red{and on-shell and off-shell states can from this perspective 
differ much more than in ordinary quantum field theory.}%
%so it should be noted that on-shell and off-shell states are 
%very different.%
%両者は大きく異なることに注意が必要である。%
\footnote{\label{footnote:extremityEOM}%{\dbltinyspace}%
The scheme of $\hslash {\negtrpltinyspace}\to{\negtrpltinyspace} 0$ 
does not necessarily make quantum theory into classical theory. 
%$\hbar {\negtrpltinyspace}\to{\negtrpltinyspace} 0$ 
%の図式で量子論が古典論になるとは限らない。
We think of this as a kind of ``extremity''.
%これは``extremity''の一種と考えてよいだろう。
}
We will discuss this in more detail later.
%詳しくは，後々述べることにしよう。

\subsubsection{Several problems in QG}\label{sec:Problems}

The partition functions 
\rf{PartitionFunQG} and \rf{PartitionFunQGfixedTopology} 
have several problems: %in the following points. 
\begin{quote}
\vspace{-2pt}\normalsize
\begin{enumerate}
\renewcommand{\labelenumi}{\bf \arabic{enumi}.}
\item
How to define the coefficients $C_{{\negtinyspace}{\cal T}}$
\item
How to define the functional form of Lagrangian density ${\cal L}$
\item
How to define and 
perform the path integral %$\int\!{\cal D}g_{mn}$
$\int\!{\cal D}g_{\mu\nu} {\cal D}A_\rho {\cal D}\psi_\alpha$
%How to define the measure of ${\cal D}g_{mn}$
\end{enumerate}
\vspace{-2pt}
\end{quote}
{\bf 1.}\ 
\red{It seems  difficult to define the topology ${\cal T}$ 
if the spacetime dimension is larger than two,} 
%is hopeless if the spacetime dimension is more than $2$. 
%{\bf 1.}は，時空の次元が$3$以上なら絶望的。
\red{since the classification of topologies in higher dimensions 
is incomplete. 
It is possible only if one restricts the class of geometries 
one wants to consider to be sufficiently nice, 
but there is presently no physical motivation for such a restriction. 
Only when the spacetime dimension is two one has a classification, 
namely the genus of the two-dimensional manifold.}
%This is because it is necessary to classify 
%all topologies ${\cal T}$ of spacetime. 
%時空の全てのトポロジー${\cal T}$を分類する必要があるからである。
%However, 
%this is possible when the dimension of spacetime is 
%less than or equal to 2. 
%ところが，
%時空が$2$次元のときはこれが可能になる。
%More on this later.
%これについては後ほど述べる。

\vspace{3pt}

\noindent
{\bf 2.}\ 
\red{There is  little freedom of choice, 
especially in the case of QG, since} 
%has little freedom of choice, 
%especially in the case of QG, 
%because it is involved in the problem of {\bf 3.}.
%{\bf 2.}は，
%特に量子重力理論の場合は%超重力理論や弦理論のように%
%{\bf 3.}の問題と絡んでくるため，選択の自由度はほとんどない。
\red{an arbitrary functional form spoils renormalizability.} 
%This is because renormalization cannot be performed 
%if the functional form is determined arbitrarily.
%関数形を好き勝手に決めると繰り込みができなくなるからである。
\red{In the case of QG} 
%However, 
it is very difficult to obtain a concrete functional form, 
so we will not discuss this problem any further %here.%
in this {\Article}.
%しかし，具体的な関数形を得るのは非常に難しいので，
%この問題は先送りすることにしよう。

\vspace{3pt}

\noindent
{\bf 3.}\ 
\red{Even if one knew how to perform the path integral, 
one would still have to face the following problems 
in the case of QG:} 
%encounters a new problem in the following point, 
%even if we know how to perform the path integral.
%{\bf 3.}は，たとえ経路積分の方法がわかったとしても，
%次の点で新たな問題に遭遇する。
\begin{quote}
\vspace{-2pt}\normalsize
\begin{enumerate}
\renewcommand{\labelenumi}{\bf \alph{enumi}.}
\item
\red{How to obtain a theory that can both be renormalized and unitary}
%The possibility of renormalization
\item
\red{How to deal with the singularity 
at the moment of the birth of our universe}
%The singularity at the moment of the birth of our universe
\end{enumerate}
\vspace{-2pt}
\end{quote}
{\bf a.}\ 
\red{It is presently unknown how to quantize 4D gravity perturbatively.} 
%is the famous problem in QG. 
%It is known that the quantization of metric is naively impossible 
%in 4D gravity. 
%On the other hand, 
%the path-integration is possible in 2D gravity. 
%We will be back to this point later. 
\red{Only in the case of 2D quantum gravity one has been able} 
%The only example, which make us possible 
to perform the path integral completely and solve 
\red{problem {\bf 1.}\ at the same 
time. This theory 
%time, is the 2D Euclidean gravity. This 
is also called ``non-critical string theory".} 
In this theory we can calculate
not only \rf{PartitionFunQGfixedTopology} 
but also \rf{PartitionFunQG}. 
We consider this as a very important clue 
\red{for obtaining a QG theory which describes our universe, 
so we will explain the 2D QG theory} 
%to obtain QG which describes our universe, 
%so we will explain the 2D Euclidean gravity 
in detail in {\Chapter} \ref{sec:TwoDimGravity}. 

\vspace{3pt}

\noindent
{\bf b.}\ 
\red{The singularity of the universe} 
is another serious problem. 
\rf{PartitionFunQGfixedTopology} only gives 
one real number after the path integral. 
So, we should not sum over all possible configurations of spacetime, 
if we want to describe the process of the birth of the universe. 
We need to separate time and space and 
\red{not integrate with respect to time.} 
%leave the time without integration. 
%\rf{PartitionFunQGfixedTopology}は1つの実数値を与えるに過ぎないので，
%宇宙誕生の過程を記述したいのなら，
%全てのconfigurationを完全に足し上げてはダメで，
%時間と空間を別々にし，時間を積分せずに残す必要がある。
\red{It is a tremendous task to do this in the quantum theory, 
and usually one falls back on the classical theory,} 
%This is an ideal analysis as quantum theory, 
%but this is a tremendous task, so an approach from classical theory 
%is usually performed.
%これが量子論としての理想の解析だが，これは困難を極める作業なので，
%通常，古典論からのアプローチが行われる。
\red{but then, as mentioned, the} 
%Then, a 
special state, 
where the universe is a point at the moment when the universe is born, 
causes a problem. 
%すると，宇宙誕生の瞬間という宇宙が点となる特殊な状態が問題を引き起こす。
\red{Let us briefly elaborate on this point.} 
%Next, let us briefly explain this. 
%次に，これについて簡単に説明しよう。
%
In modern cosmology, inflation occurred before Big Bang. 
%現代の宇宙論では，ビッグバンの前はインフレーションが起きていたとされる。
The reason for the introduction of inflation was 
to explain the observation fact that 
the cosmic microwave background (CMB) is uniform 
even in areas that exceed the event horizon. 
%インフレーションが導入された理由は，
%宇宙背景輻射が事象の地平線を超えた領域でも均一になっているという
%観測事実を説明するためであった。
However, 
the introduction of inflation has caused many new questions, 
such as 
``How did inflation begin?'', 
``What kind of era was it before inflation?''
%ところが，インフレーションの導入により，
%インフレーションはどのように始まったのか？\ 
%インフレーションの前はどんな時代だったのか？\ 
%新たな疑問が生まれることになった。
Furthermore, 
this leads to the determinism mentioned 
in {\Section} \ref{sec:Determinism}, 
which naturally raises the question, 
``Does the universe start from a point?'' 
%そしてさらには，
%1.1節で述べた決定論にも繋がるが，
%\lq\lq 宇宙は点から始まったのでは？" 
%という疑問も自然に生まれた。

However, 
if the Friedmann equation makes us possible 
to go back to the time when the universe was a point, 
such a point state has the infinite energy density, 
and it becomes spacetime singularity.%
%ところが，
%Friedmann方程式が宇宙誕生時まで遡ることができるなら，
%%物質場のエネルギー密度において深刻になる。
%%これは
%%$\nabla^\mu{\tinyspace} T_{\mu\nu} {\negtrpltinyspace}={\negtrpltinyspace} 0$
%%に起因する問題で，Friedmann方程式にも現れる。
%エネルギー保存則により宇宙誕生時の点の状態は
%先ほど述べたエネルギーが有限で密度が無限大の状態，
%いわゆる時空の特異点となる。
\footnote{%{\dbltinyspace}%
One way to remove this singularity is to assume that 
time was pure imaginary 
and the spacetime metric was Euclidean when the universe was born.\,%
\cite{ImaginaryTime}
%この特異点を除去する方法として，
%宇宙が誕生するときの時間が純虚数で時空がEuclid計量だった
%とする方法がある。
However, 
all spacetime configurations are summed over in QG, 
so Euclidaen spacetime regions appear anytime, anywhere 
if one introduces such region to the theory.
%しかし，
%量子重力理論ではありとあらゆる時空を足し上げるので，
%Euclid計量を持つ時空はいつでもどこでも現れてしまう。
In this case, 
we need to find the mechanism that 
such Euclidean spacetime region appears only when the universe was born. 
%in this model. 
%それゆえ，この時空が宇宙誕生時だけに現れる仕組みを考える必要がある。
}
We should also note that 
the topologies of the point and the space with expanse are different. 
Moreover, 
suppose that 
the time when the universe was a point state 
is the origin of time coordinate, 
there is a question whether there exists a negative time. 
%しかも，宇宙が点となった時刻をゼロとすると，
%負の時刻が存在するのかしないのか，疑問も生じる。
In other words, 
there is a new problem of 
whether time and space occurred at the same time, or 
whether space was generated after time was born. 
%つまり，時間と空間は同時に発生したのか，それとも，
%時間経過の途中で空間が発生したのかという新たな問題が現れる。
%%このときの宇宙は，大きさが点であるにもかかわらず，
%%物質の総エネルギーが有限という不思議な状態になります。
%%ここで鍵になる物理量が「時間」でしょう。
%%謎が謎を呼び続けるのだ。

\newcommand{\HawkingImaginaryTime}{%
この特異点を除去する方法の一つに，
宇宙が誕生するときの時空がEuclid計量だったとする考え方がある。
つまり，
この時代の時間は純虚数になっているため，Minkowski計量はEuclid計量となり，
時間は他の空間座標と区別は付けられず，事実上，存在しない。
したがって，
どこから宇宙が誕生したかという疑問は存在せず，特異点を解消する。
しかし，虚時間の理論は
純虚数の時間がどのタイミングで実数の時間に代わるのか問題になる。
なぜなら，量子論の視点で見れば，
Euclid計量とMinkowski計量の2つの空間を接続する場所は
ありとあらゆる場所が足し上げられるため可能で，
しかも，接続条件を満たさない状態をも
足し上げなければならないからである。
さらには，量子重力理論なので，
可能な限りのありとあらゆる時空を足し上げるため，
虚時間と実時間がいろいろな場所で斑に混在する時空の理論を構築する必要がある。
しかも，虚時間となる領域でビッグバンが起きてしまっては，
ありとあらゆる時刻と場所でビッグバンが起きることになる。
%たとえば，図？のような時空である。
しかし，これは観測的に問題がある時空なので，
虚時間となる領域はビッグバン以前の時刻に限定する理論的な理由が必要になる。
いろいろな時空を足し上げるわけだから，
宇宙誕生時のときだけ虚時間の時空が現れるメカニズムを考える必要があるのだ。
%多様体にならない点についても，虚時間と同様，
%もし導入するのなら，これが主にならないメカニズムを考えなければならない。
}%\HawkingImaginaryTime

\subsubsection{Position of string theory in QG}
\label{sec:PositionOfStringTheory}

It is known that 
the string theory makes it possible to quantize the metric 
\red{without encountering UV problems.} 
%avoiding the divergence in calculation. 
In the string theory, instead of point particles, 
1D strings are regarded as fundamental objects, 
and 
$\int\!{\cal D}g_{\mu\nu} {\cal D}A_\rho {\cal D}\psi_\alpha$
is replaced by the path integral 
\red{with respect to so-called string fields} 
%of the string field 
$\Phi$. 
Since gravitons appear in this theory, 
the string theory becomes a candidate for the theory of QG.
%and then gravitons appear and make it a candidate for QG.
%弦理論は点粒子の代わりに一次元的に拡がる弦を基本場と捉えるもので，
%$\int\!{\cal D}g_{\mu\nu} {\cal D}A_\rho {\cal D}\psi_\alpha$
%は弦の場$\Phi$の経路積分$\int\!{\cal D}\Phi$に置き換わり，
%重力子が現れるため，量子重力理論の候補となる理論になる。
Moreover, 
it is expected that 
\red{problems {\bf 1.}\ and {\bf 2.}\ will also be solved at the same time.
Problem {\bf 1.}\ 
as spacetime %now 
becomes a derived concept 
in string theory,%
% so 
%there might not be any higher-dimensional spacetime topologies 
%in a strict sense,%
\footnote{\label{footnote:Landscape}%{\dbltinyspace}%
The string field theory %(SFT) 
and the string landscape are 
among these approaches. 
However, 
in the string field theory %SFT 
it is quite difficult ot find a true vacuum, 
and 
in the %string theory 
string landscape so many vacua appear, 
and there is currently no known way 
to determine which vacuum is the most likely. 
}
and 
problem {\bf 2.}\ 
as the field interactions are almost uniquely determined 
in string theory.}%
%the problem of {\bf 2.}\ will also be solved at the same time, 
%because the field interactions are almost uniquely determined. 
%しかも，場の相互作用の形もほとんど決めてしまうため，
%{\bf 2.}の問題も同時に解決されると期待されている。
\footnote{\label{footnote:Swampland}%{\dbltinyspace}%
The swampland conjecture in string theory \cite{Swampland}
is one promising approach, 
but is not covered in this {\Article} 
because the technical approach is different. 
}

%ここから
\red{
As of today only the first quantized version of perturbation theory 
is used in string theory calculations. 
So-called string field theory (SFT) is one possibility 
to go beyond this simple perturbative approach 
and it allows us to treat off-shell string fields, 
needed to understand non-perturbative effects. 
However, so far SFT has been too complicated 
to use in any practical way.} 
%Only perturbation theory is 
%known as the analysis method in string theory. 
%The string field theory (SFT) is considered to be one possibility 
%to understand non-perturbative effects, 
%because the field theory treats off-shell status of fields easily. 
%However, we failed to understand the nonperturbative effects 
%because the obtained SFT is too complicate to handle. 
%
This situation was not changed 
after the discovery of string duality. 
%By the string duality we can understand the relationship 
%between different string theories. 
%Before the discovery of string duality, 
%there are several different string models. 
\red{String duality makes it} 
%The string duality makes us 
possible to understand that 
the different string theories are the same theory with different vacua. 
Though the duality was a very important discovery, 
\red{and allowed us 
a glimpse of non-perturbative string properties, 
real non-perturbative calculations are still impossible.} 
%but from the duality 
%we know the nonperturbative properties a little and 
%the nonperturbative calculation was still impossible. 

\section{Causality Road}% to QG which describes our Universe}
\label{sec:CausalityRoad}

\subsection{Creation of our Universe from emptiness}

If the universe starts from a point, 
it might have been generated from ``emptiness''. 
%もし宇宙が点から始まったのなら，
%宇宙は無から生成されたかもしれない。
\red{This is the topic of} 
%Let us consider this in 
this {\Section}. 
%この節ではこれについて考えてみよう。

\subsubsection{Our Universe is mathematics with causality}
\label{sec:MathematicsWithCausality}

Can our universe be born out of nothing?\ 
There is an opinion that 
\lq\lq Nothing comes from nothing". 
%Uddalaka Aruni, Parmenides, Lucretius
This is certainly true, but 
there is a teaching in the Buddhist Heart Sutra that 
\lq\lq Form is emptiness, emptiness is form." 
%現在の我々の世界が無から生じることはあり得るだろうか？\ 
%「無から実体は生まれない」という意見がある一方で，
%仏教の般若心経には「色即是空 空即是色」という教えがある。
This means that by the causality 
the insubstantial \lq\lq emptiness" becomes 
\lq\lq form" that is our universe, and vice versa.
%これは実体のない「空」が因果によって我々の世界「色」になり，
%また，その逆も起こることを意味する。
The causality gives life to \lq\lq emptiness" 
and embodies it into \lq\lq form".%
\footnote{%{\dbltinyspace}%
%The reason why the causality appears here is that 
Causality is a central teaching of Buddhism.
}
%因果が「空」に命を吹き込み実体化して，我々の世界「色」になるのである。

What is \lq\lq emptiness"? 
%How is it different from \lq\lq nothing"? 
%仏の教えによれば，emptinessは無とは異なる。
According to Buddhist teachings, 
emptiness is different from nothing.
%では，\lq\lq emptiness" とはなんだろうか？\ 
%\lq\lq nothing" とどう違うのだろうか？\ 
By the way, 
Pythagoras believed 
\lq\lq All things are numbers". 
Galileo said 
\lq\lq The universe is written in the mathematical language". 
These lead to the fact that 
the theory that describes all things, i.e.\ time, space and all matter, 
is mathematics that describes numbers. 
It should be noted here that 
mathematics has no substance and exists independently of our universe 
because it consists only of logic. 
%数学は実体がないが，論理だけで構成されるので，
%この世界と無関係に存在できるものである。
Therefore, 
if we regard the identity of emptiness as ``mathematics with causality'' 
%emptinessの正体を``因果性を持つ数学''とみなすならば，
and follow the teaching that ``Form is emptiness, and vice versa'', 
then we reach{\dbltinyspace}%
\footnote{%{\dbltinyspace}%
There are similar discussions these days,\,\cite{PhysicsIsMath}
but here 
%in this {\Article} 
we will only explain the principles necessary for 
QG that creates the universe from nothing, 
and will not go into philosophical discussions. %beyond this.
%最近でも似たような議論は存在するが，ここでは，
%無から世界を発生させる量子重力理論に必要な原理の説明だけに留め，
%これを超える哲学的な議論には踏み込まない。
}
\begin{equation}\label{MassHeartSutra}
\hbox{Our world is one of the mathematics with causality, and vice versa.}
%\hbox{因果性を持つ数学は我々の世界そのものになる。}
\end{equation}
The point is not only to interpret emptiness as mathematics, 
but also to 
\red{advocate that causality is more fundamental %primal
than} 
%prepare causality in advance in 
emptiness.
%emptinessを数学と解釈しただけでなく，
%emptinessの定義の中に予め因果性を用意するところがポイントである。
%Let us also pay attention to the fact that 
%a physical quantity with the property of time called causality 
%appears separately from space.
%causalityという時間の性質を持つ物理量が
%空間とは別に現れていることにも注目しよう。

For later discussion, 
let us introduce ``time" as a coordinate specifying causality. 
However, 
since causality can exist even if Lorentz symmetry does not exist, 
so to emphasize this point, 
we refer to it as ``causal time'' rather than just ``time''.
The ``causal time'' becomes ``normal time'' when Lorentz symmetry exists. 
%ここで後の議論のため，因果的な順序を指定する座標として時間を導入しよう。
%しかし，因果的な順序を指定するだけなら，
%Lorentz対称性が存在するかどうかについては必須ではないので，
%この点を強調するため，単なる時間ではなく，因果時間と呼ぶことにする。
%それゆえ，Lorentz対称性が存在するときの因果時間は通常の時間となる。

%By the way, 
Physics studies everything: time, space and all matter. 
The only difference between physics and mathematics is that 
physics has interpretation that connects mathematics with our universe. 
Therefore, 
the theorems established by this mathematics will become 
the physical laws of our universe, 
and a universe will emerge where life is born based on these physical laws. 
%この数学の中で成立する定理は現実世界の物理法則となり，
%さらにはこの物理法則を基にして生命が誕生する世界が現れるでしょう。
%
%This means that 
%the physics which studies everything, i.e.\ time, space and all matter, 
%is one of the fields of mathematics which studies the number. 
%これは，物理にとって数学は言語というよりは
%物理が数学そのものであるという考え方だが，
%物理と数学の違いには「自然現象に対する解釈」しかないことを踏まえると，
%これは自然な考え方と言える。

\subsubsection{Mathematics of QG is simple and extremal}
\label{sec:ExtremalMathematics}

%量子重力理論の詳細な議論に入る前に，
Let us think about the mathematics 
we should aim for in order to build QG. 
%Let us think about the mathematics that 
%QG should aim for.
%ここで%少し物理と数学の関係，そして，
%量子重力理論が目指すべき数学について考えてみよう。
%では，どのような数学が物理になるのでしょう？\ 

One mathematical theory corresponds to one world, 
and one of these worlds should be our world.
%一つの数学理論は一つの世界に対応し，
%これらの世界の一つが我々の世界になるはず。
Therefore, first, for example, 
%そこで，まず，
let us consider the universe of mathematics such as Euclidean geometry.
%たとえば，ユークリッド幾何学のような数学の世界を考えてみよう。
In this universe, there are theorems such as the Pythagorean theorem, 
and these theorems will become the laws of physics in this universe.
%この世界ではピタゴラス定理などの定理が存在し，
%これら定理はこの世界の物理法則となるだろう。
However, Euclidean geometry is not a complex theoretical system 
to describe our universe.
%しかし、ユークリッド幾何学は
%我々の世界を記述するほど複雑な理論体系ではない。
The mathematical structure of Euclidean geometry is too simple 
to describe our universe.
%ユークリッド幾何学が持つ数学的な構造は現実世界を記述するには簡単過ぎるのだ。
Conversely, mathematics complex enough to describe our universe 
is the mathematics that we physicists seek.
%逆に考えると，我々の世界が記述可能なほど複雑な数学こそが、
%私たち物理学者が求める数学と考えられます。

Gravity is a force caused by the distortion of spacetime. 
%重力は，一般相対性理論で既に理解されたように，
%時空，すなわち，時間と空間で構成される空間の歪みにより生じる力である。
Gravity is described by geometry.
%重力は幾何学によって記述されるのだ。
On the other hand, 
gauge fields and matter fields are described by 
${\rm SU}(3) {\negtrpltinyspace}\times{\negtrpltinyspace}
 {\rm SU}(2) {\negtrpltinyspace}\times{\negtrpltinyspace}
 {\rm U}(1)$
Yang-Mills gauge theory. 
%一方，ゲージ場と物質場は，
%${\rm SU}(3) \!\times\! {\rm SU}(2) \!\times\! {\rm U}(1)$
%の対称性を持つYang-Millsゲージ理論により記述される。
Given that both mathematics are a kind of geometry, 
it is conceivable that 
the mathematics that describes the universe 
will be a kind of geometry. 
%どちらの数学も幾何学の一種であることを考えると，
%ある種の幾何学こそがこの世界を記述する数学だと考えてよいだろう。
%%このことに加え，Einstein方程式には物質場やゲージ場が
%%エネルギー運動量テンソルという形で現れているという事実を考慮すると，
Therefore, 
it is natural to expect that 
QG, which unites the gravitational theory and the quantum theory, 
will become a theory that describes 
not only time and space, 
but also gauge fields and matter fields. 
%それゆえ，
%重力理論と量子論を統合する量子重力理論は，
%時間と空間だけでなく，ゲージ場と物質場を加えた全てを
%記述する理論になると期待するのは自然だと考えられる。
%%\footnote{%{\dbltinyspace}%
%%もちろん，どのような物質場やゲージ場が存在するとき，
%%重力理論の量子化が可能になるのかは非自明で，
%%沼地問題はこの考え方に則った例でしょう。
%%}
Of course, this idea is not a new point of view, 
and has been taken over by Kaluza-Klein theory in the old days 
and string theory in recent years.
%もちろんこの考えは新しい視点ではなく，古くはKaluza-Klein理論，
%最近では弦理論に引き継がれている。
Therefore, 
\red{we aim for the QG theory we will discuss below  to} 
%QG we show here is considered to 
explain everything, i.e.\ time, space, %and all matter. 
gauge fields and matter fields
%そこで，我々がここで目指す量子重力理論は，

%By the way, 
String theory is not only a candidate for QG, 
but it is also the 
\red{physics that we human beings know best} 
%only physics that we human beings know 
in terms of mathematical depth. 
%ところで，
%弦理論は，
%量子重力理論の候補となるだけでなく，
%数学的な奥行きの深さで言えば，
%現代の我々人類が知る唯一の理論である。
\red{It} 
%This theory 
is based on 2D conformal field theory and 
can be regarded as a 
\red{so-called} 
{\it critical string theory} with 
26 dimensions, % as the critical dimension, 
where 
the moduli symmetry due to 2D conformal field theory 
acts to cancel the divergences 
\red{resulting from} 
%caused by the 
perturbative calculations.
%この理論は，
%2次元共形場理論を基礎にしており，
%26次元を臨界次元とする critical string theory とみなすこともできて，
%2次元共形場理論に起因するmoduliの対称性が
%摂動論で生じる発散を消す働きをする。
\red{So-called} 
%Moreover, 
non-critical string theory 
with dimensions other than 26, %dimensions 
can be considered as 2D QG with matter fields.
%しかも，
%26次元以外の非臨界次元を持つ non-critical string theory は，
%物質場を持つ2次元量子重力理論と考えることもできる。
Therefore, %そこで，%ここでは
let us 
\red{start} 
%consider starting 
with 2D Euclidean QG without matter, 
extending it to non-critical strings with matter, 
and finally arriving at a string theory with 26 critical dimensions. 
%non-critical string theory を経由して
%%critical string theory と non-critical string theory の両者の成功を踏まえて，
%26-dimensional critical string theory
%を目標に量子重力理論を構築していくことを考えてみよう。

%ところで，
%2次元重力は，古典論では作用がEuler数に比例するため，
%運動方程式に相当する式はなく，
%計量の経路積分からLiouville作用が生まれ，
%Liouville方程式が運動方程式となる。
%計量の量子効果が強く，古典論の運動方程式は量子論のそれ
%を得るための手がかりにならないのである。
%このように，QGには通常の常識が通用しない現象がある。
%これは一種の``extremity" である。

In this discussion, 
we take not only ``simplicity'' but also ``extremity'' 
as clues for logical leaps.
The property of the equations of motion referred to in footnote 
\ref{footnote:extremityEOM} is an example of ``extremity''. 
The octonion %and the exceptional algebra 
%which will be introduced later are 
is also an example of ``extremity''.%
\footnote{%{\dbltinyspace}%
The octonion is in a marginal position where 
the associative law holds where it matters and 
fails where it does not. 
The statement ``the octonions do not satisfy the associative law." 
is correct, but misleading. 
Furthermore, 
in simple Lie groups, $E_8$ is an example of extremity, 
while 
in finite simple groups, the Monster group is an example of extremity. 
}
We believe that 
``extremity'' leads to a less simplistic and fruitful world.
%具体的には，critical string theory ではなく，
%non-critical string theory を出発点とし，
%論理的な飛躍をするときは，手掛かりとして，
%「単純さ (simplicity)」と「際 (extremity)」
%という考え方をすることで，量子重力理論を構築する。
In other words, we believe that 
``Physics is a simple and extreme theory of mathematics''.
%つまり，
%``Physics is a simple and extremal theory of mathematics'',
%と考えるのである。
$W$ algebra and Jordan algebra with octonions, 
which will be discussed later, 
are choices based on this idea.
%後に述べる八元数やalbert algebraがこの考え方に基づく選択になる。

\subsection{Proposals to solve the problems in QG}%expected to be solved by QG}
\label{sec:ProposalsToSolveProblems}

In this {\Section} 
we list several proposals to solve the problems 
\red{associated with} 
%that are expected to be solved by 
QG.
%この節では，量子重力理論が解くと期待される問題を解く提案をする。

%宇宙は点の状態から始まったと仮定する。
%Our Universe started from a point. 

\subsubsection{Our Universe started from a point state}
\label{sec:UniverseStartedFromPoint}
As we go back in time from the present to Big Bang, 
the universe gets smaller and smaller.
%現在からビッグバンと時代を遡ってゆくと，宇宙はどんどん小さくなる。
Therefore, 
%それゆえ，
%少なくともインフレーションの開始時点では
%宇宙は$10^{-36}\,\hbox{[m]}$というサイズにまで小さくなるので，
it would be a natural assumption to think that 
the universe arose from a point state.
%宇宙が点の状態から発生したと思うのは自然な仮定だろう。
Under this assumption, the following problem is solved. 
%この仮定の下では，次の問題が解決する。
\vspace{-2pt}
\begin{description}
\item[\bf\hspace{12pt}%
Problem of initial conditions of wave function of the Universe%
%世界の状態の波動関数の問題%
{$\rule[-5.5pt]{0pt}{0pt}$}]\ \\
\hspace{12pt}%
As was explained in {\Section} \ref{sec:Determinism}, 
%\ref{sec:Determinism}節で説明したように，
if the universe started as a point state, 
there would be no need for a mechanism to choose the initial state, 
since the point state is unique.
%宇宙が点の状態で始まったのなら，
%点の状態は唯一なので初期状態を選ぶメカニズムは必要はなく，
%この問題は解決される。
%%古典論は宇宙の状態を一意的に決定する。
%%一方，量子論はあらゆる状態の中から一つの状態を確率的に決定するものの，
%%その確率を与える波動関数は一意的に決定される。
%%ここで一つ問題が生じる。
%%もし宇宙が点から始まらない場合，
%%現在の宇宙の波動関数を決める原理が存在しない。
%%しかし，点から始まるのなら，この問題は解決する。
\end{description}
\vspace{-2pt}
It should also be noted that 
this assumption is not inconsistent with the existence of 
the reference frame of the CMB (the CMB rest frame) 
because the volume of space becomes finite. 
However, the following new problem arises under this assumption. 
%しかし，次の新たな問題が生じる。
\vspace{-2pt}
\begin{description}
\item[\bf\hspace{12pt}%
Problem of the singularity% of matter energy% density%
%物質エネルギー保存の問題%
{$\rule[-5.5pt]{0pt}{0pt}$}]\ \\
\hspace{12pt}%
If the matter energy is conserved retroactively to the point state, 
the point state becomes a state with finite matter energy. 
%物質エネルギーが点状態まで遡って保存されるなら，
%点状態は有限の物質エネルギーを持つ状態となる。
Even when there is a conservation law based on symmetry 
such as the charge conservation law, 
the point state is also a state with a finite conserved quantity.
%電荷保存則のような対称性に基づく保存則が存在するときも，
%同様に点状態は有限の保存量を持つ状態となる。
Therefore, 
if a conserved quantity exists, 
the point-state universe becomes a state with a matter memory.
%したがって，保存量が存在すると，
%点状態の宇宙は物質の記憶を持つ状態となる。
In terms of density, it becomes infinite, 
so it becomes a so-called spacetime singularity.
%密度で言えば，無限大になるため，いわゆる時空の特異点となる。
Such a spacetime singularity is not desirable 
from the point of view of mathematics or physics, 
but that is not the only problem. 
%このような時空の特異点は，
%数学から見ても物理学から見ても好ましいものではないが，
%問題はそれだけではない。
In a quantum gravity theory, which 
\red{integrates over} 
%sums over 
all possible spacetime, 
such a spacetime singularity 
\red{can appear everywhere in} 
%appears in all 
time and space, 
and a ``Big Bang'' 
\red{can then occur everywhere.} 
%occurs there. 
%あらゆる時空を足し上げる量子重力では，
%あらゆる時間と空間でこのような時空の特異点が現れてビッグバンが起きてしまう。
%In the quantum gravity theory, 
%the theoretical seriousness is further increased. 
%量子重力理論では，理論的な深刻度はさらに増すのである。

%しかし，
%ブラックホールがバリオン数などの物質の記憶を持たないことを考えると，
%これは不自然なことかもしれない。
\vspace{3pt}
\item[\bf\hspace{12pt}%
Problem of the topology of the Universe%
%宇宙のトポロジーの問題%
{$\rule[-5.5pt]{0pt}{0pt}$}]\ \\
\hspace{12pt}%
If the universe started from a point state, 
we need a mechanism to determine the topology of the universe. 
%宇宙が点状態から始まったのなら，
%宇宙のトポロジーを決めるメカニズムが必要になる。
Since the universe becomes a space with 
\red{extension and thus a topology
immediately after it was a point state,} 
%size in the next moment from the point state, 
it is necessary to choose and 
determine the topology of the space. 
%点状態から次の瞬間に宇宙は大きさを持つ空間になるため，
%空間のトポロジーを選択し決定することが必要になる。
One can say that a point state 
%A point state 
becomes a branching point of all topologies. 
%点状態は全てのトポロジーの分岐点になるのである。
However, 
it is necessary to theoretically determine 
the weight $C_{{\negtinyspace}{\cal T}}$ 
of each topology ${\cal T}$%. 
\red{, and as already mentioned,} 
%しかし，
%それには各トポロジー${\cal T}$のウエイト$C_{{\negtinyspace}{\cal T}}$を
%理論的に決める必要がある。
%However, in current mathematics, 
the classification of topologies in 3D or higher-dimensional space 
is incomplete.
%ところが，
%現在数学では，3次元以上の空間のトポロジーの分類は未完である。
Therefore, 
summing up all possible topologies 
is hopeless except for 1D and 2D spaces.
%それゆえ，
%可能な全てのトポロジーを足し上げることは2次元空間以外は絶望的である。
\end{description}
\vspace{-2pt}

%【ボツ】
%現在の数学の理解では，
%いろいろな時空の足し上げは2次元以外では絶望的である。
%しかも，特異点の問題は，2次元では起こらない。
%さらには時空のトポロジーを決めるときにも問題が起きにくい。
%これは無から宇宙を生み出すときに利用できる。
%時空のトポロジーの選択で2次元は有利なのである。
%
%そこで，時間を空間と切り離して特別扱いし，時空は2次元とする。
%時間を切り離さなくても宇宙が点から大きさを持つようになることは可能だが，
%無から点の状態の宇宙を生み出すのは不可能。
%
%無から時空が発生するか？\ 
%この問いは宇宙の波動関数と会い入れない。
%量子力学では，波動関数の大きさの2乗は
%その波動関数が生成される確率を表す。
%では，無から時空が発生したときの宇宙の波動関数の確率は意味があるだろうか？\ 
%もし，意味があるというのなら，初期状態として無の状態が存在し，その後，
%その確率に従った宇宙が発生することになる。しかし，このプロセスには，
%時間が存在する。時間と言う実数によって表される物理量が存在しないとしても，
%少なくとも Causality は存在する。

%【ボツ】
%宇宙マイクロ波背景輻射が発見され，
%宇宙膨張とビッグバンの存在が明らかになった。
%この観測事実から，
%宇宙が点のような状態から始まったことがわかり，
%\rf{PartitionFunQGfixedTopology}の観点から見ると，
%時空のある1点が宇宙誕生の特異点となることがわかる。
%しかし，もしある1点から宇宙，すなわち，時空が誕生したとするなら，
%この世界の波動関数$\Psi(t)$の解釈に問題が生じる。
%$\Psi(t_0)$がこの世の始まりの状態とするなら
%この状態が実現する確率は意味を持つか，という問題が生じるのである。
%確率は，一定の時間経過の後にこの状態が実現する確率という意味であるため，
%時間が誕生するプロセスでは意味を持たないからである。
%ここでは，この問題を解決する一つの方法として，
%時間は空間の誕生とは無関係と考えることにする。
%時間は空間の誕生前から存在するため，
%無から空間が誕生したときにできるこの世の状態が
%$\Psi(t_0)$となる確率が意味を持つのである。

\subsubsection{The causal time axis is placed outside spacetime}
\label{sec:IntroCausalityTime}
\red{A general discussion of this subject 
will derail our line of arguments and 
we will thus assume the setting of} 
%Since advancing a generalization on this subject 
%would obscure the focus of the discussion, 
%this {\Section} assumes 
CDT described in {\Chapter} \ref{sec:CDT}. 
%一般論を展開すると話の焦点がぼやけてしまうので，%ここでは，
%この節では%今までのような一般論ではなく，
%{\Chapter} \ref{sec:CDT} の中で述べるCDTを仮定しよう。

%時間軸の立場は？\ Minkowski時空は点から誕生したのか？

The 2D \red{spacetime} %space 
of 2D Euclidean QG has fractal structure, 
and it is known that 1D universes split, merge, and disappear 
when viewed from the viewpoint of geodesic distance 
as shown in Fig.\ \ref{fig:DTpicture}.\,%
%2次元Euclid量子重力理論の2次元空間はフラクタルな構造を持ち，
%図\ref{fig:DTpicture}のような測地距離の視点から見ると，
%1次元空間の宇宙は分裂したり融合したり消滅したりする
%ことが知られている。
\cite{FractalKKSW,FractalKKMW}
If we shift this viewpoint 
from 2D Euclidean QG to 2D Minkowskian QG, 
the geodesic distance is replaced by time, 
and 1D universes split, merge, and disappear 
under the time evolution as shown in Fig.\ \ref{fig:CDTpicture}. 
%この視点を2次元Minkowski量子重力理論へ移行すると，
%測地距離は時間に置き換わり，
%図\ref{fig:CDTpicture}のような時間発展の中で
%1次元空間の宇宙が
%分裂したり融合したり消滅したりすることになる。
Understanding the physical meaning of time is important here.
%ここでは，時間が持つ物理的意味の理解は重要である。
This is because 
Euclidean space with rotational symmetry does not become 
Minkowskian space with Lorentz symmetry 
just by translating the geodesic distance into time.
%測地距離を時間に読み替えただけでは
%回転対称性を持つEuclid空間は
%Lorentz対称性を持つMinkowski空間にならないからである。
\red{The simplest way to discuss this is 
to consider a theory that abandons Lorentz symmetry, 
but still has a time where causality is valid. 
We denote such a time {\it causal time}}.
%So, as the simplest way, 
%let us consider a theory described by time 
%that abandons Lorentz symmetry and holds only causality.
%そこで，最も単純な方法として，
%%Minkowski時空にはならないが，
%Lorentz対称性を諦めて因果関係だけを保つ時間によって記述される
%理論を考えてみよう。
%%これは，図\ref{fig:CDTpicture}のような
%%時間軸が2次元時空の外に置かれた描像になる。
%This time is the time with only causality 
%where Lorentz symmetry is not required, that is, causal time.
%このときの時間はLorentz対称性が要求されていない
%因果性だけを持つ時間，つまり，因果時間である。
%%そこで，我々は時間軸を2次元時空の外側に設けることとし，
%%これを因果を決定する物理量として採用することにしよう。
%%したがって，Euclid空間の領域は否定される。さらには，
In this picture, 
the process of space appearing as a point state 
from the empty state where space does not exist, 
and then expanding into a space with an 
\red{extension} 
%size 
is expressed from the viewpoint of 
\red{an external time axis.} 
%the time axis placed outside.
%この描像では，空間が何もない空の状態から点の状態へ，
%そして，点が大きさを持つ空間へ変化してゆくプロセスが
%外側に設けられた時間軸の視点で表される。
%
However, 
the following new problems arise 
\red{for such an external time:} 
%in the causal time introduced here.
%しかし，ここで導入された因果時間について，以下のような新たな問題が生じる。
\vspace{-2pt}
\begin{description}
\item[\bf\hspace{12pt}%
Problem of the Lorentz symmetry%
%Lorentz対称性の問題%
{$\rule[-5.5pt]{0pt}{0pt}$}]\ \\
\hspace{12pt}%
Since this time is a causal time, 
it does not 
\red{contradict} %deny 
the existence of Lorentz symmetry, 
but does not guarantee it. 
%この場合の時間は因果時間なので，
%Lorentz対称性の存在を否定するには至らないが，保証することもない。
%%したがって，
%%Lorentz対称性を保証する何かが必要になる。
\red{As a side remark,} 
%By the way, 
what is important in string theory is 
the Lorentz symmetry of target spacetime 
rather than 2D Lorentz symmetry. 
%ところで，弦理論で重要なのは，
%2次元のLorentz対称性よりは，むしろ，
%target spacetimeのLorentz対称性である。
The emergence of Lorentz symmetry requires a new perspective, 
\red{that we will discuss} 
%so let us discuss this point 
in {\Subsection} \ref{sec:EmergenceOfStringTheory}. 
%Lorentz対称性の出現には新しい視点が必要なので，
%subsection \ref{sec:EmergenceOfStringTheory} で
%この点を踏まえた議論をすることにしよう。
\vspace{3pt}
\item[\bf\hspace{12pt}%
Problem of the birth from emptiness%
%時間的閉曲線の問題%
{$\rule[-5.5pt]{0pt}{0pt}$}]\ \\
\hspace{12pt}%
When time is 
\red{viewed as external,} 
%placed outside, 
it is possible to change the state from an empty state to a point state.
%時間が外側に置かれると，
%空の状態から点の状態への変化が可能になる。
However, 
\red{since a point state in general can carry charges} 
as mentioned in a previous {\Subsection}, 
%{\bf Problem of the singularity}, 
\red{in order to make a transition from an empty state to a point state, 
one needs a mechanism which can introduce charges to an empty state.}%
%in order to make this possible, 
%a mechanism which makes to introduce 
%the memory of the substance to the point 
%is required, if the point state has the memory.%
%しかし，これを可能にするには，
%前節の{\bf Problem of the singularity}で述べたように，
%もし点状態が物質の記憶を持つならば，
%点状態がどのようにしてその記憶を得るのかを
%決めるメカニズムが必要になる。zaqここまでだよん。
\footnote{%{\dbltinyspace}%
Note that 
this problem will not be resolved even if 
the value of the physical 
\red{charge in question}
%value 
is zero when the universe is a point state. 
Whether the value is zero or 
whether the value itself is meaningless 
\red{are different questions.} 
%is different. 
%物理量の値が点状態のときにゼロになってもこの問題は解決しないので注意せよ。
%ゼロという値なのか，それとも，
%値自体が意味を持たない状態なのかは意味が違うのである。
}
Details will be discussed in {\Section} \ref{sec:purCDTbyWalgebra}, 
but this problem is solved by representing CDT by $W$ operators.%
%詳細は {\Section} \ref{sec:purCDTbyWalgebra} で述べるが，
%この問題はCDTを $W$ operator で表すことで解決される。%
\footnote{%{\dbltinyspace}%
The key is that the matter fields are path-integrated out.
%物質場が path-integrate out されていることが鍵になる。
}
\end{description}
\vspace{-2pt}
On the other hand, 
the following problems are solved as a byproduct of 
\red{the assumed existence of an external causal time:} 
%this picture. 
%しかし，この描像の副産物として，次の問題が解決される。
\vspace{-2pt}
\begin{description}
\item[\bf\hspace{12pt}%
Problem of closed timelike loop%
%時間的閉曲線の問題%
{$\rule[-5.5pt]{0pt}{0pt}$}]\ \\
\hspace{12pt}%
This picture 
\red{forbids} 
%denies 
the presence of time{\tinyspace}-closed curves 
and 
the branch of the time axis.%
%時間的閉曲線の存在や時間軸の分岐を否定する。
\footnote{%{\dbltinyspace}%
Since there is no time{\tinyspace}-closed curve in this picture, 
there is no G\"{o}del solution of general relativity.\,%
\cite{GoedelSolution}
%この描像では時間的閉曲線は存在しないので，
%一般相対性理論のG\"{o}del解は存在しない。
There is no time machine which makes us possible to return to the past.
%また，過去に戻るタイムマシンも存在しない。
}
\end{description}
\vspace{-2pt}

\subsubsection{Our Universe started as a one-dimensional space}
\label{sec:UniversWasBornAsOneDimSpace}
Assuming that 
the universe initially occurs from emptiness as a 1D space 
and has changed to a high dimensional space during the expansion, 
the following problems are solved. 
%宇宙が最初はemptinessから1次元空間として発生し，
%膨張する途中で高次元の宇宙へ変化したと仮定すると，
%次のような問題が解決する。
\vspace{-2pt}
\begin{description}
\item[\bf\hspace{12pt}%
Problem of the singularity of spacetime% of matter energy% density%
%物質エネルギー保存の問題%
{$\rule[-5.5pt]{0pt}{0pt}$}]\ \\
\hspace{12pt}%
%宇宙が点の状態から始まったのならば，
%物質エネルギーの保存により，
%このときの宇宙の物質エネルギー密度は無限大となり，
%どのように扱えばよいのか問題になる。
%ところが，
If the universe has changed 
from a 1D space to a high dimensional space 
in the middle of expansion, 
the conservation law of energy and charge of matter fields 
\red{cannot be extended further back in time than to the time 
where this change of dimension took place. 
Consequently the singularity problem disappears.} 
%can only go back to this change point, 
%so the singularity problem, 
%which is mentioned in {\Subsection} \ref{sec:UniverseStartedFromPoint}, 
%is not solved and disappears. 
%宇宙が膨張の途中で1次元から
%高次元へ変化したのなら，
%物質エネルギーや電荷のような物理量の保存は
%ここまでしか遡ることはできず，
%subsection \ref{sec:UniverseStartedFromPoint} で述べた
%特異点の問題は，
%%宇宙が点の状態まで遡ることはできないため，
%問題自体が消滅する。
\vspace{3pt}
\item[\bf\hspace{12pt}%
Problem of the topology of the Universe%
%宇宙のトポロジーの問題%
{$\rule[-5.5pt]{0pt}{0pt}$}]\ \\
\hspace{12pt}%
Since there is only one topology $S^1$ in a closed 1D space, 
there is no problem of which topology is chosen 
when the universe changes from a point state to the space with a 
\red{finite extension.} 
%size.
%閉じた1次元空間のトポロジーは円$S^1$の１つしかないので，
%点状態の宇宙から大きさを持つ状態の宇宙へ変化したときに
%どのトポロジーが選ばれるかという問題は起こらない。
\end{description}
\vspace{-2pt}

%\subsubsection{Coleman mechanism}
%相転移で生じる宇宙項が消える機構
\label{sec:VanishingCosmoConst}

\subsubsection{The emergence of critical string theory}
\label{sec:EmergenceOfStringTheory}
\red{Again, a general discussion of this vast topic 
will derail our line of arguments and we will thus assume} 
%Since advancing a generalization on this subject 
%would obscure the focus of the discussion, 
%this {\Subsection} assumes 
the model proposed in {\Chapter} \ref{sec:BasicTheory}. 
%一般論を展開すると話の焦点がぼやけてしまうので，%ここでは，
%この節では%今までのような一般論ではなく，
%{\Chapter} \ref{sec:BasicTheory} の中で提案される模型を仮定しよう。

Since $c{\negdbltinyspace}={\negdbltinyspace}0$ non-critical string theory 
with Euclidean metric is a DT without matter, that is, pure DT, 
the pure CDT obtained by replacing geodesic distance in pure DT 
with the causal time is considered to be a kind of 
$c{\negdbltinyspace}={\negdbltinyspace}0$ non-critical string theory 
with an 
\red{external} 
time axis. 
%Euclid計量で定義された
%$c{\negdbltinyspace}={\negdbltinyspace}0$ non-critical string theory 
%は物質場のないDTなので，これを pure DT と呼ぶと，
%pure DT の測地距離を因果時間に置き換えることで得られた pure CDT は
%時間軸を持つ
%$c{\negdbltinyspace}={\negdbltinyspace}0$ non-critical string theory 
%の一種と考えられる。
\red{By introducing 26 scalar fields 
(each with central charge 
$c{\negdbltinyspace}={\negdbltinyspace}1$) 
in this theory one expects to obtain 
$c{\negdbltinyspace}={\negdbltinyspace}26$ 
{\it critical} string theory,} 
%Introducing 26 kinds of scalar fields into this theory 
%can be expected to obtain 
%$c{\negdbltinyspace}={\negdbltinyspace}26$ critical string theory，
that is, 
\red{standard, ordinary string theory,} 
%the string theory, 
and 
if QG constructed in this way matches string theory, 
the following problems will be solved. 
%この理論に26種類の物質場を導入すると，
%$c{\negdbltinyspace}={\negdbltinyspace}26$ critical string theory，
%すなわち，弦理論が得られると期待できて，
%もしこのようにして構成された重力理論が弦理論に一致するなら，
%次の問題が解決する。
\vspace{-2pt}
\begin{description}
\item[\bf\hspace{12pt}%
Problem of Lorentz symmetry%
%Lorentz対称性の問題%
{$\rule[-5.5pt]{0pt}{0pt}$}]\ \\
\hspace{12pt}%
If we identify causal time with the light-like time of 
\red{appearing in the light-cone gauge in string theory,} 
%light-cone gauge, 
we automatically have Lorentz symmetry 
due to the property of critical string theory in the light-cone gauge.%
%因果時間を light-cone gauge の光的時間と同一視すると，
%光円錐ゲージの臨界弦理論の性質により，
%Lorentz対称性を自動的に持つようになる。
\footnote{%{\dbltinyspace}%
Note that 
as was explained in {\Subsection} \ref{sec:IntroCausalityTime}, 
the existence of Lorentz symmetry is not guaranteed, 
but also not excluded 
in CDT. 
}
\vspace{3pt}
\item[\bf\hspace{12pt}%
Problem of the gauge symmetry of Standard Model%
%標準理論のゲージ対称性の問題%
{$\rule[-5.5pt]{0pt}{0pt}$}]\ \\
\hspace{12pt}%
\red{In the model described in {\Chapter} \ref{sec:BasicTheory} 
the high-dimensional torus, which produces gauge symmetry, 
originates from the expansion of point states.
Thus there is no reason that 
a state of high gauge symmetry such as 
$E_8 \!\times\! E_8$ appears at the very beginning.} 
%In this model, 
%since the high-dimensional torus %toroidal compact spaces 
%which produces gauge symmetry 
%expand from a point state, 
%it does not appear that a high gauge symmetry such as 
%$E_8 \!\times\! E_8$ appears from the beginning.
%この模型では，空間は点から膨張してゆくので，
%最初から9次元空間で，
%たとえば$E_8 \!\times\! E_8$ のような高いゲージ対称性が
%登場する訳ではない。
Since each flavor of space was created from nothing 
one after another in time 
and increases its extension, 
there is a high possibility that 
only some flavors of space will expand to large extension. 
Then, this may explain why 
the space of our universe is of low dimension 3, 
and the gauge symmetry
${\rm SU}(3) \!\times\! {\rm SU}(2) \!\times\! {\rm U}(1)$
is of law rank 4. 
%無から空間の各成分が次々と登場するので，
%空間の一部の成分だけが先に膨張する可能性が高く，
%宇宙が3次元で，
%%標準模型の
%ゲージ対称性
%${\rm SU}(3) \!\times\! {\rm SU}(2) \!\times\! {\rm U}(1)$
%がランク4という
%低い次元かつ低い
%%超弦理論では標準理論の
%%${\rm SU}(3) \!\times\! {\rm SU}(2) \!\times\! {\rm U}(1)$
%ゲージ対称性であることを説明する可能性がある。
%を得る可能性が高い。
\vspace{3pt}
\item[\bf\hspace{12pt}%
Problem of background metric independence%
{$\rule[-5.5pt]{0pt}{0pt}$}]\ \\
\hspace{12pt}%
QG is a theory that creates spacetime from emptiness, 
so the definition of the theory should not depend on the metric.
%量子重力理論はemptinessから時空を生成させる理論なので，
%理論の定義は計量に依存してはならない。
In the case of SFT, 
a metric-independent expression is possible 
and this problem is solved formally.\,%
%弦の場の理論の場合，計量に依存しない表記が可能となり，
%この問題は形式的に解決される。
\cite{Phi3SFT}
However, 
there is a technical problem that 
this expression is very difficult to handle 
and practical calculation is almost impossible.
%しかし，この表記は取り扱いが非常に難しく，
%具体的な計算が不可能に近いという技術的な問題がある。
On the other hand, 
as will be explained later,
%一方，後に述べるが，
the theory expressing CDT with $W$ operators 
also has the property of not depending on the metric, 
so it solves this problem, 
including the problem of indefinite metric 
because the origins of time and space are different.
%CDT を $W$ operator で表した理論も計量に依存しない性質を持つため，
%弦理論と同様に，しかも，時間と空間の起源が異なるため，
%不定計量の問題を含めてこの問題を解決する。
\end{description}
\vspace{-2pt}

%\subsubsection{StringFieldTheory}
%\label{sec:SFT}
%弦の場の理論は以下の問題を取り扱うべきだが，
%実際に取り扱えるのか疑問。
%\begin{itemize}
%\item
%Path integral and Nonperturbative theory
%\item
%Wilson renormalization
%\end{itemize}

%%%
\begin{figure}[t]
\vspace{-3mm}\hspace{10mm}
\begin{minipage}{.38\linewidth}\vspace{6mm}
  \begin{center}\hspace*{-0mm}
    \includegraphics[width=1.2\linewidth,angle=0]
                    {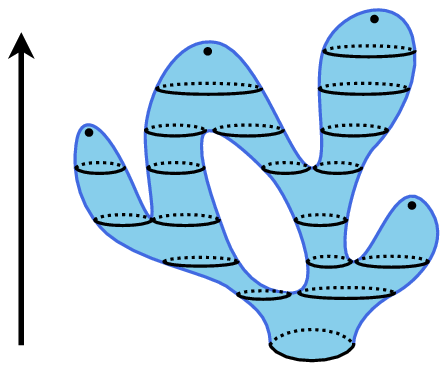}
  \end{center}
  \begin{picture}(200,0)
    \put( 32.0, 41.0){\mbox{\footnotesize$\T$}}
  \end{picture}
  \vspace*{-8mm}
  \caption[fig.1]{{One example of the DT configuration (%DTの場合，
One can reach any point in spacetime from the entrance 1D universe 
because $T$ is not time but geodesic distance.%
%$T$は時間ではなく測地距離なので
%入口の1次元宇宙から任意の時空点に辿り着くことができる。%
)}
}\vspace*{-2mm}
  \label{fig:DTpicture}
\end{minipage}
\hspace{8mm}
%\hspace{.10\linewidth}
\begin{minipage}{.38\linewidth}\vspace{6mm}
  \begin{center}\hspace*{-0mm}
    \includegraphics[width=1.2\linewidth,angle=0]
                    {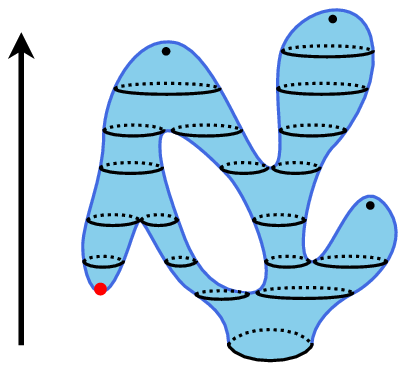}
  \end{center}
  \begin{picture}(200,0)
    \put( 32.0, 41.0){\mbox{\footnotesize$\T$}}
  \end{picture}
  \vspace*{-8mm}
  \caption[fig.2]{{One example of the CDT configuration (%CDTの場合，
One cannot reach the red dot from the entrance 1D universe 
%In order to reach the red point, 
because one cannot go backwards about time.%
%入口の1次元宇宙から赤い点に辿り着くことはできない。
%赤い点に辿り着くには時間を遡らないとならないからである。%
)}
}\vspace*{-2mm}
  \label{fig:CDTpicture}
\end{minipage}
\vspace{1mm}
%\vspace{-5mm}
\end{figure}%\vspace*{0mm}
%%%

%%%%
%\begin{figure}[t]
%  \vspace{5mm}
%  \begin{center}\hspace*{-0mm}
%    \includegraphics[width=0.75\linewidth,angle=0]
%                    {evolutionBigBang.ps}
%   %\psfig{figure=heatbath.ps,width=12.0cm,angle=90}
%  \end{center}
%  \vspace*{-0mm}\hspace{0mm}
%  \begin{picture}(300,0)
%    \put(130.0, 78.0){\mbox{\footnotesize$\T$}}
%  \end{picture}
%  \vspace*{-10.0mm}
%  \caption[fig.1]{{configuration of CDT （CDTの場合，
%入口となる閉じた弦から赤い点に辿り着くことはできない。
%赤い点に辿り着くには時刻を遡らないとならないからである。
%ところが，DTの場合ならば，測地距離なので辿り着くことができる。）}
%}\vspace*{-2mm}
%  \label{fig:CDTpicture}
%\end{figure}
%%%%

\subsection{Strategy for QG that describes our Universe}
\label{sec:Strategy}

We treat points {\bf A)} and {\bf B)} below as the most important facts.
\vspace{-2pt}
\begin{description}
\item[\bf\hspace{12pt}%
A)]
Two-dimensional Euclidean QG 
is currently the only QG theory 
that has succeeded in calculating 
%path-integrating the metric and matter, 
the path integral defined in 
\rf{PartitionFunQG} and \rf{PartitionFunQGfixedTopology}. 
\item[\bf\hspace{12pt}%
B)]
The $c {\negtrpltinyspace}={\negtrpltinyspace} 26$
critical string theory has Lorentz symmetry and includes the graviton, 
so this theory is a candidate of QG that describes our universe. 
\end{description}
\vspace{-2pt}
However, 
\red{point {\bf A)} has the problem that as a QG theory
it has a  lower spatial dimension than the three dimensions} 
%the fact {\bf A)} has the problem that 
%it has a dimension lower than the three dimension 
of our universe, 
and there is no concept of time because the metric is Euclidean. 
\red{Further, point {\bf B)} has the problem that so far
the string theory cannot describe the birth of universe. 
String theory is generally not good at 
describing the time evolution of the universe. 
Both in points} 
%The fact {\bf B)} has the problem that 
%the string theory cannot describe the birth of universe so far. 
%The string theory is not good at describing the time evolution of universe. 
%In both facts 
{\bf A)} and {\bf B)} 
``time'' seems to be the keyword. 
By the way, 
in the statement \rf{MassHeartSutra} 
``causality'' is the key. 
So, it is quite natural to think that 
``causal time'' is the key that connects 
points {\bf A)} and {\bf B)}. 
%我々は以下の2つの事実を最重要とみなし，
%{\bf 1.}\ を出発点，
%{\bf 2.}\ を終着点として，
%その間を繋ぐ経路の道標として因果時間を利用する方法を考えてみよう。
The important point here is that it is causal time, 
not Lorentzian symmetry time.
Moreover, 
only critical string theory has Lorentz symmetry. 
%and non-critical string theory does not have it. 
Given that 2D QG 
is equivalent to a non-critical string theory, 
we are led to believe that 
we can arrive at the critical string theory 
starting from 2D QG and 
using causal time as a guide. 
We will close our eyes on Lorentz symmetry for a while 
and use causal time as time that has only a causal relationship. 

The following is a strategy for getting 
the Minkowskian critical string theory 
starting from 2D Euclidean QG. 
The basic structure of this strategy is 
to arrive at the Minkowskian critical string theory 
which is described by the time-evolution picture, 
starting from 2D Euclidean QG 
and 
walking on the causality road.%
\footnote{%{\dbltinyspace}%
The causality road is the road of making use of causality.
}
\begin{quote}
\vspace{-2pt}\normalsize
\begin{enumerate}
\renewcommand{\labelenumi}{\bf \arabic{enumi})}
\setcounter{enumi}{-1}
\item
{\bf Preparation:}
We here list the theories that describe 2D Euclidean QG 
and briefly explain their relationships.%
%2次元量子重力理論を表す理論をまとめる。
%\footnote{%{\dbltinyspace}%
%ここで挙げたそれぞれの理論の詳細は
%このハンドブックの他の章を見られたい。
%}
\begin{enumerate}
\renewcommand{\labelenumii}{\bf \alph{enumii})}
\item
There are several theories that describe 2D Euclidean QG, 
for example, %and the typical ones are 
Liouville Gravity, Matrix Model, Dynamical Triangularion (DT). 
%2次元量子重力を表す理論はいくつか存在し，
%代表的なものは，Liouville重力理論，行列模型，力学的単体分割である。
\item
Liouville Gravity and Matrix Model are considered to be equivalent, 
including the theories which have matter. 
%(The proof of its equivalence has not yet been completed.)
%Liouville重力理論と行列模型は物質場が存在する理論も含め等価
%と考えられている。（その等価性の証明はまだできていない。）
\item
DT and Matrix Model are almost equivalent, 
including the theories which have matter. 
Moreover, their relationship is mostly obvious. 
%力学的単体分割と行列模型は物質場が存在する理論も含めほぼ等価。
%しかも，その関係は自明な場合がほとんどである。
\item
DT can be expressed by reduced $W$ operators, 
including the theory which has matter. 
%力学的単体分割は物質場が存在する場合も含め
%reduced $W$ operator で表すことができる。
Matter is considered to be path-integrated out 
in the expression by reduced $W$ operators. 
%since there are no operators other than the reducing $W$ operator 
%even when the theory has matter. 
%Even when there exists matter, 
%there are no operators other than reduced $W$ operators, 
%so in the expression by reduced $W$ operator, 
%the matter is considered to be path-integrated out.
%この場合，reduced $W$ operator 以外の演算子は現れないので，
%reduced $W$ operator による表記では
%物質場は path-integrated out されていると考えられる。
\end{enumerate}
\item
{\bf Pure DT by non-critical SFT:}
We study the geometrical structure of space of pure DT 
using the Hamiltonian formalism 
where the geodesic distance is treated as time. 
%力学的単体分割によって作られた空間の幾何学的構造を，
%測地距離を時間のように扱うHamilton形式で理解する。
`pure' means 
that matter does not exist in the theory, 
and then 
the conformal dimension of matter is 
$c {\negdbltinyspace}={\negdbltinyspace} 0$. 
%the theory do not have matter. 
This Hamiltonian formalism is called ``non-critical SFT'', 
and is a formalism that allows 
1D universe to expand, shrink, disappear, 
and separate and merge with other 1D universe.
%これは``non-critical SFT'' と呼ばれ，
%1次元の宇宙を膨張・収縮したり，消滅したり，
%宇宙同士を分離・融合したりする形式である。
(See Fig.\ \ref{fig:DTpicture}.)
\label{item:SFTbyPureDT}
\item
{\bf Pure DT %DT without and with matter expressed 
by $W$ operators:}
We express 
the non-critical SFT of pure DT %obtained in {\bf \ref{item:SFTbyPureDT}} 
by $W$ operators. 
We also study DT which has matter, 
and then 
the conformal dimension of matter is 
$c {\negtrpltinyspace}\neq{\negtrpltinyspace} 0$. 
%{\bf \ref{item:SFTbyPureDT}.}\ で得られた 
%DT の non-critical SFT を$W$演算子で表現する。
We find that 
the time of DT with matter is not the geodesic distance 
and the time evolution by Hamiltonian is nonlocal.\,%
\cite{StringFieldThIKmatter,EuclidWalgAW}
\item
{\bf Pure CDT by non-critical SFT:}
We change the geodesic distance of 
the non-critical SFT of pure DT %obtained in {\bf \ref{item:SFTbyPureDT}} 
to causal time. 
(See Fig.\ \ref{fig:CDTpicture}.)
%{\bf \ref{item:SFTbyPureDT}.}\ で得られた
%pure DT の non-critical SFT の測地距離を因果時間に変更する。
This theory is called ``Causal Dynamical Triangulation (CDT)'',%
\footnote{%{\dbltinyspace}%
Only 2D CDT is treated in this {\Article}. 
\red{There is a lot of numerical work on higher-dimensional
CDT, see \cite{physrep} for a review.}
}
\red{and in our notation ``pure CDT'' 
since no matter is coupled to the geometry.} 
We do not treat CDT with matter, 
because the time of Hamiltonian expressed by $W$ operators 
is nonlocal in the case of DT with matter. 
%Replacing this with causal time may be a little unreasonable. 
%$c {\negtrpltinyspace}\neq{\negtrpltinyspace} 0$ non-critical SFT 
%この理論を ``causal dynamical triangulation (CDT)" というのだが，
%%【因果性が1番目の手掛かり】
%特に，この理論は，物質場のないDT，つまり，
%pure DT から得られた理論なので，pure CDT になる。
\label{item:SFTbyPureCDT}
\item
{\bf Pure CDT by Matrix Model:}
The purpose here is to 
clarify the relationship between pure CDT and a Matrix Model 
in order to confirm that 
the CDT theoretical structure is as rich as 
the  DT theoretical structure. 
%CDTの数学的構造がDTのそれと同じくらい豊かであることを確認するのである。
We also try to find the theory equivalent to 
Liouville Gravity of DT in pure CDT. 
%我々はDTのLiouville Gravity theoryに相当する理論を
%CDTについて探し出す。
%
\item
{\bf Pure CDT by $W$ operators:}
As in pure DT, 
we express 
the non-critical SFT of pure CDT %obtained in {\bf \ref{item:SFTbyPureCDT}} 
by $W$ operators. 
%pure DT から $W$ operator による表記を得たのと同様にして，
%{\bf \ref{item:SFTbyPureCDT}.}\ で得られた
%pure CDT の non-critical SFT を$W$演算子で表現する。
%【$W$演算子が2番目の手掛かり】
\label{item:WfromPureCDT}
\item
{\bf CDT with matter by $W$ operators:}
Given that 
pure DT is $c{\negtrpltinyspace}={\negtrpltinyspace}0$ 
non-critical string theory, 
and that pure CDT is constructed based on pure DT, 
pure CDT is considered to be 
$c{\negtrpltinyspace}={\negtrpltinyspace}0$ non-critical string theory
with causal time. 
%pure DT が 
%$c{\negtrpltinyspace}={\negtrpltinyspace}0$ non-critical string theory 
%であること，
%また，
%pure CDT が pure DT を模して構成されたことを踏まえると，
%pure CDT は時間軸を持つ
%$c{\negtrpltinyspace}={\negtrpltinyspace}0$ non-critical string theory
%である。
In order to obtain 
$c{\negtrpltinyspace}={\negtrpltinyspace}26$ 
critical string theory, 
one may add $26$ bosonic scalar fields, for example. 
However, simply adding $26$ bosonic scalar fields 
does not determine the Hamiltonian uniquely.
%この理論から
%$c{\negtrpltinyspace}={\negtrpltinyspace}26$ の CDT を得るには，
%たとえば，$26$個の bosonic scalar の振動モードを加えればよいのだが，
%$26$個の bosonic scalar を単純に加えるだけでは
%Hamiltonian の形が定まらず，理論が定義できない。
So, we here fix the Hamiltonian 
based on the idea of ``extremity'' mentioned above. 
%そこで，「際」という考えを踏まえて理論を選択する。
As a result, 
$W$ operators based on the Jordan algebra with octonions will be introduced. 
%その結果，
%Jordan代数を基礎とする $W$ operator を導入する。
%【臨界次元$26$とJordan代数が3番目の手掛かり】
\label{item:WfromMatterCDT}
\item
{\bf Basic properties of CDT with matter:}
We investigate how the universe is born and 
evolve in the CDT with matter 
%25D space of string theory occurs from the 2D spacetime theory 
obtained in {\bf \ref{item:WfromMatterCDT})}.
%弦理論の25次元空間が{\bf \ref{item:WfromMatterCDT}.}\ 
%で得られた2次元時空の理論からどのように発生するのか調べる。
\label{item:WwithJmodel}
\item
{\bf Some phenomenological predictions by CDT with matter:}
We obtain 
the modified Friedmann equation 
from CDT with matter obtained in {\bf \ref{item:WfromMatterCDT})}, 
and 
investigate 
the physical phenomena %and observation facts 
that follow from this equation. 
%modified Friedmann equation を求め，
%それから導かれる現象と観測事実を突き合わせる。
%
\item
{\bf The relationship between string theory and CDT with matter:}
In addition to the number ``$26$'' which is 
the conformal dimension of scalar fields introduced to pure CDT 
in {\bf \ref{item:WfromMatterCDT})}, 
we search for other clues that CDT with matter connects to string theory.%
%臨界次元が$26$となること以外に弦理論と繋がる手がかりを探す。
\footnote{%{\dbltinyspace}%
This approach is under study. 
}
\end{enumerate}
\vspace{-2pt}
\end{quote}
It should be emphasized that in the above strategy, 
``causality'' is the most important key. 
%以上の戦略の中で因果性が論理的な飛躍をするときの鍵になっていること
%に注目されたい。
The causality road is the central axis of this strategy. 
%因果の道がこの戦略の中軸になるのである。

\section{2D Euclidean Gravity}\label{sec:TwoDimGravity}
\label{sec:DT}

%\subsection{The aim to study 2D Euclidean gravity}

%2次元の量子重力理論の成功は目に見張るものがある。
%その中でも重要な成功をいくつか列挙すると，
%\begin{itemize}
%\item
%量子重力理論では計量の経路積分が必須の課題だが，
%Euclid計量の2次元時空だけ成功している事実は非常に重要である。
%\item
%2次元量子重力理論である Liouville Gravity と行列模型の等価性は
%非常に重要である。ただし，このときの2次元時空はEuclidean計量に限る。
%\item
%行列模型は連続極限ではDTと等価になる。
%\item
%DTは，次のサブセクションで見るように，
%non-critical SFT と等価になる。
%\end{itemize}
In this {\Chapter}, 
we derive the non-critical SFT which describes DT 
using the loop equation of Matrix Models, 
and express the non-critical SFT by $W$ operators. 
%この章では，
%DTの振幅を定義し，これが満足する loop equation を利用して
%non-critical SFT を導く。
Since explanations of Liouville Gravity and Matrix Models 
are omitted in this {\Article}, 
we refer to literature for information about these. 

\subsection{$c {\negdbltinyspace}={\negdbltinyspace} 0$ 
non-critical SFT (DT version)}
\label{sec:purDTbySFT}

In this {\Section} 
we will derive the %$c {\negdbltinyspace}={\negdbltinyspace} 0$ 
non-critical SFT of pure DT.\,\cite{StringFieldThIK,StringFieldThWata}

%{\bf ここに，Disk amplitude の仮定と propagator の仮定の説明をする}
\red{Other {\Article}s in the Handbook will explain details about DT, 
so in this {\Article}} 
%Since we omit the explanation about DT in this {\Article}, 
we simply assume 
the disk amplitude 
$F_1^{(0)}{\negdbltinyspace}(L{\tinyspace};\cc)\big|_{\GG=0} 
 {\negtinyspace}=
3 {\tinyspace}
( 1 {\negtrpltinyspace}+{\negtrpltinyspace} \sqrt{\cc} L )
{\dbltinyspace}
\E^{- \sqrt{\cc} L}
{\negtinyspace}/{\tinyspace}
(4 \sqrt{\pi g}{\trehalftinyspace} L^{5/2})
$
and 
the propagator 
$\Hkin{\negtinyspace}\big(\pder{L};\cc\big) 
 {\negdbltinyspace}={\negdbltinyspace}
0
$, 
both of which are necessary for the construction of 
the non-critical SFT of pure DT. 
$\GG {\negtrpltinyspace}={\negtrpltinyspace} 0$ means 
there are no handles in the 2D spacetime. 
The Laplace transformed disk amplitude of pure DT is{\dbltinyspace}%
\footnote{%{\dbltinyspace}%
$g$ is the string coupling constant introduced in 
the Hamiltonian \rf{pureGravity_Hamiltonian}, 
not the determinant of the metric $g_{\mu\nu}$. 
}
\cite{BIPZ}
\begin{equation}\label{DT_DiskAmpConcreteExpression}
\tilde{F}_1^{(0)}{\negdbltinyspace}(\xi;\cc)\Big|_{\GG=0}
\,=\,
%%\lambda(\xi)
%%{\tinyspace}+
\frac{1}{\sqrt{g}}
\Big( \xi - \frac{\sqrt{\cc}}{2} {\trpltinyspace}\Big)
\sqrt{\xi + \sqrt{\cc}}
%%\,,
%%\qquad
%%\lambda(\xi)
%%{\dbltinyspace}={\dbltinyspace}
%%C_1 {\tinyspace}\ep^{-3/2}
%%{\dbltinyspace}-{\dbltinyspace}
%%C_2 {\dbltinyspace}\ep^{-1/2} {\tinyspace} \xi
\,,
\end{equation}
%%%where 
%%%$\ep$ is the cut-off parameter and 
%%%$C_1$ and $C_2$ are constants which depend on the regularization. 
and the Laplace transformed propagator of pure DT is{\dbltinyspace}%
%\footnote{%{\dbltinyspace}%
%$F_1^{(0)}{\negdbltinyspace}(L{\tinyspace};\cc)\Big|_{\GG=0}
% {\negtinyspace}=
%3 {\tinyspace}
%( 1 {\negtrpltinyspace}+{\negtrpltinyspace} \sqrt{\cc} L )
%{\dbltinyspace}
%\E^{- \sqrt{\cc} L}
%{\negtinyspace}/{\tinyspace}
%(4 \sqrt{\pi g}{\trehalftinyspace} L^{5/2})
%$
%%,
%%$\bar{F}_1^{(0)}(L{\tinyspace};V)\Big|_{\GG=0}
%% {\negtrpltinyspace}={\negtinyspace}
%%3 \sqrt{L}
%%{\dbltinyspace}
%%\E^{- L^2 / (4 V)}
%%/{\tinyspace}
%%(16 \pi {\tinyspace} \sqrt{g}{\trpltinyspace} V^{5/2})
%%$
%and
%$\Hkin{\negtinyspace}\big(\pder{L};\cc\big)
% {\negdbltinyspace}={\negdbltinyspace}
%0
%$. 
%}
\cite{StringFieldThIK}
\begin{equation}\label{DT_PropagatorLaplaceTransf}
\tHkin{\negtinyspace}(\xi{\tinyspace};\cc)
\,=\,
0
\,.
\end{equation}
In 2D Euclidean gravity, 
the equation of motion of the metric vanishes 
because the action is proportional to 
$\int\dd^2 x {\tinyspace}
 \sqrt{\det g_{\mu\nu}(x)} {\dbltinyspace} R^{(2)}(x)$ 
which is constant according to the Gauss-Bonnet theorem. 
This fact leads to \rf{DT_PropagatorLaplaceTransf}. 

\begin{figure}[t]
\vspace{0mm}\hspace{6mm}
\begin{minipage}{.45\linewidth}\vspace{16mm}
  \begin{center}\hspace*{-26mm}
    \includegraphics[width=0.2\linewidth,angle=0]
                    {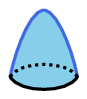}
  \end{center}
  \begin{picture}(200,0)
    \put( 35.5, 7.0){\mbox{\footnotesize$L$}}
  \end{picture}
  \vspace*{1mm}
  \caption[fig.3]{{disk topology}
  }
  \label{fig:DiskAmplitude}
\end{minipage}
\hspace{5mm}
\begin{minipage}{.45\linewidth}\vspace{16mm}
  \begin{center}\hspace*{-15mm}
    \includegraphics[width=0.2\linewidth,angle=0]
                    {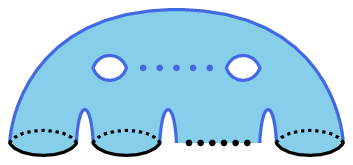}
  \end{center}
  \begin{picture}(200,0)
    \put( 26.0, 68.5){\mbox{\footnotesize$h$ handles}}
    \put(  5.7, 60.5){\mbox{\footnotesize$\overbrace{\hspace{70pt}}$}}
    \put(-17.5,  7.0){\mbox{\footnotesize$L_1$}}
    \put( 17.0,  7.0){\mbox{\footnotesize$L_2$}}
    \put( 91.5,  7.0){\mbox{\footnotesize$L_N$}}
    \put(-25.0,  5.0){\mbox{\footnotesize$\underbrace{\hspace{133pt}}$}}
    \put( 21.5, -9.5){\mbox{\footnotesize$N$ universes}}
  \end{picture}
  \vspace*{1.6mm}
  \caption[fig.4]{general topology}
  \label{fig:GeneralTopologyAmplitude}
\end{minipage}
\vspace{2mm}
%\vspace{-4mm}
\end{figure}

Let $\Psi^\dagger{\negdbltinyspace}(L)$ and $\Psi(L)$ be string operators
which creates and annihilates one 1D universe 
with length $L$, respectively.%
\footnote{%{\dbltinyspace}%
To be precise, 
$\Psi^\dagger{\negdbltinyspace}(L)$ creates a 1D universe with one marked point 
and 
$\Psi(L)$ annihilates a 1D universe with no marked point. 
The topology of a 1D universe is a circle $S^1$. 
}
The commutation relations of the string operators are 
\begin{equation}\label{CommutationRelationPsi}
\commutator{\Psi(L)}{\Psi^\dagger{\negdbltinyspace}(L')}
= \delta(L-L')
\,,
\qquad
\commutator{\Psi(L)}{\Psi(L')}
=
\commutator{\Psi^\dagger{\negdbltinyspace}(L)}
           {\Psi^\dagger{\negdbltinyspace}(L')}
=
0
\,.
\end{equation}
In the Hamiltonian formalism the disk amplitude,
$F_1^{(0)}{\negdbltinyspace}(L{\tinyspace};\cc)\big|_{\GG=0}$, i.e.\ 
the amplitude that one 1D universe annihilates into the vacuum, 
is  obtained by
\begin{equation}\label{DiskAmp}
F_1^{(0)}{\negdbltinyspace}(L{\tinyspace};\cc)\Big|_{\GG=0}
\,=\,
\lim_{\T \rightarrow \infty}
  \vac {\tinyspace} \Theta{\negtinyspace}(\T)\big|_{\GG=0} {\tinyspace}
    \Psi^\dagger{\negdbltinyspace}(L)
  \cuum
\,,
\end{equation}
where 
the vacuum state satisfies the condition, 
\begin{equation}\label{vacuumCondition_Psi}
\vacuumNorm = 1
\,,
\qquad
\vac \Psi^\dagger{\negdbltinyspace}(L) = 0
\,,
\qquad
\Psi(L) \cuum = 0
\,,
\end{equation}
and 
$\Theta{\negtinyspace}(\T)$ is the time{\tinyspace}-transfer operator, 
\begin{equation}\label{TransferOperator}
\Theta{\negtinyspace}(\T)
 \,\define\, 
\E^{-{\tinyspace} \T{\negtinyspace}\Hop}
\,.
\end{equation}
$\T$ is the proper time 
and
$\Hop$ is the Hamiltonian. 
%The Hamiltonian has several constants, for example, 
%the cosmological constant $\cc$, 
%the string coupling constant $g$, and so on. 
The amplitudes $F_N^{(h)}{\negdbltinyspace}(L_1,\ldots,L_N;\cc)$ 
which have general topologies 
are defined by
\begin{equation}\label{GeneralAmp}
\sum_{h=0}^\infty \GG^{{\tinyspace}h+N-1} {\tinyspace}
F_N^{(h)}{\negdbltinyspace}(L_1,\ldots,L_N;\cc)
\,=\,
\lim_{\T \rightarrow \infty}
  \vac {\tinyspace} \Theta{\negtinyspace}(\T) {\tinyspace}
    \Psi^\dagger{\negdbltinyspace}(L_1) \ldots \Psi^\dagger{\negdbltinyspace}(L_N)
  \cuum
\,.
\end{equation}
%with the transfer operator \rf{TransferOperator}, 
In Fig.\ \ref{fig:DiskAmplitude} and \ref{fig:GeneralTopologyAmplitude}, 
we show a typical configuration which contributes to 
the disk amplitude %defined in 
\rf{DiskAmp} 
and the general amplitudes %defined in 
\rf{GeneralAmp}, 
respectively. 
%In Fig.\ \ref{fig:GeneralTopologyAmplitude} we show a typical configuration
%which contributes to the general amplitude defined in \rf{GeneralAmp}.
%%%
%\vspace*{0mm}
%%%
The Hamiltonian $\Hop$ has the form
\begin{eqnarray}\label{pureGravity_Hamiltonian}
\Hop
&=&
-\,
\!\!\int_0^{{\tinyspace}\infty}\!\!\! \dd L
{\dbltinyspace}
\rho{\negtinyspace}(L{\tinyspace};\cc) \Psi(L)
%\hspace{170.0pt}\hspace{15pt}%\hspace{182.3pt}\hspace{15pt}
\hspace{133.2pt}\hspace{15pt}%\hspace{182.3pt}\hspace{15pt}
\raisebox{-3pt}[0ex][0ex]{%
\includegraphics[width=0.03\linewidth,angle=0]{disk.ps}}
\nonumber\\
&&
+\,%\Hkin
\!\!\int_0^{{\tinyspace}\infty}\!\!\! \dd L {\dbltinyspace}
\Psi^\dagger{\negdbltinyspace}(L)
%\Hkin{\negdbltinyspace}\Big(L,\pder{L};\cc\Big)
\Hkin{\negdbltinyspace}\Big(\pder{L};\cc\Big)
L {\tinyspace}
\Psi(L)
%\hspace{96.4pt}\hspace{4pt}%\hspace{108.6pt}\hspace{4pt}
%\hspace{82.5pt}\hspace{4pt}%
\hspace{80.0pt}\hspace{4pt}%
%\hspace{110.6pt}\hspace{4pt}%\hspace{122.8pt}\hspace{4pt}
\raisebox{-3pt}[0ex][0ex]{%
\includegraphics[width=0.08\linewidth,angle=0]{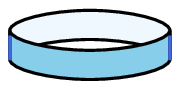}}
\nonumber\\
&&
-\,
g
\!\!\int_0^{{\tinyspace}\infty}\!\!\! \dd L_1
\!\!\int_0^{{\tinyspace}\infty}\!\!\! \dd L_2
{\dbltinyspace}
\Psi^\dagger{\negdbltinyspace}(L_1)
\Psi^\dagger{\negdbltinyspace}(L_2)
{\tinyspace}
(L_1{\negdbltinyspace}+{\negdbltinyspace}L_2)
\Psi(L_1{\negdbltinyspace}+{\negdbltinyspace}L_2)
%\hspace{14.3pt}
\hspace{14.8pt}
\raisebox{-3pt}[0ex][0ex]{%
\includegraphics[width=0.1\linewidth,angle=0]{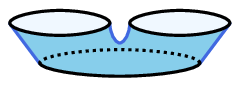}}
\nonumber\\
&&
-\,
g {\tinyspace} 
\GG
\!\!\int_0^{{\tinyspace}\infty}\!\!\! \dd L_1
\!\!\int_0^{{\tinyspace}\infty}\!\!\! \dd L_2 {\dbltinyspace}
\Psi^\dagger{\negdbltinyspace}(L_1{\negdbltinyspace}+{\negdbltinyspace}L_2)
{\tinyspace}
L_1
\Psi(L_1)
{\tinyspace}
L_2
\Psi(L_2)
\,,
\hspace{20.8pt}%{28pt}
\raisebox{-3pt}[0ex][0ex]{%
\includegraphics[width=0.1\linewidth,angle=0]{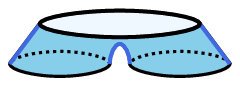}}
\end{eqnarray}
where
the constants $g$ and $\GG$ are introduced to count 
the vertices of string interaction and the number of handles of 2D space, 
respectively.%
\footnote{%{\dbltinyspace}%
$\GG$ has nothing to do with Newton constant. 
}
The last two terms in \rf{pureGravity_Hamiltonian} 
represent splitting and merging 1D universes and 
lead to the fractal structure of 2D space. 
Since the proper time $\T$ is the geodesic distance in the case of pure DT, 
the universe is never born from emptiness. 
The reason why the Hamiltonian \rf{pureGravity_Hamiltonian} satisfies 
``no big-bang condition", 
\begin{equation}\label{NoBigBangCondition}
\Hop {\dbltinyspace} \cuum  \,=\,  0
\,,
\end{equation}
comes from the fact that 
one can reach to any point in 2D Euclidean space. 
(See Fig.\ \ref{fig:DTpicture}.)
Note that 
the Hamiltonian \rf{pureGravity_Hamiltonian} has 
the time reversal symmetry 
which is the invariance under the transformation, 
\begin{equation}\label{TimeReversalTransf}
\Psi^\dagger{\negdbltinyspace}(L)
\ \leftrightarrow \ 
\GG L {\tinyspace} \Psi(L)
\,,
\end{equation}
if 
$\rho{\negtinyspace}(L{\tinyspace};\cc) {\negtrpltinyspace}={\negtrpltinyspace} 0$ 
and 
$\Hkin{\negdbltinyspace}\big(\pder{L};\cc\big)$ 
is an even function of $\pder{L}$. 
%Since $\Hkin$ is Hermitian, 
%the time evolution of the universe is invariant 
%under the time reversal. 
%%$\Hkin$ はエルミートで，その結果，宇宙の時間発展は，
%%時間反転に対して不変になる。

We here introduce the Laplace transformation defined by 
\begin{equation}\label{WaveFunLaplaceTransf}
\tilde\Psi(\eta)
\,=\,
\int_0^{{\tinyspace}\infty}\! \dd L {\dbltinyspace}
  \E^{- \eta L}{\tinyspace} \Psi(L)
\,,
\qquad\quad
\tilde\Psi^\dagger{\negdbltinyspace}(\xi)
\,=\,
\int_0^{{\tinyspace}\infty}\! \dd L {\dbltinyspace}
  \E^{- \xi L}{\tinyspace} \Psi^\dagger{\negdbltinyspace}(L)
\,.
\end{equation}
Then, the commutation relations \rf{CommutationRelationPsi} become 
\begin{equation}\label{CommutationRelationPsiLaplaceTransf}
\commutator{\tilde\Psi(\eta)}{\tilde\Psi^\dagger{\negdbltinyspace}(\xi)}
= \frac{1}{\xi+\eta}
\,,
\qquad
\commutator{\tilde\Psi(\eta)}{\tilde\Psi(\eta')}
=
\commutator{\tilde\Psi^\dagger{\negdbltinyspace}(\xi)}
           {\tilde\Psi^\dagger{\negdbltinyspace}(\xi')}
=
0
\,.
\end{equation}
The Laplace transformations of amplitudes \rf{GeneralAmp} become 
\begin{equation}\label{GeneralAmpLaplaceTransf}
\sum_{h=0}^\infty \GG^{{\tinyspace}h+N-1} {\tinyspace}
\tilde{F}_N^{(h)}{\negdbltinyspace}(\xi_1,\ldots,\xi_N;\cc)
\,=\,
\lim_{\T \rightarrow \infty}
  \vac {\tinyspace} \Theta{\negtinyspace}(\T) {\tinyspace}
    \tilde\Psi^\dagger{\negdbltinyspace}(\xi_1)
    \ldots
    \tilde\Psi^\dagger{\negdbltinyspace}(\xi_N)
  \cuum
\,,
\end{equation}
where 
the vacuum state satisfies the conditions, 
\begin{equation}\label{vacuumCondition_tildePsi}
\vac \tilde\Psi^\dagger{\negdbltinyspace}(\xi) = 0
\,,
\qquad
\tilde\Psi(\eta) \cuum = 0
\,.
\end{equation}
%and 
%$\Theta{\negtinyspace}(\T)$ is the time-transfer operator \rf{TransferOperator}. 
The Hamiltonian $\Hop$ in \rf{pureGravity_Hamiltonian} becomes 
\begin{eqnarray}\label{pureGravity_HamiltonianLaplaceTransf}
\Hop
&=&
\int_{-i\infty}^{{\tinyspace}i\infty}\!\!\! \dd \zeta
{\dbltinyspace}\bigg\{{\negqdrpltinyspace}%
-
\tilde\rho{\negtinyspace}(\zeta;\cc) \tilde\Psi(-\zeta)
{\dbltinyspace}+{\dbltinyspace}
\tilde\Psi^\dagger{\negdbltinyspace}(\zeta)
\tHkin{\negtinyspace}(\zeta{\tinyspace};\cc)
\pder{\zeta} {\tinyspace}
\tilde\Psi(-\zeta)
\nonumber\\
&&\phantom{%
\int_{-i\infty}^{{\tinyspace}i\infty}\!\!\! \dd \zeta
{\dbltinyspace}\bigg\{{\negqdrpltinyspace}%
}%
-
g
\big( \tilde\Psi^\dagger{\negdbltinyspace}(\zeta) \big)^{{\negtinyspace}2}
{\tinyspace}
\pder{\zeta} \tilde\Psi(-\zeta)
{\dbltinyspace}-{\dbltinyspace}
g {\tinyspace} \GG {\tinyspace}
\tilde\Psi^\dagger{\negdbltinyspace}(\zeta)
{\negtinyspace}
\Big( \pder{\zeta} \tilde\Psi(-\zeta) \Big)^{{\negtrpltinyspace}2}
{\dbltinyspace}\bigg\}
\,.
\end{eqnarray}
Since the operation 
$\lim_{\T \rightarrow \infty} \vac {\tinyspace} \Theta{\negtinyspace}(\T)$ 
gives a finite constant value if the volume of 2D space is finite, 
one obtains the following so-called Schwinger-Dyson equation 
from the disk amplitude, 
\begin{eqnarray}\label{DiskAmpSDeqLaplaceTransf}
0
&=&
\lim_{\T \rightarrow \infty}
  \pder{\T}
  \vac {\tinyspace} \Theta{\negtinyspace}(\T)\Big|_{\GG=0} {\tinyspace}
    \tilde\Psi^\dagger{\negdbltinyspace}(\xi)
  \cuum
=
-\!
\lim_{\T \rightarrow \infty}
  \vac {\tinyspace} \Theta{\negtinyspace}(\T)\Big|_{\GG=0} {\tinyspace}
    \commutator{\Hop}{\tilde\Psi^\dagger{\negdbltinyspace}(\xi){\negtinyspace}}
  \cuum
\nonumber\\
&=&
\tilde\rho{\negtinyspace}(\xi;\cc)
{\tinyspace}+{\tinyspace}
\pder{\xi} \Big\{
  \tilde{F}_1^{(0)}{\negdbltinyspace}(\xi;\cc)\Big|_{\GG=0}
  \tHkin{\negtinyspace}(\xi{\tinyspace};\cc)
  {\tinyspace}-{\tinyspace}
  g
  \Big(
  %\big(
    \tilde{F}_1^{(0)}{\negdbltinyspace}(\xi;\cc)\Big|_{\GG=0}
  %\big)^{{\negtinyspace}2}
  {\tinyspace}\Big)^{{\negtrpltinyspace}2}{\tinyspace}
\Big\}
\,.
\end{eqnarray}

By the way, note that 
the disk amplitude 
%$\tilde{F}_1^{(0)}{\negdbltinyspace}(\xi;\cc)\big|_{\GG=0}$ 
\rf{DT_DiskAmpConcreteExpression} 
satisfies the following loop equation, 
\begin{eqnarray}\label{DT_DiskAmpSDeqLaplaceTransf}
%g
%\Big({\negtinyspace}
  3 {\tinyspace} \xi^2
  {\dbltinyspace}-{\dbltinyspace}
  \frac{3 \cc}{4}
%\Big)
{\dbltinyspace}-{\dbltinyspace}
g{\tinyspace}
\pder{\xi}
  \Big(
  %\big(
    \tilde{F}_1^{(0)}{\negdbltinyspace}(\xi;\cc)\Big|_{\GG=0}
  %\big)^{{\negtinyspace}2}
  {\tinyspace}\Big)^{{\negtrpltinyspace}2}
\,=\,
0
\,.
\end{eqnarray}
Comparing the loop equation \rf{DT_DiskAmpSDeqLaplaceTransf} 
with the Schwinger-Dyson equation \rf{DiskAmpSDeqLaplaceTransf} 
together with 
$\tHkin {\negtrpltinyspace}={\negtrpltinyspace} 0$
\rf{DT_PropagatorLaplaceTransf}, 
one obtains 
\begin{equation}\label{DT_TadpoleLaplaceTransf}
\tilde\rho{\negtinyspace}(\xi;\cc)
\,=\,
%g
%\Big({\negtinyspace}
  3 {\tinyspace} \xi^2
  - \frac{3 \cc}{4}
%\Big)
\,.
%\qquad\quad
%\tHkin{\negtinyspace}(\xi{\tinyspace};\cc)
%\,=\,
%0
\end{equation}
%Therefore we have 
%\begin{equation}\label{DT_Tadpole}
%\rho{\negtinyspace}(L{\tinyspace};\cc)
%\,=\,
%%g
%%\Big(
%  3 {\tinyspace}\delta''(L)
%  - \frac{3 \cc}{4} {\tinyspace}\delta(L)
%%\Big)
%\,,
%%\qquad\quad
%%\Hkin{\negdbltinyspace}\Big(\pder{L};\cc\Big)
%%\,=\,
%%0
%\,.
%\end{equation}

Because of $\tHkin {\negtrpltinyspace}={\negtrpltinyspace} 0$
\rf{DT_PropagatorLaplaceTransf}, 
one can remove the parameter 
$g$ from \rf{pureGravity_HamiltonianLaplaceTransf} 
by the following rescaling. 
\begin{equation}\label{DT_Rescaling}
\Psi^\dagger{\negdbltinyspace}(L) \rightarrow 
\frac{\Psi^\dagger{\negdbltinyspace}(L)}{\sqrt{g}}
\,,
\qquad
\Psi(L) \rightarrow \sqrt{g}{\dbltinyspace} \Psi(L)
\,,
\qquad
T \rightarrow \frac{\T}{\sqrt{g}}
\,,
\qquad
\GG \rightarrow \frac{\GG}{g}
\,.
\end{equation}
So, from now on, we set 
$g {\negtrpltinyspace}={\negtrpltinyspace} 1$ 
without loss of generality. 
$\Hkin {\negtrpltinyspace}={\negtrpltinyspace} 0$ 
is linked to 
the existence of the fractal structure of 2D space. 
%フラクタル構造が存在することと
%$\Hkin {\negtrpltinyspace}={\negtrpltinyspace} 0$
%が結びついているのである。

%ところで，量子重力は計量の量子化である。
%通常の標準理論に代表される場の理論では，
%物質の生成消滅が記述され，時空はこれとは独立に存在する背景場であった。
%ところが，量子重力では，時空の幾何学的性質が量子化の対象なので，
%空間や時間の生成消滅が起こり得る理論である。
%時間の誕生とは何か？\ まずはこれを考えてみよう。
%一方，量子論は，たとえば，量子電磁気学のように，
%電磁場のような場を生成消滅させる理論です。
%それゆえ，量子重力理論は，
%時間と空間を生成消滅させる理論になると考えられます。

\subsection{The appearance of reduced $W$\hspace{1pt}algebra}
\label{sec:purDTbyWalgebra}

In this {\Section} 
we will express the non-critical SFT of pure DT 
obtained in previous {\Section} 
by reduced $W$ operators.\,\cite{EuclidWalgAW} 

%From 
%loop equation \rf{DT_DiskAmpSDeqLaplaceTransf}
%with 
%$g {\negtrpltinyspace}={\negtrpltinyspace} 1$, 
%one obtains the disk amplitude 
%%Here we use the disk amplitude obtained by Matrix Model. 
%\rf{DT_DiskAmpConcreteExpression}. 
Taylor expansion of the disk amplitude \rf{DT_DiskAmpConcreteExpression} 
around $\xi{\negtrpltinyspace}={\negtinyspace}\infty$ 
gives{\dbltinyspace}%
%最初の2項は\rf{Omega1DT}
\footnote{\label{footnote:DTdiskamp}%{\dbltinyspace}%
Note that 
the first two terms do not contribute to 
the disk amplitude with nonzero volume 
[\,$V {\negqdrpltinyspace}\neq{\negtrpltinyspace} 0$\,]
because these terms are proportional to $\delta(V)$ and $\delta'(V)$ 
after taking the inverse Laplace transformation with respect to $\mu$. 
$V$ is 2D volume. 
}
\begin{equation}\label{DT_DiskAmpConcreteExpressionExpansion}
\tilde{F}_1^{(0)}{\negdbltinyspace}(\xi;\cc)\Big|_{\GG=0}
\,=\,
%\lambda(\xi)
%\,+\,
\xi^{{\tinyspace}3/2}
\,-\,
\frac{3{\halftinyspace}\cc}{8} {\tinyspace} \xi^{{\tinyspace}-1/2}
+
\sum_{\ell=1}^\infty
  \xi^{{\tinyspace}-1/2 - \ell}{\tinyspace} f_\ell
\,.
\end{equation}
%$\Omega_1{\negtrehalftinyspace}(\xi)$ is the first several terms of disk amplitude 
%\rf{DT_DiskAmpConcreteExpression}. 
%$f_n$ [\,$n{\negtrpltinyspace}={\negtrpltinyspace}1$, $2$, \ldots]
Then, from the viewpoint of mode expansion, 
one can expect 
\begin{equation}\label{DT_CreationAnnihilationOperatorModeExpansion}
%\begin{eqnarray}
%&&
\tilde\Psi^\dagger{\negdbltinyspace}(\xi)
\,=\,
\Omega_1{\negtrehalftinyspace}(\xi)
%\mbox{(polynomial of $\xi$)}
%+
%\xi^{3/2}
%-
%\frac{3{\halftinyspace}\cc}{8}{\tinyspace}
%\xi^{-1/2}
+
\sum_{\ell=1}^\infty
  \xi^{{\tinyspace}-\ell/2 - 1}{\tinyspace} \phi^\dagger_\ell
\,,
\qquad\quad
%\label{pureDT_CreationOperatorModeExpansion}
%\\
%&&
\tilde\Psi(-\eta)
\,=\,
\sum_{\ell=1}^\infty
  \eta^{{\tinyspace}\ell/2}{\tinyspace} \phi_\ell
%\,,
%\label{pureDT_AnnihilationOperatorModeExpansion}
%\end{eqnarray}
\end{equation}
with 
\begin{equation}\label{Omega1DT}
\Omega_1{\negtrehalftinyspace}(\xi)
\,=\,
%\lambda(\xi)
%\,+\,
\xi^{{\tinyspace}3/2}
\phi^\dagger_{-5}
\,+\,
\xi^{{\tinyspace}-1/2}
\phi^\dagger_{-1}
\,.
\end{equation}
In \rf{DT_CreationAnnihilationOperatorModeExpansion}, 
the even modes of $\phi^\dagger_\ell$ and $\phi_\ell$ are introduced 
in order to obtain 
the commutation relation \rf{CommutationRelationPsiLaplaceTransf}. 
In other words, 
under this mode expansion, 
the commutation relation \rf{CommutationRelationPsiLaplaceTransf} becomes 
\begin{equation}\label{CommutationRelationPhi}
\commutator{\phi_m}{\phi^\dagger_n}
= \delta_{{\tinyspace}m,n}
\,,
\qquad
\commutator{\phi_m}{\phi_n}
=
\commutator{\phi^\dagger_m}{\phi^\dagger_n}
=
0
\,,
\qquad
\mbox{[\,$m$, $n {\negdbltinyspace}\in{\negdbltinyspace} \dbl{N}$\,]}
\,.
\end{equation}
Since $\phi_{{\tinyspace}5}$ and $\phi_1$ do not exist in this theory, 
$\phi^\dagger_{-5}$ and $\phi^\dagger_{-1}$ are not 
quantum numbers. 
To prevent 
\rf{DT_DiskAmpConcreteExpressionExpansion} and 
\rf{DT_CreationAnnihilationOperatorModeExpansion}%
-%$\cdot$%
\rf{Omega1DT} from contradiction, 
we set 
\begin{equation}\label{Omega1DTvalue}
\phi^\dagger_{-5}
=
1
\,,
\qquad
\phi^\dagger_{-1}
=
-{\dbltinyspace}
\frac{3{\halftinyspace}\cc}{8} 
\,.
\end{equation}

Using \rf{DT_CreationAnnihilationOperatorModeExpansion} with 
\rf{Omega1DT}-\rf{Omega1DTvalue}, 
the Hamiltonian $\Hop$ in \rf{pureGravity_HamiltonianLaplaceTransf} 
with 
\rf{DT_PropagatorLaplaceTransf} and \rf{DT_TadpoleLaplaceTransf} becomes 
\begin{eqnarray}\label{pureDT_HamiltonianModeExpansion}
\Hop &=&
   -\, \frac{9\cc^2}{64}{\tinyspace} \phi_{{\tinyspace}2}
 \,-\, \frac{\GG}{4}{\tinyspace} \phi_{{\halftinyspace}4}
 \,+\, \frac{3{\halftinyspace}\GG \cc}{8}{\tinyspace}
       \phi_1 \phi_{{\tinyspace}2}
 \,-\, %\GG^2{\negtinyspace}
       \frac{\GG^2}{4}
       \phi_1 \phi_1 \phi_{{\tinyspace}2}
\nonumber\\&&
 -\> \sum_{\ell=1}^\infty \phi_{\ell+1}^\dagger {\tinyspace} \ell \phi_\ell
 \,+\, \frac{3{\halftinyspace}\cc}{8}
     \sum_{\ell=4}^\infty \phi_{\ell-3}^\dagger {\tinyspace} \ell \phi_\ell
\nonumber\\&&
 -\> \frac{1}{2}
  \sum_{\ell=6}^\infty {\tinyspace} \sum_{n=1}^{\ell-5}
  \phi_n^\dagger \phi_{\ell-n-4}^\dagger {\tinyspace}
  \ell \phi_\ell
\,-\,%\nonumber\\&&
 %-\>
  \frac{\GG}{4}
  \sum_{\ell=1}^\infty {\tinyspace} \sum_{n=\max(5-\ell,1)}^\infty\!\!
  \phi_{n+\ell-4}^\dagger {\tinyspace}
  n {\halftinyspace} \phi_n {\tinyspace} \ell \phi_\ell
\,.
\end{eqnarray}
The generating function which leads to all amplitudes is 
\begin{equation}\label{GeneratingFunModeExpansion}
Z_f[j]  \ = \
\lim_{\T\rightarrow\infty}
  \vac {\tinyspace} \Theta{\negtinyspace}(\T) %{\tinyspace}
    \exp{\negtrpltinyspace}\bigg(
      \sum_{\ell=1}^\infty \phi^\dagger_\ell j_\ell
    {\negtinyspace}\bigg)
  {\negtinyspace}\cuum
\,,
\end{equation}
where 
the vacuum state satisfies the condition, 
\begin{equation}\label{vacuumCondition_phi}
\vac \phi^\dagger_\ell = 0
\,,
\qquad
\phi_\ell \cuum = 0
\,,
\qquad
\mbox{[\,$\ell {\negdbltinyspace}\in{\negdbltinyspace} \dbl{N}$\,]}
\,.
\end{equation}
%and 
%$\Theta{\negtinyspace}(\T)$ is the time-transfer operator \rf{TransferOperator}. 
Then, 
the amplitudes can be obtained by differentiation after $\ln Z_f[j]$:
\begin{eqnarray}\label{GeneralAmpModeExpansion2}
\sum_{h=0}^\infty \GG^{{\tinyspace}h+N-1}
f_N^{(h)}{\negdbltinyspace}(\ell_1,\ldots,\ell_N)
&=&
\lim_{\T\rightarrow\infty}
  \vac {\tinyspace} \Theta{\negtinyspace}(\T) {\tinyspace}
    \phi^\dagger_{\ell_1} \ldots \phi^\dagger_{\ell_N}
  \cuum
\nonumber\\&=&
\frac{\partial^{N}}
     {\partial j_{\ell_1} \, \cdots \, \partial j_{\ell_N} } 
\ln Z_f[j] \Bigg|_{j=0}
\,.
\end{eqnarray}
The direct relationship between two amplitudes 
$\tilde{F}_N^{(h)}$ in \rf{GeneralAmpLaplaceTransf} 
and 
$f_N^{(h)}$ in \rf{GeneralAmpModeExpansion2} 
is
\begin{eqnarray}\label{pureDT_GeneralAmpRelationship}
  \tilde{F}_N^{(h)}{\negdbltinyspace}(\xi_1,\ldots,\xi_N;\cc)
  &=&
  \big({\tinyspace}
  %\big\{{\negtinyspace}
    %\big(
      %\lambda(\xi_1) +
      \Omega_1{\negtrehalftinyspace}(\xi_1)
    %\big)
    {\tinyspace}\delta_{{\tinyspace}N,1}
    {\dbltinyspace}+{\dbltinyspace}
    \Omega_{{\tinyspace}2}{\neghalftinyspace}(\xi_1,\xi_2)
    {\tinyspace}\delta_{{\tinyspace}N,2}
  %\big\}
  {\tinyspace}\big)
  {\tinyspace}\delta_{{\tinyspace}h,0}
\nonumber\\
  &&+\,
  \sum_{\ell_i{\tinyspace}={\tinyspace}1,3,5,...}\!\!\!
    \xi_1^{{\tinyspace}-\ell_1/2-1} \cdots \xi_N^{{\tinyspace}-\ell_N/2-1}
    f_N^{(h)}{\negdbltinyspace}(\ell_1,\ldots,\ell_N;\cc)
\,,
\quad
\end{eqnarray}
where 
$\Omega_1{\negtrehalftinyspace}(\xi)$ is given by 
\rf{Omega1DT} and \rf{Omega1DTvalue}, 
and 
$\Omega_{{\tinyspace}2}{\neghalftinyspace}(\xi,\xi')$ is 
\begin{eqnarray}
&&
%\lambda(\xi)
%\,=\,
%C_1 {\tinyspace}\ep^{-3/2}
%{\tinyspace}-{\tinyspace}
%C_2 {\dbltinyspace}\ep^{-1/2} {\tinyspace} \xi
%\,,
%\qquad\quad
%\Omega_1{\negtrehalftinyspace}(\xi)
%\,=\,
%C_1 {\tinyspace}\ep^{-3/2}
%{\dbltinyspace}-{\dbltinyspace}
%C_2 {\dbltinyspace}\ep^{-1/2} {\tinyspace} \xi
%\,+\,
%\xi^{3/2} \,-\, \frac{3{\halftinyspace}\cc}{8} {\tinyspace} \xi^{-1/2}
%\,,
%\label{Omega1DT}
%\\
%&&
\Omega_{{\tinyspace}2}{\neghalftinyspace}(\xi,\xi')
\,=\,
\frac{1}{2 \sqrt{\xi{\tinyspace}\xi'}{\dbltinyspace}
         (\sqrt{\xi}+{\negtinyspace}\sqrt{\xi'}{\dbltinyspace})^2}
\,.
\label{Omega2}
\end{eqnarray}
%Both $\Omega_1$ and $\Omega_{{\tinyspace}2}$ contribute only on spaces with zero volume, 
%so these terms are not universal. 
Not only 
$\Omega_1{\negtrehalftinyspace}(\xi)$ and 
$\Omega_{{\tinyspace}2}{\neghalftinyspace}(\xi,\xi')$ 
but also 
$f_N^{(h)}{\negdbltinyspace}(\ell_1,\ldots,\ell_N;\cc)$ 
which has even integer $\ell$ does not have a finite volume, 
and one has 
\begin{equation}\label{purDT_VanishingEvenMode}
\pder{j_{2n}} Z_f[j]  \,=\,  0
\,,
\qquad
\mbox{[\,$n {\negdbltinyspace}\in{\negdbltinyspace} \dbl{N}$\,]}
\,.
\end{equation}

We now introduce the star operation defined by 
\begin{equation}\label{StarOpDefModeExpansion}
A_N^\star \ldots {\tinyspace} 
A_2^\star {\tinyspace} 
A_1^\star {\tinyspace} Z_f[j]
\,=\,
\lim_{\T\rightarrow\infty}
  \vac {\tinyspace} \Theta{\negtinyspace}(\T) {\tinyspace}
    A_1 A_2 \ldots A_N
    \exp{\negtrpltinyspace}\bigg(
      \sum_{\ell=1}^\infty \phi^\dagger_\ell j_\ell
    {\negtinyspace}\bigg)
  {\negtinyspace}\cuum
\,.
\end{equation}
Then, the star operation of string modes is 
\begin{equation}\label{StarOpPsiModeExpansion}
  \bigl( \phi^\dagger_\ell \bigr)^\star \,=\, \pder{j_\ell}
\,,
\qquad
  \bigl( \phi_\ell \bigr)^\star \,=\, j_\ell
\,,
\qquad
\mbox{[\,$\ell {\negdbltinyspace}\in{\negdbltinyspace} \dbl{N}$\,]}
\,.
\end{equation}
The star operation \rf{StarOpDefModeExpansion} applied to the Hamiltonian 
\rf{pureDT_HamiltonianModeExpansion} leads to 
\begin{eqnarray}\label{pureDT_HamiltonianModeExpansionStar}
 \Hopstar &=&
   -\, \frac{9\cc^2}{64}{\tinyspace} j_2
 \,-\, \frac{\GG}{4}{\tinyspace} j_4
 \,+\, \frac{3{\halftinyspace}\GG \cc}{8}{\tinyspace} j_1 j_2
 \,-\, %\GG^2{\negtinyspace}
       \frac{\GG^2}{4}{\tinyspace}
       j_1 j_1 j_2
\nonumber\\&&
 -\, \sum_{\ell=1}^\infty \ell j_\ell {\dbltinyspace} \pder{j_{\ell+1}}
 \,+\, \frac{3{\halftinyspace}\cc}{8}
     \sum_{\ell=4}^\infty \ell j_\ell {\dbltinyspace} \pder{j_{\ell-3}}
\nonumber\\&&
 -\> \half
  \sum_{\ell=6}^\infty {\tinyspace} \sum_{n=1}^{\ell-5}
  \ell j_\ell
  {\dbltinyspace} \pder{j_{n}}
  {\dbltinyspace} \pder{j_{\ell-n-4}}
\,-\,%\nonumber\\&&
 %-\>
  \frac{\GG}{4}
  \sum_{\ell=1}^\infty {\tinyspace} \sum_{n=\max(5-\ell,1)}^\infty\!\!
  n j_n {\tinyspace} \ell j_\ell
  {\dbltinyspace} \pder{j_{n+\ell-4}}
\,.
\quad
\end{eqnarray}
Corresponding to the no big-bang condition \rf{NoBigBangCondition} 
we have
\begin{equation}\label{NoBigBangConditionCurrent}
\Hopstar {\tinyspace} Z_f[j]  \,=\, 0
\,.
\end{equation}

We here introduce{\dbltinyspace}%
\footnote{%{\dbltinyspace}%
The constant 
$p {\negdbltinyspace}>{\negdbltinyspace} 0$ 
is introduced for later convenience. 
}
\begin{eqnarray}\label{DefAlphaOperatorPureDT}
\a_n \,\define\,
\left\{
\begin{array}{cl}
\displaystyle
%(\ope{a}_n^\dagger)^\star
%%{\dbltinyspace}={\dbltinyspace}
\sqrt{\frac{p}{\GG}} {\dbltinyspace} \pder{j_n}
& \hbox{[\,$n \!>\! 0$\,]}
\rule[-0pt]{0pt}{0pt}\\%\rule[-2pt]{0pt}{10pt}\\
\displaystyle
%(\ope{p})^\star
%%{\dbltinyspace}={\dbltinyspace}
\nu
& \hbox{[\,$n \!=\! 0$\,]}
\rule[-8pt]{0pt}{24pt}\\
\displaystyle
%-{\dbltinyspace} n {\tinyspace} (\ope{a}_{-n})^\star
%{\dbltinyspace}={\dbltinyspace}
-{\tinyspace} n {\tinyspace}
 \bigg(
   \lambda_{-n} + \sqrt{\frac{\GG}{p}} {\trpltinyspace} j_{-n}
 \bigg)
& \hbox{[\,$n \!<\! 0$\,]}
\end{array}
\right.
\,,
\qquad
\mbox{[\,$n {\negdbltinyspace}\in{\negdbltinyspace} \dbl{Z}$\,]}
%\,,
\end{eqnarray}
with $p {\negdbltinyspace}={\negdbltinyspace} 2$ and 
\begin{eqnarray}\label{CoherentEigenValuesPureDT}
\nu = 0
\,,
\quad
\lambda_1 = -\,\frac{3 {\halftinyspace} \cc}{4 {\sqrt{2{\tinyspace}\GG}}}
\,,
\quad
\lambda_{{\tinyspace}5} = \frac{2}{5 {\sqrt{2{\tinyspace}\GG}}}
\,,
\quad
\lambda_{{\tinyspace}n} = 0
\,,
\quad\mbox{[\,$n {\negdbltinyspace}\in{\negdbltinyspace} \dbl{N}$, 
              $n {\negtrpltinyspace}\neq{\negtrpltinyspace} 1$,\,$5$\,]}
\,,
\end{eqnarray}
then we find that 
the operators $\a_n$ satisfy the commutation relations, 
\begin{equation}\label{CommutationRelationAlpha}
\commutator{\a_m}{\a_n} \,=\, m {\tinyspace} \delta_{{\tinyspace}m+n,0}
\,,
\qquad
\mbox{[\,$m$, $n {\negdbltinyspace}\in{\negdbltinyspace} \dbl{Z}$\,]}
\,,
\end{equation}
and the Hamiltonian \rf{pureDT_HamiltonianModeExpansionStar} 
is rewritten as 
\begin{equation}\label{pureDT_HamiltonianModeExpansionStarW}
\Hopstar  \,=\,
-\,\sqrt{2{\tinyspace}\GG} \,{\dbltinyspace}
 \overline{W}^{{\dbltinyspace}(3)}_{{\negqdrpltinyspace}-2}
\,+\, Y
\,,
\end{equation}
where 
$\overline{W}^{{\dbltinyspace}(3)}_{{\negqdrpltinyspace}n}$
% [\,$n {\negdbltinyspace}\in{\negdbltinyspace} \dbl{Z}$\,] 
is a two-reduced $W^{(3)}_n$ operator and 
is defined by 
\begin{equation}
\overline{W}^{{\dbltinyspace}(3)}_{{\negqdrpltinyspace}n}
\,\define\,
%\oq (W_{2n} + \oq \tJ_{2n} ) =
\quarter \biggl( \, \onethird{\negtrpltinyspace}
\sum_{k+l+m
      {\tinyspace}={\tinyspace}2n}
{\negtrpltinyspace}{\negtrpltinyspace}{\negtrpltinyspace}
 : \a_k\a_l\a_m : 
\,+\,{\trpltinyspace}
\quarter{\tinyspace} \a_{2n} \, \biggr)
\,,
\qquad
\mbox{[\,$n$, $k$, $l$, $m {\negdbltinyspace}\in{\negdbltinyspace} \dbl{Z}$\,]}
\,,
\label{reducedW3operator}
\end{equation}
and 
\begin{equation}\label{purDT_Y}
Y
\,\define\,
\frac{1}{\sqrt{2{\tinyspace}\GG}} \Big({\dbltinyspace}
  %\frac{1}{2}{\dbltinyspace}
  {\negdbltinyspace}
  \a_6
  %\overline{W}^{{\dbltinyspace}(1)}_{\!3}
  {\dbltinyspace}-{\dbltinyspace}
  \frac{3 \mu}{4}{\dbltinyspace}
  \a_2
  %\overline{W}^{{\dbltinyspace}(1)}_{\!1}
{\negtinyspace}\Big)
\,.
\end{equation}
$Y$ is the sum of terms consisting of even operators 
$\a_{2n}$\,[\,$n {\negtrpltinyspace}\in{\negtrpltinyspace} \dbl{N}$\,] 
and essentially vanishes because 
$Y Z_f[j] {\negdbltinyspace}={\negdbltinyspace} 0$ 
obtained by \rf{purDT_VanishingEvenMode}, i.e.\ 
\begin{equation}\label{purDT_VanishingEvenMode2}
\a_{2n} {\tinyspace} Z_f[j]  \,=\,  0
\,,
\qquad
\mbox{[\,$n {\negdbltinyspace}\in{\negdbltinyspace} \dbl{N}$\,]}
\,.
\end{equation}
This fact 
\red{implies that we can drop the $Y$ term in 
$\Hopstar$ given by \rf{pureDT_HamiltonianModeExpansionStarW}.} 
%leads us that 
%$\Hopstar$ \rf{pureDT_HamiltonianModeExpansionStarW} without $Y$ 
%gives the same amplitudes. 
%$\Hopstar {\negdbltinyspace}={\negdbltinyspace}
% -2 \sqrt{G} \,{\dbltinyspace} \overline{W}^{{\dbltinyspace}(3)}_{\!-2}$
%which is 
%$\Hopstar$ \rf{pureDT_HamiltonianModeExpansionStarW} without $Y$. 

The above was 
the non-critical SFT of pure DT, 
that is, 
the $c {\negdbltinyspace}={\negdbltinyspace} 0$ non-critical SFT. 
%以上は
%$c {\negdbltinyspace}={\negdbltinyspace} 0$ non-critical SFT of DT, 
%つまり，pure DT であった。
At the end of this {\Section}, 
we give the final results of 
the $c {\negtrpltinyspace}\neq{\negtrpltinyspace} 0$ non-critical SFT of DT 
for the so-called $(p,q)$ models. 
%このサブセクションの最後に，
%$(p,q)$ model のときの
%$c {\negtrpltinyspace}\neq{\negtrpltinyspace} 0$ non-critical SFT of DT
%の最終結果をここにまとめよう。
This is DT with matter fields 
and its Hamiltonian is %defined by 
%これは物質場を持つ DT である。
\begin{equation}\label{pqHgenModeW}
\Hopstar \,\define\,
  - \, %p^{(p-1)}
    ({\tinyspace}p{\dbltinyspace}\GG)^{(p-1)/2} X \,+\, Y
\,,
\end{equation}
where 
\begin{eqnarray}
 X
 &\define&
 -\oint \! \frac{\dd z}{2 \pi i} \oint \! \frac{\dd s}{2 \pi i}\,
 s^{-p-2}{\qdrpltinyspace}
 \Norderingbig{\exp{\negtinyspace}
   \big\{{\negqdrpltinyspace}-{\negdbltinyspace}\bW(z,s) \big\}}
\nonumber\\
 &=&
 \bW^{(p+1)}_{{\negtrpltinyspace}-p}
 \,-\, \frac{1}{2}
       \sum_{k {\tinyspace}={\tinyspace} 2}^{p{\tinyspace}-1}
       \sum_{n {\tinyspace}\in{\tinyspace} \dblsmall{Z}}{\dbltinyspace}
       \Norderingbig{\bW^{(k)}_{{\negtrpltinyspace}-n} \bW^{(p-k+1)}_{{\negtrpltinyspace}n-p}}
\nonumber\\
 &&
 \,+\, \frac{1}{3!}
       \sum_{k {\dbltinyspace}\ge{\dbltinyspace} 2, ~
             l {\dbltinyspace}\ge{\dbltinyspace} 2
             \atop
             k+l+1 {\dbltinyspace}\le{\dbltinyspace} p}
       \sum_{n,m {\dbltinyspace}\in{\dbltinyspace} \dblsmall{Z}}
       \Norderingbig{\bW^{(k)}_{{\negtrpltinyspace}-n} \bW^{(l)}_{{\negtrpltinyspace}-m}
                     \bW^{(p-k-l+1)}_{{\negtrpltinyspace}n+m-p}}
 \,+\, \ldots\ldots
\label{Xoperator}
\end{eqnarray}
with
\begin{equation}\label{reducedWop}
  \bW{\negdbltinyspace}(z,s) \,\define\,
  \sum_{k=2}^\infty \bW^{(k)}{\negtinyspace}(z) {\dbltinyspace} s^k
\,,
\qquad
  \bW^{(k)}{\negtinyspace}(z) \,\define\,
  \sum_{n \in \dblsmall{Z}}
    \bW^{(k)}_{{\negtrpltinyspace}n} z^{-n-k}
\,.
\end{equation}
$\bW^{(k)}_{{\negtrpltinyspace}n}$ 
[\,$n {\negdbltinyspace}\in{\negdbltinyspace} \dbl{Z}$\,] 
is a $p${\dbltinyspace}-reduced $W^{(k)}_n$ operator. 
$\Norderingbig{\phantom{}}$\ is the normal ordering for 
$\bW^{(k)}_{{\negtrpltinyspace}n}$.
$Y$ is the polynomial of 
$\a_{pn}$\,[\,$n {\negdbltinyspace}\in{\negdbltinyspace} \dbl{N}$\,], 
and their coefficients are defined so as to satisfy 
the no big-bang condition \rf{NoBigBangConditionCurrent}. 
However, it should be noted that unlike pure DT, 
the Hamiltonian time of 
the $c {\negtrpltinyspace}\neq{\negtrpltinyspace} 0$ 
non-critical SFT is {\it not} the geodesic distance 
and the time evolution is nonlocal. 
%but that does not matter in terms of defining the theory. 
%$c {\negdbltinyspace}={\negdbltinyspace} 0$ 
%のときの pure DT とは異なり，
%この場合の$T$は測地距離にはならないが，
%理論を定義するという意味では問題はない。

%ここで，

Note that matter fields vanish 
in \rf{pqHgenModeW}-\rf{reducedWop} 
and only $\a_n$, the oscillation mode of space, remains. 
%\rf{pqHgenModeW}\,$\sim$\,\rf{reducedWop}では
%物質場が消えて空間の振動モードである$\a_n$のみが残ること
%に注目されたい。
In the case of the non-critical SFT of DT expressed by $W$ operators, 
matter fields are path-integrated out and 
\red{only appear indirectly.} 
%disappear superficially. 
%$W$演算子による表記の場合，
%物質場は経路積分されて消えてしまうのである。
This property is a hint when we introduce matter fields 
into the non-critical SFT of pure CDT.
%この性質は pure CDT に物質場を導入するときのヒントになる。

\section{2D Causal Gravity}\label{sec:TwoDimCausalGravity}
\label{sec:CDT}

The CDT introduced in this {\Chapter} is a theory 
created by replacing the geodesic distance of DT with 
causal time, as defined above. 
%この章で紹介する CDT は，
%DT の測地距離を因果時間に置き換えて作られた理論である。
In {\Section} \ref{sec:purDTbySFT} 
we derived the non-critical SFT of pure DT. 
%\ref{sec:purDTbySFT}節では DT が non-critical SFT によって
%表されることを説明した。
Conversely, 
in {\Section} \ref{sec:purCDTbySFT} 
we define %the non-critical SFT of 
pure CDT 
by replacing the geodesic distance with the causal time 
in the non-critical SFT of pure DT.
%\ref{sec:purCDTbySFT}節では，逆に，
%non-critical SFT によって表された DT の測地距離を因果時間に置き換えることで，
%CDT を定義する。
\red{A priori it is not clear that 
such a simple replacement will lead to an interesting theory. 
However, it was found that 
the theory obtained by this replacement (CDT) 
can be described by a Matrix Model} 
and a continuum theory corresponding to Lioville Gravity, 
and then 
%Of course, one cannot say that 
%the theory with excellent properties is by a simple replacement operation., 
%but
%it was found that the obtained CDT 
%can be described by a Matrix Model, 
%and 
%has a continuum theory corresponding to Lioville Gravity, 
%and 
has 
a mathematically rich structure similar to that of DT.\,%
%もちろん単なる置き換えだけでは
%よい性質の理論が作られたとは言えないが，
%得られた CDT は，
%行列模型で表すことが可能で，
%Lioville Gravityに相当する連続理論が存在し，
%DTと同じくらいに数学的に豊かな構造を持つことがわかった。
\cite{CDTandMatrixModelALWZZ,CDTandHoravaGravity}
%これは Liouville Gravity に相当する理論で，
%2D Ho\v{r}ava-Lifshitz Gravity の一種になっていた。
However, 
since this {\Article} aims at constructing high-dimensional QG, 
we omit the explanation of the relationship of CDT 
with Matrix Model and the continuum theory. 
%しかし，この{\Article}は高次元量子重力の構築を目的とするため，
%CDTの行列模型や連続理論との関係の説明は省略する。
%Ho\v{r}ava-Lifshitz Gravity
%pure CDTの遷移行列が$W$演算子で表されることを理解することにある。
This {\Chapter} is based on refs.\ 
\cite{CausalityStringFieldThALWZZ,CausalityWalg2dimAW,CausalityWJalgAW,CausalityWJalgAWfullversion}.

\subsection{$c {\negdbltinyspace}={\negdbltinyspace} 0$ 
non-critical causal SFT (CDT version)}
\label{sec:purCDTbySFT}

In this {\Section} 
we will derive the %$c {\negdbltinyspace}={\negdbltinyspace} 0$ 
non-critical SFT of pure CDT.\,%
\cite{CausalityStringFieldThALWZZ}

%{\bf ここに，Disk amplitude の仮定と propagator の仮定の説明をする}
\red{Since we omit the details about the derivation of CDT} 
in this {\Article}, 
%Since we omit the explanation about CDT 
we simply assume 
the disk amplitude 
$F_1^{(0)}{\negdbltinyspace}(L{\tinyspace};\cc)\big|_{g=0} 
 {\negtinyspace}=
\E^{- \sqrt{\cc} L}
$
and 
the propagator 
$\Hkin{\negtinyspace}\big(\pder{L};\cc\big) 
 {\negdbltinyspace}={\negdbltinyspace}
- \pdder{L} + \mu
$, 
both of which are necessary in order to construct 
the non-critical SFT of pure CDT. 
$g {\negtrpltinyspace}={\negtrpltinyspace} 0$ means 
there are no branches under the time evolution. 
The Laplace transformed disk amplitude with no branches of pure CDT is\,%
\cite{CDTkin}
\begin{equation}\label{CDT_DiskAmpConcreteExpression}
\tilde{F}_1^{(0)}{\negdbltinyspace}(\xi;\cc)\Big|_{g=0}
\,=\,
\frac{1}{\xi + \sqrt{\cc}}
\,,
\end{equation}
and the Laplace transformed propagator of pure CDT is{\dbltinyspace}%
%\footnote{%{\dbltinyspace}%
%$F_1^{(0)}{\negdbltinyspace}(L{\tinyspace};\cc)\Big|_{g=0}
% {\negtinyspace}=
%\E^{- \sqrt{\cc} L}
%$
%%,
%%$\bar{F}_1^{(0)}(L{\tinyspace};V)\Big|_{g=0}
%% {\negtrpltinyspace}={\negtinyspace}
%%L
%%{\trpltinyspace}
%%\E^{- L^2 / (4 V)}
%%/{\tinyspace}
%%(2 \sqrt{\pi}{\trpltinyspace} V^{3/2})
%%$
%and
%$\Hkin{\negtinyspace}\big(\pder{L};\cc\big)
% {\negdbltinyspace}={\negdbltinyspace}
%- \pdder{L} + \mu
%$.
%}
\cite{CDTkin}
\begin{equation}\label{CDT_PropagatorLaplaceTransf}
\tHkin{\negtinyspace}(\xi{\tinyspace};\cc)
\,=\,
%\Big({\negtinyspace}
  -\,\xi^2
  + \cc
%\Big)
\,.
\end{equation}
The propagator \rf{CDT_PropagatorLaplaceTransf} is 
related to the Friedmann equation 
without matter and with cosmological constant $\cc$.%
\footnote{%{\dbltinyspace}%
This point will be explained 
in footnote \ref{footnote:StdFriedmannEq}.
}

The discussion 
from \rf{CommutationRelationPsi} to \rf{DiskAmpSDeqLaplaceTransf} 
is the same as non-critical SFT of pure DT. 
%\rf{CommutationRelationPsi}\,$\sim$\,\rf{DiskAmpSDeqLaplaceTransf}
%の部分はDTのSFTと同じである。
On the other hand, 
the discussion after \rf{DT_DiskAmpSDeqLaplaceTransf} 
is changed, as will be described below. 
%is changed as follows.
%しかし，
%\rf{DT_DiskAmpSDeqLaplaceTransf}以降の議論については次のように変更される。

%\red{Before starting discussing the changes we note that} 
By the way, note that 
the disk amplitude 
%$\tilde{F}_1^{(0)}{\negdbltinyspace}(\xi;\cc)\big|_{g=0}$ 
\rf{CDT_DiskAmpConcreteExpression} 
satisfies the following loop equation, 
\begin{eqnarray}\label{CDT_DiskAmpSDeqLaplaceTransf}
1
{\dbltinyspace}+{\dbltinyspace}
\pder{\xi} \Big\{
  \tilde{F}_1^{(0)}{\negdbltinyspace}(\xi;\cc)\big|_{g=0}
  ({\negtrpltinyspace} - \xi^2 + \cc )
  %{\tinyspace}-{\tinyspace}
  %g \big( \tilde{F}_1^{(0)}{\negdbltinyspace}(\xi;\cc)
  %  \big)^{{\negtinyspace}2}
\Big\}
\,=\,
0
\,.
\end{eqnarray}
Comparing the loop equation \rf{CDT_DiskAmpSDeqLaplaceTransf} 
with the Schwinger-Dyson equation \rf{DiskAmpSDeqLaplaceTransf} 
together with \rf{CDT_PropagatorLaplaceTransf}, 
one obtains 
\begin{equation}\label{CDT_TadpoleLaplaceTransf}
\tilde\rho{\negtinyspace}(\xi;\cc)
\,=\,
1
\,.
%\qquad\quad
%\tHkin{\negtinyspace}(\xi{\tinyspace};\cc)
%\,=\,
%%\Big({\negtinyspace}
%  -\,\xi^2
%  + \cc
%\Big)
%\,.
\end{equation}
%Therefore we have 
%\begin{equation}\label{CDT_Tadpole}
%\rho{\negtinyspace}(L{\tinyspace};\cc)
%\,=\,
%\delta(L)
%\,,
%\qquad\quad
%\Hkin{\negtinyspace}\Big(\pder{L};\cc\Big)
%\,=\,
%-\,\pdder{L} {\dbltinyspace}+{\dbltinyspace} \cc
%\,.
%\end{equation}

\subsection{The appearance of $W$\hspace{1pt}algebra}
\label{sec:purCDTbyWalgebra}

In this {\Section} 
we will express the non-critical SFT of pure CDT 
obtained in previous {\Section} 
by $W$ operators.\,\cite{CausalityWalg2dimAW} 

%From 
%loop equation \rf{CDT_DiskAmpSDeqLaplaceTransf}
%with 
%$g {\negtrpltinyspace}={\negtrpltinyspace} 0$, 
%one obtains the disk amplitude without branching, 
%%loop equation \rf{CDT_DiskAmpSDeqLaplaceTransf}から得られる
%%%分岐がない, つまり，
%%$g {\negtrpltinyspace}={\negtrpltinyspace} 0$のときの disk amplitude は
%%Here we use the disk amplitude obtained by Matrix Model. 
%\rf{CDT_DiskAmpConcreteExpression} 
Taylor expansion of the disk amplitude \rf{CDT_DiskAmpConcreteExpression} 
around $\xi{\negtrpltinyspace}={\negtinyspace}\infty$ 
gives{\dbltinyspace}%
%最初の項は\rf{Omega1DT}
\footnote{\label{footnote:CDTdiskamp}%{\dbltinyspace}%
Note that 
the first term does not contribute to 
the disk amplitude with nonzero volume 
[\,$V {\negqdrpltinyspace}\neq{\negtrpltinyspace} 0$\,]
by the same reason 
mentioned in footnote \ref{footnote:DTdiskamp}. 
This term becomes $\delta(V)$ 
after taking the inverse Laplace transformation 
with respect to $\xi$ and $\mu$. 
$V$ is 2D volume. 
}
\begin{equation}\label{CDT_DiskAmpConcreteExpressionExpansion}
\tilde{F}_1^{(0)}{\negdbltinyspace}(\xi;\cc)\Big|_{g=0}
\,=\,
%\lambda(\xi)
%\,+\,
\xi^{{\tinyspace}-1}
+
\sum_{\ell=1}^\infty
  \xi^{{\tinyspace}-1 - \ell}{\tinyspace} f_\ell
\,.
\end{equation}
%$f_n$ [\,$n{\negtrpltinyspace}={\negtrpltinyspace}1$, $2$, \ldots]
Then, from the viewpoint of mode expansion, 
one can expect 
\begin{equation}\label{CDT_CreationAnnihilationOperatorModeExpansion}
\tilde\Psi^\dagger{\negdbltinyspace}(\xi)
\,=\,
\Omega_1{\negtrehalftinyspace}(\xi)
{\tinyspace}+{\tinyspace}
\sum_{\ell=1}^\infty
  \xi^{{\tinyspace}-\ell - 1}{\tinyspace} \phi^\dagger_\ell
\,,
\qquad\quad
\tilde\Psi(-\xi)
\,=\,
\sum_{\ell=1}^\infty
  \xi^{{\tinyspace}\ell}{\tinyspace} \phi_\ell
%\,.
\end{equation}
with 
\begin{equation}\label{Omega1CDT}
\Omega_1{\negtrehalftinyspace}(\xi)
\,=\,
%\lambda(\xi)
%\,+\,
\xi^{{\tinyspace}-1}{\tinyspace} \phi^\dagger_0
\,.
\end{equation}
Under this mode expansion, 
the commutation relation \rf{CommutationRelationPsiLaplaceTransf} becomes 
\rf{CommutationRelationPhi}. 
Since $\phi_0$ does not exist in this theory, 
$\phi^\dagger_0$ is not 
a quantum number. 
To prevent 
\rf{CDT_DiskAmpConcreteExpressionExpansion} and 
\rf{CDT_CreationAnnihilationOperatorModeExpansion}%
-%
\rf{Omega1CDT} from contradiction, 
we set 
\begin{equation}\label{Omega1CDTvalue}
\phi^\dagger_0
=1
\,.
\end{equation}
Note that 
the space with short length $L$ ``almost'' corresponds to 
the space with small mode $\ell$, 
because 
\begin{equation}\label{CDT_CreationAnnihilationOperatorModeExpansionL}
\Psi^\dagger{\negdbltinyspace}(L) %{\negtinyspace}={\negtrpltinyspace}
\,=\,
\sum_{\ell=0}^\infty \frac{L^\ell}{\ell!} \phi_\ell^\dagger
\,,
\end{equation}
which is obtained by the inverse Laplace transformation of 
\rf{CDT_CreationAnnihilationOperatorModeExpansion}. 

Using \rf{CDT_CreationAnnihilationOperatorModeExpansion} with 
\rf{Omega1CDT}-\rf{Omega1CDTvalue}, 
the Hamiltonian $\Hop$ in \rf{pureGravity_HamiltonianLaplaceTransf} 
with 
\rf{CDT_PropagatorLaplaceTransf}-\rf{CDT_TadpoleLaplaceTransf} becomes 
\begin{eqnarray}\label{pureCDT_HamiltonianModeExpansion}
 \Hop &=&
 \cc {\halftinyspace}
 %\phi^\dagger_0 {\tinyspace}
 \phi_1
 -\, 2 g {\tinyspace}
 %\phi^\dagger_0 {\tinyspace} \phi^\dagger_0 {\halftinyspace}
 \phi_{{\tinyspace}2}
 \,-\, g {\tinyspace} \GG {\tinyspace}
 %\phi^\dagger_0 {\halftinyspace}
 \phi_1 \phi_1
 %\,+\, \mbox{(polynomial of $\phi^\dagger$)}
\nonumber\\&&
 -\> \sum_{\ell=1}^\infty \phi_{\ell+1}^\dagger {\tinyspace} \ell \phi_\ell
 \,+\, \cc
     \sum_{\ell=2}^\infty \phi_{\ell-1}^\dagger {\tinyspace} \ell \phi_\ell
 \,-\, 2 g {\tinyspace} %\phi^\dagger_0
     \sum_{\ell=3}^\infty \phi_{\ell-2}^\dagger {\tinyspace} \ell \phi_\ell
\nonumber\\&&
 -\> g
  \sum_{\ell=4}^\infty {\tinyspace} \sum_{n=1}^{\ell-3}
  \phi_n^\dagger \phi_{\ell-n-2}^\dagger {\tinyspace}
  \ell \phi_\ell
\,-\,%\nonumber\\&&
 %-\>
 g {\tinyspace} \GG
  \sum_{\ell=1}^\infty {\tinyspace} \sum_{n=\max(3-\ell,1)}^\infty
  \phi_{n+\ell-2}^\dagger {\tinyspace}
  n {\halftinyspace} \phi_n {\tinyspace} \ell \phi_\ell
\,.
\end{eqnarray}
%From now on,  we set 
%$\phi^\dagger_0
% {\negdbltinyspace}={\negdbltinyspace} 
% \expect{\phi^\dagger_0}
% {\negdbltinyspace}={\negdbltinyspace} 1$ 
%because 
%the leading term of the disk amplitude \rf{CDT_DiskAmpConcreteExpression} 
%is $\xi^{-1}$ and there is no equivalent operator for $\phi_0$. 
The direct relationship between two amplitudes 
$\tilde{F}_N^{(h)}$ in \rf{GeneralAmpLaplaceTransf} 
and 
$f_N^{(h)}$ in \rf{GeneralAmpModeExpansion2} 
is
\begin{eqnarray}\label{pureCDT_GeneralAmpRelationship}
  \tilde{F}_N^{(h)} (\xi_1,\ldots,\xi_N;\cc)
  &=&
  \Omega_1{\negtrehalftinyspace}(\xi_1) {\tinyspace} \delta_{{\tinyspace}N,1}
  \delta_{{\tinyspace}h,0}
\nonumber\\
  &&+\,
  \sum_{\ell_i{\tinyspace}={\tinyspace}1,2,3,...}
    \xi_1^{{\tinyspace}-\ell_1-1} \cdots \xi_N^{{\tinyspace}-\ell_N-1}
    f_N^{(h)} (\ell_1,\ldots,\ell_N;\cc)
\,.
\end{eqnarray}
The star operation \rf{StarOpDefModeExpansion} applied to the Hamiltonian 
\rf{pureCDT_HamiltonianModeExpansion} leads to 
\begin{eqnarray}\label{pureCDT_HamiltonianModeExpansionStar}
 \Hopstar &=&
 \cc {\halftinyspace} j_1
 -\, 2 {\halftinyspace} g j_2
 \,-\, g{\tinyspace}\GG j_1 j_1
\nonumber\\&&
 -\, \sum_{\ell=1}^\infty \ell j_\ell {\dbltinyspace} \pder{j_{\ell+1}}
 \,+\, \cc \sum_{\ell=2}^\infty \ell j_\ell {\dbltinyspace} \pder{j_{\ell-1}}
 \,-\, 2 g \sum_{\ell=3}^\infty \ell j_\ell {\dbltinyspace} \pder{j_{\ell-2}}
\nonumber\\&&
 -\> g
  \sum_{\ell=4}^\infty {\tinyspace} \sum_{n=1}^{\ell-3}
  \ell j_\ell
  {\dbltinyspace} \pder{j_{n}}
  {\dbltinyspace} \pder{j_{\ell-n-2}}
\,-\,%\nonumber\\&&
 %-\>
 g{\tinyspace}\GG
  \sum_{\ell=1}^\infty {\tinyspace} \sum_{n=\max(3-\ell,1)}^\infty
  n j_n {\tinyspace} \ell j_\ell
  {\dbltinyspace} \pder{j_{n+\ell-2}}
\,.
\quad
\end{eqnarray}

It should be noted that 
three constants $g$, $\GG$ and $\cc$ appear in CDT,
while only  $\GG$ and $\cc$ appear in DT.
The reason is that in 
pure DT 
%$c {\negdbltinyspace}={\negdbltinyspace} 0$ non-critical SFT of DT 
one can remove $g$ by a rescaling \rf{DT_Rescaling} 
because $\Hkin {\negtrpltinyspace}={\negtrpltinyspace} 0$. 
This is impossible in 
pure CDT 
%$c {\negdbltinyspace}={\negdbltinyspace} 0$ non-critical SFT of CDT 
because $\Hkin {\negtrpltinyspace}\neq{\negtrpltinyspace} 0$. 

We here introduce 
$\a_n$
\rf{DefAlphaOperatorPureDT} %, 
%\rf{DefCreationAnnihilationOperatorPureDT}, 
%\rf{vacuumCondition_ap}
%\begin{equation}
%\a_0
%%(\ope{p})^\star
%%{\dbltinyspace}={\dbltinyspace}
%%\nu
%{\dbltinyspace}={\dbltinyspace}
%\frac{1}{\sqrt{\GG}}
%\end{equation}
with $p {\negdbltinyspace}={\negtrpltinyspace} 1$ and 
\begin{eqnarray}\label{CoherentEigenValuesPureCDT}
\nu = \frac{1}{\sqrt{\GG}}
\,,
\quad
\lambda_1 = -\,\frac{\cc}{2 g {\sqrt{\GG}}}
\,,
\quad
\lambda_3 = \frac{1}{6 g {\sqrt{\GG}}}
\,,
\quad
\lambda_{{\tinyspace}n} = 0
\,,
\quad\mbox{[\,$n {\negdbltinyspace}\in{\negdbltinyspace} \dbl{N}$, 
              $n {\negtrpltinyspace}\neq{\negtrpltinyspace} 1$,\,$3$\,]}
\,,
\end{eqnarray}
then we find that 
the Hamiltonian \rf{pureCDT_HamiltonianModeExpansionStar} 
is rewritten as 
\begin{equation}\label{pureCDT_HamiltonianModeExpansionStarW}
\Hopstar
\,=\,
-\,
g \sqrt{\GG}\, W^{(3)}_{-2}
\,+\,
Y
\,,
\end{equation}
where
$W^{(3)}_n$
% [\,$n {\negdbltinyspace}\in{\negdbltinyspace} \dbl{Z}$\,] 
is a standard $W^{(3)}_n$ operator and
is defined by 
\begin{equation}
W^{(3)}_n
\,\define\,
\onethird{\negtrpltinyspace}
\sum_{k+l+m
      {\tinyspace}={\tinyspace}n}
{\negtrpltinyspace}{\negtrpltinyspace}{\negtrpltinyspace}
 : \a_k\a_l\a_m : 
\,,
\qquad
\mbox{[\,$n$, $k$, $l$, $m {\negdbltinyspace}\in{\negdbltinyspace} \dbl{Z}$\,]}
\,,
\label{W3operator}
\end{equation}
and 
\begin{equation}\label{purCDT_Y}
Y
\,\define\,
\frac{1}{\sqrt{\GG}} \Big({\dbltinyspace}
  \frac{1}{4 g} {\tinyspace} \a_4
  {\dbltinyspace}-{\dbltinyspace}
  \frac{\cc}{2 g} {\tinyspace} \a_2
  {\dbltinyspace}+{\dbltinyspace}
  \a_1
{\negtinyspace}\Big)
\,+\,
\frac{\cc^2}{4 {\halftinyspace} g {\tinyspace} \GG}
\,.
\end{equation}
However, at this stage we encounter a different situation from DT. 
%ところが，我々はこの段階でDTとは異なる状況に遭遇する。
In the case of CDT, 
which uses the causal time instead of the geodesic distance, 
condition \rf{NoBigBangCondition} cannot be used 
because it loses its physical meaning, 
and we cannot eliminate from $\Hop$ 
given by \rf{pureGravity_Hamiltonian} 
the term that the universe is created from nothing. 
%測地距離の代わりに因果時間を採用するCDTの場合，
%%Hamiltonian $\Hop$ \rf{pureGravity_Hamiltonian}が満足する
%条件
%\rf{NoBigBangCondition}は物理的な意味を失うので使うことができず，
%宇宙が無から生成する項を
%%Hamiltonian 
%$\Hop$ \rf{pureGravity_Hamiltonian}の中から
%除去することができないのである。
Therefore, unlike DT, 
CDT leaves the ambiguity that any polynomial of $\Psi^\dagger$ 
might be added to the definition of $\Hop$ 
given by \rf{pureGravity_Hamiltonian}. 
%それゆえ，DTとは異なり，CDTでは，
%%Hamiltonian 
%$\Hop$ \rf{pureGravity_Hamiltonian}の定義に
%$\Psi^\dagger$の任意の多項式を付け加えてよい
%という曖昧さが残る。
%%しかし，
%%とりあえずここでは
%%\rf{DiskAmp}や\rf{GeneralAmp}で与えた振幅を曖昧なく定義するため，
%%\rf{NoBigBangCondition}を採用することにしよう。
%%ところが，DTのときの性質\rf{purDT_VanishingEvenMode2}とは異なり，
%%$Y$を物理的な理由で消去する性質はCDTにはない。
%%\rf{purCDT_Y}の右辺の最後の項は定数なので，
%%エネルギーの基準を与えることでゼロにすることが可能だが，
%%最初の3項は，それぞれ，
%%$\phi^\dagger_4$, $\phi^\dagger_{{\tinyspace}2}$, $\phi^\dagger_1$に比例し，
%%無から空間を誕生させる項になるからだ。
Let us take advantage of this ambiguity and remove $Y$ from 
%So, inversely, 
%let us take advantage of this ambiguity. 
%That is, let us remove $Y$ from 
%\rf{TransferOperator}-%
\rf{pureCDT_HamiltonianModeExpansionStarW} by hand, 
and redefine the time{\tinyspace}-transfer operator 
\rf{TransferOperator} as 
%そこで，逆にこの曖昧さを利用し，
%\rf{TransferOperator}と
%\rf{pureCDT_HamiltonianModeExpansionStarW}から$Y$を手で取り除いて，
\begin{equation}\label{TransferOperatorW}
\Theta_{\rm W}^\star(\T)
\define
\E^{-{\tinyspace} \T{\negtinyspace}\Hopstar_{\rm W}}
\,,
\qquad
\Hopstar_{\rm W}
\define
-\,
g \sqrt{\GG}\, W^{(3)}_{-2}
\,.
\end{equation}
%と定義しなおそう。
Although this definition does not satisfy the 
``no big-bang condition'' \rf{NoBigBangCondition}, 
it has the advantage that terms 
like $Y$ that do not have a mathematical role, 
do not appear, 
and as a result, 
CDT has the same %level of 
mathematical structure as DT.
%この定義は big-bang condition \rf{NoBigBangCondition}を満たさないが，
%$Y$のような数学的な役割を持たない項が現れないという長所を持ち，
%結果として，CDTはDTと同等なレベルの数学的構造を持つことになる。
The Big Bang appears from the viewpoint of mathematics!\ 
With this modification, 
the Hamiltonian is changed from $\Hop$ to $\Hop^{\phantom{\star}}_{\rm W}$ 
and becomes 
%そして，この変更により，
%Hamiltonian は$\Hop$から$\Hop^{\phantom{\star}}_{\rm W}$へ変更され，
\begin{eqnarray}\label{pureCDT_ModifiedHgeneralModeExpansion}
 \Hop^{\phantom{\star}}_{\rm W} &=&
 \Hop
 \,-\, \frac{1}{\GG} \Big({\dbltinyspace}
         \frac{1}{4 g} {\tinyspace} \phi^\dagger_4
         {\dbltinyspace}-{\dbltinyspace}
         \frac{\cc}{2 g} {\tinyspace} \phi^\dagger_{{\tinyspace}2}
         {\dbltinyspace}+{\dbltinyspace}
         \phi^\dagger_1
       {\negtinyspace}\Big)
 \,-\, \frac{\cc^2}{4 {\halftinyspace} g {\tinyspace} \GG}
\,.
\end{eqnarray}
%となる。

\subsection{The appearance of Jordan algebra}
\label{sec:purCDTbyWandJalgebra}

Since the theory in {\Section} \ref{sec:purCDTbyWalgebra} 
is derived by replacing the geodesic distance by the causal time 
in $c {\negdbltinyspace}={\negdbltinyspace} 0$ 
non-critical string theory of DT, 
this theory is considered to be a kind of 
$c {\negdbltinyspace}={\negdbltinyspace} 0$ non-critical string theory 
with causal time. 
Therefore, 
it is thought that 
one can obtain 
$c {\negtrpltinyspace}={\negtrpltinyspace} 26$ critical string theory 
by adding, for example, 26 bosonic scalar fields to this theory. 
Namely, 
%もし Section \ref{sec:purCDTbyWalgebra}で述べた理論が 
%$c {\negdbltinyspace}={\negdbltinyspace} 0$ 
%のnon-critical string theory の一種なら，
%$c {\negtrpltinyspace}={\negtrpltinyspace} 26$ 
%の critical string theory を得るには，この理論に，
%たとえば，26種類のbosonの振動モードを加えればよいと考えられる。
%つまり，
\begin{equation}
\hbox{$\a_n$ is replaced by $\a_n^0$}
\qquad\hbox{and}\qquad
\hbox{$\a_n^a$
 [\,$a {\negdbltinyspace}={\negdbltinyspace} 1$, $2$, \ldots, $26$\,] 
 are added.}
\end{equation}
Totally $27$ degrees of freedom appear as bosonic oscillation modes 
$\a_n^\mu$ where $\mu$ takes value $0$ and values $a$
 [\,$a {\negdbltinyspace}={\negdbltinyspace} 1$, $2$, \ldots, $26$\,]. 
The commutation relation of $\a_n^\mu$ is obtained 
by adding flavors to the commutation relation \rf{CommutationRelationAlpha}, 
resulting in{\dbltinyspace}%
\footnote{{\dbltinyspace}%
Since $\alpha_{{\dbltinyspace}0}^\mu$ are commutative with 
all $\alpha_{{\dbltinyspace}n}^\mu$, 
the expectation values of 
$\alpha_{{\dbltinyspace}0}^\mu$ are constants $\nu^\mu$ 
in \rf{vacuumCondition_apGeneral}. 
}
\begin{equation}\label{CommutationRelationAlphaFlavor}
\commutator{\a_m^\mu}{\a_n^\nu}
\,=\,
m {\tinyspace}\delta_{{\tinyspace}m+n,0} {\dbltinyspace}
\delta^{{\halftinyspace}\mu,\nu}
\,,
\qquad
\mbox{[\,$m$, $n {\negdbltinyspace}\in{\negdbltinyspace} \dbl{Z}$\,]}
\,.
\end{equation}
The reason why we chose Kronecker delta $\delta^{{\halftinyspace}\mu,\nu}$ 
instead of the indefinite metric 
is that we cannot introduce an operator that produces a negative norm 
because there is no gauge symmetry.
%\footnote{%{\dbltinyspace}%
%Operators with negative norms cannot be interpreted probabilistically. 
%%formally real としたのは確率解釈を可能にするためである。
%%formally real でないJordan algebra は，
%%たとえ数学的には構築可能な理論であったとしても
%%確率解釈ができないため，現実の世界にはなりえない。
%}
%不定計量ではなく Kronecker delta $\delta^{{\halftinyspace}\mu,\nu}$
%としたのは，
%ゲージ対称性が存在しないため
%負のノルムを生む演算子を導入できないからである。

\red{We now give $\a_n^\mu$ an algebraic structure, by writing} 
%Here, 
%we give $\a_n^\mu$ an algebraic structure, let 
$\a_n {\negdbltinyspace}\define
 \sum_\mu{\negdbltinyspace} E_\mu \a_n^\mu$, 
\red{where $E_\mu$ belongs to a Jordan algebra.} 
%and assume that $E_\mu$ satisfies Jordan algebra. 
%ここで，$\a_n^\mu$に代数構造を入れるのだが，
%$\a_n \define \sum_\mu E_\mu \a_n^\mu$
%とし，$E_\mu$は Jordan algebra を満たすと仮定する。
\red{The algebra of pure CDT is just $\dbl{R}$, 
which {\it is} a Jordan algebra (although a trivial one), 
and the $W$ operators appear} 
%This is because pure CDT has $\dbl{R}$ algebra, 
%which is one of Jordan algebras, 
%and $W$ operator is 
cubic in $\a_n$,%
\footnote{%{\dbltinyspace}%
Here, 
based on the idea of an extension of pure CDT,
we thought that $W^{(3)}_{-2}$ alone was a more natural extension 
than general $W^{(p)}_n$ or a mixture of them, 
but this is open for discussion. 
}
so 
\red{the concept of Jordan algebras appears naturally.}%
%Jordan algebra fits nicely.%
\footnote{%{\dbltinyspace}%
For example, 
if one tries to express the cubic expression of $\a_n$ 
by the Lie algebra product, it will be 
$\tr{\dbltinyspace}\commutator{\a_l}{\commutator{\a_m}{\a_n}}
 {\negdbltinyspace}={\negdbltinyspace} 0$, 
but 
by the Jordan algebra product, it will be 
$\tr{\dbltinyspace}\anticommutator{\a_l}{\anticommutator{\a_m}{\a_n}}
 {\negdbltinyspace}\neq{\negdbltinyspace} 0$. 
%which works well.
}
%pure CDT が Jordan algebra の一つである $\dbl{R}$ algebra を持っており，
%$W$代数が$\a_n$の3次なので，\footnote{%{\dbltinyspace}%
%ここでは，pure CDT の拡張という考えに基づき，一般的な
%$W^{(p)}_n$% [\,$p {\negtrpltinyspace}\ge{\negtrpltinyspace} 4$\,]
%やそれらを混合したものよりも
%$W^{(3)}_{-2}$単独の方を自然な拡張と考えたが，
%これについては議論の余地がある。
%}
%%Lie algebra よりも 
%Jordan algebra がしっくりくるからである。\footnote{%{\dbltinyspace}%
%たとえば，%Lie algebra の場合，
%$\a_n$の3次式をLie algebraの積で表そうとすると，
%$\tr{\dbltinyspace}\commutator{\a_l}{\commutator{\a_m}{\a_n}}
% {\negdbltinyspace}={\negdbltinyspace} 0$
%となってしまうが，
%Jordan algebra の積では
%$\tr{\dbltinyspace}\anticommutator{\a_l}{\anticommutator{\a_m}{\a_n%}}
% {\negdbltinyspace}\neq{\negdbltinyspace} 0$
%となり，うまくゆく。
%}
\red{The so-called simple Jordan algebras} 
%Though 
%the simple Jordan algebras
which have $27$ degrees of freedom are 
the ${\rm C\ell}_{26}(\dbl{R})$ algebra 
and the ${\rm H}_3(\dbl{O})$ algebra.%
\footnote{%{\dbltinyspace}%
See {\Chapter} \ref{sec:EuclideanJordanAlgebra} %Appendix B
for details on the Jordan algebra.
%Jordan代数の詳細については
%Appendix \ref{sec:EuclideanJordanAlgebra}を参照せよ。
}
We here choose the ${\rm H}_3(\dbl{O})$ algebra 
\red{by the above-mentioned principle of extremity, 
because the ${\rm H}_3(\dbl{O})$ algebra is 
the only finite-dimensional so-called {\it exceptional} algebra 
among the Jordan algebras.} 
%by extremity 
%because ${\rm H}_3(\dbl{O})$ algebra is the only exceptional algebra 
%in the Jordan algebra. 
%ここでは
%${\rm H}_3(\dbl{O})$ algebra
%が%Albert代数と呼ばれる
%Jordan代数唯一の例外代数であることから，
%extremityにより
%${\rm H}_3(\dbl{O})$ algebra
%を選択する。
%%pure CDT が $\dbl{R}$ algebra を Jordan algebra 
%として持つことに注目されたい。

%次の章では，
%こうして作られた理論の詳細を説明しよう。

\section{Basic of Theory}
\label{sec:BasicTheory}

In this {\Chapter}, we will explain the details of the theory constructed 
in {\Section} \ref{sec:purCDTbyWandJalgebra}. 
This {\Chapter} is based on refs.\ 
\cite{CausalityWJalgAW,CausalityWJalgAWfullversion}.

\subsection{Definition of $W$\hspace{1pt}and Jordan algebra gravity}
\label{sec:DefWJGravity}

Our theory is defined by the following transfer operator, 
\begin{equation}\label{WJgravityDef}
\Theta^\star%{\negtinyspace}(\T)
\,\define\,
\E^{\cW_{-2}^{(3)}}
%\,=\,
%\exp{\negdbltinyspace}\Big\{
%  {\rm res}_{z=0}{\dbltinyspace}
%  \Tr\Nordering{\onethird \big( J(z) \big)^3}
%\!\Big\}
\,,
\end{equation}
where 
$\cW_n^{(3)}$
% [\,$n {\negdbltinyspace}\in{\negdbltinyspace} \dbl{Z}$\,] 
is a $W^{(3)}_n$ operator with flavors and
is defined by 
%$J(z) = \sum_\mu E_\mu J^\mu(z)$
%\begin{eqnarray}\label{WJgravityDefW}
%&&
%\cW^{(3)}(z)
%\,=\,
%\sum_{n{\tinyspace}\in{\tinyspace}\dblsmall{Z}} \frac{\cW_n^{(3)}}{z^{n+3}}
%\,=\,
%%c_2 
%\Tr\Nordering{\onethird \big( J(z) \big)^3}
%\,,
%\end{eqnarray}
\begin{eqnarray}\label{WJgravityDefW}
&&
\cW^{(3)}_n
\,\define\,
\onethird {\trpltinyspace}\Tr{\negtrpltinyspace}
\sum_{k+l+m
      {\tinyspace}={\tinyspace}n}
{\negtrpltinyspace}{\negtrpltinyspace}{\negtrpltinyspace}
 : \a_k\a_l\a_m : 
\,,
\qquad
\mbox{[\,$n$, $k$, $l$, $m {\negdbltinyspace}\in{\negdbltinyspace} \dbl{Z}$\,]}
\,,
\end{eqnarray}
where
%\begin{eqnarray}
%&&
%J(z)
%{\tinyspace}={\tinyspace}
%\sum_{n{\tinyspace}\in{\tinyspace}\dblsmall{Z}} \frac{\a_n}{z^{n+1}}
%%\,=\,
%%\sum_\mu E_\mu J^\mu(z)
%\,,
%\qquad
%\a_n
%{\tinyspace}={\tinyspace}
%\sum_\mu E_\mu {\tinyspace} \a_n^\mu
%\,,
%\qquad
%J^\mu(z)
%\,=\,
%\sum_{n{\tinyspace}\in{\tinyspace}\dblsmall{Z}} \frac{\a_n^\mu}{z^{n+1}}
%\,,
%\\
%
%\qquad
%%&&
%\commutator{\a_m^\mu}{\a_n^\nu}
%{\tinyspace}={\tinyspace}
%m {\tinyspace}\delta_{{\tinyspace}m+n,0} {\dbltinyspace}
%\delta^{{\halftinyspace}\mu,\nu}
%%\qquad\hbox{[\,$m$, $n \in \dbl{Z}$\,]}
%\qquad
%\end{eqnarray}
$\a_n {\negtrpltinyspace}\define
 \sum_\mu{\negtrpltinyspace} E_\mu \a_n^\mu$ 
is the bosonic scalar current on the complex plane $z$ 
together with the set of matrices $E_\mu$ with flavors $\mu$. 
%(for example, the set of $3 \!\times\! 3$ Hermitian octonion matrices). 
%$\Nordering{\phantom{}}$ is the normal ordering for the operator $J(z)$.
$\Tr$ is the trace for the flavor matrix. 
%${\rm res}_{z=0}$ picks up the residue at $z \!=\! 0$. 
%The correspondence of the operator \rf{WJgravityDef} in the Standard Model is
%\begin{equation}\label{StandardModelDef}
%A \,=\,
%\exp\!\bigg\{
%  i \!\int\!\dd^4 x\, \Tr{\dbltinyspace}\bigg(
%    \quarter F_{\mu\nu} F^{\mu\nu}
%    + \bar\psi \gamma^\mu (i \partial_\mu + A_\mu ) \psi
%    + \ldots
%  \bigg)\!
%\bigg\}
%\,.
%\end{equation}
The set of matrices $E_\mu$ depends on Jordan algebra of the model. 
%Jordan algebra appears naturally in this model. 
In order to avoid negative probability, 
we consider 
the commutation relation \rf{CommutationRelationAlphaFlavor} 
that produces only positive norm states, 
and then, the algebra is 
the formally real Jordan algebra, i.e.\ 
Euclidean Jordan algebra. 
The partition function and any other correlation functions (amplitudes) 
are obtained by taking the expectation values of $\Theta^\star$. 
%\footnote{%{\dbltinyspace}%
%The transfer operator $\Theta^\star$ is related to 
%the partition function $Z$ by 
%$Z \define \tr( P {\dbltinyspace} \Theta^\star )$, 
%where $P$ is the projection operator to a physical state 
%such as $\cuum \vac$, 
%and $\tr$ is the trace of the Fock space of $\a_n^\mu$. 
%}
It should be noted that 
this model has no parameters and 
\red{does not refer to any underlying background geometry.} 
%is independent of background. 
The definition of our model is based on 2D conformal field theory 
and has no geometric meaning. 

Replacing $\a_n$ as 
$\a_n \to (g\sqrt{\GG}\,\T)^{{\tinyspace}-n/2} \a_n$ 
does not change the commutation relation of $\a_n^\mu$ and 
replaces $\cW^{(3)}_{-2}$ as 
$\cW^{(3)}_{-2} \to g\sqrt{\GG}\,\T{\tinyspace}\cW^{(3)}_{-2}$
[\,$\GG {\negdbltinyspace}>{\negdbltinyspace} 0$\,]. 
%%\rf{WJgravityDefW}において
%$\a_n^\mu \to (g\sqrt{\GG}\,\T)^{{\tinyspace}-n/2} \a_n^\mu$
%の置き換えをすると，
%$\a_n^\mu$の交換関係は変わらないまま
%$\cW^{(3)}_{-2} \to g\sqrt{\GG}\,\T{\tinyspace}\cW^{(3)}_{-2}$となる。
As a result, 
%その結果，
%%Transfer operator 
$\Theta^\star$ given by \rf{WJgravityDef} becomes 
\rf{TransferOperatorW}, that is, 
\begin{equation}\label{TransferOperatorGeneralW}
\Theta^\star
\to
\Theta_{\rm W}^\star(\T)
\define
\E^{-{\tinyspace} \T{\negtinyspace}\Hopstar_{\rm W}}
\,,
\qquad
\Hopstar_{\rm W}
\define
-\,
g \sqrt{\GG}\, \cW^{(3)}_{-2}
\,.
\end{equation}
Time $\T$ appears in transfer operator $\Theta^\star$. 
%Transfer operator $\Theta^\star$に時間$\T$が現れる。%\rf{TransferOperator}
The origin of time $\T$ is scale. 
%時間$\T$の起源は%振動モード$\a_n$の
%スケールなのである。
Conversely, 
\rf{TransferOperatorGeneralW} can be used 
instead of \rf{WJgravityDef} as the starting point of the theory, 
and in this case the theory is invariant 
under the following scale transformation, 
%逆に，
%\rf{WJgravityDef}の代わりに
%\rf{TransferOperatorGeneralW}を理論の出発点とすることもできて，
%このときの理論にはスケール変換
\begin{equation}\label{WopScaleTransformation}
z \to c {\dbltinyspace} z
\,,
\qquad
\a_n \to c^{{\tinyspace}n} {\negtinyspace} \a_n
\,,
\qquad
\T \to c^{{\tinyspace}2} {\tinyspace} \T
\,.
\end{equation}
%による対称性が存在することになる。
Thus, we can let $\T$ be, 
say, $\T{\negtrpltinyspace}={\negtrpltinyspace}1$, 
and find that time $\T$ is effectively non-existent. 
%したがって，
%$\T$をたとえば$\T{\negtrpltinyspace}={\negtrpltinyspace}1$%
%とすることができて，
%時間$\T$は事実上存在しないことがわかる。
In order for $\T$ to have the physical meaning of time, 
the vacuum $\cuum$ must break this symmetry, and thus time is born.%
\footnote{%{\dbltinyspace}%
The Higgs potential does not appear here, 
but it is the same as the Higgs mechanism in the sense that 
the change of a vacuum state causes symmetry breaking.
}
%$\T$が時間という物理的な意味を持つためには，
%真空$\cuum$がこの対称性を破る必要があり，
%これにより時間が誕生するのである。\footnote{%{\dbltinyspace}%
%%理論が対称性を壊すわけではなく，真空が対称性を壊す。
%Higgs potential はここでは登場しないが，
%真空の状態変化が対称性の破れを引き起こす
%という意味では，Higgs機構と同じである。
%}
Note also that 
from a statistical mechanics point of view 
$T$ corresponds to the inverse temperature.%
%\footnote{%{\dbltinyspace}%
%$T{\negtrpltinyspace}={\negtrpltinyspace}0$ and 
%$T{\negtrpltinyspace}={\negtinyspace}\infty$ 
%corresponds to infinite and zero temperature, respectively. 
%}

Next, 
let us consider the definition of the vacuum $\cuum$ 
that breaks the symmetry of the scale transformation.
%次に，
%スケール変換\rf{WopScaleTransformation}の対称性を破る
%真空$\cuum$の定義を考えてみよう。
First, we introduce 
operators, 
$\ope{a}_n^{{\tinyspace}\mu\dagger}$, 
$\ope{p}^{{\tinyspace}\mu}$, 
and 
$\ope{a}_n^{{\tinyspace}\mu}$ 
as 
%\rf{DefAlphaOperatorPureDT} with flavors, that is, 
%まずはflavorを持つときの\rf{DefAlphaOperatorPureDT}，つまり，
\begin{eqnarray}\label{DefAlphaOperatorGeneralCDT}
\a_n^\mu \,=\,
\left\{
\begin{array}{cl}
\displaystyle
(\ope{a}_n^{{\tinyspace}\mu\dagger})^\star
%{\dbltinyspace}={\dbltinyspace}
%\frac{1}{\sqrt{\GG}} {\dbltinyspace} \pder{j_n}
& \hbox{[\,$n \!>\! 0$\,]}
\rule[-2pt]{0pt}{10pt}\\
\displaystyle
({\dbltinyspace}\ope{p}^{{\tinyspace}\mu})^\star
%{\dbltinyspace}={\dbltinyspace}
%\nu
& \hbox{[\,$n \!=\! 0$\,]}
\rule[-12pt]{0pt}{30pt}\\
\displaystyle
-{\dbltinyspace} n {\tinyspace} (\ope{a}_{-n}^{{\tinyspace}\mu})^\star
%{\dbltinyspace}={\dbltinyspace}
%- n {\tinyspace} ( \lambda_{-n} + \sqrt{\GG} {\dbltinyspace} j_{-n} )
& \hbox{[\,$n \!<\! 0$\,]}
\end{array}
\right.
\,,
\qquad
\mbox{[\,$n {\negdbltinyspace}\in{\negdbltinyspace} \dbl{Z}$\,]}
\,.
\end{eqnarray}
%を導入する。
The commutation relations of 
these operators are 
%operators %introduced here 
%on the rhs of \rf{DefAlphaOperatorGeneralCDT} 
%satisfy 
%ここで導入された演算子の交換関係は
%\begin{equation}\label{CreationAnnihilationOperatorCommutatorPureDT}
%\commutator{\ope{a}_m}{\ope{a}_n^\dagger}
%=
%\delta_{{\tinyspace}m,n}
%\,,
%\end{equation}
\begin{eqnarray}\label{CreationAnnihilationOperatorCommutationRelation}
&&
\commutator{\ope{a}_m^{{\tinyspace}\mu}{\negdbltinyspace}}
           {\ope{a}^{{\negtinyspace}\nu\dagger}_n}
=
\delta_{{\tinyspace}m,n}
{\tinyspace}
\delta^{{\halftinyspace}\mu,\nu}
\,,
\quad
\commutator{\ope{a}_m^{{\tinyspace}\mu}{\negdbltinyspace}}
           {\ope{a}_n^{\nu}}
=
\commutator{\ope{a}_m^{{\tinyspace}\mu\dagger}{\negdbltinyspace}}
           {\ope{a}^{{\negtinyspace}\nu\dagger}_n}
=
0
\,,
\nonumber\\
&&
\commutator{\ope{p}^{{\tinyspace}\mu}{\negdbltinyspace}}
           {\ope{p}^{{\negtinyspace}\nu}}
=
\commutator{\ope{p}^{{\tinyspace}\mu}{\negdbltinyspace}}
           {\ope{a}^{{\negtinyspace}\nu\dagger}_n}
=
\commutator{\ope{p}^{{\tinyspace}\mu}{\negdbltinyspace}}
           {\ope{a}_n^{\nu}}
=
0
\,,
\qquad
\mbox{[\,$m$, $n {\negdbltinyspace}\in{\negdbltinyspace} \dbl{N}$\,]}
\,.
\end{eqnarray}
%を満たす。
Using these operators, 
%そして，これらの演算子を利用して
{\it the absolute vacuum} $|0\rangle$ 
is defined as follows:
%を次のように定義する。
\begin{equation}\label{DefVacuum}
\ope{a}_n^{{\tinyspace}\mu} {\tinyspace} |0\rangle
 {\tinyspace}={\tinyspace} 
\ope{p}^{{\tinyspace}\mu} {\trehalftinyspace} |0\rangle
 {\tinyspace}={\tinyspace} 
0
%\,,
%\qquad\quad\hbox{[\,$n \!=\!  1$, $2$, \ldots\,]}
\,,
\qquad
\mbox{[\,$n {\negdbltinyspace}\in{\negdbltinyspace} \dbl{N}$\,]}
\,.
\end{equation}
Then, 
the absolute vacuum $\ket{0}$ satisfies{\dbltinyspace}%
\footnote{%{\dbltinyspace}%
We here chose $W^{(3)}_{-2}$ out of 
$W^{(3)}_n$\ [\,$n {\negdbltinyspace}\in{\negdbltinyspace} \dblsmall{Z}$\,]
based on mathematical grounds, 
but this is also considered a kind of extremity 
because the absolute vacuum $\ket{0}$ satisfies \rf{NoBigBangConditionW}. 
%$W^{(3)}_n$\ [\,$n {\negtrpltinyspace}\in{\negtrpltinyspace} \dbl{Z}$\,]
%の中から$W^{(3)}_{-2}$を選んだのは，
%数学的な根拠に基づいたことだが，
%絶対真空$\ket{0}$と\rf{NoBigBangConditionW}の関係があるので，
%これも一種のextremityと考えられる。
}
\begin{equation}\label{NoBigBangConditionW}
\Hop^{\phantom{\star}}_{\rm W} {\tinyspace} |0\rangle  \,=\,  0
%\,,
\end{equation}
for $\Hop^{\phantom{\star}}_{\rm W}$ in \rf{TransferOperatorGeneralW}, 
and %then, 
is invariant under the scale transformation \rf{WopScaleTransformation}. 
%スケール変換\rf{WopScaleTransformation}に対する不変性を持つ真空となる。
On the other hand, 
we introduce the physical vacuum $\cuum$ which satisfies 
%\rf{vacuumCondition_ap}
%\begin{eqnarray}\label{vacuumCondition_ap}
%\ope{a}_n \cuum = \lambda_n {\halftinyspace}\cuum
%\,,
%\qquad
%\ope{p} {\trehalftinyspace}\cuum = \nu {\tinyspace}\cuum
%\,.
%\end{eqnarray}
%with flavors, 
%i.e.\ 
%the following condition is satisfied. 
%これに対し，
%現実の世界の真空$\cuum$は
%flavorを持つときの\rf{vacuumCondition_ap}，すなわち，
\begin{eqnarray}\label{vacuumCondition_apGeneral}
\ope{a}_n^{{\tinyspace}\mu} {\tinyspace}\cuum
 = 
\lambda_{{\tinyspace}n}^{{\negtrehalftinyspace}\mu} {\trehalftinyspace}\cuum
\,,
\qquad
\ope{p}^{{\tinyspace}\mu} {\trehalftinyspace}\cuum = \nu^\mu {\tinyspace}\cuum
\,,
\qquad
\mbox{[\,$n {\negdbltinyspace}\in{\negdbltinyspace} \dbl{N}$\,]}
\,.
\end{eqnarray}
%を満たすとする。
%$\cuum$は明らかにスケール変換\rf{WopScaleTransformation}を破る。
The physical vacuum $\cuum$ breaks 
the scale symmetry under the transformation \rf{WopScaleTransformation}, 
and has the relationship with the absolute vacuum $\ket{0}$ as 
%この真空$\cuum$は，
%スケール変換\rf{WopScaleTransformation}を破り，
%絶対真空$\ket{0}$と
\begin{equation}\label{CondensationOfUniverse}
\cuum =
\bigg\{
{\negdbltinyspace}
\prod_n %V(\lambda_n)
{\negtrpltinyspace}
\exp\!{\negtinyspace}\bigg({\negtinyspace}
  \sum_\mu %{\negtrpltinyspace}
    %\Big({\negtinyspace}
        \lambda_{{\tinyspace}n}^{{\negtrehalftinyspace}\mu}
        \ope{a}_n^{{\tinyspace}\mu\dagger}
      %- \frac{|\lambda_{{\tinyspace}n}^{{\negtrehalftinyspace}\mu}|^2}{2}
    %\Big)
{\negtrpltinyspace}
\bigg)
{\negtrpltinyspace}
\bigg\}
{\dbltinyspace}
\e^{{\tinyspace}i
  \sum_\mu {\negdbltinyspace} \nu^\mu \ope{q}^{{\tinyspace}\mu}
}
{\tinyspace} |0\rangle
\,,
\end{equation}
%という関係にある。
where $q^\mu$ is an operator 
which satisfies the following commutation relations. 
%ただし，
%$q^\mu$は次の交換関係を満たす演算子である。
\begin{eqnarray}\label{CreationAnnihilationOperatorCommutationRelationQ}
%&&
%\ope{a}_0^\mu = 
%\frac{1}{\sqrt{2}}
%  ( \ope{q}^{{\tinyspace}\mu} {\negdbltinyspace}+{\negdbltinyspace}
%    i{\tinyspace}\ope{p}^{{\tinyspace}\mu} )
%\,,
%\quad
%\ope{a}_0^{\mu\dagger} = 
%\frac{1}{\sqrt{2}}
%  ( \ope{q}^{{\tinyspace}\mu} {\negdbltinyspace}-{\negdbltinyspace}
%    i{\tinyspace}\ope{p}^{{\tinyspace}\mu} )
%\,,
%\nonumber\\
&&
\commutator{\ope{q}^{{\tinyspace}\mu}{\negdbltinyspace}}
           {\ope{p}^{{\negtinyspace}\nu}}
=
i{\tinyspace}
\delta^{{\halftinyspace}\mu,\nu}
\,,
\quad
\commutator{\ope{q}^{{\tinyspace}\mu}{\negdbltinyspace}}
           {\ope{q}^{{\neghalftinyspace}\nu}}
=
\commutator{\ope{q}^{{\tinyspace}\mu}{\negdbltinyspace}}
           {\ope{a}^{{\negtinyspace}\nu\dagger}_n}
=
\commutator{\ope{q}^{{\tinyspace}\mu}{\negdbltinyspace}}
           {\ope{a}^{{\negtinyspace}\nu}_n}
=
0
\,,
\qquad
\mbox{[\,$n {\negdbltinyspace}\in{\negdbltinyspace} \dbl{N}$\,]}
\,.
\end{eqnarray}
%\footnote{%{\dbltinyspace}%
Since $\ope{a}_n^{{\tinyspace}\mu\dagger}$ is 
an operator that creates universes, 
$\cuum$ can be thought of as a kind of coherent state 
made up of condensed mini-universes. 
%$\ope{a}_n^{{\tinyspace}\mu\dagger}$は宇宙を生成する演算子なので，
%$\cuum$は小さな宇宙が凝縮してできた一種のcoherent stateと考えることができる。
%}
\footnote{%{\dbltinyspace}%
On the other hand, 
$\vac = \langle 0| {\tinyspace}
  \e^{{\tinyspace}-{\tinyspace}i{\halftinyspace}
    \sum_\mu {\negdbltinyspace} \nu^\mu \ope{q}^{{\tinyspace}\mu}
  }$. 
Then, we have $\vacuumNorm = \langle 0|0\rangle = 1$.
}

For convenience, we here introduce the operators, 
$\phi_n^{{\tinyspace}\mu\dagger}$ and $\phi_n^{{\tinyspace}\mu}$, 
by 
%\begin{equation}\label{DefCreationAnnihilationOperatorPureDT}
%\ope{a}_n^\dagger
%=
%\frac{1}{\sqrt{\GG}} {\dbltinyspace} \phi^\dagger_n
%\,,
%\quad
%\ope{a}_n
%=
%\lambda_n
%{\negtinyspace}+{\negtinyspace} 
%\sqrt{\GG} {\dbltinyspace} \phi_n
%\,,
%\end{equation}
\begin{equation}\label{DefCreationAnnihilationOperatorGeneral}
\ope{a}_n^{{\tinyspace}\mu\dagger}
=
\frac{1}{\sqrt{\GG}} {\dbltinyspace} \phi^{{\tinyspace}\mu\dagger}_n
\,,
\quad
\ope{a}_n^{{\tinyspace}\mu}
=
\lambda_{{\tinyspace}n}^{{\negtrehalftinyspace}\mu}
{\negtinyspace}+{\negtinyspace} 
\sqrt{\GG} {\dbltinyspace} \phi_n^{{\tinyspace}\mu}
\,,
\qquad
\mbox{[\,$n {\negdbltinyspace}\in{\negdbltinyspace} \dbl{N}$\,]}
\,.
\end{equation}
Then, the commutation relations of these operators become 
\begin{equation}\label{CommutationRelationPhiGeneralCDT}
\commutator{\phi^{{\tinyspace}\mu}_m}{\phi^{{\negtinyspace}\nu\dagger}_n}
= 
\delta_{{\tinyspace}m,n}
{\tinyspace}
\delta^{{\halftinyspace}\mu,\nu}
\,,
\qquad
\commutator{\phi^{{\tinyspace}\mu}_m}{\phi^{{\negtinyspace}\nu}_n}
=
\commutator{\phi^{{\tinyspace}\mu\dagger}_m}{\phi^{{\negtinyspace}\nu\dagger}_n}
=
0
\,,
\qquad
\mbox{[\,$m$, $n {\negdbltinyspace}\in{\negdbltinyspace} \dbl{N}$\,]}
\,,
\end{equation}
and the vacuum satisfies 
\begin{equation}\label{vacuumCondition_phiGeneral}
\vac \phi^{{\tinyspace}\mu\dagger}_\ell = 0
\,,
\qquad
\phi^{{\tinyspace}\mu}_\ell \cuum = 0
\,,
\qquad
\mbox{[\,$\ell {\negdbltinyspace}\in{\negdbltinyspace} \dbl{N}$\,]}
\,.
\end{equation}

\newcommand{\MathStructure}{%
\subsection{Mathematical structure}
\label{sec:MathStructureOfWJGravity}

The Jordan product is defined by 
\begin{equation}
A(z) {\negtinyspace}\circ{\negtinyspace} B(w)
\,=\,
\half \anticommutator{A(z)}{B(w)}
%\qquad
%\anticommutator{A(z)}{B(w)}
%\,=\,
%A(z) B(w) + B(w) A(z)
\end{equation}
The exponentiation of $A(z)$ is defined by 
\begin{eqnarray}
&&
\big( A(z) \big)^2
\,=\,
A(z) {\negtinyspace}\circ{\negtinyspace} A(z)
\\
&&
\big( A(z) \big)^3
\,=\,
A(z) {\negtinyspace}\circ{\negtinyspace}
\big(
  A(z) {\negtinyspace}\circ{\negtinyspace} A(z)
\big)
\\
&&
\big( A(z) \big)^n
\,=\,
\underbrace{%
A(z) {\negtinyspace}\circ{\negtinyspace}
\big( \cdots \big(
  A(z) {\negtinyspace}\circ{\negtinyspace}
  \big(
    A(z) {\negtinyspace}\circ{\negtinyspace} A(z)%
}_{\mbox{\scriptsize $n$ pieces of $A(z)$}}%
{\negdbltinyspace}\big)
\big) \cdots \big)
\qquad\mbox{[\,$n {\negtrpltinyspace}\ge{\negtrpltinyspace} 4$\,]}
\end{eqnarray}

\begin{eqnarray}
&&
A(z) {\negtinyspace}\circ{\negtinyspace} B(w)
\,=\,
B(w) {\negtinyspace}\circ{\negtinyspace} A(z)
\\
&&
A(z) {\negtinyspace}\circ{\negtinyspace}
\big(
  \big( A(z) \big)^2 {\negtinyspace}\circ{\negtinyspace} B(w)
\big)
\,=\,
\big( A(z) \big)^2 {\negtinyspace}\circ{\negtinyspace}
\big(
  A(z) {\negtinyspace}\circ{\negtinyspace} B(w)
\big)
\end{eqnarray}
}%\MathStructure	%End of newcommand

\subsection{Birth of Spaces and THT expansion}
\label{sec:BirthSpace}

As an example of a vacuum state $\cuum$ 
given by \rf{CondensationOfUniverse}, 
let us assume{\dbltinyspace}%
\footnote{%{\dbltinyspace}%
%Since 
%%$\Psi^\dagger{\negdbltinyspace}(L) {\negtinyspace}={\negtrpltinyspace}
%% \sum_{\ell=0}^\infty \frac{L^\ell}{\ell!} \phi_\ell^\dagger$, 
%the space with short length $L$ almost corresponds to 
%the space with small mode $l$, as was explained by using 
%\rf{CDT_CreationAnnihilationOperatorModeExpansionL}, 
This assumption means that 
the vacuum $\cuum$ has momentum $\nu$ and 
is the coherent state of tiny spaces created by 
$\ope{a}^\dagger_1$ and $\ope{a}^\dagger_3$. 
%Note that 
%the small mode state almost corresponds to 
%the short lengh state as was explained around eq.\ 
%\rf{CDT_CreationAnnihilationOperatorModeExpansionL}. 
}
%真空$\cuum$ \rf{CondensationOfUniverse} の一例として，
%\rf{CoherentEigenValuesPureCDT}を拡張した
\begin{eqnarray}\label{CoherentEigenValuesGeneralCDT}
\nu = \frac{\omega}{\sqrt{\GG}}
\,,
\quad
\lambda_1 = -\,\frac{\cc}{2 g {\sqrt{\GG}}}
\,,
\quad
\lambda_{{\tinyspace}3} = \frac{\sigma}{6 g {\sqrt{\GG}}}
\,,
\quad
\lambda_{{\tinyspace}n} = 0
\,,
\quad\mbox{[\,$n {\negdbltinyspace}\in{\negdbltinyspace} \dbl{N}$, 
              $n {\negtrpltinyspace}\neq{\negtrpltinyspace} 1$,\,$3$\,]}
\,,
\ 
\end{eqnarray}
%そして，
%\begin{equation}
%\nu = \sum_\mu E_\mu \nu^\mu
%\,,
%\qquad
%\cc = \sum_\mu E_\mu \cc^\mu
%\,,
%\qquad
%\sigma = \sum_\mu E_\mu \sigma^\mu
%\,,
%\end{equation}
which is an extension of \rf{CoherentEigenValuesPureCDT}. 
%を仮定する。
%Then, similar to pure CDT, 
%\rf{DefCreationAnnihilationOperatorGeneral} introduces 
%$\phi_n^\dagger$ and $\phi_n$. 
%そして，pure CDT と同様，
%\rf{DefCreationAnnihilationOperatorPureDT}により
%$\phi_n^\dagger$% = 
%% \sum_\mu {\negdbltinyspace} E_\mu {\tinyspace} \phi_n^{\mu\dagger}$
%と
%$\phi_n$% = 
%% \sum_\mu {\negdbltinyspace} E_\mu {\tinyspace} \phi_n^\mu$
%を導入する。
$\omega$, $\cc$, $\sigma$ 
%, $\phi_n^\dagger$, $\phi_n$ 
are constant matrices %or operators 
which can be expanded on the $E_\mu$ matrices%
%$\omega$, $\cc$, $\sigma$, 
%$\phi_n^\dagger$, $\phi_n$は，
%いずれも行列$E_\mu$を持つ定数と演算子である。%
\footnote{%{\dbltinyspace}%
$\omega
 {\negtinyspace}={\negqdrpltinyspace}
 \sum_\mu{\negdbltinyspace} E_\mu \omega^\mu$, 
$\mu
 {\negtinyspace}={\negqdrpltinyspace}
 \sum_\mu{\negdbltinyspace} E_\mu \mu^\mu$, 
$\sigma
 {\negtinyspace}={\negqdrpltinyspace}
 \sum_\mu{\negdbltinyspace} E_\mu \sigma^\mu$.
%, 
%$\phi_n^\dagger
% {\negtinyspace}={\negqdrpltinyspace}
% \sum_\mu{\negdbltinyspace} E_\mu \phi_n^{\mu\dagger}$, 
%$\phi_n
% {\negtinyspace}={\negqdrpltinyspace}
% \sum_\mu{\negdbltinyspace} E_\mu \phi_n^\mu$. 
}.
Note that 
the scale of $\omega$ and $\sigma$ are meaningless 
because $\sqrt{\GG}$ and $g$ express their scales. 
Then, 
%$\phi_n \cuum {\negtrpltinyspace}={\negtrpltinyspace} 0$
%%[\,$n {\negtrpltinyspace}={\negtrpltinyspace} 1$, $2$, \ldots\,]
%[\,$n {\negdbltinyspace}\in{\negdbltinyspace} \dbl{N}$\,]
%holds from \rf{vacuumCondition_apGeneral}, 
%すると，
%\rf{vacuumCondition_apGeneral}より
%$\phi_n \cuum {\negtrpltinyspace}={\negtrpltinyspace} 0$
%[\,$n {\negtrpltinyspace}={\negtrpltinyspace} 1$, $2$, \ldots\,]
%が成り立ち，
%and 
$\Hop^{\phantom{\star}}_{\rm W}$ in \rf{TransferOperatorGeneralW}
becomes 
\begin{equation}\label{JordanWalgebraHamiltonianAfterSSB}
\Hop^{\phantom{\star}}_{\rm W}
\,=\,
\Hop_{\rm birth}
\,+\,
\Hop_{\rm death}
\,+\,
\Hop_{\rm kin}
\,+\,
\Hop_{\rm int}
\,-\,
\Tr{\dbltinyspace}\bigg\{\!
 \frac{\mu {\halftinyspace} \mu {\halftinyspace} \omega}
      {4 g {\tinyspace} \GG}\,
\!\bigg\}
\,,
\end{equation}
where
\begin{eqnarray}%\label{JordanWalgebraHamiltonianAfterSSB}
\hspace{-10pt}&&\hspace{-10pt}
\Hop_{\rm birth}
\,\define\,
%c_2 
\Tr{\dbltinyspace}\bigg\{
 \frac{\sigma}{\GG}
\bigg(\!\!
  - \frac{\sigma}{4 g} {\tinyspace} \phi^{\dagger}_4
  + \frac{\mu}{2 g} {\tinyspace} \phi^{\dagger}_{{\tinyspace}2}
  - \omega {\tinyspace} \phi^{\dagger}_1
\!\;\bigg)\!
\bigg\}
\,,
\label{JordanWalgebraHamiltonianBirthTerm}
\\
\hspace{-10pt}&&\hspace{-10pt}
\Hop_{\rm death}
\,\define\,
%c_2 
\Tr{\dbltinyspace}\Big\{
    \omega \big(
      \mu {\halftinyspace} \phi_1
    - 2 g {\tinyspace} \omega {\tinyspace} \phi_{{\tinyspace}2}
    - g {\tinyspace} \GG {\tinyspace} \phi_1 \phi_1
    {\tinyspace}\big)
\!\Big\}
\,,
\label{JordanWalgebraHamiltonianDeathTerm}
\\
\hspace{-10pt}&&\hspace{-10pt}
\Hop_{\rm kin}
\,\define\,
%c_2 
\Tr{\dbltinyspace}\bigg\{\!{\negdbltinyspace}
-{\negtinyspace}
    \sigma \sum_{\ell=1}^\infty \phi^{\dagger}_{\ell+1} {\tinyspace}
    \ell \phi_\ell
+
    \mu \sum_{\ell=2}^\infty \phi^{\dagger}_{\ell-1} {\tinyspace}
    \ell \phi_\ell
- 2 g {\halftinyspace}
    \omega \sum_{\ell=3}^\infty \phi^{\dagger}_{\ell-2} {\tinyspace}
    \ell \phi_\ell
\bigg\}
\,,
\label{JordanWalgebraHamiltonianKineticTerm}
\\
\hspace{-10pt}&&\hspace{-10pt}
\Hop_{\rm int}
\,\define\,
-\,g\,
%c_2 
\Tr{\dbltinyspace}\bigg\{\!
    \sum_{\ell=4}^\infty {\tinyspace} \sum_{n=1}^{\ell-3}
      \phi^{\dagger}_n \phi^{\dagger}_{\ell-n-2} {\tinyspace}
      \ell \phi_\ell
%\nonumber\\&&\phantom{
%\Tr\bigg\{
%}
{\dbltinyspace}+{\dbltinyspace}
  \GG
    \sum_{\ell=1}^\infty {\tinyspace} \sum_{n=\max(3-\ell,1)}^\infty\!\!
      \phi^{\dagger}_{n+\ell-2} {\tinyspace}
      n {\halftinyspace} \phi_n {\tinyspace} \ell \phi_\ell
\bigg\}
\,.
\quad
\label{JordanWalgebraHamiltonianTreeTerm}
\end{eqnarray}
The last term on the rhs of \rf{JordanWalgebraHamiltonianAfterSSB} 
only gives the origin of Hamiltonian and has no physical meaning. 
%\rf{JordanWalgebraHamiltonianAfterSSB}の右辺の最後の項は
%Hamiltonianの基準を与えるに過ぎず，物理な意味を持たない。
$\Hop_{\rm birth}$ is the Hamiltonian which creates tiny spaces by 
$\phi^\dagger_1$, $\phi^\dagger_{{\tinyspace}2}$ and $\phi^\dagger_4$, 
and 
$\sigma{\negtrpltinyspace}\neq{\negtrpltinyspace}0$ 
is a necessary condition. 
On the other hand, 
$\Hop_{\rm death}$ is the Hamiltonian which annihilates tiny spaces by 
$\phi_1$ and $\phi_{{\tinyspace}2}$, 
and 
$\omega{\negtrpltinyspace}\neq{\negtrpltinyspace}0$ 
is a necessary condition.
%$\Hop_{\rm birth}$は空間を生成する項で，
%特に$\sigma{\negtrpltinyspace}\neq{\negtrpltinyspace}0$%
%であることが必要条件，
%$\Hop_{\rm death}$は空間を消滅する項で，
%特に$\omega{\negtrpltinyspace}\neq{\negtrpltinyspace}0$%
%であることが必要条件である。
%\footnote{%{\dbltinyspace}%
%Note that 
%the space with short length $L$ almost corresponds to 
%the space with small mode $l$, as was explained around eq.\ 
%\rf{CDT_CreationAnnihilationOperatorModeExpansionL}. 
%}
The Hamiltonians 
$\Hop_{\rm birth}$ and $\Hop_{\rm death}$ 
and the third term in $\Hop_{\rm kin}$ 
break the symmetry 
under the time reversal transformation \rf{TimeReversalTransf}. 
For later convenience, 
we introduce the following scale transformation 
which changes the physical constants as{\dbltinyspace}%
\footnote{%{\dbltinyspace}%
Since the physical constants have been changed 
under the scale transformation \rf{WopScaleTransformationParameter}, 
this transformation does not lead to the symmetry, 
but leads to the so-called dimensional analysis. 
}
\begin{eqnarray}\label{WopScaleTransformationParameter}
&&
\phi_\ell \to c^{-\ell} \phi_\ell
\,,
\quad
\phi_\ell^\dagger \to c^{{\tinyspace}\ell} \phi_\ell^\dagger
\,,
\quad
\xi \to c {\trpltinyspace} \xi
\,,
\quad
L \to c^{-1} {\negtinyspace} L
\,,
\quad
\T \to c^{-1} {\tinyspace} \T
\,,
\nonumber
\\
&&
\GG \to \GG
\,,
\qquad\hspace{7.45pt}%
g \to c^3 g
\,,
\qquad\hspace{0.30pt}%
\omega \to \omega
\,,
\qquad\hspace{-7.20pt}%
\cc \to c^{{\halftinyspace}2} \cc
\,,
\qquad\hspace{-7.30pt}%
\sigma \to \sigma
\,.
\end{eqnarray}
The transfer operator $\Theta$ is invariant under the scale transformation 
\rf{WopScaleTransformationParameter}. 
The combination $g\sqrt{\GG}\,\T$ have the same scale transformation 
as $\T$ in \rf{WopScaleTransformation}, 
so \rf{WopScaleTransformationParameter} is consistent with 
\rf{WopScaleTransformation}. 
The scale transformation \rf{WopScaleTransformationParameter} 
gives the %following 
dynamical timescales %, which have the dimension of time, 
%(This is the so-called dimensional analysis.)
\begin{equation}\label{TimeScaleCCg}
t_\cc \,\define\, |\cc|^{-1/2}
\,,
\qquad
t_g \,\define\, |g|^{-1/3}
\,,
\end{equation}
for constants $\cc$ and $g$, respectively. 
The coefficient of each equation is of order of $1$ 
but is not determined because this is the so-called dimensional analysis. 

%ところで，
%Jordan代数の構造定数$d_{\mu\nu\rho}$の中で非零になるものは，
%singlet を$S$, multipletを$M$とすると，
%%\begin{equation}\label{JordanStructureConstantSM}
%$d_{SSS}$, 
%%\,, \quad
%$d_{SMM}$, 
%%\,, \quad
%$d_{MM'M''}$
%%\end{equation}
%である。

Since $\sigma$, $\cc$ and $\omega$ 
introduced in \rf{CoherentEigenValuesGeneralCDT} 
still have many components, 
the properties of the physical states they create 
vary greatly with their values. 
%\rf{CoherentEigenValuesGeneralCDT}で導入した
%$\sigma$, $\cc$, $\omega$はたくさんの成分を持ち，
%それらが作る物理的な状態の性質はそれら成分の値で大きく異なる。
Therefore, in this {\Article}, 
we will introduce a relatively simple model among these nontrivial models.
%それゆえ，
%このレビューでは，
%非自明な模型の中から比較的単純な模型を紹介することにしよう。
This is a model in which Jordan algebra is ${\rm H}_3(\dbl{O})$,%
\footnote{%{\dbltinyspace}%
See {\Chapter} \ref{sec:EuclideanJordanAlgebra} %Appendix B
for more specifics. 
}
and the vacuum is such that 
only 
%これは
%Jordan algebra が
%${\rm H}_3(\dbl{O})$ algebra となるもので，%
%\footnote{%{\dbltinyspace}%
%%\rf{JordanStructureConstantSM}の
%より具体的なことについては
%Appendix \ref{sec:EuclideanJordanAlgebraH3Def}を参照せよ。
%}
%かつ，
\begin{equation}\label{H3realmodel}
\sigma^0
=
%{\negtrpltinyspace}={\negtrpltinyspace}
\frac{1}{\CoeffSinglet}
%\sqrt{\frac{3}{2}}
\,,
\qquad
\cc^0
=
%{\negtrpltinyspace}={\negtrpltinyspace}
\frac{\cc_0}{\CoeffSinglet}
%\sqrt{\frac{3}{2}}{\dbltinyspace}\cc_0
\,,
\qquad
\omega^0
=
%{\negtrpltinyspace}={\negtrpltinyspace}
\frac{1}{\CoeffSinglet}
%\sqrt{\frac{3}{2}}
\,,
\qquad
\cc^8
=
%{\negtrpltinyspace}={\negtrpltinyspace}
\sqrt{3}{\dbltinyspace}\bar{\cc}
\,,
\qquad
\cc^3
=
%{\negtrpltinyspace}={\negtrpltinyspace}
\cc'
\,,
\end{equation}
is nonzero.%
\footnote{%{\dbltinyspace}%
The $0$ component flavor was set to match pure CDT 
(this model has a simple \dblsmall{R} algebra.). 
%$0$成分のフレーバーについては pure CDT ($\dbl{R}$ algebra を持つ模型) 
%に一致する値にした。
Since there is no firm policy on the definition of vacuum, 
we chose a vacuum that realize a geometric picture. 
%真空の定義については確固たる方針がないので，
%幾何学的な描像が実現できる真空を選択したのである。
The geometry appears from the algebraic structure 
after the birth of space in this theory. 
}
($\CoeffSinglet = \sqrt{\frac{2}{3}}${\dbltinyspace})\,
In this case, 
%このとき，
$\Hop_{\rm kin}$ \rf{JordanWalgebraHamiltonianKineticTerm}
becomes 
%singlet と multiplet を分離して扱うと，
\begin{equation}\label{WalgebraHamiltonianKineticTermSM}
\Hop_{\rm kin}
\,=\,
\Hop_{\rm kin}^{[{\rm A}]}
+
\Hop_{\rm kin}^{[{\rm B}]}
\end{equation}
with 
\begin{eqnarray}
\Hop_{\rm kin}^{[{\rm A}]}
&\define&
-
    \sum_{\ell=1}^\infty
      \phi^{\oIndex\dagger}_{\ell+1} \ell \phi^{\oIndex}_\ell
+
  \cc^{\oIndex}{\negdbltinyspace}
    \sum_{\ell=2}^\infty
      \phi^{\oIndex\dagger}_{\ell-1} \ell \phi^{\oIndex}_\ell
- 2 g %{\tinyspace} \phi^\dagger_0
    {\negdbltinyspace}
    \sum_{\ell=3}^\infty
      \phi^{\oIndex\dagger}_{\ell-2} \ell \phi^{\oIndex}_\ell
\nonumber\\
&&
-
    \sum_{\ell=1}^\infty
      \phi^{\pIndex\dagger}_{\ell+1} \ell \phi^{\pIndex}_\ell
+
  \cc^{\pIndex}{\negdbltinyspace}
    \sum_{\ell=2}^\infty
      \phi^{\pIndex\dagger}_{\ell-1} \ell \phi^{\pIndex}_\ell
- 2 g %{\tinyspace} \phi^\dagger_0
    {\negdbltinyspace}
    \sum_{\ell=3}^\infty
      \phi^{\pIndex\dagger}_{\ell-2} \ell \phi^{\pIndex}_\ell
\nonumber\\
&&
-
    \sum_{\ell=1}^\infty
      \phi^{\mIndex\dagger}_{\ell+1} \ell \phi^{\mIndex}_\ell
+
  \cc^{\mIndex}{\negdbltinyspace}
    \sum_{\ell=2}^\infty
      \phi^{\mIndex\dagger}_{\ell-1} \ell \phi^{\mIndex}_\ell
- 2 g %{\tinyspace} \phi^\dagger_0
    {\negdbltinyspace}
    \sum_{\ell=3}^\infty
      \phi^{\mIndex\dagger}_{\ell-2} \ell \phi^{\mIndex}_\ell
\,,
\label{WalgebraHamiltonianKineticTermS}
\\
%%%% Irregular Pagebreak (Temporal)
&&\phantom{X}\nonumber\\
&&\phantom{X}\nonumber\\
%%%%
\Hop_{\rm kin}^{[{\rm B}]}
&\define&
\sum_i{\negdbltinyspace}
\bigg\{\!{\negdbltinyspace}
-{\negtrpltinyspace}
    {\negdbltinyspace}
    \sum_{\ell=1}^\infty
      \phi^{i\dagger}_{\ell+1} \ell \phi^{i}_\ell
+
  \cc^{\prime\oIndex}{\negdbltinyspace}
    \sum_{\ell=2}^\infty
      \phi^{i\dagger}_{\ell-1} \ell \phi^{i}_\ell
- 2 g %{\tinyspace} \phi^\dagger_0
    {\negdbltinyspace}
    \sum_{\ell=3}^\infty
      \phi^{i\dagger}_{\ell-2} \ell \phi^{i}_\ell
\bigg\}
\nonumber\\
&&
+\,
\sum_I{\negdbltinyspace}
\bigg\{\!{\negdbltinyspace}
-{\negtrpltinyspace}
    {\negdbltinyspace}
    \sum_{\ell=1}^\infty
      \phi^{I\dagger}_{\ell+1} \ell \phi^{I}_\ell
+
  \cc^{\prime\pIndex}{\negdbltinyspace}
    \sum_{\ell=2}^\infty
      \phi^{I\dagger}_{\ell-1} \ell \phi^{I}_\ell
- 2 g %{\tinyspace} \phi^\dagger_0
    {\negdbltinyspace}
    \sum_{\ell=3}^\infty
      \phi^{I\dagger}_{\ell-2} \ell \phi^{I}_\ell
\bigg\}
\nonumber\\
&&
+\,
\sum_{\tilde{I}}{\negdbltinyspace}
\bigg\{\!{\negdbltinyspace}
-{\negtrpltinyspace}
    {\negdbltinyspace}
    \sum_{\ell=1}^\infty
      \phi^{{\tilde{I}}\dagger}_{\ell+1} \ell \phi^{{\tilde{I}}}_\ell
+
  \cc^{\prime\mIndex}{\negdbltinyspace}
    \sum_{\ell=2}^\infty
      \phi^{{\tilde{I}}\dagger}_{\ell-1} \ell \phi^{{\tilde{I}}}_\ell
- 2 g %{\tinyspace} \phi^\dagger_0
    {\negdbltinyspace}
    \sum_{\ell=3}^\infty
      \phi^{{\tilde{I}}\dagger}_{\ell-2} \ell \phi^{{\tilde{I}}}_\ell
\bigg\}
\,,
\label{WalgebraHamiltonianKineticTermM}
\end{eqnarray}
where 
\begin{eqnarray}\label{RealModelField}
\hspace{-24pt}
&&
\phi^{\oIndex\dagger}_\ell
\define
%-\,
  \frac{1}{\sqrt{3}}\,\phi^{0\dagger}_\ell
- \sqrt{\frac{2}{3}}\,\phi^{8\dagger}_\ell
\,,
\qquad\hspace{39.8pt}
%\qquad\hspace{48.7pt}
%\qquad\hspace{39.4pt}
%\qquad\hspace{45pt}
\phi^{\oIndex}_\ell
\,\define\,
%-\,
  \frac{1}{\sqrt{3}}\,\phi^0_\ell
- \sqrt{\frac{2}{3}}\,\phi^8_\ell
\,,
\nonumber\\
\hspace{-24pt}
&&
\phi^{\pmIndex\dagger}_\ell
\define
%\pm
%\Big(
  \frac{1}{\sqrt{3}}\,\phi^{0\dagger}_\ell
  +%{\negdbltinyspace}+{\negtinyspace}
  \frac{1}{\sqrt{6}}\,\phi^{8\dagger}_\ell
%\Big){\negtinyspace}
\pm
  \frac{1}{\sqrt{2}}\,\phi^{3\dagger}_\ell
\,,
\qquad
%\ 
%\qquad\hspace{9.3pt}
\phi^{\pmIndex}_\ell
\define
%\pm
%\Big(
  \frac{1}{\sqrt{3}}\,\phi^0_\ell
  +%{\negdbltinyspace}+{\negtinyspace}
  \frac{1}{\sqrt{6}}\,\phi^8_\ell
%\Big){\negtinyspace}
\pm
  \frac{1}{\sqrt{2}}\,\phi^3_\ell
\,,
\end{eqnarray}
and
\begin{eqnarray}\label{RealModelCC}
&&
\cc^{\oIndex} \define
\cc_0 {\negtinyspace}-{\negtinyspace} 2{\tinyspace}\bar{\cc}
\,,
\qquad\hspace{0pt}
\cc^{\pmIndex} \define
\cc_0 {\negtinyspace}+{\negtinyspace} \bar{\cc}
      {\negtinyspace}\pm{\negtinyspace} \cc'
\,,
\nonumber\\
&&
\cc^{\prime\oIndex} \define
\cc_0 {\negtinyspace}+{\negtinyspace} \bar{\cc}
\,,
\qquad\hspace{4.42pt}
\cc^{\prime\pmIndex} \define
\cc_0 {\negtinyspace}+{\negtinyspace}
  \frac{- \bar{\cc} {\negtinyspace}\pm{\negtinyspace} \cc'}{2}
\,.
\end{eqnarray}
Note that 
the time evolution by Hamiltonian 
\rf{WalgebraHamiltonianKineticTermSM} with 
\rf{WalgebraHamiltonianKineticTermS}%
-%
\rf{WalgebraHamiltonianKineticTermM}
%による時間発展は
is independent for each flavor, 
%各フレーバー毎に独立になっており，
and moreover, 
%しかも，
%pure CDT にも現れた 
the ``basic'' Hamiltonian
\begin{eqnarray}\label{WalgebraHamiltonianKineticTerm}
\bar{\Hop}_{\rm kin}
&\define&
-{\negtrpltinyspace}
    \sum_{\ell=1}^\infty \phi^{\dagger}_{\ell+1} {\tinyspace} \ell \phi_\ell
+
    \mu
    {\negdbltinyspace}
    \sum_{\ell=2}^\infty \phi^{\dagger}_{\ell-1} {\tinyspace} \ell \phi_\ell
- 2 g %{\tinyspace} \phi^\dagger_0
    {\negdbltinyspace}
    \sum_{\ell=3}^\infty \phi^{\dagger}_{\ell-2} {\tinyspace} \ell \phi_\ell
\,,
\end{eqnarray}
which also appeared in pure CDT, is common to all flavors.
%が全てのフレーバーに共通していることに注目されたい。

Now let us assume that 
%ここでさらに，
$g$ is negligible but nonzero, i.e.\ 
``small nonzero $|g|$''.%
%を仮定してみよう。
\footnote{%{\dbltinyspace}%
Note that 
small nonzero $|g|$ is possible but 
$g{\negtrpltinyspace}={\negtrpltinyspace}0$
is impossible. 
The reason is 
to let the kinetic term dominate 
in a theory defined only by 
the interactions of the three universes \rf{WJgravityDefW}. 
%３宇宙の相互作用\rf{WJgravityDefW}を出発点とする理論において，
%Kinetic termを主に考えようとしているのだから，
%$g=0$が無理なことは注意せよ。
}
The condition ``small nonzero $|g|$''
is consistent with the fact that 
the splitting and merging of the universe has not been observed 
from the big bang to the present. 
%この条件は，宇宙の分裂や融合が観測されない観測事実に
%辻褄合わせをしたという意味がある。
Using the dynamical timescale $t_g$ \rf{TimeScaleCCg}, 
this observed fact leads us to 
\begin{equation}\label{gScaleCondition}
t_0 \,\lesssim\, t_g
\ \ \Longrightarrow \ \ 
|g|^{1/3} {\tinyspace}={\tinyspace} \frac{1}{t_g}  \,\lesssim\, \frac{1}{t_0}
\,,
\end{equation}
where $t_0$ is the present time. 
The dynamical timescale $t_g$ can be considered as 
``the effective lifetime of our universe'' 
because 
the universe will enter the chaos {\period} after $t_g$.%
\footnote{%{\dbltinyspace}%
%The reason why the expression ``effective'' is used is that 
%after $t_g$ the state of universe will enter the chaos {\period}. 
%(
See {\Chapter} \ref{sec:Overview} 
about the chaos {\period}.
%)
}
For ``small nonzero $|g|$'', 
%$g {\negtrpltinyspace}\sim{\negtrpltinyspace} 0$, 
%すると，
the terms proportional to $g$ in 
$\Hop_{\rm kin}^{[{\rm A}]}$ given by \rf{WalgebraHamiltonianKineticTermS}
and in 
$\Hop_{\rm kin}^{[{\rm B}]}$ given by \rf{WalgebraHamiltonianKineticTermM}, 
%proportional to $g$, i.e.\ 
%the third term of 
%$\bar{\Hop}_{\rm kin}$ \rf{WalgebraHamiltonianKineticTerm}
are negligible, and 
$\Hop_{\rm int}$ \rf{JordanWalgebraHamiltonianTreeTerm}
is also negligible. 
If $g$ cannot be ignored, 
$\Hop_{\rm int}$ in \rf{JordanWalgebraHamiltonianTreeTerm} 
is dominant %the main player
and a non-perturbative calculation is inevitable. 
%$g$が無視できないときは
%$\Hop_{\rm int}$ \rf{JordanWalgebraHamiltonianTreeTerm}
%が主役になるため，摂動計算ができなくなるからである。
When $g {\negtrpltinyspace}\to{\negtrpltinyspace} 0$, 
%$g {\negtrpltinyspace}\to{\negtrpltinyspace} 0$ のときの
$\Hop_{\rm kin}
 {\negdbltinyspace}+{\negdbltinyspace}
 \Hop_{\rm int}$
thus becomes the sum of 
%は
%\rf{WalgebraHamiltonianKineticTerm}において
%$g {\negtrpltinyspace}={\negtrpltinyspace} 0$
%とした
basic Hamiltonian, 
$\bar{\Hop}_{\rm kin}\big|_{g=0}$, 
defined by \rf{WalgebraHamiltonianKineticTerm}. 
%が基本のHamiltonianになるのである。
%Therefore, 
%そこで，
Here, 
introducing 
$\Psi^\dagger{\negdbltinyspace}(L)$ and $\Psi(L)$ 
by \rf{CDT_CreationAnnihilationOperatorModeExpansion} 
and \rf{WaveFunLaplaceTransf} 
%\rf{CDT_CreationAnnihilationOperatorModeExpansion}と
%\rf{WaveFunLaplaceTransf}によって
%$\Psi^\dagger{\negdbltinyspace}(L)$と$\Psi(L)$を導入し，
and using $\bar{\Hop}_{\rm kin}$ from \rf{WalgebraHamiltonianKineticTerm}, 
we define Green function as 
%$\bar{\Hop}_{\rm kin}$ \rf{WalgebraHamiltonianKineticTerm}を利用して
%Green関数を
\begin{equation}\label{GreenFun}
G(L_{{\tinyspace}0},L{\tinyspace};T)
\,\define\,
\vac \Psi(L)
 {\dbltinyspace}\E^{- T {\negtinyspace} \bar{\Hop}_{\rm kin}}{\tinyspace}
 \Psi^\dagger{\negdbltinyspace}(L_{{\tinyspace}0}) \cuum
\,.
\end{equation}
%と定義する。
Then, 
we can calculate 
the expected value and fluctuation of the size of the universe. 
Especially when 
$g {\negtrpltinyspace}={\negtrpltinyspace} 0$, 
these are 
%すると，宇宙の大きさの期待値と揺らぎが計算できて，
%特に，
%$g {\negtrpltinyspace}={\negtrpltinyspace} 0$
%のときは，
\begin{equation}\label{THTexpansion}
\expect{L}_{{\negtinyspace}(0,T)}{\negtinyspace}\big|_{g=0}
=
\sqrt{
  \expect{L^2}_{{\negtinyspace}(0,T)}{\negtinyspace}\big|_{g=0}
  -
  \big(
    \expect{L}_{{\negtinyspace}(0,T)}{\negtinyspace}\big|_{g=0}
  \big)^{{\negtinyspace}2}
}
=
\left\{
\begin{array}{cl}
\displaystyle
\frac{1}{\sqrt{\cc}}
 \tanh{\negtrpltinyspace}\big( \sqrt{\cc}\,T {\dbltinyspace}\big)
& \hbox{[\,$\cc \!>\! 0$\,]}
\\
\rule[0pt]{0pt}{24pt}
\displaystyle
\frac{1}{\sqrt{-\cc}}
 \tan{\negtrpltinyspace}\big( \sqrt{-\cc}{\dbltinyspace}T {\dbltinyspace}\big)
& \hbox{[\,$\cc \!<\! 0$\,]}
\end{array}
\right.
\!,
\end{equation}
where the definition of the average is 
\begin{equation}
\expect{f(L)}_{{\negtinyspace}(L_{{\tinyspace}0},T)}
\,\define\,
\frac{
  \int_0^\infty\! \frac{\dd L}{L} f(L) G(L_{{\tinyspace}0},L{\tinyspace};T)
     }
     {
  \int_0^\infty\! \frac{\dd L}{L} G(L_{{\tinyspace}0},L{\tinyspace};T)
     }
\,.
\end{equation}
The expansion of the universe varies greatly 
with the sign of the cosmological constant $\cc$. 
%宇宙の膨張は宇宙定数$\cc$の符号で大きく変わり，
The universe expands as hyperbolic tangent when 
$\cc {\negdbltinyspace}>{\negdbltinyspace} 0$
and as tangent when 
$\cc {\negdbltinyspace}<{\negdbltinyspace} 0$. 
%宇宙は，
%$\cc {\negdbltinyspace}>{\negdbltinyspace} 0$
%のときは hyperbolic tangent expansion, 
%$\cc {\negdbltinyspace}<{\negdbltinyspace} 0$
%のときは tangent expansion 
%をする。
From now on, we will call this ``{\dbltinyspace}THT expansion''. 
%以後は，これを ``THT expansion'' と呼ぶことにする。
From \rf{THTexpansion} one can also understand that 
the fluctuation of length $\sqrt{\expect{L^2} - \expect{L}^{2}}$ 
is equal to the expectation of length $\expect{L}$. 
Spatial fluctuations are quite large. 
%また，
%\rf{THTexpansion}から
%揺らぎの大きさ$\sqrt{\expect{L^2} - \expect{L}^{2}}$
%は大きさの期待値$\expect{L}$に等しく，
%空間の揺らぎが大きいこともわかる。

Spaces with various flavors are generated 
one after another from the vacuum $\cuum$ 
by $\Hop_{\rm birth}$ given by \rf{JordanWalgebraHamiltonianBirthTerm}, 
and then THT expansion occurs for each space 
by $\Hop_{\rm kin}\big|_{g=0}$. 
%but depending on the cosmological constant values of each flavor, 
When $\cc {\negtrpltinyspace}<{\negdbltinyspace} 0$ 
the tangent expansion creates a huge space, 
and leads to a kind of inflation. 
On the other hand, 
when $\cc {\negdbltinyspace}>{\negdbltinyspace} 0$ 
the hyperbolic tangent expansion creates 
a torus of size 
$t_\cc
 {\negdbltinyspace}={\negdbltinyspace}
 1/{\negdbltinyspace}\sqrt{\cc}$ 
\rf{TimeScaleCCg},%
\footnote{%{\dbltinyspace}%
The scale $t_\cc$ is considered to give the Planck size $t_{\rm pl}$, 
i.e.\ 
$t_{\rm pl} {\negdbltinyspace}\sim{\negdbltinyspace} t_\cc$ 
by the hyperbolic tangent expansion, 
and is also the dynamical timescale in THT expansion. 
}
\red{and this torus should have a gauge symmetry,} 
%leads to the gauge symmetry 
%いろいろなフレーバーを持つ空間が真空$\cuum$から次々と発生し，
%その後 THT expansion してゆくのだが，
%フレーバーが持つ宇宙定数の値によって，
%tangent expansion なら巨大な空間が，
%hyperbolic tangent expansion ならプランクスケールのトロイダル空間が生まれる。
%The former gives rise to inflation, 
%the latter gives rise to gauge symmetry 
if the theory should agree with string theory. 
%前者はインフレーションとなり，%って地平線問題を解決し，
%後者は，この理論が弦理論に一致するなら，ゲージ対称性を生むのである。
For example, 
%たとえば，
in the case that 
the values of $\cc$'s in \rf{RealModelCC} satisfy 
%\rf{RealModelCC}の$\cc$の値が，
\begin{equation}\label{RealPhysicalValues}
\underbrace{\hspace{3pt}%
\cc^{\mIndex} < \cc^{\prime\oIndex}
}_{\mbox{\scriptsize tangent expansion}}
\hspace{-4pt}<{\tinyspace} 0 {\tinyspace}<\hspace{1pt}
\underbrace{\hspace{3pt}%
\cc^{\pIndex} < \cc^{\prime\mIndex} < \cc^{\prime\pIndex} < \cc^{\oIndex}
}_{\mbox{\scriptsize hyperbolic tangent expansion}}
\,.
\end{equation}
%となる場合は，
9D space with $\mIndex$ and $i$ flavors 
expands tangentially, 
and 
the rest 18D space 
expands hyperbolic tangentially. 
%expands into a huge space by tangent expansion
%and the rest 18D space expand by hyperbolic tangent expansion. 
%$\mIndex$と$i$成分のフレーバーを持つ9次元空間は
%tangent expansionにより巨大な空間へ膨張をする。
$\cc_0$ plays the role of 
shifting the origin of %giving the reference point in 
\rf{RealPhysicalValues}. 
%$\cc_0$は\rf{RealPhysicalValues}の原点をずらす役割をしている。
%インフレーションの条件
%\begin{description}
%\item[\ 1.]
%ビッグバンのエネルギーを説明する。
%\item[\ 2.]
%[horizon problem] 地平線を超えて均一になっていることを説明する。
%\item[\ 3.]
%[flatness problem]
%$K \!\neq\! 0$のときに限り，$\Omega_K$が非常に小さいことを説明する。
%\item[\ 4.]
%[monopole problem]
%GUT scale のエネルギーが存在する場合に限り，
%monopoleなどの残留物が観測できないほど薄まることを説明する。
%\end{description}
%
%\begin{description}
%\item[\ 1.]
%GUTが存在しない場合，
%THT-expansion がインフレーションの代わりをし，地平線問題を解決する。
%\item[\ 2.]
%GUTが存在する場合，モノポール問題が生じるため，
%GUTの対称性が破れた後に第二のインフレーションが起きる必要がある。
%（再出なので整理する必要あり）
%\end{description}

\subsection{Higher-dimension enhancement}
\label{sec:HDenhancement}

%In the last part of {\Section} \ref{sec:BirthSpace}, 
%we consider 
%$\Hop_{\rm kin}$ \rf{JordanWalgebraHamiltonianKineticTerm} 
%and 
%$\Hop_{\rm int}$ \rf{JordanWalgebraHamiltonianTreeTerm} 
%for very small $g$. 
In this {\Section}, 
let us consider the physical phenomenon caused by $g$ 
assuming that the value of $g$ is very small but nonzero. 
So, $\Hop_{\rm int}$ given by \rf{JordanWalgebraHamiltonianTreeTerm}
is not negligible. 

In the case of \rf{H3realmodel}, 
%$\cuum$が\rf{H3realmodel}となるとき，
$\Hop_{\rm int}$ \rf{JordanWalgebraHamiltonianTreeTerm}
becomes 
%singlet と multiplet を分離して扱うと，
\begin{equation}
\Hop_{\rm int}
\,=\,
\Hop_{\rm int}^{[{\rm A}]}
+
\Hop_{\rm int}^{[{\rm B}]}
+
\Hop_{\rm int}^{[{\rm C}]}
+
\Hop_{\rm int}^{[{\rm D}]}
\end{equation}
with 
\begin{eqnarray}
\Hop_{\rm int}^{[{\rm A}]}
&\define&
-\,\sqrt{2}\>\!g{\negtinyspace}
    \sum_{\ell,\,n}{\negdbltinyspace}
  \big\{{\negdbltinyspace}
      \phi^{\oIndex\dagger}_n \phi^{\oIndex\dagger}_{\ell-n-2}
      \ell \phi^{\oIndex}_\ell
    +
    \GG
%    \sum_{\ell,\,m}{\negdbltinyspace}
      \phi^{\oIndex\dagger}_{n+\ell-2}
      n \phi^{\oIndex}_n \ell \phi^{\oIndex}_\ell
  \big\}
\nonumber\\&&
-\,\sqrt{2}\>\!g{\negtinyspace}
    \sum_{\ell,\,n}{\negdbltinyspace}
  \big\{{\negdbltinyspace}
      \phi^{+\dagger}_n \phi^{\pIndex\dagger}_{\ell-n-2}
      \ell \phi^{+}_\ell
    +
    \GG
%    \sum_{\ell,\,m}{\negdbltinyspace}
      \phi^{+\dagger}_{n+\ell-2}
      n \phi^{\pIndex}_n \ell \phi^{\pIndex}_\ell
  \big\}
\nonumber\\&&
-\,\sqrt{2}\>\!g{\negtinyspace}
    \sum_{\ell,\,n}{\negdbltinyspace}
  \big\{{\negdbltinyspace}
      \phi^{-\dagger}_n \phi^{\mIndex\dagger}_{\ell-n-2}
      \ell \phi^{-}_\ell
    +
    \GG
%    \sum_{\ell,\,m}{\negdbltinyspace}
      \phi^{\mIndex\dagger}_{n+\ell-2}
      n \phi^{\mIndex}_n \ell \phi^{\mIndex}_\ell
  \big\}
\,,
\label{WalgebraHamiltonianTreeTerm3S}
\\
\Hop_{\rm int}^{[{\rm B}]}
&\define&
-\,\sqrt{2}\>\!g{\negtinyspace}
  \sum_{i}{\negdbltinyspace}
      \sum_{\ell,\,n}{\negdbltinyspace}
    \big\{{\negdbltinyspace}
        (
          \phi^{\pIndex\dagger}_n
          {\negdbltinyspace}+{\negtinyspace}
          \phi^{\mIndex\dagger}_n
        )
          \phi^{i\dagger}_{\ell-n-2}
          {\dbltinyspace}
          \ell \phi^{i}_\ell
%\nonumber\\&&\phantom{%
%-\,\sqrt{2}\>\!g{\negtinyspace}
%  \sum_{i}{\negdbltinyspace}
%    \big\{{\negdbltinyspace}
%}
  + \GG
%      \sum_{\ell,\,m}{\negdbltinyspace}
        \phi^{i\dagger}_{\ell+n-2}
        {\tinyspace}
        \ell \phi^{i}_\ell
        {\tinyspace}
        n (
          \phi^{\pIndex}_n
          {\negdbltinyspace}+{\negtinyspace}
          \phi^{\mIndex}_n
        )
    \big\}
\nonumber\\&&
-\,\sqrt{2}\>\!g{\negtinyspace}
  \sum_{I}{\negdbltinyspace}
      \sum_{\ell,\,n}{\negdbltinyspace}
    \big\{{\negdbltinyspace}
        (
          \phi^{\oIndex\dagger}_n
          {\negdbltinyspace}+{\negtinyspace}
          \phi^{\pIndex\dagger}_n
        )
        \phi^{I\dagger}_{\ell-n-2}
        {\dbltinyspace}
        \ell \phi^{I}_\ell
%\nonumber\\&&\phantom{%
%-\,\sqrt{2}\>\!g{\negtinyspace}
%  \sum_{I}{\negdbltinyspace}
%    \big\{{\negdbltinyspace}
%}
  + \GG
%      \sum_{\ell,\,m}{\negtrpltinyspace}
        \phi^{I\dagger}_{\ell+n-2}
        {\tinyspace}
        \ell \phi^{I}_\ell
        n (
          \phi^{\oIndex}_n
          {\negdbltinyspace}+{\negtinyspace}
          \phi^{\pIndex}_n
        )
    \big\}
\nonumber\\&&
-\,\sqrt{2}\>\!g{\negtinyspace}
  \sum_{\tilde{I}}{\negdbltinyspace}
      \sum_{\ell,\,n}{\negdbltinyspace}
    \big\{{\negdbltinyspace}
        (
          \phi^{\oIndex\dagger}_n
          {\negdbltinyspace}+{\negtinyspace}
          \phi^{\mIndex\dagger}_n
        )
        \phi^{\tilde{I}\dagger}_{\ell-n-2}
        {\dbltinyspace}
        \ell \phi^{\tilde{I}}_\ell
%\nonumber\\&&\phantom{%
%-\,\sqrt{2}\>\!g{\negtinyspace}
%  \sum_{\tilde{I}}{\negdbltinyspace}
%    \big\{{\negdbltinyspace}
%}
  + \GG
%      \sum_{\ell,\,m}{\negtrpltinyspace}
        \phi^{\tilde{I}\dagger}_{\ell+n-2}
        {\tinyspace}
        \ell \phi^{\tilde{I}}_\ell
        n (
          \phi^{\oIndex}_n
          {\negdbltinyspace}+{\negtinyspace}
          \phi^{\mIndex}_n
        )
    \big\}
\,,
\label{WalgebraHamiltonianTreeTermSM}
\\
%
%\end{eqnarray}
%\begin{eqnarray}
%
\Hop_{\rm int}^{[{\rm C}]}
&\define&
-\,\sqrt{2}\>\!g{\negtinyspace}
  \sum_{i}{\negdbltinyspace}
      \sum_{\ell,\,n}{\negdbltinyspace}
    \big\{{\negdbltinyspace}
        \phi^{i\dagger}_n
        \phi^{i\dagger}_{\ell-n-2}
        {\dbltinyspace}
        \ell (
          \phi^{\pIndex}_\ell
          {\negdbltinyspace}+{\negtinyspace}
          \phi^{\mIndex}_\ell
      )
%\nonumber\\&&\phantom{%
%-\,\sqrt{2}\>\!g{\negtinyspace}
%  \sum_{i}{\negdbltinyspace}
%    \big\{{\negdbltinyspace}
%}
  + \GG
%      \sum_{\ell,\,m}{\negdbltinyspace}
        (
          \phi^{\pIndex\dagger}_{n+\ell-2}
          {\negdbltinyspace}+{\negtinyspace}
          \phi^{\mIndex\dagger}_{n+\ell-2}
        )
        {\dbltinyspace}
        n \phi^{i}_n
        {\dbltinyspace}
        \ell \phi^{i}_\ell
    \big\}
\nonumber\\&&
-\,\sqrt{2}\>\!g{\negtinyspace}
  \sum_{I}{\negdbltinyspace}
      \sum_{\ell,\,n}{\negdbltinyspace}
    \big\{{\negdbltinyspace}
        \phi^{I\dagger}_n
        \phi^{I\dagger}_{\ell-n-2}
        {\dbltinyspace}
        \ell
        (
          \phi^{\oIndex}_\ell
          {\negdbltinyspace}+{\negtinyspace}
          \phi^{\pIndex}_\ell
        )
%\nonumber\\&&\phantom{%
%-\,\sqrt{2}\>\!g{\negtinyspace}
%  \sum_{I}{\negdbltinyspace}
%    \big\{{\negdbltinyspace}
%}
  + \GG
%      \sum_{\ell,\,m}{\negtrpltinyspace}
        (
          \phi^{\oIndex\dagger}_{n+\ell-2}
          {\negdbltinyspace}+{\negtinyspace}
          \phi^{\pIndex\dagger}_{n+\ell-2}
        )
        {\dbltinyspace}
        n \phi^{I}_n
        {\tinyspace}
        \ell \phi^{I}_\ell
    \big\}
\nonumber\\&&
-\,\sqrt{2}\>\!g{\negtinyspace}
  \sum_{\tilde{I}}{\negdbltinyspace}
      \sum_{\ell,\,n}{\negdbltinyspace}
    \big\{{\negdbltinyspace}
        \phi^{\tilde{I}\dagger}_n
        \phi^{\tilde{I}\dagger}_{\ell-n-2}
        {\dbltinyspace}
        \ell
        (
          \phi^{\oIndex}_\ell
          {\negdbltinyspace}+{\negtinyspace}
          \phi^{\mIndex}_\ell
        )
%\nonumber\\&&\phantom{%
%-\,\sqrt{2}\>\!g{\negtinyspace}
%  \sum_{\tilde{I}}{\negdbltinyspace}
%    \big\{{\negdbltinyspace}
%}
  + \GG
%      \sum_{\ell,\,m}{\negtrpltinyspace}
        (
          \phi^{\oIndex\dagger}_{n+\ell-2}
          {\negdbltinyspace}+{\negtinyspace}
          \phi^{\mIndex\dagger}_{n+\ell-2}
        )
        n \phi^{\tilde{I}}_n
        {\tinyspace}
        \ell \phi^{\tilde{I}}_\ell
    \big\}
\,,
\qquad\ 
\label{WalgebraHamiltonianTreeTermSMp}
\end{eqnarray}
and
\begin{eqnarray}
\Hop_{\rm int}^{[{\rm D}]}
&\define&
-{\dbltinyspace}
2 {\tinyspace} g{\negdbltinyspace}
\sum_{i,\,I,\,\tilde{I}}{\negdbltinyspace}
 d_{iI\tilde{I}}{\negtinyspace}
    \sum_{\ell,n}{\negdbltinyspace}%\sum_{\ell=4}^\infty \sum_{n=1}^{\ell-3}
\big\{{\negdbltinyspace}
      \phi^{I\dagger}_n \phi^{\tilde{I}\dagger}_{\ell-n-2}
      {\dbltinyspace}
      \ell \phi^{i}_\ell
  {\tinyspace}+{\tinyspace}
  \GG
%    \sum_{\ell,m}%\sum_{\ell=1}^\infty \sum_{m=\max(3-\ell,1)}^\infty\!\!\!\!
      \phi^{i\dagger}_{n+\ell-2}
      {\tinyspace}
      n \phi^{I}_n \ell \phi^{\tilde{I}}_\ell
\nonumber\\&&\phantom{%
-{\dbltinyspace}
2 {\tinyspace} g{\negdbltinyspace}
\sum_{i,\,I,\,\tilde{I}}{\negdbltinyspace}
 d_{iI\tilde{I}}{\negtinyspace}
    \sum_{\ell,n}{\negdbltinyspace}%\sum_{\ell=4}^\infty \sum_{n=1}^{\ell-3}
\big\{{\negdbltinyspace}
}\hspace{-1.4pt}%\hspace{-2.7pt}
+
      \phi^{i\dagger}_n \phi^{\tilde{I}\dagger}_{\ell-n-2}
      {\dbltinyspace}
      \ell \phi^{I}_\ell
  {\tinyspace}+{\tinyspace}
  \GG
%    \sum_{\ell,m}%\sum_{\ell=1}^\infty \sum_{m=\max(3-\ell,1)}^\infty\!\!\!\!
      \phi^{I\dagger}_{n+\ell-2}
      {\tinyspace}
      n \phi^{i}_n \ell \phi^{\tilde{I}}_\ell
\nonumber\\&&\phantom{%
-{\dbltinyspace}
2 {\tinyspace} g{\negdbltinyspace}
\sum_{i,\,I,\,\tilde{I}}{\negdbltinyspace}
 d_{iI\tilde{I}}{\negtinyspace}
    \sum_{\ell,n}{\negdbltinyspace}%\sum_{\ell=4}^\infty \sum_{n=1}^{\ell-3}
\big\{{\negdbltinyspace}
}\hspace{-1.4pt}%\hspace{-2.7pt}
+
      \phi^{i\dagger}_n \phi^{I\dagger}_{\ell-n-2}
      {\dbltinyspace}
      \ell \phi^{\tilde{I}}_\ell
  {\tinyspace}+{\tinyspace}
  \GG
%    \sum_{\ell,m}%\sum_{\ell=1}^\infty \sum_{m=\max(3-\ell,1)}^\infty\!\!\!\!
      \phi^{\tilde{I}\dagger}_{n+\ell-2}
      {\tinyspace}
      n \phi^{i}_n \ell \phi^{I}_\ell
\big\}
\,.
\label{WalgebraHamiltonianTreeTerm3M}
\end{eqnarray}
%Note that 
%although \rf{WalgebraHamiltonianTreeTerm3M} is 
%the interaction of three different multiplets, 
%it also contains zero terms.
%ただし，
%\rf{WalgebraHamiltonianTreeTerm3M}は
%3個の異なるmultipletによる相互作用なのだが，
%%\footnote{%{\dbltinyspace}%
%%\rf{WalgebraHamiltonianTreeTerm3M}は非零だけでなく，
%ゼロとなる項も含まれているので，注意されたい。
%%}

\begin{figure}[t]
\vspace{10mm}\hspace{5mm}
\begin{minipage}{.45\linewidth}\vspace{0mm}
  \begin{center}\hspace*{-12mm}
    \includegraphics[width=0.5\linewidth,angle=0]
                    {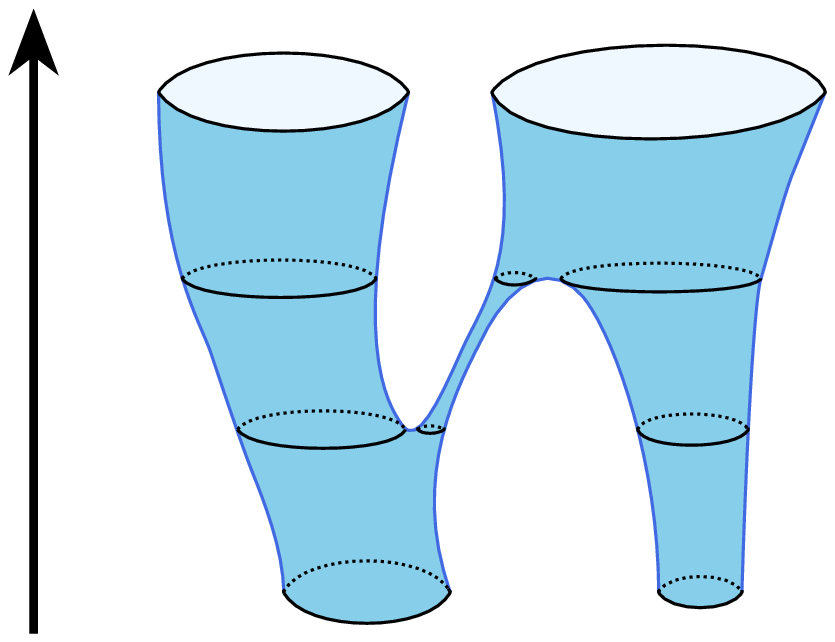}
  \end{center}
  \begin{picture}(200,0)
    \put( 12.0, 45.0){\mbox{\footnotesize$\T$}}
    \put( 50.0,  9.0){\mbox{\footnotesize$L_{{\tinyspace}1}$}}
    \put( 81.0,  9.0){\mbox{\footnotesize$L_{{\trehalftinyspace}2}$}}
    \put( 43.0, 78.0){\mbox{\footnotesize$L_{{\tinyspace}1}'$}}
    \put( 77.0, 78.0){\mbox{\footnotesize$L_{{\trehalftinyspace}2}'$}}
    \put(133.0, 43.0){\mbox{$\Longrightarrow$}}
  \end{picture}
  \vspace*{-6mm}
  \caption[fig.5]{{wormhole interaction}
  }
  \label{fig:WormholeInteractionA}
\end{minipage}
\hspace{6mm}
\begin{minipage}{.45\linewidth}\vspace{0mm}
  \begin{center}\hspace*{-12mm}
    \includegraphics[width=0.5\linewidth,angle=0]
                    {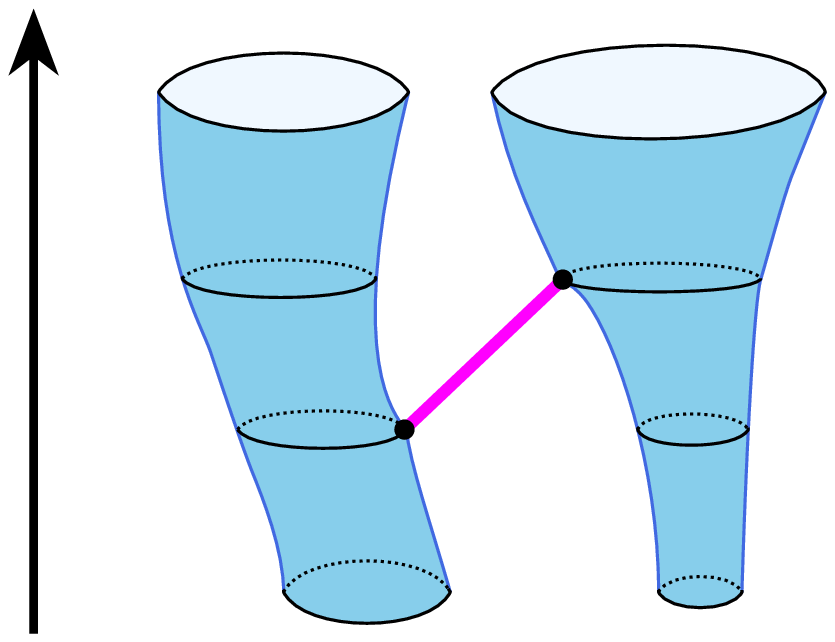}
  \end{center}
  \begin{picture}(200,0)
    \put( 12.0, 45.0){\mbox{\footnotesize$\T$}}
    \put( 50.5,  9.0){\mbox{\footnotesize$L_{{\tinyspace}1}$}}
    \put( 81.0,  9.0){\mbox{\footnotesize$L_{{\trehalftinyspace}2}$}}
    \put( 43.0, 78.0){\mbox{\footnotesize$L_{{\tinyspace}1}'$}}
    \put( 78.0, 78.0){\mbox{\footnotesize$L_{{\trehalftinyspace}2}'$}}
  \end{picture}
  \vspace*{-6mm}
  \caption[fig.6]{simplified diagram of Fig.\,\ref{fig:WormholeInteractionA}}
  \label{fig:WormholeInteractionB}
\end{minipage}
\vspace{-1mm}
%\vspace{-4mm}
\end{figure}

Now let us consider wormholes created by $\Hop_{\rm int}$. 
%ここで，$\Hop_{\rm int}$ が作り出す wormhole について考えてみよう。
This is a thin space that connects two spaces as shown in 
Fig.\ \ref{fig:WormholeInteractionA}, 
%これは，
%Fig.\ \ref{fig:WormholeInteractionA}のように
%2つの空間を繋げる細い空間のことで，
and from now on, this is shown as in 
Fig.\ \ref{fig:WormholeInteractionB}, for simplicity. 
%以後は簡単のため，
%Fig.\ \ref{fig:WormholeInteractionB}のように描く。
%%%
%\vspace*{0mm}
%%%
By the way, 
the Green function $G(L_{{\tinyspace}0},L{\tinyspace};T)$ given by \rf{GreenFun} becomes 
$\delta(L-{\negdbltinyspace}L_{{\tinyspace}0})$ 
for $T {\negdbltinyspace}\to{\negdbltinyspace} 0$ 
and in the region 
$T {\negtrpltinyspace}\sim{\negdbltinyspace} 0$ 
one finds 
\begin{equation}\label{GreenFunSmallT}
G(L_{{\tinyspace}0},L{\tinyspace};T)
\,\sim\,
\frac{1}{\sqrt{4 \pi {\tinyspace} L_{{\tinyspace}0} T}}\,
 e^{-\,\frac{(L {\tinyspace}-{\tinyspace} L_{{\tinyspace}0})^2}
            {4 {\tinyspace} L_{{\tinyspace}0} T\rule{0pt}{5pt}}}
\,,
\qquad\hbox{[ $T {\negtinyspace}\sim{\negtinyspace} 0$ ]}
\,.
\end{equation}
%For $T \to 0$, 
%the Green function \rf{GreenFunSmallT} becomes 
%a natural result, the delta function $\delta(L-L_{{\tinyspace}0})$. 
When $L = L_{{\tinyspace}0}$, 
the Green function \rf{GreenFunSmallT} becomes 
\begin{equation}\label{GreenFunSmallTsameL}
G(L,L{\tinyspace};T)
\,\sim\,
\frac{1}{\sqrt{4 \pi {\tinyspace} L {\tinyspace} T}}
\,,
\qquad\hbox{[ $T {\negtinyspace}\sim{\negtinyspace} 0$ ]}
\,.
\end{equation}
$V_{\rm wh} {\negtrpltinyspace}\define{\negtrpltinyspace}
 L {\tinyspace} T$ 
is the 2D volume of wormhole. 
Let us call the wormhole with a tiny volume 
$V_{\rm wh}$ a ``tiny wormhole". 

According to \rf{GreenFunSmallTsameL}, 
the smaller the wormhole size $V_{\rm wh}$ is, 
the larger the amplitude is. 
%ワームホールの大きさ$V_{\rm wh}$は小さければ小さいほど大きな振幅を与える。
Therefore, 
%したがって，
even if $g$ is very small, 
%たとえ$g$が非常に小さくとも，
if $V_{\rm wh}$ is small enough to satisfy 
$g^2\GG /{\negdbltinyspace} \sqrt{V_{\rm wh}} \gg 1$, 
%$g^2\GG /{\negdbltinyspace} \sqrt{V_{\rm wh}} \gg 1$
%が成り立つほど$V_{\rm wh}$が小さければ，
the interaction by Green function $G(L,L{\tinyspace};T)$ 
given by \rf{GreenFunSmallTsameL} 
cannot be ignored, 
and spaces with different flavors are connected 
by such tiny wormholes. % as shown in Fig.\ \ref{torusknitting}. 
%この相互作用は無視できなくなり，
%異なるフレーバーを持つ空間は図\ref{torusknitting}のように，
%極小ワームホールで結ばれてゆく。
%At this time, 
\red{The interactions are such that spaces with different flavors do not merge 
and the tiny wormhole which can connect them will remain. 
On the other hand, 
spaces with the same flavor {\it can} merge into one space 
and the corresponding tiny wormholes connecting them will then disappear.} 
%Spaces with different flavors do not merge 
%and the tiny wormhole remains, 
%on the other hand, 
%spaces with the same flavor can merge into one
%and the tiny wormhole disappears. 
%この際，
%同じフレーバーを持つ空間同士の場合は1つに融合するのだが，
%異なるフレーバーを持つ空間同士の場合は融合は起こらず，
%tiny wormhole は小さいまま留まる。
\newcommand{\KnittingFigiure}{%
%%%
\begin{figure}[h]
\vspace{-60pt}
%\centerline{\includegraphics{GreenFun.ps}}
  \begin{center}
    \includegraphics[width=180pt]{Knitting.ps}
  \end{center}
  \begin{picture}(200,0)
    \put(113, 92){\mbox{\footnotesize$\T$}}
    \put(150, 21){\mbox{\footnotesize$x_1$}}
    \put(284, 21){\mbox{\footnotesize$x_2$}}
    \put(210,161){\mbox{\footnotesize$x_1$}}
    \put(221,161){\mbox{\footnotesize$x_2$}}
  \end{picture}
\vspace{-20pt}
\caption[torusknitting]{{\footnotesize%\small
%Points of two 1D universes with 
%different flavors are identified via knitting,
%merging 
The two coordinates $x_1$ and $x_2$ 
of two 1D %$T^1$ 
spatial universes 
are connected by tiny wormholes and knitted 
into one coordinate $(x_1,x_2)$ 
of 2D %$T^2$ 
spatial universe. 
}}
\label{torusknitting}
\vspace{-1mm}
%\vspace{-4mm}
\end{figure}%
%%%
}%\KnittingFigiure%The end of \KnittingFigiure
A tiny wormhole gives a large amplitude, 
so all points of 1D space are connected by such tiny wormholes. 
%By the way, 
%ところで，
Only amplitudes that satisfy the condition 
$g^2\GG /{\negdbltinyspace} \sqrt{V_{\rm wh}} \gg 1$ survive. 
Other interactions 
which do not contribute to these amplitudes in 
\rf{WalgebraHamiltonianTreeTerm3S}-\rf{WalgebraHamiltonianTreeTerm3M}
can be ignored when $g$ is very small. 
%ここでは Fig.\ \ref{fig:WormholeInteractionA} 
%に寄与する相互作用だけを考えたが，
%\rf{WalgebraHamiltonianTreeTerm3S}\,$\sim$\,\rf{WalgebraHamiltonianTreeTerm3M}
%の他の相互作用は$g$が非常に小さいときは無視できることに注意しよう。
%Only those amplitudes that satisfy the condition 
%$g^2\GG /{\negdbltinyspace} \sqrt{V_{\rm wh}} \gg 1$ survive. 
%条件$g^2\GG /{\negdbltinyspace} \sqrt{V_{\rm wh}} \gg 1$を
%満たす振幅だけが生き残るのである。
We call this phenomenon ``the knitting mechanism''. 
%This is the essential property in the knitting mechanism.

\begin{figure}[b]
\vspace{1mm}\hspace{10mm}
\begin{minipage}{.45\linewidth}\vspace{6mm}
  \begin{center}\hspace*{0mm}
    \includegraphics[width=0.3\linewidth,angle=0]
                    {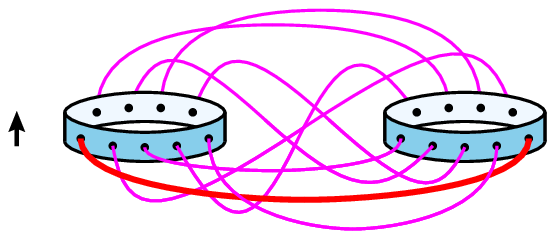}
  \end{center}
  \begin{picture}(200,0)
    \put(-17.0, 25.0){\mbox{\scriptsize $\Delta T$}}
    \put( 10.5, 15.5){\mbox{\scriptsize $x_1$}}
    \put(137.0, 15.5){\mbox{\scriptsize $x_2$}}
    \put(160.0, 25.0){\mbox{$\Longrightarrow$}}
  \end{picture}
  \vspace*{-3mm}
  \caption[fig.7]{{A typical knitting interaction 
                   to form a $T^2${\negdbltinyspace} manifold
                   in a short amount of time $\Delta T$\\
    (The red wormhole connects two coordinates $x_1$ and $x_2$ in each space.)}
  }
  \label{fig:KnittingInteraction}
\end{minipage}
\hspace{6mm}
\begin{minipage}{.45\linewidth}\vspace{9.0mm}
  \begin{center}\hspace*{-18mm}
    \includegraphics[width=0.1\linewidth,angle=0]
                    {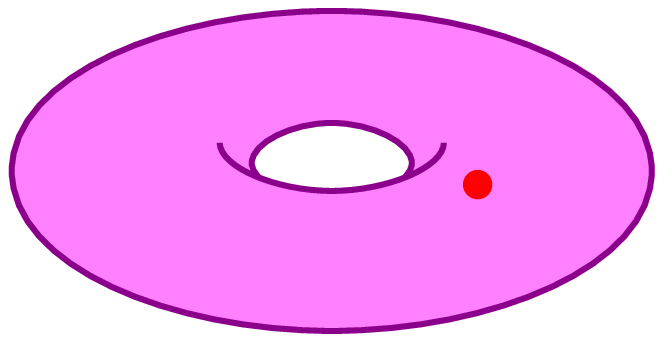}
  \end{center}
  \begin{picture}(200,0)
    \put( 54.0, 14.3){\mbox{\scriptsize $(x_1,x_2)$}}
  \end{picture}
  \vspace*{-1mm}
  \caption[fig.8]{A knitted $T^2${\negdbltinyspace} manifold\\
    (The red dot is the red wormhole in 
     \\
     the left figure, and is the coordinate 
     \\
     $(x_1,x_2)$ on $T^2${\negdbltinyspace} manifold.)}
  \label{fig:Torus}
\end{minipage}
\vspace{-1mm}
%\vspace{-4mm}
\end{figure}

Next, 
let us explain the knitting mechanism in detail again 
using two 1D spaces.
%次に，
%2次元空間を利用し，編み上げ機構についてもう一度詳しく説明しよう。
Suppose that these are two 1D spaces with different flavors. 
%ここに，2種類の異なるフレーバーを持つ1次元空間があるとする。
Let the coordinates in the 1D spaces be $x_1$ and $x_2$, respectively, 
and let the Green function of the wormhole 
connecting the two points $x_1$ and $x_2$ be 
$G_{\rm wh}(x_1,x_2)$. 
%その座標をそれぞれ$x_1$, $x_2$とし，
%2点$x_1$と$x_2$を繋ぐワームホールの
%グリーン関数を$G_{\rm wh}(x_1,x_2)$とする。
Then, 
%すると，
a lot of wormholes connecting any two points in the two 1D spaces 
appear as shown in 
Fig.\ \ref{fig:KnittingInteraction}, 
%あらゆる2点を繋ぐワームホールが
%図\ref{fig:KnittingInteraction}のように現れるのだが，
and 
tiny wormholes give a large amplitude 
when the knitting condition 
$g^2\GG /{\negdbltinyspace} \sqrt{V_{\rm wh}} \gg 1$
is satisfied, 
and the set of $(x_1,x_2)$ that gives a large amplitude forms a 2D torus 
as shown in 
Fig.\ \ref{fig:Torus}.
%編み上げ条件$g^2\GG /{\negdbltinyspace} \sqrt{V_{\rm wh}} \gg 1$
%が成り立つときは，
%極小のワームホールは大きな振幅を与え，
%%図\ref{torusknitting}のように極小になると，
%大きな振幅を与える$(x_1,x_2)$の集合は2次元トーラスを構成するようになる。
The distance on this 2D torus inherits the distance from the 1D spaces 
and becomes the metric distance in 2D Euclidean space. 
%この2次元トーラス上の距離は，1次元空間の距離を引き継ぎ，
%2次元Euclid空間の計量の距離になる。
For similar reasons, 
the scales of 1D spaces become the scale of the 2D space.%
%同様な理由で，
%1次元空間のスケールは
%2次元空間のスケールになる。%
\footnote{%{\dbltinyspace}%
This point is used in the derivation of 
modified Friedmann equation 
in {\Subsection} \ref{sec:MFEderivation}. 
}
Also, 
%また，
the value ${g^2 \GG}/{\sqrt{4 \pi V_{\rm wh}}}$ is put 
on each point in 2D space 
and becomes the cosmological constant of 2D space. 
%空間の各点には
%$\frac{g^2 \GG}{\sqrt{V_{\rm wh}}}$
%の値が乗り，2次元空間の宇宙定数を与える。

\begin{figure}[h]
\vspace*{12mm}\hspace{0mm}
%\vspace{-90pt}
%\centerline{\includegraphics{GreenFun.ps}}
  \begin{center}
    \includegraphics[width=0.1\linewidth,angle=0]
                    {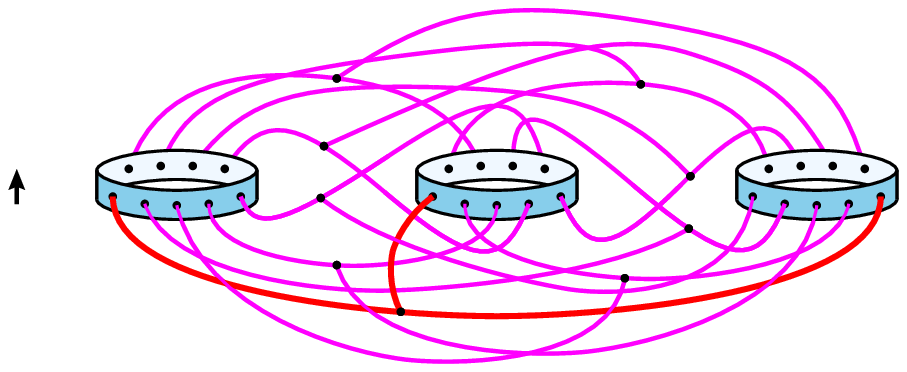}
  \end{center}
  \begin{picture}(200,0)
    \put( 55.0, 20.5){\mbox{\scriptsize $\Delta T$}}
    \put( 83.0, 13.5){\mbox{\scriptsize $x_1$}}
    \put(143.5, 24.0){\mbox{\scriptsize $x_2$}}
    \put(245.0, 13.5){\mbox{\scriptsize $x_3$}}
  \end{picture}
\vspace{6mm}
\caption[fig.9]{{A typical knitting interaction 
to form a $T^3${\negdbltinyspace} manifold
in a short amount of time $\Delta T$\\
(The red wormhole connects three coordinates $x_1$, $x_2$ and $x_3$
 in each space %\hfill\break\hspace*{5pt}%
and becomes the coordinate $(x_1,x_2,x_3)$ 
on $T^3${\negdbltinyspace} manifold.)}
}
\label{fig:ToroidalSpace}
\vspace{-1mm}
%\vspace{-4mm}
\end{figure}

In the case of three 1D spaces with different flavors, 
%3種類の異なるフレーバーの場合，
wormholes are connected by the three{\tinyspace}-point interaction 
$\Hop_{\rm int}^{[{\rm A}]}$ 
given by \rf{WalgebraHamiltonianTreeTerm3S}, 
as shown in Fig.\ \ref{fig:ToroidalSpace}, 
and these become tiny wormholes 
\red{that will mediate the formation of a 3D torus.} 
%to form a 3D torus.
%ワームホールは3点相互作用
%$\Hop_{\rm int}^{[{\rm A}]}$ \rf{WalgebraHamiltonianTreeTerm3S}
%により，
%図\ref{fig:ToroidalSpace}のように繋がれ，
%これらが極小ワームホールになることで，3次元トーラスを構成する。
Similarly, if there are many more flavors, 
the set of tiny wormholes forms a torus of general dimension.
%さらに多くの種類のフレーバーが存在する場合も同様で，
%極小ワームホールの集合は
%一般的な次元のトロイダルの空間を構成する。
%%%
%\vspace*{0mm}
%%%
%
%%%
%%%
In other words, 
%つまり，
high-dimensional space is composed of tiny wormholes,%
\footnote{%{\dbltinyspace}%
Each point of the torus is one wormhole, 
so the spacetime of our universe is discrete.
%空間の各点は１つのwormholeになっており，それゆえ，
%空間は離散的になっていると言える。
}
%宇宙空間の各点は極小ワームホールで構成され，
and has a toroidal topology. 
Note that 
the knitting mechanism continues to preserve the high-dimensional space 
even after the space becomes high-dimensional.%
\footnote{%{\dbltinyspace}%
We call the phenomenon of changing the spatial dimension 
from 1D to high dimension ``dimension enhancement''. 
}
%
%\begin{equation}
%\mbox{Higher-dimensional space formed toroidal topology.}
%\end{equation}
\red{We thus conclude that 
the knitting mechanism in a natural way results in a universe with
toroidal topology, and thus allows a universe with zero spatial curvature.} 
%This concludes that by the knitting mechanism 
%the topology of our universe becomes toroidal 
%and then 
%その結果，%よって$K {\negtrpltinyspace}={\negtrpltinyspace} 0$となり，
%the universe with flat curvature is realized. 
%宇宙は平坦となる。
\red{We can summarize the situation as follows:
since the set of $n$ labeling} 
%Since a set of $n$ of 
$\a_n$ is a set of integers $\dbl{Z}$, 
the inverse Laplace transformation creates the 1D loop spaces $S^1$.
%$\a_n$の$n$は整数であることから，
%この逆Laplace変換により1次元のloopが現れる。
\red{Then, by the knitting mechanism, 
a direct product of loop spaces is formed,} 
%By the knitting mechanism, it forms a direct product of loop spaces, 
i.e.\ a high-dimensional torus $T^n$. 
%そして，編み上げ機構によりloopの直積として高次元トーラス$T^n$となる。
The concept of ``neighborhood'' in spacetime 
is introduced into the knitted space. 
\red{This is how the dynamics of the} 
%This is how this 
theory determines
the spatial topology and the spatial distances of our universe.
%これがこの理論が宇宙の空間のトポロジー，
%そして，空間の距離を決める仕組みである。

\red{The knitting mechanism \red{not only solves} 
with problems of the emergence of topology and metrics, 
but also the problem of the energy conservation of our universe.} 
%The knitting mechanism is consistent 
%not only with problems of metric and topology emergence, 
%but also with problems of the energy conservation law. 
%編み上げ機構は，
%計量やトポロジー創発の問題だけでなく，エネルギー保存則の問題とも整合性がよい。
If the current 4D spacetime started from a point, 
as mentioned in {\Subsection}s 
\ref{sec:UniverseStartedFromPoint} and \ref{sec:UniversWasBornAsOneDimSpace},
%現在の4次元時空が点から発生した場合，
%\ref{sec:UniversWasBornAsOneDimSpace}節で述べたように
problems arise due to the energy conservation law, but
%エネルギー保存則に起因する問題が現れるが，
in our theory, the current 4D spacetime appears 
after the knitting mechanism, 
so the energy conservation law can only be traced back to the time 
when 4D spacetime appeared, 
and in this way the singularity of energy density does not occur.
%我々の理論では現在の4次元時空は編み上げ機構のときに出現するため，
%エネルギー保存則は4次元時空が出現した時点までしか遡ることはできず，
%エネルギー密度が発散するという問題は現れない。
In addition, even the 
%ところで，
%note also that 
2D energy density $L_{{\tinyspace}0}$ is not conserved 
in the period when the universe is a 1D space, 
because 
$\commutator{\Hopstar_{\rm W}}{L_{{\tinyspace}0}} \propto
 -\commutator{W_{-2}}{L_{{\tinyspace}0}} =
 2{\tinyspace}W_{-2} \neq 0$. 
%宇宙が1次元空間の時代は，
%$\commutator{\Hopstar_{\rm W}}{L_{{\tinyspace}0}} \propto
% -\commutator{W_{-2}}{L_{{\tinyspace}0}} =
% 2{\tinyspace}W_{-2} \neq 0$
%が成り立つため，2次元エネルギー$L_{{\tinyspace}0}$は保存されないことにも注目されたい。
Even in this period 
\red{a potential singularity is not caused by energy conservation.} 
%the singularity caused by the energy conservation law does not appear. 
%ここでもエネルギー保存則に起因する問題は起こらないのである。

At the end of this {\Section}, 
we make several comments on 
the phenomena of the concrete model \rf{RealPhysicalValues}. 
By using the interactions 
$\Hop_{\rm int}^{[{\rm A}]}$ given by \rf{WalgebraHamiltonianTreeTerm3S} 
and 
$\Hop_{\rm int}^{[{\rm B}]}$ given by \rf{WalgebraHamiltonianTreeTermSM}, 
%$\Hop_{\rm int}^{[{\rm C}]}$ given by \rf{WalgebraHamiltonianTreeTermSMp}, 
%により，
spaces with the flavors of $\pIndex$ and $\OIndex$ %components 
act as wormholes that cause the knitting mechanism, 
%$\pIndex$成分と$\OIndex$成分のフレーバーを持つ空間が
%編み上げ機構を引き起こすwormholeの役割をする。
and 1D spaces with the rest of 25 flavors of 
$i$, $I$, $\tilde{I}$, $\mIndex$ %components 
form a high-dimensional space. 
%そして，25個の$i$, $I$, $\tilde{I}$, $\mIndex$が高次元空間を形成するのである。

Finally, let us comment on the birth and the death of spaces. 
%The birth and death of spaces is as follows: 
Firstly, 
%まず，
$\Hop_{\rm birth}$ in eq.\ \rf{JordanWalgebraHamiltonianBirthTerm}, i.e.\ 
\begin{eqnarray}\label{WalgebraHamiltonianBirthTerm3S}
\Hop_{\rm birth}
&=&
\frac{1}{\sqrt{2}{\tinyspace} %g {\tinyspace}
         \GG}
\bigg\{\!{\negdbltinyspace}
  -{\tinyspace}
    \frac{
      \phi^{\oIndex\dagger}_4
      {\negtrpltinyspace}+{\negtinyspace}
      \phi^{\pIndex\dagger}_4
      {\negtrpltinyspace}+{\negtinyspace}
      \phi^{\mIndex\dagger}_4
    }{4 {\halftinyspace} g}
  {\tinyspace}+{\tinyspace}
    \frac{
      \cc^{\oIndex} \phi^{\oIndex\dagger}_2
      {\negtrpltinyspace}+{\negtinyspace}
      \cc^{\pIndex} \phi^{\pIndex\dagger}_2
      {\negtrpltinyspace}+{\negtinyspace}
      \cc^{\mIndex} \phi^{\mIndex\dagger}_2
    }{2 {\halftinyspace} g}
\nonumber\\
&&\phantom{%
\frac{1}{\sqrt{2}{\tinyspace} %g {\tinyspace}
         \GG}
\bigg\{\!{\negdbltinyspace}
}%{\tinyspace}
  -{\tinyspace} %g
    (
      \phi^{\oIndex\dagger}_1
      {\negtinyspace}+
      \phi^{\pIndex\dagger}_1
      {\negtinyspace}+
      \phi^{\mIndex\dagger}_1
    )
\bigg\}
\end{eqnarray}
creates spaces with three singlets 
$\OIndex$, $\pIndex$, and $\mIndex$. 
%3つのsinglet $\OIndex$, $\pIndex$, $\mIndex$ を持つ空間が誕生する。
Secondly, 
%そして，
$\Hop_{\rm int}^{[{\rm C}]}$ in eq.\ \rf{WalgebraHamiltonianTreeTermSMp}
creates spaces with three multiples 
$i$, $I$, and $\tilde{I}$ 
from the three singlets 
$\OIndex$, $\pIndex$, and $\mIndex$. 
%\footnote{%{\dbltinyspace}%
%Note that 
%the leading term of this creation is zeroth order of $g$ 
%because $g$'s in 
%\rf{WalgebraHamiltonianBirthTerm3S} and 
%\rf{WalgebraHamiltonianTreeTermSMp} are cancelled. 
%}
%3つのsingletから3つのmultiplet $i$, $I$, $\tilde{I}$ を持つ空間が誕生する。
The annihilation of spaces by 
$\Hop_{\rm death}$ \rf{JordanWalgebraHamiltonianDeathTerm}, i.e.\ 
\begin{equation}\label{WalgebraHamiltonianDeathTerm3S}
\Hop_{\rm death}
=
\frac{1}{\sqrt{2}}
\big\{{\tinyspace}
      \cc^{\oIndex} \phi^{\oIndex}_1
      {\negtinyspace}+{\negtinyspace}
      \cc^{\pIndex} \phi^{\pIndex}_1
      {\negtinyspace}+{\negtinyspace}
      \cc^{\mIndex} \phi^{\mIndex}_1
  - 2 {\halftinyspace} g {\tinyspace}
    (
      \phi^{\oIndex}_2
      {\negtinyspace}+
      \phi^{\pIndex}_2
      {\negtinyspace}+
      \phi^{\mIndex}_2
    )
{\negtinyspace}\big\}
  - g {\tinyspace} \GG
    \sum_\mu{\negtinyspace} \phi^\mu_1 \phi^\mu_1
\end{equation}
follows almost the same process as the creation of spaces. 
Here it should be noted that 
``small nonzero $|g|$'' 
is a necessary condition for the kinetic term to be dominant 
because the definition of the theory 
\rf{WJgravityDef}-\rf{WJgravityDefW}
has only three-point interactions.

\subsection{Vanishing the cosmological constant and Big Bang}
\label{sec:BigBang}
%\subsection{End of inflation}
%\label{sec:EndInflation}

%\subsubsection{Vanishing the cosmological constant}

\begin{figure}[t]
\vspace{-2mm}\hspace{-0mm}
%\centerline{\includegraphics{GreenFun.ps}}
  \begin{center}
    \includegraphics[width=60pt]{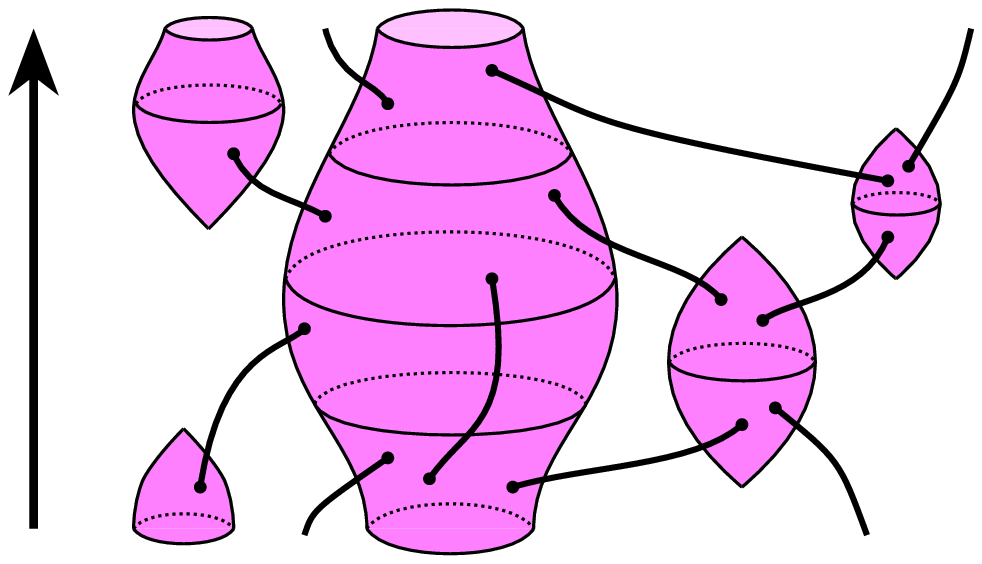}
  \end{center}
  \begin{picture}(200,0)
    \put(125, 51){\mbox{\footnotesize$\T$}}
    %\put(157, 75){\mbox{\footnotesize$\T$}}
  \end{picture}
\vspace{-4mm}
\caption[vanishingcosmologicalconstant]{%
%We show 
A typical configuration of 
Coleman mechanism. 
Many universes are connected by many thin tube wormholes. 
(Black curves are thin tube wormholes.)
}
\label{fig:vanishingcosmologicalconstant}
\vspace{-1mm}
%\vspace{-5mm}
\end{figure}

In the knitting mechanism, 
even when ``small nonzero $|g|$'', 
%$g {\negtrpltinyspace}\sim{\negtrpltinyspace} 0$, 
the tiny wormholes give a large amplitude, 
and the knitting mechanism forms a high-dimensional space 
and produces a huge cosmological constant. 
%Knitting mechanism では，
%$g {\negtrpltinyspace}\sim{\negtrpltinyspace} 0$
%のときでさえ
%tiny wormhole が大きな振幅を与えることで
%編み上げ機構により高次元空間が形成され，
%巨大な宇宙定数が生み出される。
%
On the other hand, it is known that 
even when many spaces are connected 
by wormholes with finite length as shown in 
Fig.\ \ref{fig:vanishingcosmologicalconstant}, 
it gives a large amplitude. 
%実は，
%Fig.\ \ref{fig:vanishingcosmologicalconstant}のように
%多くの空間が有限の長さの wormhole で結ばれたときも
%大きな振幅を与えることが知られている。
This is called the Coleman mechanism 
and works 
\red{in such a way that the cosmological term becomes zero.}%
%to make the cosmological term zero.%
%これは Coleman機構と呼ばれ，
%宇宙項をゼロにする働きをする。%
\footnote{%{\dbltinyspace}%
The Coleman mechanism occurs 
not only in two dimensions but also in high dimensions.
%このメカニズムは，2次元だけでなく，高次元でも起きる。
}
\cite{ColemanMecha,ColemanMechaKO,ColemanMecha2D,ColemanMechaHKK}
%仕組みはほとんど Knitting mechanism と同じだが，
%唯一異なるのが次元である。
%Knitting mechanism が2次元時空で発生したのに対し，
%Coleman mechanism は2次元時空では発生せず？
%高次元時空で発生する。
%したがって，発生の順序は，
%Knitting mechanism の後，Coleman mechanism となる。
%%%
%%%
%
%\subsubsection{Big Bang}
%
By the Coleman mechanism, 
the cosmological term vanishes 
and its energy is converted into a uniform and huge energy.%
\footnote{%{\dbltinyspace}%
%As a detailed mechanism, it is thought that 
%some field %such as the dilaton field 
%transfers energy from the cosmological term, i.e.\ 
%the vacuum energy to matter energy. 
It is 
\red{believed that this
energy is extracted from the vacuum by some} 
%thought that 
%energy is extracted from the vacuum by 
mechanisms, 
%Mechanisms which extracts energy from the vacuum, 
\red{e.g.\  via 
the blackbody radiation or via interactions with dilaton fields.} 
%for example, 
%the blackbody radiation, the interactions via such as dilaton fields. 
%and so on, 
%are thought to work. 
}
%Coleman mechanism によって宇宙項は消え，
%そのエネルギーは均一かつ巨大なエネルギーに転化される。
This is the Big Bang.
%ビッグバンである。
%\footnote{%{\dbltinyspace}%
Since THT expansion is considered a kind of inflation, 
the entropy of the universe at the beginning of the Big Bang 
is approximately equal to 
%of the order of the length of universe $t_\cc$, 
the value when \lq\lq THT inflation" began, 
thus resulting in a low-entropy state. 
%THT膨張はインフレーションの一種と考えられて，
%ビッグバン開始時の宇宙のエントロピーは，
%THTインフレーションが始まった時代の$t_\cc$ のオーダーの値，つまり，
%低いエントロピー状態になる。
%}
%
The wormholes are considered not to transmit energy 
because there exist no transverse modes 
in 2D thin tube spacetime of wormholes.
%\footnote{%{\dbltinyspace}%
%Rigorous explanation is under study. %necessary here. 
%}
Then, the energy is conserved. 
It is also noted that 
the vanishing cosmological constant induced by the Coleman mechanism 
determines the reference point for potential energy, 
which cannot be done in the Standard Model of particle physics.%
\footnote{%{\dbltinyspace}%
The vanishing cosmological constant 
helps to realize %will make 
supersymmetry (SUSY) smoothly. %possible smoothly. 
However, 
we need more delicate and detailed discussions about this. 
}

The Coleman mechanism works as long as wormholes exist.%
\footnote{%{\dbltinyspace}%
This property is the same as the knitting mechanism. 
}
Therefore, it begins 
when the size of the universe becomes bigger than 
the size of wormholes{\dbltinyspace}%
\footnote{%{\dbltinyspace}%
Note that the size of wormholes is up to 
$1{\negtinyspace}/{\negdbltinyspace}\sqrt{\cc}$ 
because they expand hyperbolic tangentially 
according to \rf{THTexpansion}. 
}
and does not end %its role 
when the Big Bang occurs.
%Coleman mechanism はワームホールがあれば常に働くメカニズムで，
%しかも，ビッグバンの発生でその役割を終えることはない。
It works even after the Big Bang and even now. 
%ビッグバン発生後も，そして今現在も働くのである。
For example, 
%たとえば，
SSB of ${\rm SU}(2) \!\times\! {\rm U}(1)$ changes 
the vacuum energy, 
and it gives a nonzero cosmological term, 
but the Coleman mechanism returns it to zero. 
%電弱相互作用の対称性の破れによるSSBが起きたときは，
%真空のエネルギーは変化するが，これは非零の宇宙項になるため，
%Coleman機構により宇宙項は零に戻る。
The Coleman mechanism works 
not only on SSB, but also on all phenomena 
that make the cosmological term nonzero, 
and makes the cosmological term zero in the end.%
%Coleman機構は，SSBに限らず，宇宙項を非零にするような現象すべてに働き，
%宇宙項を零にするのである。%
\footnote{%{\dbltinyspace}%
\red{Dark energy cannot exist according to the Coleman mechanism.}
%Coleman mechanism denies the existence of dark energy.
%したがって，ダークエネルギーを否定する。
}

\red{Let us end this section with the following remark. 
If our model is equivalent to string theory, 
%it is guaranteed that 
high gauge symmetry like GUT/SUSY exists in the theory 
%GUTのような高いゲージ対称性が理論の背景にあることは弦理論により保証される。
and 
the gauge symmetry starts from nothing and will only grow higher. 
However, 
if the energy density of the universe does not reach 
to that realized by the GUT/SUSY symmetry, 
GUT/SUSY does not appear and then any consequences triggered 
by the spontaneous symmetry breaking (SSB) of GUT/SUSY 
will be irrelevant in a cosmological context. 
For instance 
there will then not be the cosmological monopole problem.} 

%By the way, 
%if our model is equivalent to string theory, 
%%it is guaranteed that 
%high gauge symmetry like GUT exists behind this theory 
%%GUTのような高いゲージ対称性が理論の背景にあることは弦理論により保証される。
%and 
%the gauge symmetry starts from nothing and grows higher. 
%However, 
%if the energy density of the universe does not reach 
%to that realize the GUT symmetry, 
%GUT does not appear and then 
%residues triggered by the spontaneous symmetry breaking (SSB) of GUT 
%such as monopoles does not. 
%
%この際，
%宇宙のエネルギー密度がGUTのエネルギー密度を超えないとき，
%または，
%一瞬だけしか超えないときは，
%GUTの対称性が現れない。
%In this case, 
%the symmetry breaking of GUT does not occur, 
%and residues such as monopoles does not appear.%
%それゆえ，その後，宇宙の温度が低くなったとしても，
%GUTの対称性の破れが起きることはなく，
%モノポールなどの残留物は現れない。
%\footnote{%{\dbltinyspace}%
%On the other hand, 
%since symmetry gradually increases from emptiness, 
%there is no phenomenon that 
%a large amount of monopoles will be generated 
%when the high gauge symmetry is broken
%in this case. 
%, 
%if the matter energy density does not reach the energy density 
%at which high gauge symmetry is realized, 
%and even if it is reached, 
%there is no time to realize high gauge symmetry. 
%%if the energy density begins to decrease 
%%before high gauge symmetry is realized.%
%%その一方，対称性は無から徐々に高くなってゆくので，
%%高い対称性が実現するまえにエネルギーが低くなると，
%%対称性の破れの際にモノポールが多量発生するという現象は現れない。
%}

\section{Modified Friedmann Equation}
\label{sec:MFE}

\subsection{Expansion of our Universe}
\label{sec:ExpandingUniv}

When $g$ is very small, 
we can ignore the splitting and merging of universes in principle. 
%$g$が非常に小さいときは，宇宙の分離融合を無視することができる。
However, 
the coupling constant $g$ plays an important role 
in the knitting mechanism and the Coleman mechanism 
even if it is a small value. 
In this {\Section}, we will show another phenomenon 
in which such a small effect of $g$ appears 
\red{in a surprising way.}%
%slightly but directly.
%\cite{ModFriedmannEqAW,ModFriedmannEqHubbleConstAW}
\footnote{%{\dbltinyspace}%
In previous {\Chapter}, 
%前sectionでは，
except for the special effects, 
the knitting mechanism and the Coleman mechanism, 
%KMやCMのような例外的な効果を除けば，
the assumption ``small nonzero $|g|$'' 
makes us possible to ignore 
the interactions of three universes. 
%``small nonzero $|g|$'' という仮定は３宇宙の相互作用を無視することに成功した。
However, 
this interaction is fundamental 
from the viewpoint of the theory definition \rf{WJgravityDefW}. 
%しかし，無視をした３宇宙の相互作用は，
%理論の定義\rf{WJgravityDefW}から明らかなように，
%本来は基本的なものである。
Thus, we hope to proceed our study to the phenomena by $\Hop_{\rm int}$. 
%そこで，$\Hop_{\rm int}$による３宇宙の相互作用に立ち入りたいのだが，
But before doing this study, 
we study the effect by 
the interaction appeared in $\Hop_{\rm kin}$, i.e.\ 
the term which is proportional to $g$ in $\bar{\Hop}_{\rm kin}$ 
in this {\Chapter}. 
%その前に，$\Hop_{\rm kin}$の中に現れる３宇宙の相互作用，つまり，
%$g$に比例する項の物理的な効果を調べてみようというのがこのsectionの目的である。
}
%
%Let us consider a physical phenomenon 
%in which such a small effect of $g$ appears slightly. 
%そんな小さな$g$の効果が僅かに表れる物理現象を考えてみよう。
This {\Section} is based on refs.\ 
\cite{CDTandHoravaGravity,ModFriedmannEqAW,ModFriedmannEqHubbleConstAW}.

\subsubsection{Derivation of a Modified Friedmann Equation}
\label{sec:MFEderivation}

%\subsubsubsection{Two-dimensional modified Friedmann equation}

Here, we restore the lapse function $N(t)$ of metric 
that has been integrated out by the path integral.%
%ここでは，経路積分によって積分してしまった計量の
%lapse関数$N(t)$を復活させる。%
\footnote{%{\dbltinyspace}%
Let us remember that the matter fields are not non-existent, 
but path-integrated out. 
%物質場は存在しないのではなく，
%path-integral out されていることを思い出そう。
}
Then, 
%すると，
using the Hamiltonian 
$\bar{\Hop}_{\rm kin}$ \rf{WalgebraHamiltonianKineticTerm}
which is common to all flavors, 
we get the following classical Hamiltonian.%
\footnote{%{\dbltinyspace}%
Note that 
the Hamiltonian \rf{ClassicalHamiltonian} 
does not have the spatial curvature term. 
The fact that the universe is flat 
is consistent with 
the fact that the space created by the knitting mechanism is a torus. 
Algebraic structure and geometric structure are consistent. 
}
%全てのフレーバーで共通な構造の
%$\bar{\Hop}_{\rm kin}$ \rf{WalgebraHamiltonianKineticTerm}より，
%これに相当する古典的なHamiltonianとして
\begin{equation}\label{ClassicalHamiltonian}
\Hc  \,=\, 
N L {\tinyspace}\Big(
 \Pi^2 - \mu + \frac{2 g}{\Pi}
{\dbltinyspace}\Big)
\,,
\qquad
%\end{equation}
%where
%\begin{equation}
\Pbracket{{\negtinyspace}L}{\Pi{\tinyspace}} \,=\, 1
%\qquad
%\commutator{L}{L} \,=\, 
%\commutator{\Pi}{\Pi} \,=\, 0
\,,
\end{equation}
%を得る。
where 
$\Pbracket{{\trpltinyspace}}{{\trpltinyspace}}$ 
is the Poisson bracket. 
%$N(t)$ is the lapse function. 
$L(t)$ is the length of the 1D universe 
before the dimension enhancement, 
and is a physical quantity representing the length scale 
after the dimension enhancement. 
%$L(t)$は編み上げ機構が働く前なら1D universe で，
%編み上げ機構が働いたのちは長さのスケールを表す物理量である。
$\Pi(t)$ is the conjugate momentum of $L(t)$. 
The derivative of $\Hc$ \rf{ClassicalHamiltonian} 
with respect to $N(t)$ %, i.e.\ 
%$\frac{\partial{\dbltinyspace} \Hc}{\partial{\halftinyspace} N(t)}
% {\negdbltinyspace}={\negdbltinyspace} 0$
gives the Friedmann equation
\begin{equation}\label{ModifiedFriedmannEquation0}
\Pi^2 - \mu + \frac{2 g}{\Pi}
\,=\, 0
\,.
\end{equation}
This corresponds to 
$\Hc {\negtrpltinyspace}={\negtrpltinyspace} 0$ 
and means 
$\Pi(t)$ is constant. 
The equations of motion derived from 
%the classical Hamiltonian 
$\Hc$ \rf{ClassicalHamiltonian} are 
\begin{eqnarray}
\dot{L} &=& 
-\,\Pbracket{{\negtinyspace}\Hc}{L{\tinyspace}}
\,=\,
2 N {\negtinyspace} L {\dbltinyspace}
\Big( \Pi - \frac{g}{\Pi^2} \Big)
\,,
\label{ClassicalEquationL}
\\
\dot{\Pi} &=& 
-\,\Pbracket{{\negtinyspace}\Hc}{\Pi{\tinyspace}}
\,=\,
-\,N {\tinyspace}\Big(
 \Pi^2 - \mu + \frac{2 g}{\Pi}
{\dbltinyspace}\Big)
\,.
\label{ClassicalEquationPi}
\end{eqnarray}
The first equation \rf{ClassicalEquationL} becomes 
\begin{equation}\label{CubicEquationF}
F^2 - F^3 \,=\, x
\,,
\end{equation}
where
\begin{equation}\label{FunctionsFxH}
F  \,\define\,  \frac{2 {\tinyspace} \Pi}{H}
\,,
\qquad
x \,\define\, -{\dbltinyspace}\frac{8 {\tinyspace} g}{H^3}
\,,
\qquad
H  \,\define\,  \frac{\dot{L}}{N L}
\,.
\end{equation}
%\rf{CubicEquationF} tells us that 
%$F$ is a function of $x$.
%\rf{FunctionsFxH} tells us that 
%$x$ and $\Pi$ are functions of $H$.
%Therefore, \rf{ModifiedFriedmannEquation0} means 
%$\Pi$ and $H$ are constant. 
%Eq.\ \rf{ModifiedFriedmannEquation0} gives $\Pi$ is constant.  
%So, it is consistent with the second equation \rf{ClassicalEquationPi}. 
The second equation \rf{ClassicalEquationPi} gives 
no new information 
because $\Pi(t)$ is the solution of \rf{ModifiedFriedmannEquation0}. 
\rf{CubicEquationF} means $F$ is the function of $x$, i.e.\ $F(x)$. 
Eq.\ \rf{ModifiedFriedmannEquation0} is rewritten as 
\begin{equation}\label{ModifiedFriedmannEquation}
H^2
\,=\,
4 \mu
+ \frac{x H^2 \big( 1 + 3 {\tinyspace} F(x) \big)}
       {\big( F(x) \big)^{{\negtinyspace}2}}
\,.
\end{equation}
%By the knitting mechanism, 
%the 2D Friedmann equation \rf{ModifiedFriedmannEquation} 
%will be unchanged. 
%So, \rf{ModifiedFriedmannEquation} becomes 
%the 4D Friedmann equation. 

By the knitting mechanism, 
the tiny wormholes which connect 
every point of 1D universes with different flavors 
give the cosmological constant $\Lambda$ 
to every point of high-dimensional space. 
By the Coleman mechanism, 
the cosmological constant $\Lambda$ will be replaced by 
matter energy density $\rho(t)$,
%\footnote{%{\dbltinyspace}%
%The energy conservation law is assumed here. 
%As a detailed mechanism, it is thought that 
%some field %such as the dilaton field 
%transfers energy from the cosmological term, i.e.\ 
%the vacuum energy to matter energy. 
%Mechanisms which extracts energy from the vacuum, 
%such as blackbody radiation, 
%interaction by through dilaton field, %and so on, 
%are thought to work. 
%}
i.e.\ 
\begin{equation}\label{ReplacementbyKMandCM}
4 \mu
\,\rightarrow\,  
\frac{\Lambda}{3}
\,\rightarrow\,  
\frac{\kappa \rho(t)}{3}
\quad\mbox{in \rf{ModifiedFriedmannEquation}}
\,,
\end{equation}
where $\kappa {\negtrpltinyspace}={\negtrpltinyspace} 8 \pi G$. 
($G$ is Newton constant.)
Then, 
the replacement \rf{ReplacementbyKMandCM} changes 
the Friedmann equation \rf{ModifiedFriedmannEquation} 
into 
the Friedmann equation in higher-dimensional space, 
\begin{equation}\label{ModifiedFriedmannEquation4D}
H^2
\,=\,
\frac{\kappa \rho}{3}
+ \frac{x H^2 \big( 1 + 3 {\tinyspace} F(x) \big)}
       {\big( F(x) \big)^{{\negtinyspace}2}}
\,.
\end{equation}
Finally, 
taking the $N(t){\negtrpltinyspace}={\negtrpltinyspace}1$ gauge, 
the Hubble parameter $H(t)$ in \rf{FunctionsFxH} becomes 
$H{\negtrpltinyspace}={\negtrpltinyspace}\dot{L}/L$, 
and then 
the equation \rf{ModifiedFriedmannEquation4D} becomes 
a kind of 
``the modified Friedmann equation''.%
%that will be used in the rest of this {\Article}.%
%最後に$N(t)=1$ gauge を取ると，$H=\dot{L}/L$ となり，
%\rf{ModifiedFriedmannEquation4D}は以後で扱う
%modified Friedmann equation となる。
\footnote{%{\dbltinyspace}%
Note that 
%the spatial curvature of the universe is zero 
the spatial geometry of the universe is flat 
in the modified Friedmann equation \rf{ModifiedFriedmannEquation4D}. 
%The fact that the universe is flat 
%is consistent with 
%the fact that the space created by the knitting mechanism is a torus. 
%because the shape of the universe is toroidal in this model. 
}
If the second term on the rhs of \rf{ModifiedFriedmannEquation4D} 
is replaced by $\Lambda/3$, 
the equation becomes the standard Friedmann equation with 
the cosmological constant $\Lambda$. %and zero spatial curvature of universe. 

\subsubsection{Accelerating Expansion of our Universe}
\label{sec:ExpandingUnivAcceleratingExpansion}

We examine the model which obeys 
the modified Friedmann equation \rf{ModifiedFriedmannEquation4D}, i.e.\ 
\begin{equation}\label{ModifiedFriedmannEquationRealModel}
\Omega_{\rm m}
{\tinyspace}+{\tinyspace}
\Omega_B
\,=\,
1
\,,
\qquad\quad
\Omega_{\rm m}
{\dbltinyspace}\define{\dbltinyspace}
\frac{\kappa \rho}{3 H^2}
\,,
\quad
\Omega_B
{\dbltinyspace}\define{\dbltinyspace}
\frac{x {\tinyspace} \big( 1 + 3 {\tinyspace} F(x) \big)}
     {\big( F(x) \big)^{{\negtinyspace}2}}
\,,
\end{equation}
where 
\begin{equation}\label{ModifiedFriedmannEquationHxB}
H  \,\define\,  \frac{\dot{L}}{L}
\,,
\qquad
x \,\define\, \frac{B}{H^3}
\,,
\qquad
B \,\define\, -{\dbltinyspace}8 g
\,.
\end{equation}
$x$ has to be less than or equal to 4/27 and 
one has to choose the solution $F(x)$ 
to the third order equation \rf{CubicEquationF} 
which is larger than or equal to 2/3. 
%We here set $N {\negdbltinyspace}={\negdbltinyspace} 1$. 
The modified Friedmann equation has 
only one physical constant $B$. 

\begin{figure}[t]
\vspace{1mm}\hspace{-0mm}
%\vspace{-90pt}
%\centerline{\includegraphics{GreenFun.ps}}
  \begin{center}
    \includegraphics[width=100pt]{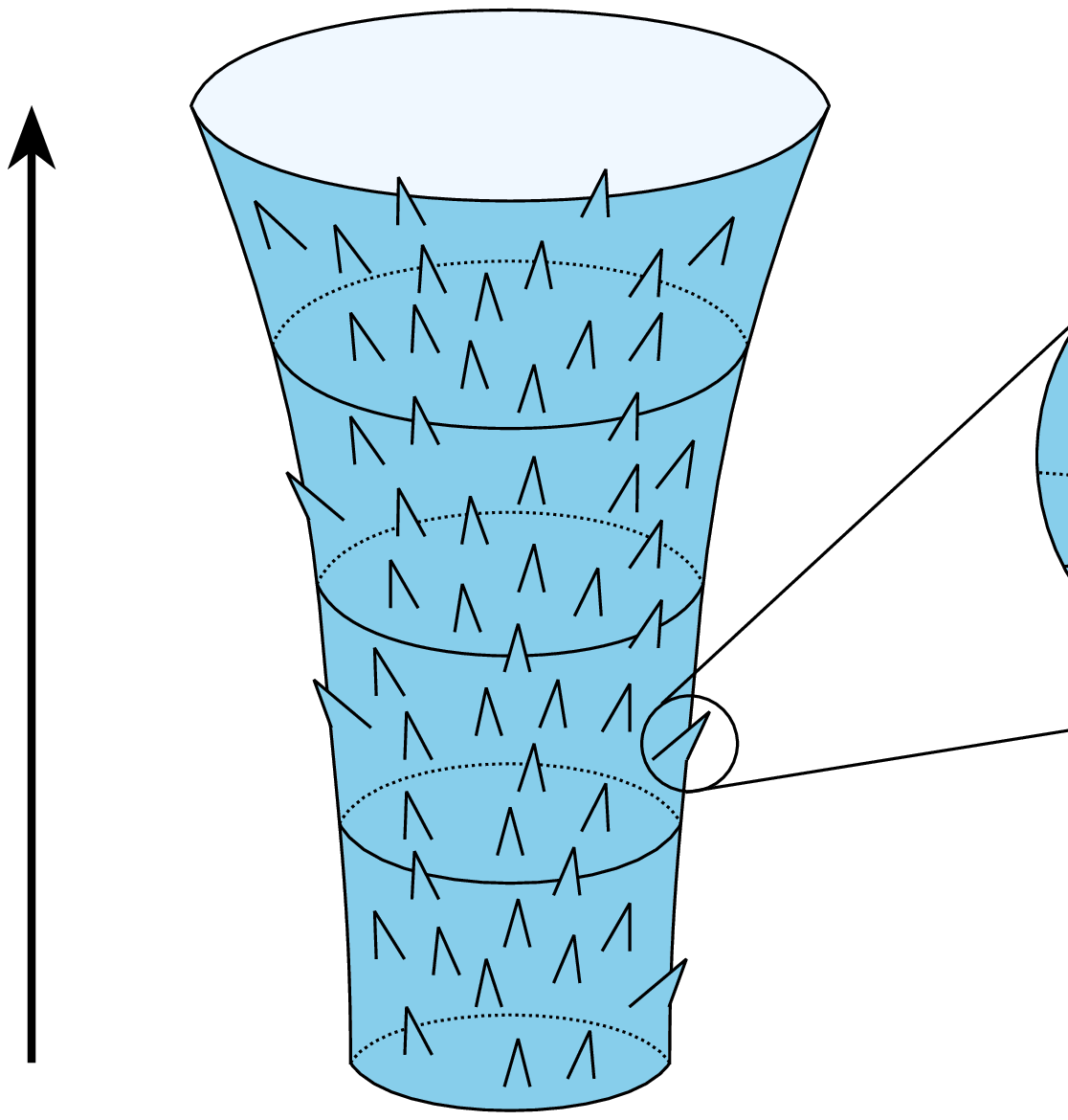}
  \end{center}
  \begin{picture}(200,0)
    \put(113, 35){\mbox{\footnotesize$\T$}}
    \put(221, 61){\mbox{\footnotesize$F^{\rm (baby)}{\negdbltinyspace}(L{\tinyspace};V)
                                      = \delta(V)$}}
    %\put(268, 72){\mbox{\tiny$\phi^\dagger_0 = 1$}}
    \put(227, 40){\mbox{\footnotesize$g = - \frac{B}{8}$}}
  \end{picture}
\vspace{1mm}
\caption[AccelExpandUniv]{%
A typical configuration of Porcupinefish expansion by baby universes. 
(Even if the space is a high-dimensional space 
formed by the knitting mechanism, 
each spine represents the production of a 1D baby universe 
with the amplitude 
$F^{\rm(baby)}{\negdbltinyspace}(L{\tinyspace};V)
 {\negtrpltinyspace}={\negtrpltinyspace}
 \delta(V)$. 
%たとえ空間が編み上げ機構後の高次元空間であったとしても，
%それぞれの棘は1次元空間の baby universe の生成になる。
%(Note that the positive $g$ gives the decelerating expansion of universe. 
%The shape of the expanding universe is similar to \lq\lq Porcupinefish".
%ハリセンボン%
%パイナップル%
%cactus（サボテン）%
One can see that 
the phenomenon of accelerating/decelerating expansion of the universe 
appears uniformly in space.%
)}
\label{fig:AccelExpandUniv}
\vspace{-2mm}
%\vspace{-7mm}
\end{figure}

The modified Friedmann equation 
\rf{ModifiedFriedmannEquation4D}-%
\rf{ModifiedFriedmannEquationHxB} 
describes the expansion of the universe, 
but before doing this analysis 
let us consider the geometric meaning of each term in this equation. 
The first and second terms on the rhs of \rf{WalgebraHamiltonianKineticTerm}, 
i.e.\ 
the fourth and fifth terms on the rhs of \rf{pureCDT_HamiltonianModeExpansion} 
come from 
the second term in the parenthesis of 
\rf{pureGravity_HamiltonianLaplaceTransf}. 
Then, 
it turns out that 
$\Pi^2 {\negdbltinyspace}-{\negdbltinyspace} \cc$ 
in \rf{ClassicalHamiltonian} 
correspond to 
$\tHkin{\negtinyspace}(\xi{\tinyspace};\cc)$
\rf{CDT_PropagatorLaplaceTransf}.%
\footnote{\label{footnote:StdFriedmannEq}%{\dbltinyspace}%
%Note that 
%$\Pi^2 {\negdbltinyspace}-{\negdbltinyspace} \cc$ 
%in \rf{ModifiedFriedmannEquation0} 
%comes from 
%$\Pi^2 {\negdbltinyspace}-{\negdbltinyspace} \cc$ 
%in \rf{ClassicalHamiltonian} 
%and then %comes from 
%the equivalent terms in 
%\rf{WalgebraHamiltonianKineticTerm}, 
%and the origin of these terms is 
%$\tHkin{\negtinyspace}(\xi{\tinyspace};\cc)$
%\rf{CDT_PropagatorLaplaceTransf}. 
%Thus, 
The propagator 
\rf{CDT_PropagatorLaplaceTransf} 
is related to the standard 
Friedmann equation 
without matter and with cosmological constant $\cc$ 
in this way. 
}
Also note that both $\xi$ and $\Pi$ are conjugate momentums of $L$. 
Since 
$\tHkin{\negtinyspace}(\xi{\tinyspace};\cc)$
\rf{CDT_PropagatorLaplaceTransf} 
comes from the local geometric structure of 2D spacetime \cite{CDTkin}, 
and so does 
$\Pi^2 {\negdbltinyspace}-{\negdbltinyspace} \cc$ 
in \rf{ClassicalHamiltonian}. 
On the other hand, 
the third term on the rhs of \rf{WalgebraHamiltonianKineticTerm}, 
i.e.\ 
the sixth term on the rhs of \rf{pureCDT_HamiltonianModeExpansion} 
comes from 
the third term in the parenthesis of 
\rf{pureGravity_HamiltonianLaplaceTransf}. 
$2 g / \Pi$ in \rf{ClassicalHamiltonian} 
comes from a different geometric origin, 
and is {\it the quantum effect of quantum gravity}.
Using 
$\phi^\dagger_0 {\negdbltinyspace}={\negdbltinyspace} 1$
\rf{Omega1CDTvalue}, 
the third term on the rhs of \rf{WalgebraHamiltonianKineticTerm} 
is rewritten as 
%\rf{WalgebraHamiltonianKineticTerm}の第3項は，
%$\phi^\dagger_0 {\negdbltinyspace}={\negdbltinyspace} 1$
%\rf{Omega1CDTvalue}を利用すると，
\begin{equation}\label{PorcupinefishTerm}
-\, 2 g %{\tinyspace} \phi^\dagger_0
    {\negdbltinyspace}
    \sum_{\ell=3}^\infty \phi^\dagger_{\ell-2} {\tinyspace} \ell \phi_\ell
\,=\,
-\, 2 g {\tinyspace} \phi^\dagger_0
    \sum_{\ell=3}^\infty \phi^\dagger_{\ell-2} {\tinyspace} \ell \phi_\ell
\,,
\qquad
\hbox{[\,$\phi^\dagger_0 {\negtrpltinyspace}={\negtrpltinyspace} 1$\,]}
\,.
\end{equation}
%と書き直すことができる。
Since 
$\phi^\dagger_0$
appears in %the first term 
$\Omega_1{\negtrehalftinyspace}(\xi)$ which is the first term of 
the wave function $\tilde\Psi^\dagger{\negdbltinyspace}(\xi)$ 
\rf{CDT_CreationAnnihilationOperatorModeExpansion}, 
the term \rf{PorcupinefishTerm} causes time evolution 
with branching of baby universes as shown in 
Fig.\ \ref{fig:AccelExpandUniv}. 
%$\phi^\dagger_0$は
%disk amplitude \rf{CDT_DiskAmpConcreteExpressionExpansion}の右辺第1項なので，
%\rf{PorcupinefishTerm}は
%図\ref{fig:AccelExpandUniv}のような
%baby universe の分岐を持つ時間発展を引き起こす。
%As was mentioned in footnote \ref{footnote:CDTdiskamp}, 
The first term $\xi^{{\tinyspace}-1}$ of 
the disk amplitude $\tilde{F}_1^{(0)}{\negdbltinyspace}(\xi;\cc)\big|_{g=0}$
\rf{CDT_DiskAmpConcreteExpressionExpansion} 
becomes $\delta(V)$ 
after taking the inverse Laplace transformation 
with respect to $\xi$ and $\mu$.%
\footnote{%{\dbltinyspace}%
Note also that 
the term $\xi^{{\tinyspace}-1}$ is the only term that 
destroys the square integrability.% property. 
}
Then, 
$F^{\rm(baby)}{\negdbltinyspace}(L{\tinyspace};V)
 {\negtrpltinyspace}={\negtrpltinyspace}
 \delta(V)$, 
which is a part of 
the disk amplitude $F_1^{(0)}{\negdbltinyspace}(L{\tinyspace};V)\big|_{g=0}$, 
is the amplitude of baby universe. 
Therefore, 
the branching of universe makes the size of universe smaller 
but matter fields and their energy do not flow into the baby universe 
because the baby universe has its size 
[\,$L {\negtrpltinyspace}\neq{\negtrpltinyspace} 0$\,]
but no volume 
[\,$V {\negtrpltinyspace}={\negtrpltinyspace} 0$\,]. 
Then, 
$2 g / \Pi$
in \rf{ClassicalHamiltonian}
%the third term on the rhs of \rf{WalgebraHamiltonianKineticTerm} 
causes a decelerating expansion of the universe. 
%分岐は宇宙を小さくするので，
%\rf{WalgebraHamiltonianKineticTerm}の第3項は
%宇宙の減速膨張を引き起こす。
Namely, 
%つまり，
if 
$g {\negdbltinyspace}={\negtrpltinyspace}
 - \frac{B}{8} {\negdbltinyspace}>{\negdbltinyspace} 0$, 
the expansion of the universe decelerates. 
%なら宇宙の膨張速度は減速するのである。
Conversely, 
%逆に考えれば，
if 
$g {\negdbltinyspace}={\negtrpltinyspace}
 - \frac{B}{8} {\negdbltinyspace}<{\negdbltinyspace} 0$, 
the conclusion is the opposite, that is, 
the expansion of the universe accelerates. 
%ならその逆，つまり，宇宙の膨張速度は加速することになる。
Since 
the production of baby universes 
is independent of 
the interactions by wormholes, 
the accelerating/decelerating expansion 
is independent of 
the vanishing cosmological constant which is caused by the Coleman mechanism. 
%%%
%%%

%From now on, 
In the rest of this {\Article} 
we always assume the cold dark matter (CDM) for all models. 
Let us call 
the model which obeys 
the standard Friedmann equation 
with the cosmological constant $\Lambda$ 
and zero spatial curvature of the universe 
``the $\Lambda$CDM model'', 
and also 
call the $\Lambda$CDM model which is created by the CMB data 
observed by the Planck satellite \lq\lq the Planck $\Lambda$CDM model". 
The list of input constants is %as follows:
\begin{eqnarray}
t_0^{({\rm CMB})}
&=&
13.8 %\times 10^{9}
\ [\,{\rm Gyr}\,]
\,,
\label{PresentTimePlanck}
\\
H_0^{({\rm CMB})}
&=&
67.3 \pm 0.6
\ [\,{\rm km}\,{\rm s}^{-1}\!\;{\rm Mpc}^{-1}\,]
\,,
\label{HubbleConstantPlanck}
\\
z_{{\tinyspace}\rm LS}^{({\rm CMB})}
&=&
1089.95
\,,
\label{RedShiftAtLastScatteringPlanck}
\end{eqnarray}
where 
$t_0$, 
$H_0 {\negtrpltinyspace}\define{\negtrpltinyspace} H(t_0)$ 
and 
$z_{{\tinyspace}\rm LS}$
are
the present time, 
the Hubble parameter at $t_0$ 
and 
the redshift at $t_{{\tinyspace}\rm LS}$. 
$t_{{\tinyspace}\rm LS}$ is 
the time of last scattering of CMB. 
The superscript ``(CMB)'' 
means the CMB data observed by 
the Planck satellite \cite{planck}. %[{\tinyspace}Planck 2018]. 
Let $L_\Lambda(t)$ and $H_\Lambda(t) \define \dot{L}(t)/L(t)$ be 
the scale length and the Hubble parameter 
in $\Lambda$CDM model, respectively. 
$t_{{\tinyspace}\rm LS}^{({\rm CMB})}$ and $\Lambda^{({\rm CMB})}$ 
are determined by{\dbltinyspace}%
\footnote{%{\dbltinyspace}%
In the determination of $t_{{\tinyspace}\rm LS}^{({\rm CMB})}$ 
the effect from radiation cannot be ignored. 
However, one can neglect this effect 
because this effect %is small and 
can be incorporated by shifting the origin of time $t$. 
%
%Calculations ignoring the radiation field are sufficient here.
%ここでは輻射場を無視した計算で十分である。
%When the radiation field is considered, 
%the origin of the time $t$ when ignoring the radiation field is shifted, 
%but this shift is very small compared to $t_0$, 
%so the value of $t_0$ does not change. 
%輻射場を考慮するときは輻射場を無視したときの
%時刻$t$の原点をずらすことになるが，
%このずれは$t_0$に比べて非常に小さいため$t_0$の値を変えることはない。
}
\begin{equation}\label{LastScatteringRedShift_HubbleConst_CMB}
\frac{L_{\Lambda^{({\rm CMB})}}{\negdbltinyspace}(t_0^{({\rm CMB})})}
     {L_{\Lambda^{({\rm CMB})}}{\negdbltinyspace}(t_{{\tinyspace}\rm LS}^{({\rm CMB})})}
\,=\,
1 + z_{{\tinyspace}\rm LS}^{({\rm CMB})}
\,,
\qquad\quad
H_{\Lambda^{({\rm CMB})}}{\negdbltinyspace}(t_0^{({\rm CMB})})
\,=\,
H_0^{({\rm CMB})}
\,.
\end{equation}

Here, 
we use another input parameter obtained by direct observation, 
\begin{eqnarray}\label{HubbleConst_SC}
H_0^{({\rm SC})}
&=&
73.0 \pm 1.0 \, [\,{\rm km}\,{\rm s}^{-1}\!\;{\rm Mpc}^{-1}\,]
\,.
\end{eqnarray}
%where 
%$H_0^{({\rm SC})}$ 
%is
%the Hubble parameter at the present time $t_0$ 
The superscript ``(SC)'' 
means the data observed by 
using the Standard Candles %Cosmic Distance Ladder 
(SC) \cite{latestshoes}.
%[{\tinyspace}SH0ES 2021]. 
%
If we assume 
$z_{{\tinyspace}\rm LS}^{({\rm CMB})}\!$, 
$t_{{\tinyspace}\rm LS}^{({\rm CMB})}$ and $H_0^{({\rm SC})}$, 
then, in the case of $\Lambda$CDM model, 
$t_0^{({\rm SC})}$ and $\Lambda^{({\rm SC})}$ are determined by 
\begin{equation}\label{LastScatteringRedShift_HubbleConst_SC}
\frac{L_{\Lambda^{({\rm SC})}}{\negdbltinyspace}(t_0^{({\rm SC})})}
     {L_{\Lambda^{({\rm SC})}}{\negdbltinyspace}(t_{{\tinyspace}\rm LS}^{({\rm CMB})})}
\,=\,
1 + z_{{\tinyspace}\rm LS}^{({\rm CMB})}
\,,
\qquad\quad
H_{\Lambda^{({\rm SC})}}{\negdbltinyspace}(t_0^{({\rm SC})})
\,=\,
H_0^{({\rm SC})}
\,.
\end{equation}
The result is 
\begin{eqnarray}\label{ValuePresentTimeLambda_SC}
%&&
t_0^{({\rm SC})}
\,=\,
13.31 \pm 0.05 \ [\,{\rm Gyr}\,]
\,,
%\\
\qquad\quad
%&&
\big( t_0^{({\rm SC})} \big)^{{\negtinyspace}2} {\tinyspace}
\Lambda^{({\rm SC})}
\,=\,
2.17 \pm 0.06
\,.
\end{eqnarray}
We call this model ``the late time $\Lambda$CDM model''. 
In the case of our model with CDM, 
$t_0$ and $B$ are determined by 
\begin{equation}\label{LastScatteringRedShift_HubbleConst}
\frac{L(t_0)}
     {L(t_{{\tinyspace}\rm LS}^{({\rm CMB})})}
\,=\,
1 + z_{{\tinyspace}\rm LS}^{({\rm CMB})}
\,,
\qquad\quad
H(t_0)
\,=\,
H_0^{({\rm SC})}
\,.
\end{equation}
The result is 
\begin{eqnarray}\label{ValuePresentTimeB}
%&&
t_0
\,=\,
13.89 \pm 0.06 \ [\,{\rm Gyr}\,]
\,,
%\\
\qquad\quad
%&&
t_0^3 {\tinyspace} B
\,=\,
0.149 \pm 0.006
\,.
\end{eqnarray}
Using the dynamical timescale $t_g$ %\rf{TimeScaleCCg} 
and the result \rf{ValuePresentTimeB}, 
the inequality \rf{gScaleCondition} becomes 
\begin{equation}\label{EffectiveLifetimeUniverse}
t_0 \,\lesssim\, t_g %|g|^{-1/3}
% \,\sim\, 3.77 {\dbltinyspace} t_0
 \,\sim\, 3.8 {\dbltinyspace} t_0
% \,\sim\, 52.4 \ [\,{\rm Gyr}\,]
 \,\sim\, 52 \ [\,{\rm Gyr}\,]
\,.
\end{equation}
It should be noted that 
the result \rf{ValuePresentTimeB} satisfies %is consistent with 
the inequality \rf{gScaleCondition}.%
\footnote{%{\dbltinyspace}%
Since the dynamical timescale $t_g$ in \rf{TimeScaleCCg} is obtained 
by the dimensional analysis, 
the inequality `$<$' of `$\lesssim$' 
is less important than 
the approximate equality `$\sim$' of `$\lesssim$' 
in this case. 
}
The fact that 
the splitting and merging of the universe has not been observed 
%from the big bang to the present 
leads to a very small constant $B$, 
which represents the accelerating expansion of the universe. 
%現在までに宇宙の分裂が観測されていないという事実から
%宇宙の加速膨張を表す定数$B$の大きさが非常に小さいことが
%導かれるのである。
From the perspective of the {\it anthropic principle}, 
the fact that the value of $t_0$ %\rf{ValuePresentTimeB} 
satisfies %the inequality 
$t_0 {\negdbltinyspace}\lesssim t_g$ 
\rf{EffectiveLifetimeUniverse} 
suggests that 
human beings 
will encounter the era of $t_0 {\negdbltinyspace}\sim t_g$, but 
will find it difficult to survive beyond $t_g$. 
%Moreover, 
%it is natural that the scale of $t_0$ is comparable to that of $t_g$.
%There are observation data about $(z,H)$.
Fig.{\dbltinyspace}\ref{fig:HubbleConstantPresentTime} shows the graph of 
$t_0$-$H_0$. 
%Fig.{\dbltinyspace}\ref{fig:HubbleParameter} shows the graph of 
%$H(t)$. 
Fig.{\dbltinyspace}\ref{fig:HubbleParameterOverZ} shows the graph of 
$H(z) / (1 {\negdbltinyspace}+{\negdbltinyspace} z)^{3/2}$.%
\footnote{%{\dbltinyspace}%
See ref.\ \cite{ModFriedmannEqHubbleConstAW} 
for the list of observation data 
used in Fig.{\dbltinyspace}\ref{fig:HubbleParameterOverZ}. 
}
The chi-square $( \chi_{\rm red} )^2$ of 
$H(z) / (1 {\negdbltinyspace}+{\negdbltinyspace} z)^{3/2}$ 
for each graph is{\dbltinyspace}%
\footnote{%{\dbltinyspace}%
%The definition of $( \chi_{\rm red} )^2$ is
$( \chi_{\rm red} )^2 \define
 \sum_{i=1}^N ( (x - x_i) / \sigma_i )^2 / N$, 
where $x_i$ are data with error bars $\sigma_i$. 
}
\begin{equation}\label{chisquareHz}
( \chi_{\rm red} )^2
= 1.8
\,,
\qquad
\big( \chi_{\rm red}^{\rm (SC)} \big)^2
= 3.5
\,,
\qquad
\big( \chi_{\rm red}^{\rm (CMB)} \big)^2
= 5.4
\,.
\end{equation}
Observation data seems to support our CDM model more than 
the $\Lambda$CDM model.
%Observation data seems to support our CDM model over the $\Lambda$CDM model.
\newcommand{\HubbleParameterTimeDependence}{%
\begin{figure}[b]
\vspace{-8mm}\hspace{-0mm}
%\begin{minipage}{.45\linewidth}\vspace{6mm}
%  \begin{center}\hspace*{0mm}
%    \includegraphics[width=1.0\linewidth,angle=0]
%                    {Graph_q_t.eps}
%  \end{center}
%  \begin{picture}(200,0)
%    \put(2.0,68.5){\mbox{\tiny$0$}}
%  \end{picture}
%  \vspace*{-10mm}
%  \caption[fig.1]{{$\tau${\dbltinyspace}-{\tinyspace}$%
%\big({\negqdrpltinyspace}-{\negdbltinyspace}q(\tau)\big)$}
%  }
%  \label{fig:AcceleratingUniverse}
%\end{minipage}
\begin{minipage}{.45\linewidth}\vspace{6mm}
  \begin{center}\hspace*{0mm}
    \includegraphics[width=1.0\linewidth,angle=0]
                    {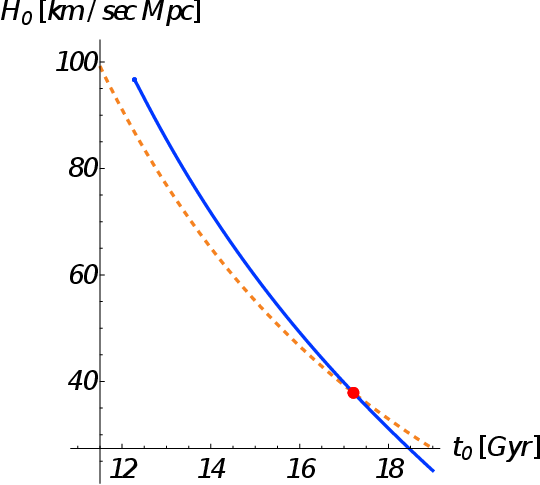}%eps}
  \end{center}
  \begin{picture}(200,0)
    \put(64.0,37.0){\mbox{\tiny
      $\Lambda{\negdbltinyspace}={\negdbltinyspace}B{\negdbltinyspace}={\negdbltinyspace}0$ model}}
  \end{picture}
  \vspace*{-7mm}
  \caption[fig.2]{%{$t_0$-$H_0$}
The blue curve is {$t_0$-$H_0$} 
for the modified Friedmann equation, 
while 
the orange dashed curve is {$t_0$-$H_0$} 
based on $\Lambda$CDM model. 
The red dot is {$t_0$-$H_0$} 
for Friedmann equation without $\Lambda$ and $B$. 
  }
  \label{fig:HubbleConstantPresentTime}
\end{minipage}
\hspace{6mm}
\begin{minipage}{.45\linewidth}\vspace{6mm}
  \begin{center}\hspace*{0mm}
    \includegraphics[width=1.0\linewidth,angle=0]
                    {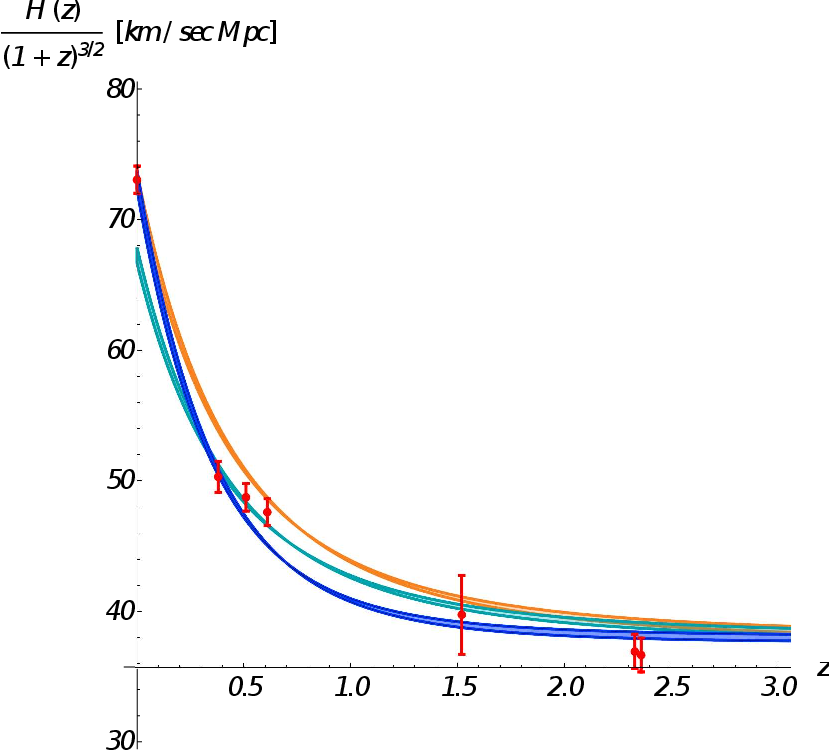}%eps}
  \end{center}
  \begin{picture}(200,0)
    %\put(-18.0,90.0){\mbox{\tiny$H_0^{\rm (SC)}$}}
  \end{picture}
  \vspace*{-8mm}
  \caption[fig.2]{%{$z${\tinyspace}-$\frac{H(z)}{(1+z)^{3/2}}$}
$z$ versus 
$H(z)/(1{\negdbltinyspace}+{\negdbltinyspace}z)^{3/2}$ 
for three models: 
our model (blue), 
the Planck $\Lambda$CDM model (green) 
and 
the late time $\Lambda$CDM model (orange). 
%The blue line shows 
%$H(z)/(1{\negdbltinyspace}+{\negdbltinyspace}z)^{3/2}$ 
%for the modified Friedmann equation, 
%while 
%the orange line shows 
%$H_{\Lambda^{({\rm SC})}}(z)/(1{\negdbltinyspace}+{\negdbltinyspace}z)^{3/2}$, 
%%based on $\Lambda$CDM model 
%%with $H_0^{\rm (CMB)}$ and $\Lambda^{\rm (CMB)}$ taken from \cite{planck}. 
%and 
%the green line shows 
%$H_{\Lambda^{({\rm CMB})}}(z)/(1{\negdbltinyspace}+{\negdbltinyspace}z)^{3/2}$.
The red dots with error bars 
are observation data. 

%based on $\Lambda$CDM model 
%with $H_0^{\rm (SC)}$ and $\Lambda^{\rm (SC)}$ as described in the main text.
  }
  \label{fig:HubbleParameterOverZ}
\end{minipage}
%\hspace{6mm}
%\begin{minipage}{.45\linewidth}\vspace{6mm}
%  \begin{center}\hspace*{0mm}
%    \includegraphics[width=1.0\linewidth,angle=0]
%                    {Graph_w_t.eps}
%  \end{center}
%  \vspace*{-0mm}
%  \caption[fig.2]{{$\tau${\dbltinyspace}-{\tinyspace}$w(\tau)$}
%  }
%  \label{fig:wParameter}
%\end{minipage}
\vspace{-4mm}
%\vspace{-4mm}
\end{figure}%\vspace*{0mm}
%%%
}\HubbleParameterTimeDependence

\newcommand{\ObservablesModelIndependence}{%
The Planck value of $\Omega_{\rm{m},0} H_0^2$ is entirely determined by physics before 
$t_{{\tinyspace}\rm LS}$ since we have 
\begin{equation}\label{omegaH2}
\frac{\Omega_{\rm m}{\negtinyspace}(t_{{\tinyspace}\rm LS}) \big( H(t_{{\tinyspace}\rm LS}) \big)^{{\negtinyspace}2}
    }{\big( 1 + z(t_{{\tinyspace}\rm LS}) \big)^{{\negtinyspace}3}
    }
\,=\,
\Omega_{\rm m}{\negtinyspace}(t_0) \big( H(t_0) \big)^{{\negtinyspace}2}
\,=\,
\Omega_{{\rm m},0} H_0^2.
\end{equation}
Our model will also satisfy eq.\ \rf{omegaH2} and since our value of $H(t_{{\tinyspace}\rm LS})$ by assumption agrees
with the Planck value at $z(t_{{\tinyspace}\rm LS})$ we will have the same value of $\Omega_{{\rm m},0} H_0^2$. However,
by construction we have  a larger 
value for $H_0$ than Planck collaboration since we used the $H_0$ measured at late time to determine 
our value of the constant $B$. Consequently we have a lower value 
of $\Omega_{{\rm m},0}$. 
}%\ObservablesModelIndependence

\newcommand{\OtherObservables}{%
Our model also gives 
\begin{equation}
\Omega_{\rm m}{\negtinyspace}(t_0)
=
0.27 \pm 0.01
\,,
%\\
\quad
-{\tinyspace}q(t_0)
%\define 
%\frac{\ddot{L}(t)}{\big( H(t) \big)^2 L(t)}\bigg|_{t=t_0}
=
0.82 \pm 0.01
\,.
%\\
\quad
w(t_0)
\,=\,
-1.206 \pm 0.006
\,,
\end{equation}
where 
\begin{equation}
-{\tinyspace}q(t) 
\define 
\frac{\ddot{L}(t)}
     {\big( H(t) {\negtinyspace}\big)^{{\negtinyspace}2} L(t)}
\,,
\qquad
w(t) 
\define 
-\,\frac{1 - 2 {\tinyspace} q(t)}
        {3 {\tinyspace} \Omega_B{\negtinyspace}(t)}
\,.
\end{equation}
}\OtherObservables

\subsection{Large-Scale Structure of our Universe}\label{sec:LSS}

%"real model" and "real-model independent"

In this {\Section} 
we study the large-scale structure of our universe. 
%predicted by the model which obeys the modified Friedmann equation 
%\rf{ModifiedFriedmannEquationRealModel}. 
%As with the previous {\Subsection}, 
%the constant $B$ can be ignored in the previous period of $t_{{\tinyspace}\rm LS}$. 
In the period before $t_{{\tinyspace}\rm LS}$, 
the influences of both 
$\Lambda$ in $\Lambda$CDM model and $B$ in our model are small, 
and not only that, but 
the differences in all physical observables between both models 
are so small that they can be ignored. 
We will use this property in this {\Section}.%
\footnote{%{\dbltinyspace}%
We have used this property in the previous {\Section}. 
$t_{{\tinyspace}\rm LS}$ and $z_{{\tinyspace}\rm LS}$ are such observables. 
%When $t {\negtrpltinyspace}<{\negtrpltinyspace} t_{{\tinyspace}\rm LS}$
%the influences of both $\Lambda$ and $B$ are very small, 
%and not only that, 
%but the difference between two models is so small 
%that it can be ignored.
}
This {\Section} is based on ref.\ 
\cite{ModFriedmannEqLargeScaleStructureAW}.

%\begin{itemize}
%\item
%Baryon Acoustic Oscillation (BAO) and $D_{\rm V}/r_{\rm s}$
%\item
%Redshift Space Distortions (RSD) and $f_{\rm m} \sigma_8$
%\end{itemize}

\subsubsection{Baryon Acoustic Oscillations (BAO)}

The baryon acoustic oscillations (BAO) are 
fluctuations in the baryon density 
caused by the baryon acoustic oscillation in the early universe. 
In BAO the inverse of the angular diameter distance 
$D_V{\negtinyspace}(z) / r_{\rm s}$ 
%\begin{equation}
%\frac{1}{\theta(z)} \,\define\, \frac{D_V{\negtinyspace}(z)}{r_{\rm s}}
%\end{equation}
is defined 
where 
\begin{equation}
D_V{\negtinyspace}(z)
 \define
\sqrt[3]{z {\dbltinyspace} D_H{\negtinyspace}(z) \big( D_M{\negtinyspace}(z) \big)^{2}}
\,,
\quad
z {\dbltinyspace} D_H{\negtinyspace}(z)
 \define \frac{%c {\dbltinyspace} 
               z}{H(z)}
\,,
\quad
D_M{\negtinyspace}(z)
 \define 
%(1+z) D_A(z) \,=\,
\int_0^{{\tinyspace}z}\! %\dd z' {\tinyspace}
  \frac{%c {\dbltinyspace} 
        \dd z'}{H(z')}
\,.
\end{equation}
$D_V{\negtinyspace}(z)$ is a kind of average distances with three directions
%\begin{equation}
%z D_H{\negtinyspace}(z) \,\define\, \frac{c {\dbltinyspace} z}{H(z)}
%\,,
%\qquad\quad
%D_M{\negtinyspace}(z) \,\define\, %(1+z) D_A(z) \,=\,
%\int_0^{{\tinyspace}z}\! %\dd z' {\tinyspace}
%  \frac{c {\dbltinyspace} \dd z'}{H(z')}
%\,.
%%\qquad\quad
%%D_A(z) \,\define\,
%%\frac{1}{1+z} \!\int_0^{{\tinyspace}z}\! \dd z' {\tinyspace}
%%  \frac{c}{H(z')}
%\end{equation}
and 
$r_{\rm s}$ defined by 
%which is the end of baryon drag and considered before $t_{{\tinyspace}\rm LS}$
%[\,$t_{\rm drag} {\negdbltinyspace}<{\negdbltinyspace} t_{{\tinyspace}\rm LS}$\,], 
\begin{equation}\label{rsDef}
r_{\rm s}%(z_{\rm drag})
%%&=&
%%\frac{1}{1 + z(t_{\rm drag})}
%%\!\int_{z_{\rm drag}}^{{\tinyspace}\infty}\! \dd z {\dbltinyspace}
%%\frac{c_{\rm s}(z)}{H(z)}
{\dbltinyspace}\define{\dbltinyspace}
\int_0^{{\dbltinyspace}t_{\rm drag}}\!
  \dd{\trehalftinyspace}t {\dbltinyspace}
  \frac{L(t_{\rm drag})}{L(t)}
{\dbltinyspace} c_{\rm s}(t)
%%\nonumber\\&=&
%%\frac{2 {\tinyspace} c {\dbltinyspace} a_{\rm drag}}
%%     {a_0 \sqrt{3 R(t_0) {\tinyspace} \Omega_{{\rm m},0} H_0^2}}
%%\log\!\bigg(
%%  \frac{\sqrt{1 + R(t_{\rm drag})} + \sqrt{R(t_{\rm EQ}) + R(t_{\rm drag})}}
%%       {1 + \sqrt{R(t_{\rm EQ})}}
%%{\dbltinyspace}\bigg)
%\,,
%\qquad
%c_{\rm s}(t)
%{\dbltinyspace}\define{\dbltinyspace}
%\frac{c}{\sqrt{3 \big( 1 + \frac{3 {\tinyspace} \rho_{{\tinyspace}\rm b}(t)}
%                                {4 {\tinyspace} \rho_{\gamma}(t)} \big)}}
\end{equation}
is the comoving sound horizon at $t_{\rm drag}$, 
where $c_{\rm s}(t)$ is the speed of sound. 
%$\rho_{{\tinyspace}\rm b}(t)$ is the baryon density 
%and 
%$\rho_{\gamma}(t)$ is the photon density. 
%
%\begin{equation}
%\frac{\rho_{\rm m}{\negtinyspace}(t_{\rm EQ})}
%     {\rho_{\rm r}(t_{\rm EQ})}
%\,=\,
%\frac{\Omega_{\rm m}{\negtinyspace}(t_{\rm EQ})}
%     {\Omega_{\rm r}(t_{\rm EQ})}
%\,=\,
%1
%\end{equation}
%
Since $r_{\rm s}$ \rf{rsDef} is determined by 
the physical phenomena before $t_{\rm drag}$, i.e.\ before $t_{{\tinyspace}\rm LS}$, 
the value $r_{\rm s}$ of the model which obeys the modified Friedmann equation 
\rf{ModifiedFriedmannEquation4D}-%
\rf{ModifiedFriedmannEquationHxB}
%\rf{ModifiedFriedmannEquationRealModel} 
coincides with that of Planck $\Lambda$CDM model. 
So, 
using the data by Planck Mission, 
we have 
\begin{equation}\label{rsValueDrag}
r_{\rm s}
\,\sim\,
r_{\rm s}^{\rm (SC)}
\,\sim\,
r_{\rm s}^{\rm (CMB)}
\,=\,
147.05 \pm 0.30
\ [\,{\rm Mpc}\,]
\,.
\end{equation}

Fig.{\dbltinyspace}\ref{fig:DvRs} shows the graph of 
$D_V{\negtinyspace}(z) / r_{\rm s}$.%
\footnote{%{\dbltinyspace}%
See ref.\ \cite{ModFriedmannEqLargeScaleStructureAW} 
for the list observation data 
used in Fig.{\dbltinyspace}\ref{fig:DvRs}. 
}
%Fig.{\dbltinyspace}\ref{fig:DvRsDiff} shows the graph of 
%%$1/\theta$ 
%$D_V{\negtinyspace}(z) / r_{\rm s}$ 
%divided by the Planck $\Lambda$CDM graph 
%%$1/\theta^{\rm (CMB)}$. 
%$D_V^{\rm (CMB)}{\negdbltinyspace}(z) / r_{\rm s}^{\rm (CMB)}$. 
The chi-square $( \chi_{\rm red} )^2$ of 
%$1/\theta$ 
$D_V{\negtinyspace}(z) / r_{\rm s}$ 
for each graph is 
\begin{equation}\label{chisquareInvTheta}
( \chi_{\rm red} )^2
= 1.0
\,,
\qquad
\big( \chi_{\rm red}^{\rm (SC)} \big)^2
= 4.7
\,,
\qquad
\big( \chi_{\rm red}^{\rm (CMB)} \big)^2
= 1.6
\,.
\end{equation}
The reason why 
$\chi_{\rm red}^{\rm (CMB)}$
is very small compared to 
$\chi_{\rm red}^{\rm (SC)}$
is that 
the Planck Mission analysis 
ignores the Hubble constant 
$H_0^{({\rm SC})}$ %\rf{HubbleConst_SC}
but 
takes into account the fluctuation of CMB 
which is directly related to BAO. 
%$\chi_{\rm red}^{\rm (SC)}$に比べ
%$\chi_{\rm red}^{\rm (CMB)}$が非常に小さいのは，
%Planck mission の解析がHubble定数\rf{HubbleConst_SC}を無視し，
%BAOと直結するCMBの揺らぎを考慮したせいである。
On the other hand, in our model, 
%一方，我々の模型では，
although the redshift of CMB is taken into account, 
the effect of the fluctuation of CMB is ignored, 
and instead the Hubble constant 
$H_0^{({\rm SC})}$ %\rf{HubbleConst_SC} 
is taken into account. 
%but it is adjusted to the Hubble constant \rf{HubbleConst_SC} 
%ignoring the effect of the fluctuation of CMB. 
%CMBの赤方偏移は考慮するが，CMBの揺らぎの効果を無視して
%Hubble定数\rf{HubbleConst_SC}に合わせている。
Nevertheless, 
%それにもかかわらず，
the result of our model is better than that of Planck Mission. 
%Planck mission による解析結果よりも
%よりよい結果となっていることに注目されたい。
%%%%
%\begin{figure}[ht]
%\vspace{-4mm}\hspace{-0mm}
%\begin{minipage}{.45\linewidth}\vspace{6mm}
%  \begin{center}\hspace*{0mm}
%    \includegraphics[width=1.0\linewidth,angle=0]
%                    {Graph_Dv_z.eps}
%  \end{center}
%  \begin{picture}(200,0)
%    \put(1.2,23.0){\mbox{\tiny$0$}}
%  \end{picture}
%  \vspace*{-14mm}
%  \caption[fig.1]{{$z${\dbltinyspace}-{\tinyspace}$D_V{\negtinyspace}(z)/r_{\rm s}$}
%  }
%  \label{fig:DvRs}
%\end{minipage}
%\hspace{6mm}
%\begin{minipage}{.45\linewidth}\vspace{6mm}
%  \begin{center}\hspace*{0mm}
%    \includegraphics[width=1.0\linewidth,angle=0]
%                    {Graph_Dv_z_diff.eps}
%  \end{center}
%  \begin{picture}(200,0)
%    \put(4.6,50.0){\mbox{\tiny$0$}}
%  \end{picture}
%  \vspace*{-14mm}
%  \caption[fig.2]{{$z${\dbltinyspace}-{\tinyspace}$\frac{D_V{\negtinyspace}(z)/r_{\rm s}}{D_V^{\rm (CMB)}(z)/r_{\rm s}^{\rm (CMB)}}$}
%  }
%  \label{fig:DvRsDiff}
%\end{minipage}
%\vspace{-2mm}
%\end{figure}%\vspace*{0mm}
%%%%

\subsubsection{The growth of fluctuations in linear theory}
%\subsubsection{Redshift Space Distortions (RSD)}

The matter density fluctuation 
$\delta_{{\tinyspace}\rm m}{\negtinyspace}(\Bf{x},t)$ 
is defined by 
\begin{equation}\label{MatterDensityFluctuationDef}
\delta_{{\tinyspace}\rm m}{\negtinyspace}(\Bf{x},t) \,\define\, 
\frac{\rho_{\rm m}{\negtinyspace}(\Bf{x},t)}
     {\bar\rho_{\rm m}{\negtinyspace}(t)} - 1
\qquad\hbox{and}\qquad
\bar\rho_{\rm m}{\negtinyspace}(t) \,\define\,
\frac{\int\!\dd^3 x {\dbltinyspace}\rho_{\rm m}{\negtinyspace}(\Bf{x},t)}
     {\int\!\dd^3 x}
\,.
\end{equation}
%In the linear approximation, 
%the matter density fluctuation $\delta_{{\tinyspace}\rm m}{\negtinyspace}(\Bf{x},t)$ 
%obeys the differential equation 
%%(See eq.(8.1.12) in Weinberg's book.)
%\begin{equation}\label{MatterDensityFluctuationDiffeq}
%\frac{\dd}{\dd t}\Big(
%  \big( L(t) \big)^2 \frac{\dd}{\dd t} \delta_{{\tinyspace}\rm m}{\negtinyspace}(\Bf{x},t)
%\Big)
%\,=\,
%\frac{\kappa}{2} \big( L(t) \big)^2
%  \bar\rho_{\rm m}{\negtinyspace}(t) {\tinyspace}
%  \delta_{{\tinyspace}\rm m}{\negtinyspace}(\Bf{x},t)
%{\dbltinyspace}+{\dbltinyspace}
%{\cal O}(B)
%\end{equation}
In the linear approximation, %Then, 
we have the differential equation of 
$\delta_{{\tinyspace}\rm m}{\negtinyspace}(\Bf{x},t)$ 
%growth factor $F_{\rm m}{\negtinyspace}(t)$ 
as
%is the solution of (See eq.(8.1.15) in Weinberg's book.)
\begin{equation}\label{MatterDensityFluctuationFtDef}
\frac{\dd}{\dd{\trehalftinyspace}t}\bigg(
  \big( L(t) \big)^2 \frac{\dd}{\dd{\trehalftinyspace}t}
  \delta_{{\tinyspace}\rm m}{\negtinyspace}(\Bf{x},t)
\Big)
\,\sim\,
\frac{\kappa}{2} \big( L(t) \big)^{{\negtinyspace}2}
\bar\rho_{\rm m}{\negtinyspace}(t) {\tinyspace} 
\delta_{{\tinyspace}\rm m}{\negtinyspace}(\Bf{x},t)
%{\dbltinyspace}+{\dbltinyspace} {\cal O}(B)
\,.
\end{equation}
%Here, we have assumed 
%the contribution from the production of baby universes is negligible. 
%So, we delete ${\cal O}(B)$ from \rf{MatterDensityFluctuationFtDef} 
%in later calculations. 
We here assume that 
$\delta_{{\tinyspace}\rm m}{\negtinyspace}(\Bf{x},t)$ is decoupled as 
%(See eq.(8.1.13) in Weinberg's book.)
\begin{equation}\label{MatterDensityFluctuationDecoupling}
\delta_{{\tinyspace}\rm m}{\negtinyspace}(\Bf{x},t) \,=\, F_{\rm m}{\negtinyspace}(t) \Delta_{\rm m}{\negtinyspace}(\Bf{x})
\,.
\end{equation}
%where $F_{\rm m}{\negtinyspace}(t)$ is the growth factor 
%%which represents the growth of fluctuation 
%and 
%$\Delta_{\rm m}{\negtinyspace}(\Bf{x})$ is the fluctuation at a certain time. 
The boundary conditions of the growth factor $F_{\rm m}{\negtinyspace}(t)$ are 
\begin{equation}\label{MatterDensityFluctuationFtCondition}
F_{\rm m}{\negtinyspace}(t_0) \,=\, 1
\qquad\hbox{and}\qquad
f_{\rm m}{\negtinyspace}(t_{{\tinyspace}\rm LS})
\,=\,
1
\,,
\quad\hbox{where}\ \ 
f_{\rm m}{\negtinyspace}(t) \define \frac{\dd \log F_{\rm m}{\negtinyspace}(t)}{\dd \log L(t)}
\,.
\end{equation}
%where 
%$f_{\rm m}{\negtinyspace}(t)$ is the growth rate defined by 
%\begin{equation}
%f_{\rm m}{\negtinyspace}(t) \,\define\, \frac{\dd \log F_{\rm m}{\negtinyspace}(t)}{\dd \log L(t)}
%\,.
%\end{equation}
%$t_{{\tinyspace}\rm LS}$ is the time of the last scattering of photons. 
%The first condition means we set the initial condition of fluctuation as 
%$\Delta_{\rm m}{\negtinyspace}(\Bf{x}) = \delta_{{\tinyspace}\rm m}{\negtinyspace}(\Bf{x},t_0)$.

Here, one introduces the average of fluctuation 
inside the sphere with radius $R$ as 
%(See eq.(8.1.44) in Weinberg's book.)
\begin{equation}\label{MatterDensityFluctuationAverage}
\sigma_R{\neghalftinyspace}(t)
\,\define\,
\sqrt{
\bigg<\!
  \bigg(
    \frac{1}{\frac{4}{3} \pi R^3}
    \!\!\int_{|\Bfscript{x}| < R}\!\!\!\dd^3 x {\dbltinyspace}
    \delta_{{\tinyspace}\rm m}{\negtinyspace}(\Bf{x},t)
  \bigg)^{{\negtrpltinyspace}2}
{\trpltinyspace}\bigg>
}
%
%\,=\,
%F_{\rm m}{\negtinyspace}(t)
%\sqrt{
%\bigg<\!
%  \bigg(
%    \frac{1}{\frac{4}{3} \pi R^3}
%    \!\!\int_{|\Bfscript{x}| < R}\!\!\!\dd^3 x {\dbltinyspace}
%    \Delta_{\rm m}{\negtinyspace}(\Bf{x})
%  \bigg)^{{\negtrpltinyspace}2}
%{\trpltinyspace}\bigg>
%}
\,.
\end{equation}
\newcommand{\sigmaRcal}{%
\rf{MatterDensityFluctuationAverage} is written as 
\begin{equation}\label{MatterDensityFluctuationAverage0powerspectrum}
\sigma_R{\neghalftinyspace}(t_0)
\,=\,
\sqrt{
  \int_0^\infty\frac{\dd k {\tinyspace} k^2}{2 \pi^2}
  \big( f(k R) \big)^{{\negtinyspace}2}
  P(k)
}
\end{equation}
where $f(k R)$ is the window function 
\begin{eqnarray}\label{WindowFunDef}
f(k R) &\define&
\frac{1}{\frac{4}{3} \pi R^3}
\!\int_{|\Bfscript{x}| < R}\!\!\!\dd^3 x\,
\E^{i \Bfscript{k}\cdot \Bfscript{x}}
\,=\,
\frac{3 j_1(k R)}{k R}
\end{eqnarray}
and $P(k)$ is the matter power spectrum 
\begin{equation}\label{MatterPowerSpectrumDef}
P(k)
\,\define\,
(2 \pi)^3
\big| \tilde\Delta_{\rm m}{\negtinyspace}(k) \big|^2
\end{equation}
The derivation of \rf{MatterDensityFluctuationAverage0powerspectrum} 
can be done using the Fourier transfomation 
\begin{equation}\label{DeltaFourierTransf}
\Delta_{\rm m}{\negtinyspace}(\Bf{x})
\,=\,
  \!\int\!\!\dd^3 q
  {\dbltinyspace}
  \E^{i \Bfscript{q}\cdot\Bfscript{x}}
  \alpha(\Bf{q}) \tilde\Delta_{\rm m}\big(|\Bf{q}|\big)
\end{equation}
and the expectation value of fluctuation
\begin{equation}
  \big<
    \alpha(\Bf{q}) {\tinyspace} \alpha^\ast(\Bf{q'})
  \big>
\,=\,
  \delta^{(3)}( \Bf{q} - \Bf{q'} )
\end{equation}
}%\sigmaRcal %End of \sigmaRcal
The decoupling \rf{MatterDensityFluctuationDecoupling} 
gives us an important information, i.e.\ 
the ratio $\sigma_R{\neghalftinyspace}(t) / F_{\rm m}{\negtinyspace}(t)$ is independent of $t$. 
Since $t$ is an arbitrary time, if this is before $t_{{\tinyspace}\rm LS}$, 
it turns out 
using 
$F_{\rm m}{\negtinyspace}(t_0) {\negtrpltinyspace}={\negtrpltinyspace} 1$
\rf{MatterDensityFluctuationFtCondition} 
that $\sigma_R{\neghalftinyspace}(t_0)$ depends only on 
the physical phenomena before $t_{{\tinyspace}\rm LS}$, 
and then, is independent of the present time $t_0$. 
Therefore, 
we find{\dbltinyspace}%
\footnote{%{\dbltinyspace}%
Here we avoid using $\sigma_R$ with different $R$ 
because we need $n_s$ to compute their differences. 
Not only %exponential expansion (
the standard inflation but also THT inflation 
gives $n_s {\negtrpltinyspace}={\negtrpltinyspace} 1$ 
because of the scale invariance of matter field equations 
under the Robertson-Walker metric. 
However, about this we have to do a delicate discussion %about this, 
including the effects of the knitting mechanism and the Coleman mechanism.
}
\begin{equation}\label{sigmaRatCMB}
\sigma_R{\neghalftinyspace}(t_0)
\,\sim\,
\sigma_R^{\rm (SC)}{\negdbltinyspace}(t_0^{({\rm SC})})
\,\sim\,
\sigma_R^{\rm (CMB)}{\negdbltinyspace}(t_0^{({\rm CMB})})
\,.
\end{equation}
%if the model is consistent with the Planck $\Lambda$CDM model
%in the period before $t_{{\tinyspace}\rm LS}$. 

%Fig.{\dbltinyspace}\ref{fig:fm} shows the graph of 
%$f_{\rm m}{\negtinyspace}(z)$.
Fig.{\dbltinyspace}\ref{fig:fmsigma} shows the graph of 
$f_{\rm m}{\negtinyspace}(z)\sigma_8{\neghalftinyspace}(z)$,%
\footnote{%{\dbltinyspace}%
See ref.\ \cite{ModFriedmannEqLargeScaleStructureAW} 
for the list observation data 
used in Fig.{\dbltinyspace}\ref{fig:fmsigma}. 
}
where $\sigma_8{\neghalftinyspace}(z)$ is $\sigma_R{\neghalftinyspace}(t)$ at 
$R{\negdbltinyspace}={\negdbltinyspace}
%800{\dbltinyspace}{\rm km}{\trpltinyspace}{\rm s}^{-1}
%{\negdbltinyspace}/ H_0%
%{\negdbltinyspace}={\negdbltinyspace}
8{\tinyspace}{\rm Mpc}/h%
$, 
%\footnote{%{\dbltinyspace}%
%$
%8{\dbltinyspace}{\rm Mpc}/h
%=
%800{\dbltinyspace}{\rm km}{\trpltinyspace}{\rm s}^{-1}
%{\negdbltinyspace}/ H_0
%$
%depends on the present Hubble parameter $H_0$. 
%Since $\sigma_R{\neghalftinyspace}(t)$ 
%is independent of the present time $t_0$, 
%the dependence of $t_0$ is artificially introduced in $\sigma_8{\neghalftinyspace}(t)$. 
%So, we here compare the observation data which use the same $h$ 
%observed in Planck satellite. 
%}
[\,$h{\negtrpltinyspace}\sim{\negtrpltinyspace}0.68$\,]
and satisfies %the properties 
%\begin{equation}\label{MatterDensityFluctuationAverage8}
%\sigma_8{\neghalftinyspace}(t)
%%\,\define\,
%%\sigma_R{\neghalftinyspace}(t)\Big|_{R{\dbltinyspace}={\dbltinyspace}8{\rm Mpc}/h}
%%\,=\,
%%F_{\rm m}{\negtinyspace}(t) {\tinyspace}
%%\sigma_R{\neghalftinyspace}(t_0)\Big|_{R{\dbltinyspace}={\dbltinyspace}8{\rm Mpc}/h}
%\,=\,
%F_{\rm m}{\negtinyspace}(t) {\tinyspace}
%\sigma_8{\neghalftinyspace}(t_0)
%\end{equation}
%and 
\begin{equation}\label{sigma8atCMB}
\sigma_8{\neghalftinyspace}(t_0)
\,\sim\,
\sigma_8^{\rm (SC)}{\negdbltinyspace}(t_0^{({\rm SC})})
\,\sim\,
\sigma_8^{\rm (CMB)}{\negdbltinyspace}(t_0^{({\rm CMB})})
\,=\,
0.8120 \,\pm 0.0073%{{+0.034}\atop{-0.040}}
\,.
\end{equation}
\rf{sigma8atCMB} is derived from 
%\rf{MatterDensityFluctuationAverage} and 
\rf{sigmaRatCMB} and %with 
the data by Planck Mission. 
The chi-square $( \chi_{\rm red} )^2$ of 
$f_{\rm m}{\negtinyspace}(t) \sigma_8{\neghalftinyspace}(t)$
for each graph is 
\begin{equation}\label{chisquarefsigma8}
( \chi_{\rm red} )^2
= 0.49
\,,
\qquad
\big( \chi_{\rm red}^{\rm (SC)} \big)^2
= 0.26
\,,
\qquad
\big( \chi_{\rm red}^{\rm (CMB)} \big)^2
= 0.29
\,.
\end{equation}
Since the chi-squares %$\chi^2$ 
of all models are less than $1$, 
these are consistent with current observations, 
and one cannot judge which is better model from \rf{chisquarefsigma8}.%
%どの模型の$\chi^2$も$1$以下なので
%こちらは現在の観測精度と矛盾はなく，甲乙は付けられない。
\footnote{%{\dbltinyspace}%
There is another value %$S_8$ defined by 
$S_8 \define
 \sigma_8{\neghalftinyspace}(t_0)
 ( \Omega_{\rm m}{\negtinyspace}(t_0) / 0.3 )^{0.5}$. 
The chi-square%s $( \chi_{\rm red} )^2$ for each model is 
s are 
%We also calculate 
%$( \chi_{\rm red} )^2$ of 
%$S_8 \define
% \sigma_8{\neghalftinyspace}(t_0)
% ( \Omega_{\rm m}{\negtinyspace}(t_0) / 0.3 )^{0.5}$, 
%\begin{equation}
$( \chi_{\rm red} )^2
{\negtrpltinyspace}={\negtrpltinyspace}
%= 0.56
%\,,
%\qquad
\big( \chi_{\rm red}^{\rm (SC)} \big)^2
{\negtrpltinyspace}={\negtrpltinyspace}
%= 
0.56$
and 
%\,,
%\qquad
$\big( \chi_{\rm red}^{\rm (CMB)} \big)^2
{\negtrpltinyspace}={\negtrpltinyspace}
%= 
11.2$.
%\,.
%\end{equation}
The result of our model is much better than that of Planck Mission. 
}
%%%
\newcommand{\FluctuationTimeDependence}{%
\begin{figure}[h]
\vspace{-4mm}\hspace{-0mm}
\begin{minipage}{.45\linewidth}\vspace{6mm}
  \begin{center}\hspace*{0mm}
    \includegraphics[width=1.0\linewidth,angle=0]
                    {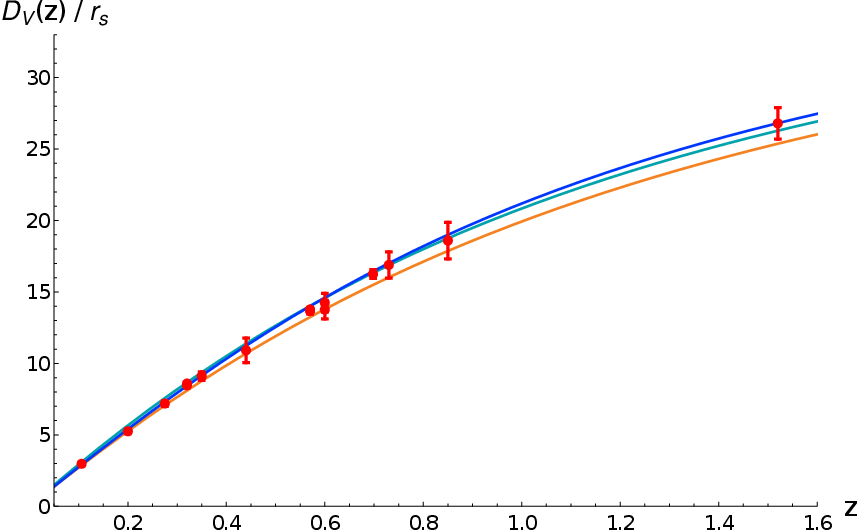}%eps}
  \end{center}
  \begin{picture}(200,0)
    %\put(4.6,50.0){\mbox{\tiny$0$}}
  \end{picture}
  \vspace*{-6mm}
  \caption[fig.2]{%{$z${\dbltinyspace}-{\tinyspace}$D_V{\negtinyspace}(z)/r_{\rm s}$}
$z$ versus 
$D_V{\negtinyspace}(z)/r_{\rm s}$ 
for three models: 
our model (blue), 
the Planck $\Lambda$CDM model (green) 
and 
the late time $\Lambda$CDM model (orange). 
%The blue line shows 
%$D_V{\negtinyspace}(z)/r_{\rm s}$ 
%for the modified Friedmann equation, 
%while 
%the orange line shows 
%$D_V^{({\rm SC})}{\negdbltinyspace}(z)/r_{\rm s}^{({\rm SC})}$, 
%and 
%the green line shows 
%$D_V^{({\rm CMB})}{\negdbltinyspace}(z)/r_{\rm s}^{({\rm CMB})}$. 
The red dots %with error bars 
are observation data. 
  }
  \label{fig:DvRs}
\end{minipage}
%\begin{minipage}{.45\linewidth}\vspace{6mm}
%  \begin{center}\hspace*{0mm}
%    \includegraphics[width=1.0\linewidth,angle=0]
%                    {Graph_Dv_z_diff.eps}
%  \end{center}
%  \begin{picture}(200,0)
%    \put(4.6,50.0){\mbox{\tiny$0$}}
%  \end{picture}
%  \vspace*{-14mm}
%  \caption[fig.2]{{$z${\dbltinyspace}-{\tinyspace}$\frac{D_V{\negtinyspace}(z)/r_{\rm s}}{D_V^{\rm (CMB)}(z)/r_{\rm s}^{\rm (CMB)}}$}
%  }
%  \label{fig:DvRsDiff}
%\end{minipage}
%\begin{minipage}{.45\linewidth}\vspace{6mm}
%  \begin{center}\hspace*{0mm}
%    \includegraphics[width=1.0\linewidth,angle=0]
%                    {Graph_f_smallz.eps}
%  \end{center}
%  \vspace*{-5mm}
%  \caption[fig.1]{{$z${\dbltinyspace}-{\tinyspace}$f_{\rm m}{\negtinyspace}(z)$}
%  }
%  \label{fig:fm}
%\end{minipage}
\hspace{6mm}
\begin{minipage}{.45\linewidth}\vspace{11mm}
  \begin{center}\hspace*{0mm}
    \includegraphics[width=1.0\linewidth,angle=0]
                    {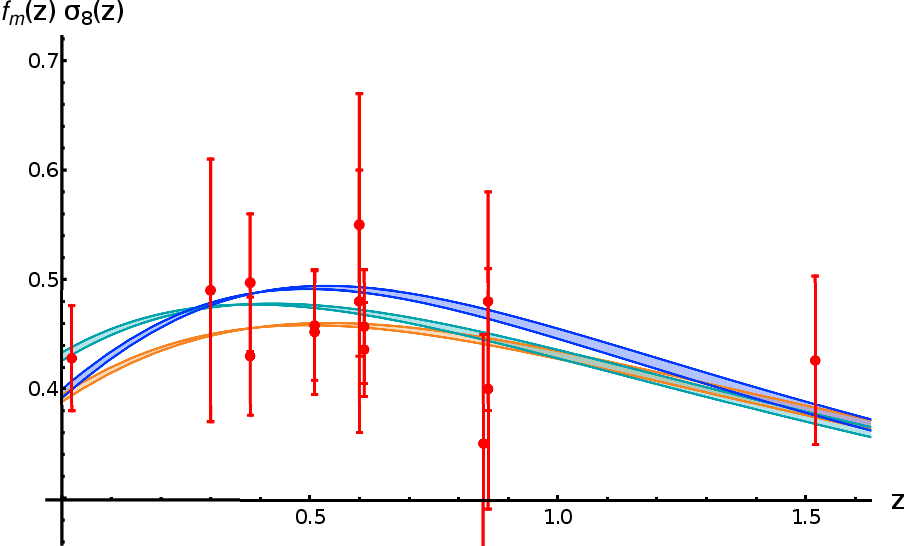}%eps}
  \end{center}
  \vspace*{0mm}
  \caption[fig.2]{%{$z${\dbltinyspace}-{\tinyspace}$f_{\rm m}{\negtinyspace}(z) \sigma_8{\neghalftinyspace}(z)$}
$z$ versus 
$f_{\rm m}{\negtinyspace}(z) \sigma_8{\neghalftinyspace}(z)$ 
for three models: 
our model (blue), 
the Planck $\Lambda$CDM model (green) 
and 
the late time $\Lambda$CDM model (orange). 
%The blue line shows 
%$f_{\rm m}{\negtinyspace}(z) \sigma_8{\neghalftinyspace}(z)$ 
%for the modified Friedmann equation, 
%while 
%the orange line shows 
%$f_{\rm m}^{({\rm SC})}{\negdbltinyspace}(z) \sigma_8^{({\rm SC})}{\negdbltinyspace}(z)$, 
%and 
%the green line shows 
%$f_{\rm m}^{({\rm CMB})}{\negdbltinyspace}(z) \sigma_8^{({\rm CMB})}{\negdbltinyspace}(z)$. 
The red dots %with error bars 
are observation data. 
  }
  \label{fig:fmsigma}
\end{minipage}
\vspace{-1mm}%zaq
\end{figure}%\vspace*{0mm}
%%%
}\FluctuationTimeDependence

\section{Change of vacuum and birth of time}

%物理学では，運動方程式やHamilton形式のような時間発展の記述が主である。
Before discussing the birth of time, 
let us summarize the words in which the concept of time is hidden.
%時間誕生の議論をする前に，時間の概念が隠れている言葉をまとめておこう。
The words that seem to be related to QG are 
``causality'', 
``fluctuation'', 
``conservation law'', 
``spontaneous symmetry breaking (SSB)'', and 
``pregeometry''. 
%量子重力に関係すると思われる言葉だけを挙げると，
%「causality」
%%「probability」
%「揺らぎ」
%「conservation law」
%「spontaneous symmetry breaking (SSB)」
%「pregeometry」
%がある。
They do not have the word ``time'', 
but the concept of ``causal time'' is inseparably placed in their background. 
%これらは時間という言葉を持たないが，
%因果時間がこれらの背景に切り離せないように入っている。
Now let us return to QG. 
%ここで量子重力に話を移そう。
It has been pointed out that 
when the universe was born, 
the spacetime fluctuated, repeated birth and death, 
and the possibility that the current large universe was born 
after overcoming the potential barrier. 
%宇宙誕生時は，時空は揺らぎ，誕生や死滅を繰り返し，
%その中からポテンシャル障壁を乗り越えて
%現在の大きな宇宙が誕生したという可能性が指摘されている。
However, 
the words and concepts mentioned above 
appear here in a way that contradicts the birth of time.
%しかし，ここには先ほど述べた言葉や概念が組み込まれており，
%時間誕生と矛盾する。
%The birth of time is not physical problem.
%It is mathematical problem because we cannot observe it.

In our model, 
the origin of space was the oscillation of 2D scalar fields, 
whereas the origin of time was the scale of the system. 
%我々の模型では，
%空間の起源が2次元スカラー場の振動であったのに対し，
%時間の起源は系のスケールであった。
They have different origins. 
%起源が異なるのである。
This makes it possible to separate the birth of time and space.
%このことは，時間と空間の誕生を切り離すことを可能にする。
Specifically, 
the following process that breaks the scale transformation 
produces the birth of time, 
%具体的には，スケール変換を壊す
%次のプロセスが時間誕生になる。
\begin{equation}\label{CreationOfTime}
|0\rangle
\,\to\,
\cuum
% =
%  \bigg|
%    \lambda_3=\frac{\sigma}{6 g {\sqrt{\GG}}},
%    \lambda_1=-\,\frac{\cc}{2 g {\sqrt{\GG}}},
%    \nu=\frac{\omega}{\sqrt{\GG}}
%\bigg\rangle
\,.
\end{equation}
%
%\begin{equation}\label{jk8}
%| \lambda \rangle\!\rangle
% \,=\,
%\exp{\negqdrpltinyspace}\bigg(
%  \sum_{n=0}^\infty \lambda_n a_n^\dagger
%  - \half \sum_{n=0}^\infty \lambda_n^2
%\bigg)
%| 0 \rangle\!\rangle
%\,,
%\end{equation}
%where the state $| 0 \rangle\!\rangle$ is defined by 
%\beq\label{jk9}
%a_n | 0 \rangle\!\rangle
%\,=\,
%0 \quad {\rm for}\quad n=0,1,2,\ldots .
%\eeq
In fact 
the vacuum may not have changed directly. 
%しかし，実際には，\rf{CreationOfTime}がいきなり起きたのではなく，
Instead of \rf{CreationOfTime}, 
the following process might have been happened 
if the sum of $\lambda_{{\tinyspace}n}$ %[\,$n \in \dbl{Z}$\,] 
is %respectively 
conserved for each $n$.%
%もし $\lambda_{{\tinyspace}n}$ [\,$n \in \dbl{Z}$\,] 
%の和がそれぞれ保存するなら，%
\footnote{%{\dbltinyspace}%
There is a detailed discussion of this in 
ref.\ \cite{CausalityWJalgAWfullversion}. 
%これについては ref.\ \cite{CausalityWJalgAWfullversion} に詳しい議論がある。
}
\begin{eqnarray}
\big(
  |0,0,\nu\rangle
  {\negtinyspace}\otimes{\negtinyspace}
  |0,\lambda_1,0\rangle{\negdbltinyspace}
\big) {\negtinyspace}\otimes{\negtinyspace}
|\lambda_{{\tinyspace}3},0,0\rangle
\to
|0,\lambda_1,\nu\rangle
{\negtinyspace}\otimes{\negtinyspace}
|\lambda_{{\tinyspace}3},0,0\rangle
\to
|\lambda_{{\tinyspace}3},\lambda_1,\nu\rangle
= \cuum
.
\hspace{15pt}
\end{eqnarray}
%のような相互作用が過去にあったかもしれない。%可能性がある。
There are many unknowns here. 
These theoretical analyses require 
an understanding of the mathematics behind the $W$ and Jordan algebras. 
%これらの理論的解析は$W$代数の背景にある数学の理解が必要で，
%未知の要素が多分にある。
In any case, 
it is a process that creates time, 
so it is highly likely that it occurs kinematically, not dynamically.
%いずれにしても時間が誕生するプロセスなので，
%力学的ではなく，運動学的な発生の可能性が高い。

%\begin{itemize}
%\item
%The origin of time is the scale of $J(z)$.
%\item
%The origin of space is the integer $n$ which comes from the modes $\a_n$.
%\end{itemize}
%The appearance of coherent states leads us to the birth of time. 
%We will discuss this phenomena later. 

\section{Overview}\label{sec:Overview}

\subsection{Cosmic age division}

%We discuss here the history %{\age}s and periodization 
%of our universe.
%
We divide the time into several {\age}s and {\period}s as follows:
\begin{quote}
\vspace{-2pt}
\begin{description}
\item[{\bf Pre- and Post-world {\age}}] ~\\
This is a world where 
no spaces are yet created from a vacuum 
or 
all spaces have disappeared to a vacuum, 
such as 
the absolute vacuum 
$\ket{0}$ 
or 
$\cuum$ with 
$\sigma {\negdbltinyspace}={\negdbltinyspace} 0$. 
%(the Pre-world {\age}), 
%or where all spaces have disappeared (Post-world {\age}). 
%たとえば $\ket{0}$ 
%や
%$\sigma {\negdbltinyspace}={\negdbltinyspace} 0$のときの$\cuum$
%のように，真空から空間が発生しない世界である。
The dynamics in this world 
%both in the pre-world {\age} and in the post-world {\age} 
will be studied by mathematics, not physics. 

\item[{\bf Cosmic Dawn {\age}}] ~\\
\vspace{-13pt}
\begin{description}
\item[\bf Space{\tinyspace}-birth {\period}] ~\\
At the beginning of this {\period}, 
the present physical vacuum is established. 
In this {\period}, many 1D universes are created from the physical vacuum. 
\item[\bf Wormhole {\period}] ~\\
This {\period} starts by dimension enhancement and 
ends by the vanishing cosmological constant. 
During this {\period}, many wormholes play important roles, 
\red{and %During this period 
our universe expands exponentially,
%(the $\tan{\negdbltinyspace}\sqrt{\mu}{\trpltinyspace}T$ expansion 
%discussed above), 
i.e.\ the standard inflation appears, 
but without the need of an ``inflaton''.} 
%Between two phenomena, 
%the universe expands exponentially, i.e.\ the standard inflation appears.%
%\footnote{%{\dbltinyspace}%
%So-called ``inflaton'' does not exist in our theory. 
%}
(%In this {\period} 
Our universe is high-dimensional space and 
the Friedmann equation is the modified Friedmann equation
after 
$4\cc {\negdbltinyspace}\to{\negdbltinyspace} \Lambda{\negtrehalftinyspace}/3$ 
%substitution 
and 
before 
$\Lambda {\negdbltinyspace}\to{\negdbltinyspace} \kappa \rho$%
%substitution%
.)
\end{description}
\item[{\bf Cosmic Growth {\age}}] ~\\
%\item[Single Universe {\age}]　\\
\vspace{-13pt}
\begin{description}
\item[\bf Big-bang {\period}] ~\\
This {\period} starts by the vast matter energy 
produced by the vanishing cosmological constant 
at the end of the wormhole {\period} 
($\Lambda {\negdbltinyspace}\to{\negdbltinyspace} \kappa \rho$).
\item[\bf Transition {\period}] ~\\
%\item[Accelerating expansion {\period}]　\\
%\item[Accelerating-expansion {\period}]　\\
The direct effect of small $g$ starts to appear. 
However, 
the perturbative calculations are still possible in this {\period}. 
The accelerating expansion of universe is one of them. 
Present time is in this {\period}. 
%and is also at the entrance to the Chaos {\period}.
%
\end{description}
\item[{\bf Cosmic Dusk {\age}}] ~\\
\vspace{-13pt}
\begin{description}
\item[\bf Chaos {\period}] ~\\
%\item[Non-perturbation {\period}]　\\
The effects by $g$ appear in several phenomena. 
Among them 
the fifth contribution of constant $g$ is drastic.%
\footnote{%{\dbltinyspace}%
The first one is the creation of many 1D spaces. 
The second and third ones are 
the dimension enhancement by the knitting mechanism and 
the vanishing cosmological constant by the Coleman mechanism, respectively. 
The fourth one is the accelerating expansion of the universe. 
}
All Hamiltonians 
$\Hop_{\rm int}^{[{\rm A}]}$, 
$\Hop_{\rm int}^{[{\rm B}]}$, 
$\Hop_{\rm int}^{[{\rm C}]}$ and 
$\Hop_{\rm int}^{[{\rm D}]}$ given by 
\rf{WalgebraHamiltonianTreeTerm3S}-\rf{WalgebraHamiltonianTreeTerm3M} 
participate in. 
Though 
``small nonzero $|g|$'' 
is a necessary condition for the kinetic term to be dominant, 
this situation is broken for large $T$.%
\footnote{%{\dbltinyspace}%
The larger $|g|$ is, 
the earlier the Chaos {\period} begins. 
The dynamical timescale of phenomena in Chaos {\period} 
is $t_g$ \rf{TimeScaleCCg}. 
}
For example, 
the non-perturbative effects by the three-universe interaction 
cannot be negligible in this {\period}.%
\footnote{%{\dbltinyspace}%
Note that 
the high-dimensional space created by the knitting mechanism 
does not split into two, %straight away, 
but 
\red{one flavor space splits in two flavors.} 
The space splitting phenomena occure in one coordinate. 
%splits in one direction. 
%編み上げ機構によって形成された空間は，
%そのまま２つに分裂することはなく，
%１方向に分裂することに注意されたい。
}
As a result, 
the interactions \red{between} %by 
universes are dominant 
and the present universe starts to be chaotic 
because not only large spaces take part but also the small spaces 
which \red{led to} %give 
the gauge symmetry. 
\item[\bf Doomsday {\period}] ~\\
Entropy increases monotonically and reaches to a maximum value. 
Then, the physical time ends at a certain critical time $T_{\rm c}$. 
%
%\item[Post-world {\age}]　\\
%The present physical vacuum changes. 
%
\end{description}
\end{description}
\vspace{-2pt}
\end{quote}

Little is known about the mechanism by which the vacuum $\cuum$ changes. 
So, with our current understanding, 
we cannot say when the current physical vacuum $\cuum$ will change.

\newcommand{\WthreeSdual}{%
The end of the world is described by our model with 
$T {\negtrpltinyspace}\sim{\negtrpltinyspace} \infty$ 
will be larger than $T_{\rm c}$. 
Since the large $T$ is equivalent to the strong coupling constant, 
the model which describes the end of the world 
is a kind of S-dual of 
$\Theta_{\rm W}^\star(\T)$ in \rf{TransferOperatorGeneralW}. 
In this period, 
time goes backwards,%
%時間は逆行し，%
\footnote{%{\dbltinyspace}%
This is because 
the direction of time coincides with the direction in which entropy increases. 
%時間の向きはエントロピーが増大する方向になるからである。
}
and as a result, 
a description similar to the above may be reproduced, 
however if $\cuum$ is different, 
the situation will be very different, 
so reexamination is necessary.
%結果として，上記と似たような記述が再現するかもしれないが，
%$\cuum$が異なれば状況は大きく異なるので再検討が必要である。。
%問題{\bf 1.}の解決も難しい。
%非常にたくさんの真空が候補となってしまい，
%どの真空が現実世界の真空として選ばれるべきなのか分からないのです。
%
%The candidates of the dual model are 
%${\rm C\ell}_{26}(\dbl{R})$ algebra 
%and 
%${\rm H}_3(\dbl{O})$ algebra 
%with different physical vacuum. 
}%\WthreeSdual

%\subsection{Predictions}

\subsection{Future problems}

From \rf{EffectiveLifetimeUniverse}, 
there appears a problem 
``Why is the ratio of 
  the effective lifetime of our universe $t_g$ and 
  the Planck time $t_{\rm pl}$, 
  i.e.\ 
  $t_g /{\tinyspace} t_{\rm pl}
   {\negtinyspace}\sim{\negtinyspace}
   3 \!\times\!{\negtinyspace} 10^{{\tinyspace}61}$ so large?'' 
Abnormally large value of 
$t_g /{\tinyspace} t_\cc
 {\negtinyspace}={\negtinyspace}
 |\cc|^{1/2} {\negtrpltinyspace}/ |g|^{1/3}$ 
is an unsolved problem in our model.%
\footnote{%{\dbltinyspace}%
This problem is essentially the same as 
the cosmological constant hierarchy problem in $\Lambda$CDM model 
and is almost equivalent to the hierarchy problem 
between $t_0$ and $t_{\rm pl}$ 
because of \rf{EffectiveLifetimeUniverse}. 
}

It is also an unsolved problem whether $T_{\rm c}$ is finite or infinite. 
If $T_{\rm c}$ is finite, 
the phase transition appears at 
$T {\negtrpltinyspace}={\negtrpltinyspace} T_{\rm c}$. 
However, 
we do not know what the theory is for regions 
$T {\negdbltinyspace}>{\negdbltinyspace} T_{\rm c}$ in this case. 

This {\Article} showed one possibility, 
\red{namely the one realized by the choice of physical constants in} 
\rf{RealPhysicalValues} as a physical vacuum, 
but the real vacuum may be slightly different. 
In addition, 
there are many unclear points about 
how the knitting mechanism and the Coleman mechanism specifically occur. %, 
%and numerical simulations may be necessary to understand 
%\red{the mechanisms in more detail.} %them.
%また，
%Knitting mechanism and Coleman mechanism 
%が具体的にどのように起きているかは不明な点が多く，
%これに関する理解にはシミュレーションが必要だろう。
It should also be noted that 
the knitting mechanism, which identifies different space points, 
is a suitable mechanism for creating 
not only a torus but also a torus orbifold. 
%Knitting mechanism は orbifold を作り出すのに適している。
%さらには，Calabi-Yau のような空間を作ることも可能かもしれない。
%Since 
%the inequality \rf{gScaleCondition} is derived from the fact that 
%the splitting and merging of the universe has not been observed 
%until the present time $t_0$ 
%and 
%is independent of the observation of the accelerating expansion, 
%this coincidence means that 
%the problem of the scale of cosmological constant in $\Lambda$CDM model 
%is translated to 
%the problem of the scale of the present time $t_0$, i.e.\ 
%``Why is 
% $t_0 / t_{\rm pl} {\negtinyspace}\sim{\negtinyspace}
%  8 \!\times\!{\negtinyspace} 10^{60}$ so large?'' 
%in our model. 

%\footnote{%{\dbltinyspace}%
%This 
%is similar to %may have something to do with 
%the holographic principle.
%}
%However, 
So far, 
only the dimensions have been confirmed 
to be consistent with string theory, 
but finding other properties in common with string theory 
is a future task. 
%今のところ弦理論との一致は次元だけしか確認されていないが，
%これ以外に弦理論と共通する性質はないのだろうか？\ 
%Calabi-Yao manifold や orbifold のような空間をどのように作り出すか？\ 
%今後の大きな課題である。
%This {\Article} showed the possibility of a physical vacuum 
%by choosing a physics constant at one possibility, 
%that is, \rf{RealPhysicalValues} as a vacuum, 
%but the actual vacuum may be slightly different.
%
If our model is equivalent to string theory, 
our model 
%will be obtained by path-integrating 
%all the dynamical modes of string theory. 
%Our model 
corresponds to the head of QG 
and 
the so-called ordinary string theory to the body of QG. 
%because 
Our model is talkative in phenomenology. 
Little is known about the neck 
which connects the head to the body. 

%宇宙の加速膨張以外に small $g$ の影響を与える現象はないのだろうか？

\appendix

\section{Appendix: Octonions}

The octonions are numbers defined by
\begin{eqnarray}\label{CartesianCoordinateOctonion}
&&
z  
\,=\,%&=&
x {\dbltinyspace}+{\dbltinyspace} y{\trehalftinyspace}i
  {\dbltinyspace}+{\dbltinyspace} u{\halftinyspace}j
  {\dbltinyspace}+{\dbltinyspace} v{\halftinyspace}k
  {\dbltinyspace}+{\dbltinyspace} \bar{x}{\tinyspace}\ell
  {\dbltinyspace}+{\dbltinyspace} \bar{y}{\dbltinyspace}\bar{i}
  {\dbltinyspace}+{\dbltinyspace} \bar{u}{\tinyspace}\bar{j}
  {\dbltinyspace}+{\dbltinyspace} \bar{v}{\tinyspace}\bar{k}
\,=\,%\nonumber\\&=&
\xi {\negtinyspace}+{\negtinyspace}
 \eta{\tinyspace}\ell
\,,
\end{eqnarray}
where
$x$,{\negtinyspace} $y$,{\negtinyspace}
$u$,{\negtinyspace} $v$,{\negtinyspace} $\bar{x}$,{\negtinyspace}
$\bar{y}$,{\negtinyspace} $\bar{u}$,{\negtinyspace} 
$\bar{v} {\negdbltinyspace}\in{\negdbltinyspace} \dbl{R}$
are all real numbers, 
and
\begin{eqnarray}
&&
\xi 
\,=\,%&=&
x {\dbltinyspace}+{\dbltinyspace} y{\trehalftinyspace}i
  {\dbltinyspace}+{\dbltinyspace} u{\halftinyspace}j
  {\dbltinyspace}+{\dbltinyspace} v{\halftinyspace}k
\in \dbl{H}
\,,\qquad%\nonumber\\
\eta 
\,=\,%&=&
\bar{x}
  {\dbltinyspace}+{\dbltinyspace} \bar{y}{\trehalftinyspace}i
  {\dbltinyspace}+{\dbltinyspace} \bar{u}{\halftinyspace}j
  {\dbltinyspace}+{\dbltinyspace} \bar{v}{\halftinyspace}k
\in \dbl{H}
\end{eqnarray}
are all quaternions, 
%$\xi$, $\eta {\negdbltinyspace}\in{\negdbltinyspace} \dbl{H}$.
and 
$i$, $j$, $k$, $\ell$ satisfy
\begin{eqnarray}\label{OctonionUnitSquare}
&&
i^2 \,=\, j^2 \,=\, k^2 \,=\,
\ell^2 \,=\,
-1
\,,
\nonumber\\
&&
i{\tinyspace}j \,=\, - j{\tinyspace}i \,=\, k
\,,
\qquad
j{\tinyspace}k \,=\, - k{\tinyspace}j \,=\, i
\,,
\qquad
k{\tinyspace}i \,=\, - i{\tinyspace}k \,=\, j
\,,
\nonumber\\
&&
\bar{i} \,\define\,
i{\tinyspace}\ell \,=\, - \ell{\tinyspace}i
\,,
\qquad
\bar{j} \,\define\,
j{\tinyspace}\ell \,=\, - \ell{\tinyspace}j
\,,
\qquad
\bar{k} \,\define\,
k{\tinyspace}\ell \,=\, - \ell{\tinyspace}k
\,.
\end{eqnarray}
The multiplication of two octonions is defined by
\begin{equation}\label{OctonionMultiplication}
( \xi {\negtinyspace}+{\negtinyspace}
  \eta{\tinyspace}\ell {\tinyspace})
( \xi' {\negtinyspace}+{\negtinyspace}
  \eta'{\negtinyspace}\ell {\tinyspace})
\,=\,
\xi \xi' {\negtinyspace}-{\negtinyspace} \eta^{\prime\ast} \eta
+ (
    \eta' \xi {\negtinyspace}+{\negtinyspace}
    \eta {\tinyspace} \xi^{\prime\ast}
  ){\trehalftinyspace}\ell
\,,
\end{equation}
where $\xi$, $\eta$, $\xi'$, $\eta'$ are quaternions.

\section{Appendix: Formally real Jordan algebra}
\label{sec:EuclideanJordanAlgebra}

\subsection{Definition and Properties}
\label{sec:EuclideanJordanAlgebraDef}

The Jordan product \lq\lq$\circ$" is defined by 
\begin{equation}
A {\negtinyspace}\circ{\negtinyspace} B
\,\define\,
\half \anticommutator{A}{B}
\,,
\qquad\quad
\anticommutator{A}{B}
\,\define\,
A B + B {\tinyspace} A
\,.
\end{equation}
The exponentiation of $A$ is defined by 
\begin{equation}
A^1
{\dbltinyspace}\define{\dbltinyspace}
A
\,,
\qquad\quad
A^2
{\dbltinyspace}\define{\dbltinyspace}
A {\negtinyspace}\circ{\negtinyspace} A
\,,
\qquad\quad
A^n
{\dbltinyspace}\define{\dbltinyspace}
A {\negtinyspace}\circ{\negtinyspace} A^{n-1}
\,,
\quad\mbox{[\,$n {\negtrpltinyspace}={\negtrpltinyspace} 3$, $4$, \ldots\,]}
\,.
\end{equation}
The Jordan product satisfies the following properties:
\begin{eqnarray}
&&
A {\negtinyspace}\circ{\negtinyspace} B
\,=\,
B {\negtinyspace}\circ{\negtinyspace} A
\,,
\\
&&
A^m {\negdbltinyspace}\circ{\negtinyspace}
(
  A^n {\negdbltinyspace}\circ{\negtinyspace} B
)
\,=\,
A^n {\negdbltinyspace}\circ{\negtinyspace}
(
  A^m {\negdbltinyspace}\circ{\negtinyspace} B
)
\,,
\qquad\mbox{[\,$m$, $n {\negtrpltinyspace}={\negtrpltinyspace} 1$, $2$, \ldots\,]}
\,.
\end{eqnarray}
%where $A^1 \define A$.

The formally real Jordan algebra, 
which is also called the Euclidean Jordan algebra, 
is a kind of Jordan algebra which satisfies the condition 
\begin{equation}
X^2 + Y^2 = 0
\quad\Longrightarrow\quad
X = Y = 0
\,.
\end{equation}
The element $X$ of the formally real Jordan algebra 
is expressed by 
\begin{equation}
X
\,=\,
\sum_\mu E_\mu {\tinyspace} X^\mu
\,=\,
E_0 {\tinyspace} X^0 + \sum_a E_a {\tinyspace} X^a
\,,
\end{equation}
where 
$X^0$ and $X^a$ are real numbers. 
The formally real Jordan algebra is the direct sum of 
the following simple algebras.%
\footnote{%{\dbltinyspace}%
The index `$\mu{\negtrpltinyspace}={\negtrpltinyspace}0$' 
does not mean the negative norm, 
and it is a singlet in Jordan algebra. 
}
\begin{itemize}
\item
$\dbl{R}$ algebra\,:\ 
singlet
\\
\quad
$E_0 {\negdbltinyspace}={\negdbltinyspace} 
 \CoeffSinglet$
\item
${\rm C\ell}_n(\dbl{R})$ algebra\,:\ 
%$\{ E_\mu \} = \{ 1 ,\, \gamma_1 ,\, \gamma_2 ,\, \ldots ,\, \gamma_{n-1} \}$
%\qquad
%$\anticommutator{\gamma_\mu}{\gamma_\nu} = 2 \delta_{{\tinyspace}\mu,\nu}$
%$\dbl{R} {\negdbltinyspace}\oplus{\negdbltinyspace} \dbl{R}^{n-1}$
singlet and 
$n$D %-dimensional
 gamma matrices
\quad\hbox{[\,$n {\negdbltinyspace}\ge{\negdbltinyspace} 2$\,]}
\\
\quad
$E_0 {\negdbltinyspace}={\negdbltinyspace} 
 \CoeffSinglet {\tinyspace}
 I$,\quad
$E_a {\negdbltinyspace}={\negdbltinyspace} \gamma_a$\ 
[\,$a {\negtrpltinyspace}={\negtrpltinyspace} 1$, \ldots, $n$\,]
\item
${\rm H}_n(\dbl{R})$ algebra\,:\ 
%$\{ E_\mu \} = \{ \lambda_1 ,\, \ldots ,\, \lambda_{n^2-1} \}$
$n$D %-dimensional
 symmetric matrices
\quad\hbox{[\,$n {\negdbltinyspace}\ge{\negdbltinyspace} 3$\,]}
\\
\quad
$E_0 {\negdbltinyspace}={\negdbltinyspace}
 \CoeffSinglet {\tinyspace}
% \sqrt{\frac{2}{n}}{\tinyspace}
 I$,\quad
$E_a {\negdbltinyspace}={\negdbltinyspace} \lambda_{{\tinyspace}a}$\ 
[\,$a {\negtrpltinyspace}={\negtrpltinyspace} 1$, \ldots,
   $\frac{n(n+1)}{2} {\negtinyspace}-{\negdbltinyspace} 1$\,]
\item
${\rm H}_n(\dbl{C})$ algebra\,:\ 
$n$D %-dimensional
 Hermitian matrices
\quad\hbox{[\,$n {\negdbltinyspace}\ge{\negdbltinyspace} 3$\,]}
\\
\quad
$E_0 {\negdbltinyspace}={\negdbltinyspace}
 \CoeffSinglet {\tinyspace}
% \sqrt{\frac{2}{n}}{\tinyspace}
 I$,\quad
$E_a {\negdbltinyspace}={\negdbltinyspace} \lambda_{{\tinyspace}a}$\ 
[\,$a {\negtrpltinyspace}={\negtrpltinyspace} 1$, \ldots,
   $n^2 {\negdbltinyspace}-{\negdbltinyspace} 1$\,]
\item
${\rm H}_n(\dbl{H})$ algebra\,:\ 
$n$D %-dimensional
 Hermitian quaternion matrices
\quad\hbox{[\,$n {\negdbltinyspace}\ge{\negdbltinyspace} 3$\,]}
\\
\quad
$E_0 {\negdbltinyspace}={\negdbltinyspace}
 \CoeffSinglet {\halftinyspace}
% \sqrt{\frac{2}{n}}{\tinyspace}
 I$,\quad
$E_a {\negdbltinyspace}={\negdbltinyspace} \lambda_{{\tinyspace}a}$\ 
[\,$a {\negtrpltinyspace}={\negtrpltinyspace} 1$, \ldots,
   $n(2n{\negtrpltinyspace}-{\negtrpltinyspace}1)
    {\negdbltinyspace}-{\negdbltinyspace} 1$\,]
\item
${\rm H}_{n=3}(\dbl{O})$ algebra (Albert algebra)\,:\ 
%$(n{\negtrpltinyspace}={\negtrpltinyspace}3)$-dimensional 
$3$D %-dimensional
 Hermitian octonion matrices
\\
\quad
$E_0 {\negdbltinyspace}={\negdbltinyspace}
 \CoeffSinglet {\halftinyspace}
% \sqrt{\frac{2}{3}}{\tinyspace}
 I$,\quad
$E_a {\negdbltinyspace}={\negdbltinyspace} \lambda_{{\tinyspace}a}$\ 
[\,$a {\negtrpltinyspace}={\negtrpltinyspace} 1$, \ldots,
   $26$\,]
\end{itemize}
$\gamma_a$ are 
$N \!\times\! N$ 
%$2^{{\halftinyspace}[n/2]} \!\times\! 2^{{\halftinyspace}[n/2]}$ 
gamma matrices which satisfy 
$(\gamma_a)^\dag = \gamma_a$, 
$\tr{\tinyspace}(\gamma_a) = 0$, 
$\anticommutator{\gamma_a}{\gamma_{{\tinyspace}b}} =
 2 {\tinyspace}\delta_{{\tinyspace}a,b}$. 
$\lambda_{{\tinyspace}a}$ are 
$N \!\times\! N$ 
%$n \!\times\! n$ 
traceless Hermitian matrices which satisfy 
$(\lambda_{{\tinyspace}a})^\dag = \lambda_{{\tinyspace}a}$, 
$\tr{\tinyspace}(\lambda_{{\tinyspace}a}) = 0$. 
$N$ is the dimension of the matrix, 
$I$ is the unit matrix. 
In the case of $\dbl{R}$ algebra,
we define 
$\CoeffSinglet {\negdbltinyspace}={\negdbltinyspace}
 N {\negdbltinyspace}={\negdbltinyspace}
 1$. 
In the case of the spin factor type, i.e.\ 
${\rm C\ell}_n(\dbl{R})$ algebra,
we define 
\begin{equation}
\CoeffSinglet = 1
\,,
\qquad
%c_1 = c_2 = \frac{1}{N}%\frac{1}{2^{[n/2]}}
%\,,
%\qquad
N = 2^{{\halftinyspace}[n/2]}
\,.
\end{equation}
In the case of the Hermitian matrix types, i.e.\ 
${\rm H}_n(\dbl{R})$, 
${\rm H}_n(\dbl{C})$,  
${\rm H}_n(\dbl{H})$, and 
${\rm H}_{n=3}(\dbl{O})$ algebras,
we define 
\begin{equation}
\CoeffSinglet = 
\sqrt{\frac{2}{n}}
\,,
\qquad
%c_1 = \frac{1}{\sqrt{2 {\halftinyspace} N}}
%\,,
%\qquad
%c_2 = \frac{1}{2}
%\,,
%\qquad
N = n
\,.
\end{equation}

In the case of 
%the spin factor type and the Hermitian matrix types, 
all types above, 
the generator $E^\mu$ satisfies
\begin{equation}
E_\mu {\negdbltinyspace}\circ{\negtinyspace} E_\nu
\,=\,
\sum_\rho d_{\mu\nu\rho} {\dbltinyspace} E_\rho
\,,
\end{equation}
and 
\begin{equation}
\Tr(
E_\mu
)
=
\frac{1}{\CoeffSinglet}{\dbltinyspace}\delta_{{\tinyspace}\mu,0}
%\,=\,
%\frac{c_2}{c_1}{\tinyspace}\delta_{{\tinyspace}\mu,0}
\,,
\qquad
\Tr(
E_\mu {\negdbltinyspace}\circ{\negtinyspace} E_\nu
)
=
\delta_{{\tinyspace}\mu,\nu}
\quad\hbox{with}\quad
\Tr
 {\negtinyspace}\define{\negtinyspace}
 \frac{1}{\CoeffSinglet^2 N}{\dbltinyspace}\tr%
% {\negtrpltinyspace}\define{\negtrpltinyspace}
% (c_2/c_1)^2 \frac{1}{N}\tr%
\,,
\end{equation}
%where %The definition of $\Tr$ is 
%$
%\Tr
% {\negtrpltinyspace}\define{\negtrpltinyspace}
% \frac{1}{\CoeffSinglet^2 N}\tr%
%$, 
where 
%$N$ is the dimension of the matrix, 
$\tr$ is the standard trace to the matrix. 
%Especially, 
%$E_0
% {\negtrpltinyspace}={\negtrpltinyspace}
% \CoeffSinglet {\tinyspace} I%
%% {\negtrpltinyspace}={\negtrpltinyspace}
%% \frac{c_1}{c_2} I%
%$. 
Then, one finds 
\begin{equation}
d_{\mu\nu\rho}  \,=\,
\Tr\big(
E_\rho {\negtrehalftinyspace}\circ{\negtinyspace} (
E_\mu {\negdbltinyspace}\circ{\negtinyspace} E_\nu
)
\big)
\,,
\qquad\quad
d_{0\mu\nu} \,=\,
\CoeffSinglet{\trpltinyspace}
\Tr(
E_\mu {\negdbltinyspace}\circ{\negtinyspace} E_\nu
)
%\,=\,
%\frac{c_1}{c_2}{\dbltinyspace}
%\Tr(
%E_\mu {\negdbltinyspace}\circ{\negtinyspace} E_\nu
%)
\,=\,
\CoeffSinglet{\dbltinyspace}\delta_{{\tinyspace}\mu,\nu}
%\,=\,
%\frac{c_1}{c_2}{\tinyspace}\delta_{{\tinyspace}\mu,\nu}
\,.
\end{equation}
One also finds that 
$d_{\mu\nu\rho}$ is a real number and completely symmetric. 
\newcommand{\JalgCoefficientDerivation}{
\begin{eqnarray*}
\Tr(
E_\mu {\negdbltinyspace}\circ{\negtinyspace} E_\nu
)
&=&
\sum_\rho d_{\mu\nu\rho} \Tr( E_\rho )
\,=\,
\sum_\rho d_{\mu\nu\rho} {\dbltinyspace}
  \frac{1}{c_1} {\tinyspace} \delta_{{\tinyspace}\rho,0}
\\&=&
d_{\mu\nu0} {\tinyspace} \frac{1}{c_1}
\,=\,
\frac{c_1}{c_2}{\tinyspace}\delta_{{\tinyspace}\mu,\nu} {\tinyspace} \frac{1}{c_1}
\,=\,
\frac{1}{c_2} {\tinyspace} \delta_{{\tinyspace}\mu,\nu}
\\
\Tr\big(
E_\rho {\negtrehalftinyspace}\circ{\negtinyspace} (
E_\mu {\negdbltinyspace}\circ{\negtinyspace} E_\nu
)
\big)
&=&
\Tr\big(
E_\rho {\negtrehalftinyspace}\circ{\negtinyspace} (
\sum_\sigma d_{\mu\nu\sigma} E_\sigma
)
\big)
\,=\,
\sum_\sigma d_{\mu\nu\sigma}
\Tr(
E_\rho {\negtrehalftinyspace}\circ{\negtinyspace} E_\sigma
)
\\&=&
\sum_\sigma d_{\mu\nu\sigma} {\tinyspace}
\frac{1}{c_2} {\tinyspace} \delta_{{\tinyspace}\rho,\sigma}
\,=\,
\frac{1}{c_2} {\tinyspace} d_{\mu\nu\rho}
\end{eqnarray*}
}%\JalgCoefficientDerivation

\subsection{${\rm H}_3(\Dbl{O})$ algebra (Albert algebra)}
\label{sec:EuclideanJordanAlgebraH3Def}

The generators of ${\rm H}_3(\dbl{O})$ algebra are 
a set of octonion Hermitian $3 \!\times\! 3$ matrices. 
This algebra is exceptional, so 
this is a kind of ``extremity''. 
The generators consist of one identity matrix 
$\CoeffSinglet {\halftinyspace} I$ 
($\CoeffSinglet {\negdbltinyspace}={\negdbltinyspace}
  \sqrt{\frac{2}{3}}${\dbltinyspace})
and 
the following 26 traceless matrices,
\begin{eqnarray}
&&
\hbox{%
  $\lambda_1 = S_{xy}$,\quad
  $\lambda_{{\tinyspace}4} = S_{yz}$,\quad
  $\lambda_{{\tinyspace}6} = S_{zx}$,\quad
  $\lambda_{{\tinyspace}3} = S_{x^2-y^2}$,\quad
  $\lambda_{{\tinyspace}\eight} = S_{z^2}$,
}
\nonumber\\
&&
\hbox{%
  $\lambda_{{\tinyspace}2} = i A_{xy}$,\quad
  $\lambda_{{\tinyspace}2'} = j A_{xy}$,\quad
  $\lambda_{{\tinyspace}2''} = k A_{xy}$,\quad
  $\lambda_{{\tinyspace}2^\circ} = \ell A_{xy}$,
}
\nonumber\\&&
\hbox{%
  $\lambda_{{\tinyspace}\bar{2}} = \bar{i} A_{xy}$,\quad
  $\lambda_{{\tinyspace}\bar{2}'} = \bar{j} A_{xy}$,\quad
  $\lambda_{{\tinyspace}\bar{2}''} = \bar{k} A_{xy}$,
}
\nonumber\\
&&
\hbox{%
  $\lambda_{{\tinyspace}7} = i A_{yz}$,\quad
  $\lambda_{{\tinyspace}7'} = j A_{yz}$,\quad
  $\lambda_{{\tinyspace}7''} = k A_{yz}$,\quad
  $\lambda_{{\tinyspace}7^\circ} = \ell A_{yz}$,
}
\nonumber\\&&
\hbox{%
  $\lambda_{{\tinyspace}\bar{7}} = \bar{i} A_{yz}$,\quad
  $\lambda_{{\tinyspace}\bar{7}'} = \bar{j} A_{yz}$,\quad
  $\lambda_{{\tinyspace}\bar{7}''} = \bar{k} A_{yz}$,
}
\nonumber\\
&&
\hbox{%
  $\lambda_{{\tinyspace}5} = -i A_{zx}$,\quad
  $\lambda_{{\tinyspace}5'} = -j A_{zx}$,\quad
  $\lambda_{{\tinyspace}5''} = -k A_{zx}$,\quad
  $\lambda_{{\tinyspace}5^\circ} = -\ell A_{zx}$,
}
\nonumber\\&&
\hbox{%
  $\lambda_{{\tinyspace}\bar{5}} = -\bar{i} A_{zx}$,\quad
  $\lambda_{{\tinyspace}\bar{5}'} = -\bar{j} A_{zx}$,\quad
  $\lambda_{{\tinyspace}\bar{5}''} = -\bar{k} A_{zx}$,
}
\end{eqnarray}
where $S$ and $A$ are symmetric-antisymmetric matrices defined by
\begin{eqnarray}\label{SU3SpinMatrixS}
&&
S_{xy} \define
  \left(\!\begin{array}{ccc}
           0   &    1   &    0   \\
           1   &    0   &    0   \\
           0   &    0   &    0
  \end{array}\!\right)
,
\quad
S_{yz} \define
  \left(\!\begin{array}{ccc}
           0   &    0   &    0   \\
           0   &    0   &    1   \\
           0   &    1   &    0
  \end{array}\!\right)
,
\quad
S_{zx} \define
  \left(\!\begin{array}{ccc}
           0   &    0   &    1   \\
           0   &    0   &    0   \\
           1   &    0   &    0
  \end{array}\!\right)
,
\nonumber\\
&&
S_{x^2-y^2} \define
  \left(\!\begin{array}{ccc}
           1   &    0   &    0   \\
           0   &   -1   &    0   \\
           0   &    0   &    0
  \end{array}\!\right)
,
\quad
S_{z^2} \define
  \frac{1}{\sqrt{3}}\!
  \left(\!\begin{array}{ccc}
           1   &    0   &    0   \\
           0   &    1   &    0   \\
           0   &    0   &   -2
  \end{array}\!\right)
,
\nonumber\\
&&
A_{xy} \define
  \left(\!\begin{array}{ccc}
           0   &   -1   &    0   \\
           1   &    0   &    0   \\
           0   &    0   &    0
  \end{array}\!\right)
,
\quad
A_{yz} \define
  \left(\!\begin{array}{ccc}
           0   &    0   &    0   \\
           0   &    0   &   -1   \\
           0   &    1   &    0
  \end{array}\!\right)
,
\quad
A_{zx} \define
  \left(\!\begin{array}{ccc}
           0   &    0   &    1   \\
           0   &    0   &    0   \\
          -1   &    0   &    0
  \end{array}\!\right)
.
\qquad\ 
\end{eqnarray}
$\lambda_{{\tinyspace}a}$ are the extended Gell-Mann matrices{\dbltinyspace}%
\footnote{%{\dbltinyspace}%
$\lambda_1$, $\lambda_{{\tinyspace}2}$, \ldots, $\lambda_{{\tinyspace}8}$ 
are Gell-Mann matrices.
}
which satisfies 
\begin{equation}
%2\lambda_{{\tinyspace}a}
%{\negtinyspace}\circ{\negtinyspace}
%\lambda_{{\tinyspace}b}
%\,=\,
\anticommutator{\lambda_{{\tinyspace}a}}{\lambda_{{\tinyspace}b}}
\,=\,
\frac{4}{3}{\tinyspace} \delta_{{\tinyspace}a,b} {\dbltinyspace} I
\,+\,
2 \sum_c d_{abc} {\tinyspace} \lambda_{{\tinyspace}c}
\,.
\end{equation}
The generators of 
${\rm H}_3(\dbl{R})$, 
${\rm H}_3(\dbl{C})$, 
${\rm H}_3(\dbl{H})$  algebra are obtained 
by truncating several $\lambda_{{\tinyspace}a}$ from
the generators of 
${\rm H}_3(\dbl{O})$. 
The indices of ${\rm H}_3(\dbl{O})$ are classified as 
\begin{eqnarray}
&&
\{ \mu \} {\negdbltinyspace}={\negdbltinyspace} \{ 0, a \}
\,,
\qquad
\{ a \} {\negdbltinyspace}={\negdbltinyspace}
\{
 8,3,i,I,\tilde{I}{\tinyspace}
\}
\,,
\qquad
\{ i \} {\negdbltinyspace}={\negdbltinyspace}
\{
 1,2,2',2'',2^\circ,\bar{2},\bar{2}',\bar{2}''
\}
\,,
\qquad
\nonumber\\
&& 
\{ I{\tinyspace} \}
 =
\{
 4,5,5',5'',5^\circ,\bar{5},\bar{5}',\bar{5}''
\}
\,,
\qquad
\{
 \tilde{I}{\tinyspace} \}
 =
\{
 6,7,7',7'',7^\circ,\bar{7},\bar{7}',\bar{7}''
\}
\,,
\end{eqnarray}
and 
\begin{equation}
\{ S{\tinyspace} \}
=
\{{\tinyspace}
 0{\tinyspace},{\qdrpltinyspace}
 8{\tinyspace},{\qdrpltinyspace}
 3 {\tinyspace}\}
\,,
\qquad\quad
\{ M{\tinyspace} \}
=
\{{\tinyspace}
 i{\tinyspace},{\qdrpltinyspace}
 I{\tinyspace},{\qdrpltinyspace}
 \tilde{I} {\femhalftinyspace}\}
\,.
\end{equation}
$\{ S{\tinyspace} \}$ are three singlets 
and 
$\{ M{\tinyspace} \}$ are three multiplets (octets).%
\footnote{%{\dbltinyspace}%
Note that 
singlets and multiplets 
are generally linear combinations of several components. 
(For example, $\oIndex$, $\pIndex$, $\mIndex$ introduced in 
\rf{RealModelField} and \rf{RealModelCC} 
in the case of \rf{H3realmodel} 
are such three singlets.)
}
The nonzero $d_{\mu\nu\rho}$ are, up to permutation, 
\begin{eqnarray}
&&
\{ d_{SSS} \}
\,=\,
\{{\tinyspace}
d_{000}
\,|{\qdrpltinyspace}
d_{888}
{\tinyspace}\}
\,,
\nonumber\\
&&
\{ d_{SMM} \}
\,=\,
\{{\tinyspace}
d_{0aa}
\,,{\qdrpltinyspace}
d_{833}
\,,{\qdrpltinyspace}
d_{8ii}
\,,{\qdrpltinyspace}
d_{3II}
\,|{\qdrpltinyspace}
d_{8II}
\,,{\qdrpltinyspace}
d_{8\tilde{I}\tilde{I}}
\,,{\qdrpltinyspace}
d_{3\tilde{I}\tilde{I}}
{\tinyspace}\}
\,,
\nonumber\\
&&
\{ d_{MM'M''} \}
\,=\,
\{{\tinyspace}
\hbox{some of}\ 
d_{iI\tilde{I}}
{\tinyspace}\}
\,.
\end{eqnarray}
The values of $d_{\mu\nu\rho}$ on the lhs of $|$ are positive, 
and 
the values of $d_{\mu\nu\rho}$ on the rhs of $|$ are negative, 
i.e.\ 
$\{\hbox{positive $d$'s}{\tinyspace}|{\tinyspace}\hbox{negative $d$'s}\}$. 
The nonzero $d_{\mu\nu\rho}$ 
[\,$\mu$, $\nu$, $\rho \!=\! 0$, $1$, \ldots\,]
are in the following:
\begin{eqnarray}\label{SOoct3antistructureconstant2}
&&
d_{000} \,=\, 
d_{0aa} \,=\, \sqrt{\frac{2}{3}}
\quad
\hbox{[\,for all $a${\dbltinyspace}]}
\,,
\qquad
d_{888} \,=\, -{\dbltinyspace}\frac{1}{\sqrt{3}}
\,,
\qquad
d_{833} \,=\, \frac{1}{\sqrt{3}}
\,,
\nonumber\\
&&
d_{811} \,=\, 
d_{822} \,=\, 
d_{82'2'} \,=\, 
d_{82''2''} \,=\, 
d_{82^\circ2^\circ} \,=\, 
d_{8\bar{2}\bar{2}} \,=\, 
d_{8\bar{2}'\bar{2}'} \,=\, 
d_{8\bar{2}''\bar{2}''} \,=\, 
%d_{833} \,=\, 
\frac{1}{\sqrt{3}}
\,,
\nonumber\\
&&
d_{844} \,=\, 
d_{855} \,=\, 
d_{85'5'} \,=\, 
d_{85''5''} \,=\, 
d_{85^\circ5^\circ} \,=\, 
d_{8\bar{5}\bar{5}} \,=\, 
d_{8\bar{5}'\bar{5}'} \,=\, 
d_{8\bar{5}''\bar{5}''} \,=\, 
\nonumber\\&&
d_{866} \,=\, 
d_{877} \,=\, 
d_{87'7'} \,=\, 
d_{87''7''} \,=\, 
d_{87^\circ7^\circ} \,=\, 
d_{8\bar{7}\bar{7}} \,=\, 
d_{8\bar{7}'\bar{7}'} \,=\, 
d_{8\bar{7}''\bar{7}''} \,=\, 
-{\dbltinyspace}\frac{1}{2\sqrt{3}}
\,,
\nonumber\\
&&
d_{344} \,=\, 
d_{355} \,=\, 
d_{35'5'} \,=\, 
d_{35''5''} \,=\, 
d_{35^\circ5^\circ} \,=\, 
d_{3\bar{5}\bar{5}} \,=\, 
d_{3\bar{5}'\bar{5}'} \,=\, 
d_{3\bar{5}''\bar{5}''} \,=\, 
\frac{1}{2}
\,,
\nonumber\\
&&
d_{366} \,=\, 
d_{377} \,=\, 
d_{37'7'} \,=\, 
d_{37''7''} \,=\, 
d_{37^\circ7^\circ} \,=\, 
d_{3\bar{7}\bar{7}} \,=\, 
d_{3\bar{7}'\bar{7}'} \,=\, 
d_{3\bar{7}''\bar{7}''} \,=\, 
-{\dbltinyspace}\frac{1}{2}
\,,
\nonumber\\
&&
d_{146} \,=\, 
d_{157} \,=\, 
d_{15'7'} \,=\, 
d_{15''7''} \,=\, 
d_{15^\circ7^\circ} \,=\, 
d_{1\bar{5}\bar{7}} \,=\, 
d_{1\bar{5}'\bar{7}'} \,=\, 
d_{1\bar{5}''\bar{7}''} \,=\, 
\nonumber\\&&
d_{256} \,=\, 
d_{2'5'6} \,=\, 
d_{2''5''6} \,=\, 
d_{2^\circ5^\circ6} \,=\, 
d_{\bar{2}\bar{5}6} \,=\, 
d_{\bar{2}'\bar{5}'6} \,=\, 
d_{\bar{2}''\bar{5}''6} \,=\, 
%\frac{1}{2}\,,
%
\nonumber\\
&&
d_{25'7''} \,=\, 
d_{2'5''7} \,=\, 
d_{2''57'} \,=\, 
d_{\bar{2}'\bar{5}7''} \,=\, 
d_{\bar{2}''\bar{5}'7} \,=\, 
d_{\bar{2}\bar{5}''7'} \,=\, 
\nonumber\\&&
d_{\bar{2}'5\bar{7}''} \,=\, 
d_{\bar{2}''5'\bar{7}} \,=\, 
d_{\bar{2}5''\bar{7}'} \,=\, 
d_{2'\bar{5}\bar{7}''} \,=\, 
d_{2''\bar{5}'\bar{7}} \,=\, 
d_{2\bar{5}''\bar{7}'} \,=\, 
\nonumber\\&&
d_{\bar{2}57^\circ} \,=\, 
d_{\bar{2}'5'7^\circ} \,=\, 
d_{\bar{2}''5''7^\circ} \,=\, 
d_{25^\circ\bar{7}} \,=\, 
d_{2'5^\circ\bar{7}'} \,=\, 
d_{2''5^\circ\bar{7}''} \,=\, 
\nonumber\\&&
d_{2^\circ\bar{5}7} \,=\, 
d_{2^\circ\bar{5}'7'} \,=\, 
d_{2^\circ\bar{5}''7''} \,=\, 
\frac{1}{2}
\,,
\nonumber\\
&&
d_{247} \,=\, 
d_{2'47'} \,=\, 
d_{2''47''} \,=\, 
d_{2^\circ47^\circ} \,=\, 
d_{\bar{2}4\bar{7}} \,=\, 
d_{\bar{2}'4\bar{7}'} \,=\, 
d_{\bar{2}''4\bar{7}''} \,=\, 
%-{\dbltinyspace}\frac{1}{2}\,,
%
\nonumber\\
&&
d_{2'57''} \,=\, 
d_{2''5'7} \,=\, 
d_{25''7'} \,=\, 
d_{\bar{2}\bar{5}'7''} \,=\, 
d_{\bar{2}'\bar{5}''7} \,=\, 
d_{\bar{2}''\bar{5}7'} \,=\, 
\nonumber\\&&
d_{2\bar{5}'\bar{7}''} \,=\, 
d_{2'\bar{5}''\bar{7}} \,=\, 
d_{2''\bar{5}\bar{7}'} \,=\, 
d_{\bar{2}5'\bar{7}''} \,=\, 
d_{\bar{2}'5''\bar{7}} \,=\, 
d_{\bar{2}''5\bar{7}'} \,=\, 
\nonumber\\&&
d_{2\bar{5}7^\circ} \,=\, 
d_{2'\bar{5}'7^\circ} \,=\, 
d_{2''\bar{5}''7^\circ} \,=\, 
d_{2^\circ5\bar{7}} \,=\, 
d_{2^\circ5'\bar{7}'} \,=\, 
d_{2^\circ5''\bar{7}''} \,=\, 
\nonumber\\&&
d_{\bar{2}5^\circ7} \,=\, 
d_{\bar{2}'5^\circ7'} \,=\, 
d_{\bar{2}''5^\circ7''} \,=\, 
-{\dbltinyspace}\frac{1}{2}
\,.
\end{eqnarray}
As usual we here omit $d_{\mu\nu\rho}$ 
obtained by the permutation of $\mu$, $\nu$, $\rho$.

%\section{$W$ operators}
%
%\subsection{$p$ reduced $W$ operators}
%
%\subsection{$W$ operators}

\newcommand{\bibAuthor}[1]{#1,}%{#1,\\}
\newcommand{\bibTitle}[1]{}%{``#1'',\\}
\newcommand{\bibJournal}[4]{#1#2#3 (#4)}
\newcommand{\bibDOI}[1]{doi:\,#1}
\newcommand{\bibArXiv}[1]{[arXiv:\,#1]}

\section{Note: References and Citations}

\end{document}